\documentstyle[rmp,eqsecnum,aps]{revtex}
\begin{document}

\title{PHASE TRANSITIONS IN QUANTUM CHROMODYNAMICS}

\draft
\author{Hildegard Meyer-Ortmanns}
\address{Institut f\"ur Theoretische Physik, Universit\"at Heidelberg,
D-69120 Heidelberg, Federal Republic of Germany \\
h.meyer-ortmanns@thphys.uni-heidelberg.de}
\date{\today}
\maketitle
%
%
\begin{abstract}
  The current understanding of finite temperature phase transitions in
  QCD is reviewed. A critical discussion of refined phase transition
  criteria in numerical lattice simulations and of analytical tools
  going beyond the mean-field level in effective continuum models for
  QCD is presented. Theoretical predictions about the order of the
  transitions are compared with possible experimental manifestations
  in heavy-ion collisions. Various places in phenomenological
  descriptions are pointed out, where more reliable data for QCD's
  equation of state would help in selecting the most realistic
  scenario among those proposed. Unanswered questions are raised about
  the relevance of calculations which assume thermodynamic
  equilibrium. Promising new approaches to implement nonequilibrium
  aspects in the thermodynamics of heavy-ion collisions are described.
\end{abstract}


\widetext
\tableofcontents

%
%
\section{INTRODUCTION}
\label{sec:I}
Phase transitions are found over a wide range of temperatures. They
start near absolute zero with the Mott transitions (Mott, 1968) from
superconductor to insulator, and are common between $10^2$ K and
$10^3$ K. Nine orders of magnitude higher one finds the phase
transitions discussed here. At a temperature typical of the QCD scale
($\Lambda_{\overline{MS}}\sim 200$~MeV) nuclear matter melts. At this
temperature quarks and gluons cease to be confined inside hadrons, and
begin moving freely. Three orders of magnitude higher is the next
phase transition, at which the electromagnetic and weak interactions
are unified.

These two transitions may seem exotic, but according to the big bang
theory have occurred at least once in the universe. The electroweak
transition is predicted to have taken place at $10^{-10}\rm{s}$, and
the deconfinement transition at $10^{-6}\rm{s}$ after the big bang.
The deconfinement transition occurred, when the temperature of the
universe dropped to the order of tera degrees, i.e. to a scale
$T_{c}\sim(2.32\pm 0.6)\cdot 10^{12}\rm{K}$ or $200\pm 50$~MeV.  (The
conversion from natural energy units [MeV] to degrees Kelvin [K] is
determined by the Boltzmann constant $k_{B}\sim 8.6\cdot
10^{-5}\rm{[eV/K]}$ or $(100~\rm{MeV}\sim 1.16\cdot 10^{12}K)$.)
These temperatures with time scales of $10^{-23}\rm{s}$ and typical
distances of 1fm may now be achievable in the laboratory. At CERN one
reaches energy densities around $1.5-2.5~\rm{GeV/fm}^{3}$, and higher
densities are expected when RHIC (Relativistic Heavy Ion Collision)
experiments come on-line in Brookhaven (Stachel, 1994).

We have glibly mentioned the QCD transition, as if it were unique.
However, there are really two, and they take place if either the
temperature or the density crosses some critical point.

Each transition is well defined in some limit of QCD. If the quark
masses are infinite, one has the {\it deconfinement} transition. If
the quark masses are zero, one has the {\it chiral} (symmetry
restoring) transition. It is not yet clear whether these transitions
persist for physical quark masses nor whether they will occur
together. If they take place at different temperatures, one expects
the chiral transition to occur after the deconfinement transition
(Banks and Casher, 1980; Campbell et al., 1990). Lattice results
indicate, however, that both transitions coincide at the same critical
point. One of them can then be regarded as driving the other, and the
nature of the transition will be determined by the driving symmetry.
Here it is assumed that both happen simultaneously, unless stated
otherwise.

As mentioned above, a transition is expected for both increasing
density as well as temperature. This review concentrates on the
effects of temperature, as this is the area in which the lattice is
strong and many partial or preliminary results are available.

One usually has many length, or equivalently energy, scales playing a
role at a phase transition. This makes it impossible, in most cases,
to find a suitable expansion parameter for a perturbative treatment.
In addition, QCD is asymptotically free in the UV-limit, but has a
coupling $g(T)\approx 1$ near the phase transition. So a
nonperturbative approach must be used; the lattice regularization may
be the only means of calculating the properties of the phase
transition from first principles.

One expects, naively, a phase transition in QCD if the density rises
to the point at which the hadrons begin to overlap. This will be
around $0.5~\rm{GeV/fm}^{3}$, the energy density in a proton.

Similarly, for temperatures above the QCD scale parameter,
$T>\Lambda_{\overline{MS}}=200~\rm{MeV}$, one expects a phase
transition. At extremely high temperatures, $T\gg
\Lambda_{\overline{MS}}$, with typical momenta $Q\approx T$, the
coupling $Q^{2}/\Lambda^{2}_{\overline{MS}}$ becomes weak, and
resummed finite temperature perturbation theory (Braaten and Pisarski,
1990 and 1992) may be applied.

There are three natural length scales in the QCD plasma, $l_{p}={1
  \over T}<l_{e}={1 \over gT}<l_{m}={1 \over g^{2}T}$ (DeGrand and
DeTar, 1986). The first gives the range over which perturbation theory
may be applied, the second corresponds to the scale set by the
chromoelectric mass, and the third to the scale set by the
chromomagnetic mass. The latter two scales reflect the increasingly
nonperturbative nature of long-range interactions in the plasma, and
the lack of the screening in the magnetic sector.

Although these long-range interactions are strong, and may lead to a
type of dynamical confinement in the plasma, the bulk properties, the
equation of state etc., are still dominated by short-range
interactions. So perturbation theory may well be valid at high
temperatures for these quantities.

More evidence that the chromomagnetic, long-range sector of QCD may
have a complicated, nonperturbative structure comes from the
phenomenon of dimensional reduction. In $3+1$ dimensional Euclidean
field theory the imaginary time dimension is proportional to the
inverse temperature, $l_{\tau}\propto 1/T$. Thus the theory reduces,
in the high temperature limit, to three dimensional QCD (Reisz, 1992;
Reisz and Petersson, 1991; K\"arkk\"ainen et al., 1992, 1993 and
1994).

Deriving the properties of the QCD phase transition is clearly
difficult. After all, it is analogous to deriving the properties of
boiling water directly from the quantum Hamiltonian of water
molecules. So the thermodynamics of the phases, and the order of
the transition between them, have only recently been settled for pure
gauge theory, despite nearly two decades of computer simulations.

The {\it order of a phase transition} is one of the basic
thermodynamic classifications. A phase transition is set to be of {\it
  first order}, if there is at least one finite gap in the first
derivatives of a suitable thermodynamic potential in the thermodynamic
limit. A finite latent heat goes often along with a gap in the order
parameter. A transition is set to be of {\it second order} if there is
a power like singularity in at least one of 
the second derivatives of the potential. If
the thermodynamic potential is analytic over the whole temperature
range (for temperature driven transitions), the phase conversion is
called a {\it crossover phenomenon}.

In general the order of the QCD transitions depends on the number of
colors ($N_{c}$), the number of flavors ($N_{f}$), the current quark
masses, and on more subtle effects related to the IR- and UV-cut-offs.
The deconfinement transition is now believed to be of first order for
three colors (Fukugita et al., 1989; Gavai et al., 1989), and the
equation of state is known in the limit of infinite quark masses
(Engels et al., 1990b; Laermann et al., 1995). Similarly the chiral
transition is believed to be of second order for 2 massless flavors
(Pisarski and Wilczek, 1984; Karsch, 1994) and of first order for
three or more massless flavors (Pisarski and Wilczek, 1984; Gavai and
Karsch, 1985; Gavai et al., 1987; Brown et al., 1990; Iwasaki, 1995;
Kanaya et al., 1995). These results do not contradict each other. They
just indicate the sensitivity of the order to the involved
approximations and should be taken as a warning not to jump to
conclusions.

Ultimately one is interested in the physically relevant case of 2
light quarks (up and down), and one heavier quark (the strange quark).
This case is still under discussion. It is an open question as to
whether the chiral transition and the deconfinement transition persist
for realistic quark masses.

\vskip12pt The order of a phase transition has far-reaching
phenomenological consequences. Phenomenological implications of QCD
transitions are supposed to be visible in relativistic heavy-ion
collisions and astrophysics. For a first order transition one expects
metastabilities with latent heat, interfaces, supercooling and
overheating etc.. The experimental consequences of a first order
transition make it relatively easy to see, especially if the plasma
``explodes'' into the hadronic phase (Cleymans et al., 1986). A second
order transition, lacking a jump in the energy density, may be less
easy to see experimentally. However, it has divergent correlation
lengths in the thermodynamic limit. These may also lead to observable
consequences (Bialas and Peschanski, 1988; Wilczek, 1992; Rajagopal
and Wilczek, 1993; Bjorken et al., 1993; Gavin et al., 1994a and b),
in analogy to the well-known phenomenon of critical opalescence.

Perhaps most likely is a smooth crossover, as in the transition to an
electric plasma from a molecular or atomic gas. One will have a smooth
transition from a pion gas at low temperatures to a quark gluon plasma
at high temperatures, with a highly nontrivial mixture of excitations
in the neighborhood of the crossover. 
If one has a sharp crossover phenomenon with a rapid
change in thermodynamic quantities over a small temperature interval
(say of the order of 10~MeV), there is still some chance for
measurable effects in experiments. For example, double $\Phi$-peaks 
in the dilepton invariant mass spectrum are still predicted as a signature
for the phase conversion, as long as the crossover phenomenon is rapid
enough (Ko and Asakawa, 1994; Ko, 1995).  

The situation if further complicated by the nature of experiments. The
ultrarelativistic heavy-ion collisions have a finite volume expanding
in time. The finite volume smooths out the non-analyticities in the
free energy. The high temperature matter created in the collisions may
not reach equilibrium either, and does have different processes
dominating in different regions of phase space. It becomes a
nontrivial task to find ``thermometers'', to tune and measure the
temperature of the transient hot plasma. There is thus no clear signal
of the phase transition known, although a number of signals has been
proposed, and the quantities measured so far can usually be modelled
by both a hot hadron gas and a quark-gluon-plasma.

In the early universe we have different competing scales. The
expansion of the universe is rather slow in units of QCD. If a typical
time scale of QCD is taken as $1/T_{c}$ ($\sim 1\rm{fm}/c \sim
10^{-23}\rm{s}$), the Hubble time is of the order of $10^{19}/T_{c}$.
Thus there is enough time for equilibration before and after the
transition. The spatial volume $V$ of the universe, $10^{-6}\rm{s}$
after the big bang, appears as almost infinite in units of QCD.  In
units of $T_{c}$ it is given as $V\cdot T_{c}^{3}=7.1\times
10^{55}({200~\rm{MeV} \over T_{c}})^{3}$, if the physical correlation
length $\xi$ is taken as $\xi(T_{c})\cdot T_{c}=1.38\pm 0.24$ in the
deconfinement phase.  This leads to a rather small upper bound on the
amount of supercooling during a (hypothetical) first order
deconfinement transition (Banerjee and Gavai, 1992). The effect of
supercooling is an important ingredient in the reasoning, why one
should see remnants of the early QCD transition even nowadays. A first
order deconfinement transition might have led to inhomogeneities in the
baryon number density in the early universe. If these inhomogeneities
had survived until the epoch of primordial nucleosynthesis, they may
have influenced the light element abundances and led to deviations
from values obtained in the standard scenario (without a phase
transition) (Schramm et al., 1992; Fuller et al., 1988; Applegate et
al., 1987). It is these deviations in the light element abundances
which have been predicted as a visible remnant nowadays. Meanwhile this
prediction has become rather questionable. Initial inhomogeneities which
are compatible with the rather small supercooling are most likely
insufficient to induce well separated proton and neutron rich regions.
Pronounced inhomogeneities in the proton and neutron distributions are
a precondition for a considerable change in the initial conditions of
primordial nucleosynthesis.

\vskip12pt Lattice Monte Carlo simulations get rather time consuming,
when one tries to simulate the realistic situation with small but
nonzero quark masses. The CPU-time of a typical simulation with
dynamical quarks is of the order of months or even years. From the
viewpoint of one less familiar with the lattice approach, it is
difficult to understand, why, for example, extensive Monte Carlo
calculations have been performed to simulate a strongly interacting
system with eight nearly massless flavors, whereas only two flavors
are light in nature. Without a deeper understanding of the lattice
methods one is left with the impression that the field of QCD
computer simulations has decoupled from the field of heavy-ion
collisions. The connection between both areas of research seems
sometimes to be lost.

It would be overly ambitious to attempt, in this article, to close the
gaps between the various approaches to understanding the finite
temperature transitions of QCD. This review will focus on the progress
made in determining the nature of the phase transitions, and the
development of sophisticated tools for this purpose. The complexity of
the task, and the time needed to turn the hints of today into firm
answers, will be explained.

\vskip12pt
The organization of the paper is as follows. Section \ref{sec:II.A}
gives the background from statistical physics. We summarize the main
steps in a renormalization group approach, both in the infinite and
finite volume, and recall some concepts to distinguish first and
second order transitions, independently of the specific QCD dynamics.
In section \ref{sec:II.B} we deal with a renormalization group
analysis applied to QCD.  Such an analysis provides a guide for a more
detailed investigation of the phase diagram.

In view of the anticipated nonperturbative nature of the transition
and the need for simulating QCD rather than a simplified model, it is
natural to start with the lattice approach (section \ref{sec:III}).
From numerous papers we select a few to explain the type of progress
which has been made. Section \ref{sec:III.A} serves mainly to give the
basic ideas, to fix the notations, and to introduce some calculational
tools of lattice gauge theory.

Section III.B deals with the pure gauge theory without matter fields.
We describe the controversial discussion about the order of the
deconfinement transition in the $SU(3)$ gauge theory (section
\ref{sec:II.B.1}).  Thermodynamics on the lattice is the topic of
section \ref{sec:III.B.2}. Over a long time it has been taken for
granted that the effect of matter on the transition is only to weaken
the strength of its order. Hence various interface or surface tensions
have been calculated as further characteristics of the transition
dynamics. Interface tensions refer to free energies associated with
interfaces between different realizations of the plasma phase, between
the phases of broken and restored chiral symmetry, or the phases of
deconfinement (plasma phase) and confinement (hadronic phase). A
reliable estimate of their values at the transition point could play a
selective role in the phase transition scenarios. A measurement of the
interface tension in the $SU(3)$ gauge theory is presented in section
\ref{sec:III.B.3}.
Intimately related to the topic of interface tensions are
phenomenological implications of the QCD transition in the early
universe. We summarize the line of arguing for finding relics of the
early QCD transition in section \ref{sec:III.B.4}.

Section \ref{sec:III.C} is devoted to the inclusion of dynamical
fermions. Here we restrict the discussion mostly to the staggered
fermion formulation. In section \ref{sec:III.C.1} we list the pitfalls
on the lattice, including infrared, ultraviolet and finite-mass
artifacts. We describe a finite size scaling analysis in the presence
of fermions in section \ref{sec:III.C.2}. Finite-mass artifacts
(analyzed in a finite-mass scaling analysis) refer to the necessity of
extrapolating results to the chiral limit from finite mass
calculations (section \ref{sec:III.C.3}). Examples for the subtleties
of UV-artifacts are given in section \ref{sec:III.C}.3 and 4. In
section \ref{sec:III.C.5} we present results for the cases of two and
three light fermion flavors, which come closest to the physical
current quark masses.  Recent results on the equation of state for
two-flavor QCD are explained in section \ref{sec:III.C.6}. Wilson
fermions are the topic of section \ref{sec:III.C.7}.

A brief review of lattice simulations at finite baryon density is
presented in section \ref{sec:III.D}.

The lattice approach plays an important role for QCD. It has the
advantage of starting from first principles. But even if lattice
calculations finally succeed in providing reliable predictions, there
is some need for further alternatives. In a typical Monte Carlo
calculation QCD's full partition function is simulated at once (if we
disregard the involved approximations for a moment). From such a
simulation one does not get an intuitive insight why a result comes
out in a particular way. Effective models try to fill this lack.  They
are usually not derived from QCD in a strict sense, but share some
important symmetries with QCD. Accordingly we consider models for
quark and gluonic degrees of freedom separately (section
\ref{sec:IV}.A and B).

In section \ref{sec:IV.A} we concentrate on theoretical tools which go
beyond the mean-field level: the renormalization group approach
(section \ref{sec:IV.A.1}), chiral perturbation theory (section
\ref{sec:IV.A}.2 and 3), and a $1/N$-expansion (section
\ref{sec:IV.A.4}). The renormalization group leads to specific
predictions for the critical indices and the temperature dependence of
meson masses close to $T_{c}$. These predictions may serve as working
hypothesis for lattice simulations (Wilczek, 1992; Rajagopal and
Wilczek, 1993).

Chiral perturbation theory is well established for describing QCD at
low energies and small temperatures. We outline the derivation of
chiral condensates as a function of temperature in section
\ref{sec:IV.A.2}. Of particular interest are the influence of finite
current quark masses and an estimate of the contribution of heavier
mesons to thermodynamic quantities. Heavier mesons turn out to be
nonnegligible in the phase transition region in contrast one's naive
guess (Gerber and Leutwyler, 1989).  Although chiral perturbation
theory fails in the vicinity of the transition region, it leads to an
upper bound for the latent heat, when Clausius-Clapeyron relations are
applied to QCD in a hypothetical first order transition (Leutwyler,
1992) (section \ref{sec:IV.A.3}). The bound excludes strong
supercooling scenarios, thus it may be of particular interest for
phenomenologists.

In section \ref{sec:IV.A.4} we apply a $1/N$-expansion to the linear
$SU(3)\times SU(3)$ sigma model. Here large $N$ refers to the number
of flavors. The topics are the mass sensitivity of the chiral
transition and the thermodynamics for physical values of
(pseudo)-scalar meson masses. An attempt is made to locate the phase
boundary between first order chiral transitions and crossover
phenomena in meson mass space (Meyer-Ortmanns and Schaefer, 1995).

As an example for a simplified physical picture of the interaction
dynamics in a pure gauge theory we consider Patel's color flux tube
models (Patel 1984a; 1984b) in section \ref{sec:IV.B.1}. In the flux
tube models the $SU(3)$-deconfinement transition occurs, when the
network of color flux tubes gets infinitely connected. The models
abandon the symmetry breaking as a driving mechanism for the
deconfinement transition.  Section \ref{sec:IV.B.2} deals with dual
Ginzburg-Landau models, explaining the confinement property of QCD in
terms of a dual Meissner effect.

If quark and gluonic degrees of freedom are included in an effective
Lagrangian for QCD, their interplay in the transition dynamics may be
investigated. We give an outlook for such attempts, and further
directions in section \ref{sec:IV.B.3}.

Section \ref{sec:V} is devoted to heavy-ion collisions. Although the
application of a thermodynamic concept is not fully established,
throughout the review we adopt the viewpoint that a thermodynamic
description of nucleus nucleus collisions is meaningful.  After a
recapitulation of basic concepts in relativistic hydrodynamics
(section \ref{sec:V.B}) we point out possibilities for measuring
thermodynamical observables (section \ref{sec:V.C.1}). Substructures
in dilepton spectra and enhanced strangeness production, which could
be sensitive to the underlying transition dynamics, are the topic of
sections \ref{sec:V.C}.2 and 3. Pion interferometry, measurements of
multiplicity fluctuations, and intermittency are powerful experimental
tools in heavy-ion collisions.  They may be also utilized for
inferring the order of the QCD transition (section
\ref{sec:V.C}.4,5,6). The common difficulty to all of these
experimental devices is to find a {\it unique} signature for a certain
type of transition.  Thus a flattening in the $\langle
p_T\rangle$-distribution of charged pions as a function of
multiplicity distributions in rapidity space is compatible with a
first order transition, but lack of available phase space serves as an
alternative and much simpler explanation as well (van Hove, 1985).

We will not cover the issue of preequilibrium production, although it
is likely to be relevant in various aspects. Part \ref{sec:V.D}
concerns with off-equilibrium aspects. Estimates of the nucleation
rate and transport coefficients are two examples for attempts to
incorporate the knowledge of equilibrium QCD in situations slightly
off-equilibrium.  An alternative off-equilibrium process to nucleation
is large domain coarsening. Under certain conditions it is more likely
the relevant mechanism for phase separation at the transition point
(section \ref{sec:V.D.3}).  A plasma evolution far out of equilibrium
is the topic of the final section \ref{sec:V.D.4}.  Methods of
nonequilibrium physics and the powerful concept of dynamical
universality classes become applicable, if the hot plasma is quenched
via almost instantaneous cooling (Rajagopal and Wilczek, 1993). Large
correlation volumes may be created in spite of a small equilibrium
correlation length, which is hampered in growing due to the finite
pion mass. In realistic heavy-ion experiments the hypothetical plasma
presumably cools neither adiabatically nor instantaneously, and the
physical masses are likely not light enough for inducing large
correlation volumes with the aid of nonequilibrium amplification.
Nevertheless, we conclude section V in sketching ideas about dynamical
universality as it may be challenging and stimulating to consider a
quench of a hot quark gluon plasma as some kind of
gedanken-experiment.

In section \ref{sec:VI} we summarize the most important points of the
preceding sections. We list the main results, the open questions, and
interesting perspectives for further studies.

\vskip12pt It has been unavoidable to select papers out of numerous
others related to the same topic to keep the article readable. If I do
not refer to a paper in spite of its relevance, this may be a result
of ignorance.

%
%
\section{GENERALITIES}
\label{sec:II}
\subsection{Phase transitions in statistical systems}
\label{sec:II.A}
\subsubsection{First and second order transitions
  in the infinite volume limit}
\label{sec:II.A.1}
The order of a phase transition is one
of the basic thermodynamic classifications. It concerns the
thermodynamic potential and its derivatives at the transition. The
thermodynamic potential $\Omega$ is the free energy $F$ in case of a
ferromagnet or the Gibbs free energy $G = F+pV$ for a fluid. In Eqs.\ 
(\ref{2.1}) we recall the basic thermodynamic formulas for a magnet.
\begin{mathletters}
  \label{2.1}
  \begin{equation}
    Z(T,H,V) = \int {\cal D}U e^{-\beta {\cal H}(U)}
    \label{2.1a}
  \end{equation}
  \begin{equation}
    F = -T \ln Z
    \label{2.1b}
  \end{equation}
  \begin{equation}
    E=-{\partial\ln Z\over \partial\beta}
    \quad
    S=-\left({\partial F\over \partial T}\right)_H
    \quad
    M=-\left({\partial F\over \partial H}\right)_T
    \label{2.1c}
  \end{equation}
  \begin{equation}
    c_H=\left({\partial E\over \partial T}\right)_H
    \quad
    c_{H,M}=T\left({\partial S\over \partial T}\right)_{H,M}
    \quad
    \chi_T={1 \over V} \left({\partial M\over\partial H}\right)_T
    .
    \label{2.1d}
  \end{equation}
\end{mathletters}
Here ${\cal H}$ denotes the spin Hamiltonian, $\beta$ is the inverse
temperature, the Boltzmann constant $k_B$ has been set 1 everywhere,
$H$ is an external magnetic field, $V$ is the volume and $\int{\cal
  D}U$ stands for the sum over all spin configurations $\{U\}$,
weighted with the Boltzmann factor $\exp\left\{-\beta {\cal
    H}\right\}$. Eq.\ (\ref{2.1b}) defines the free energy $F$,
depending on $T$, $H$, and $V$, of a system in a finite volume $V$.
In the {\it large} volume limit the free energy density $f$ is given
as $f=\lim_{V\to\infty} F(V)/V$ with $F(V)\equiv \int_V d^d x f(x)$ in
$d$ space-dimensions.  The free energy is assumed to depend on the
scaling fields $T$ and $H$.In view of a finite size scaling analysis
we keep also the dependence on $V$. The parameters $T$ and $H$ are on
an equal footing, both may drive a phase transition. One speaks of a
{\it temperature} or a {\it field driven} transition. In QCD we focus
our discussion on temperature driven transitions (although density- or
``mass-driven'' transitions may be considered as well). First
derivatives of the free energy with respect to $T$ or $H$ lead to the
internal energy $E$, the entropy $S$ or the magnetization $M$
according to Eq.\ (\ref{2.1c}). The magnetization is the conjugate
variable to the external field and plays the role of an order
parameter in the case of a magnet.  More generally it plays the role
of an order parameter, if the external field {\it explicitly} breaks
the symmetry which may be spontaneously broken at a phase transition.

On the second level of derivatives (Eqs.\ (\ref{2.1d})) we have the
specific heat c (at constant $H$ and~/ or $M$, respectively) and the
isothermal susceptibility $\chi_T$.  For a fluid the isothermal
susceptibility would be replaced by the compressibility $\kappa$.

In the infinite volume limit a phase transition is signalled by a
singularity (in the sense of non-analyticity) in the thermodynamical
potential $\Omega$. If there is a finite discontinuity in at least one
of the first derivatives of $\Omega$, the transition is called {\it
  first order}. In the case of a ferromagnet there is a jump in the
magnetization, if one passes through the transition temperature from
the phase of broken symmetry to the symmetric phase. This gives $M$
the name of an {\it order parameter}, $M$ indicates the order of
spins. This way it tells us the phase in which the system is
encountered at a given temperature.

The remaining first derivatives of $F$ with respect to $T$ (the
internal energy and entropy) usually also show a discontinuity at the
transition point. A gap in the entropy is associated with a finite
latent heat $\Delta Q=T_c\cdot\Delta S$, but there need not be such a
gap. (Consider a transition in a ferromagnet between states of
magnetization opposite in sign but equal in magnitude. The latent heat
would vanish in this case, while the magnetization jumps between
values of opposite sign.) Vice versa there may be a finite latent heat
without a gap in the order parameter at the transition point.

The second derivatives of the thermodynamic potential at a first order
transition are typically $\delta$-function singularities
(corresponding to the discontinuities in the first derivatives) or
finite.

According to the original Ehrenfest classification of phase
transitions, {\it n-th order transitions} are defined by the
occurrence of discontinuities (rather than divergences) in the n-th
order derivative of the appropriate thermodynamical potential. In
M.E.Fisher's terminology one distinguishes between first order and
{\it continuous} (or higher order) transitions. In continuous
transitions the first derivatives of $\Omega$ are continuous, whereas
second derivatives are either discontinuous or {\it divergent}. In a
{\it second order transition} at least one of the second derivatives
of $\Omega$ is divergent. (If there are at most finite discontinuities
in the second derivatives, the transition is of higher than second
order.) Hence the order parameter $M$ will vanish continuously at the
transition point.

The susceptibility $\chi_T$ and the specific heat $c$ typically both
diverge in a second order transition. (Again it is not necessary that
both of them diverge.) Here the divergences are power law
singularities. They are characterized by {\it critical indices}. For
example, in the infinite volume limit the susceptibility scales close 
to $T_c$ according to $\chi_T
\equiv -(\partial^2 f/ \partial H^2)_T \propto {|1-T/T_c|}^{-\gamma}$
with an index $\gamma$, $f$ as defined above, 
and the specific heat according to $c_H \equiv
-T({\partial^2}F/\partial T^2)_H \propto {|1-T/T_c|}^{-\alpha}$ with
critical index $\alpha$.

\vskip12pt\noindent {\it Landau's free energy.} A first criterion on
the order of the phase transition is given by Landau's theory (Landau
and Lifschitz, 1958; Aizu, 1970; Michel, 1980; Toledano, 1981). It
consists in an expansion of the free energy in powers of the order
parameter. The allowed terms in this expansion are further selected by
symmetry arguments. Phase transitions can be classified according to
their transformation behavior of the order parameter under a symmetry
transformation. In this introductory section we discuss only an order
parameter described by a scalar field $\phi$. In QCD applications the
scalar $\phi$ will be replaced by an $O(N)$-vector with $N$ components
or an $SU(3)$ matrix parametrized by two independent directions of
possible ``ordering'' (see section~\ref{sec:IV}).

The ansatz of a free energy functional for a scalar order parameter
$\phi$ in $d$ space-dimensions is given as
\begin{eqnarray}
  F\left\{\phi(x)\right\}=
  \int d^dx \left\{a\left(\nabla\phi(x)\right)^2+{r\over 2}
    \phi^2(x) \right. \nonumber\\
  \left. +{\lambda\over 4}\phi^4(x)-h\phi(x)\right\}.
  \label{2.2}
\end{eqnarray}
For vanishing $h$ this is the simplest form that admits spontaneous
symmetry breaking. Although $F$ is reflection invariant (if $h=0$),
the ground state need not be so. $F$ may take its
minimum for nonvanishing values $\pm\phi_0\neq 0$ (later denoted as
$\left<\phi\right>$), depending on the values of $a$, $r$, and
$\lambda$.  The ``couplings'' $a$, $r$, $\lambda$ and $h$ should be
considered as parameters, where $a$, $\lambda>0$. In the example of a
magnet the condition $\lambda>0$ corresponds to the physical condition
that the magnetization is bounded. Later $r$ has the meaning of a mass
squared, $\lambda$ of a coupling strength of the interaction and $h$
of an external field. In the vicinity of a second order transition the
order parameter is small (more generally, if fluctuations in the field
are allowed, its average expectation value is small), hence one drops
higher powers of $\phi$. A further assumption is that $\phi$ is slowly
varying in space (thus there are no higher derivative terms than
$(\nabla\phi)^2$). A $\phi^3$-term is missing, if a symmetry under
sign inversion $\phi\to -\phi$ is required for vanishing $h$.  For
$h=0$ it is easily verified that Eq.\ (\ref{2.2}) predicts a second
order phase transition. For $r<0$ two stable states are predicted with
magnetizations $\pm\phi_0$. The condition $r=0$ defines the critical
temperature, thus one may write for $r$
\begin{equation}
  r=\tilde r~(T-T_c)~.
  \label{2.3}
\end{equation}
\noindent
Figs.~\ref{fig:1}(a) and \ref{fig:1}(b) display the typical signatures
of a second order transition. The nontrivial minima at $\pm\phi_0$
move continuously inwards as the temperature is increased towards
$T_c$, where the $Z(2)$-symmetry is restored (Fig.~\ref{fig:1}(a)).
Fig.~\ref{fig:1}(b) illustrates the vanishing of the order parameter
as function of $T$ and the power law divergence in the susceptibility.

For fixed temperature $T<T_c$ a field driven transition can be
considered as function of $h$. In this case the transition is of first
order, $\phi$ jumps from $\sqrt{(-r/\lambda)}$ to
$-\sqrt{(-r/\lambda)}$ as $h$ changes sign.

Although the ansatz (\ref{2.2}) has been originally proposed for
second order transitions, temperature driven first order transitions
can be described as well. We mention two possibilities.

\begin{itemize}
\item $\lambda<0$ in Eq.\ (\ref{2.2}).  The coupling $\lambda$ can
  play the role of a renormalized coupling. In later applications of
  QCD $\lambda$ occurs as a renormalized coupling in an effective
  description of QCD.  It varies as a function of the value which is
  chosen for the strange quark mass $m_s$. For certain values of
  $m_s$, $\lambda$ becomes negative. If $\lambda<0$, one has to
  include a term $\propto \phi^6$ with positive coefficient to
  stabilize the free energy functional (see section \ref{sec:IV.A.1}).
  Adding a term $(\kappa/6)\cdot\phi^6(x)$ to Eq.\ (\ref{2.2})
  $(\kappa>0)$, $F$ has now two local minima over a certain
  temperature interval $T_0<T<T_1$, where $T_0$ and $T_1$ have the
  meaning of stability limits of the disordered phase in the ordered
  and the ordered in the disordered phase, respectively. At $T_c$ the
  minima are equally deep, the order parameter jumps from
  $\phi_0=\pm(3\lambda/4\kappa)^{1/2}$ to zero.
  
  In general $\lambda$ may change its sign as function of an external
  parameter $P$ ($P\equiv m_s$ in our example of section
  \ref{sec:IV}).
  
  When $\lambda$ changes its sign at some value $P^*$, a line $T_c(P)$
  of second order transitions ends at a so called {\it tricritical
    point} $T_t=T_c(P^*)$ and continues as a line of first order
  transitions. Tricritical behavior is predicated to occur for QCD in
  certain limiting cases (see section \ref{sec:IV}).
  
\item a cubic term in Eq.\ (\ref{2.2}). If such a term is not
  suppressed by a symmetry argument, it admits a first order
  temperature driven transition. For $\lambda>0$, $F$ has two minima
  at $\phi_0=0$ and at $\phi_0\neq 0$ over a temperature interval
  $T_0<T<T_1$. If the cubic term is written as $(\alpha/3)\phi^3(x)$,
  the nontrivial minimum at $T_c$ occurs for $\phi_0=-9r/\alpha$.
  
  The very existence of a cubic term in an effective potential for the
  electroweak phase transition has been under much debate in the last
  few years. In the simplest models for the electroweak transition the
  order parameter is an $O(N)$-vector field, $r$, $\lambda$ and
  $\alpha$ are renormalized parameters.
  
  Here we add a warning.  A cubic term in the classical potential of
  Eq.\ (\ref{2.2}) does not guarantee the first order of the
  transition. If the transition is weakly first order in the sense
  that the maximum between zero and the nontrivial minimum is not
  high, the transition may be washed out by fluctuations of the order
  parameter field. Such fluctuations will be discussed below.
\end{itemize}
\noindent
Landau's concept for the free energy and the criterion for the order
of the transition is widely used in applications of particle physics.
Two caveats should be mentioned at least.  The first one concerns
convexity properties, the second the validity range of the mean-field
approximation.

\vskip12pt\noindent $\bullet$ {\it Convexity properties}.  It is known
from general thermodynamic principles that thermodynamic potentials in
thermal equilibrium are convex functions of their variables.  The
non-convex shape of Fig.~\ref{fig:1} and the coexisting minima for a
first order transition are obviously at odds with the general
expectation. Landau's free energy is a macroscopic concept. Its
non-convex and physical realization may be understood as a {\it coarse
  grained free energy} arising in an intermediate step from a
microscopic to a macroscopic scale.

To be specific, let us start with a microscopic spin Hamiltonian
${\cal H}(\{s_i\})$ depending on spin variables $s_i$ associated to
sites $i$ of a hypercubic lattice. As a first step short-wavelength
fluctuations are eliminated by dividing the lattice into cells of
linear dimensions $L$ and introducing new variables $\phi(x)$ in a
block spin transformation according to
\begin{equation}
  \phi(x)={1 \over L^d} \sum_{i\in L^d(x)} s_i .
  \label{2.4}
\end{equation}
Here $x$ is a site on the block lattice, the block spin variable
$\phi$ is identified with an order parameter field as it enters the
ansatz (\ref{2.2}). This way we make the connection to Landau's ``mean
field'' free energy $F=F_{mf}$. The dynamics of the coarse grained
block lattice with field variables $\phi(x)$ is determined by the so
called coarse grained Hamiltonian ${\cal H}_{CG}$.  It is obtained as
\begin{equation}
  e^{-\beta {\cal H}_{CG} [\phi(x)] } =
  \int {\cal D} s_i P\left([\phi(x)], [s_i]\right)
  \cdot e^{-\beta {\cal H}[s_i]} ,
  \label{2.5}
\end{equation}
where the path integral extends over all spin configurations $[s_i]$
under the constraint that a particular configuration $[\phi(x)]$ on
the block lattice (determined according to Eq.\ (\ref{2.4})) is kept
fixed.  The constraint is denoted by $P(\ldots)$.  Such a block spin
transformation may be iterated.  The general folklore is then
(although it is extremely hard to be proven rigorously) that after a
sufficient number of iterations, ${\cal H}_{CG}[\phi(x)]$ has the form
of Landau's free energy $F=F_{mf}$, if one is close to a second order
phase transition. At such a transition the correlation length $\xi$
diverges, the condition $L\ll\xi$ on the linear cell size $L$ is
easily satisfied.

The bulk free energy $F$ of the system which is governed by the
Hamiltonian $H_{CG}$ is then obtained upon the remaining path
integration over all configurations $[\phi]$ according to
\begin{equation}
  F=-{1\over\beta} \ln Z = -{1\over\beta}
  \ln \int {\cal D}\phi\exp
  \left\{-\beta F_{mf} [\phi(x)] \right\}.
  \label{2.6}
\end{equation}
If it is justified to evaluate Eq.\ (\ref{2.6}) in a saddle point
approximation, i.e. to drop $\int{\cal D}\phi$ and take $F_{mf}\{ ~
\}$ at its minimum $\phi_0$, the result is
\begin{equation}
  F=F_{mf} ,
  \label{2.7}
\end{equation}
the free energy coincides with its mean-field value, i.e.  Landau's
free energy. Landau's free energy is inherently a mean-field
approximation. For a spin system, the order parameter is the
magnetization. Landau's free energy is an expansion in terms of mean
values of spins. It does not include a summation over all spin
configurations according to
\begin{equation}
  F=-{1\over\beta} \ln \int{\cal D}s_i
  \exp \left\{-\beta {\cal H}[s_i]\right\}.
  \label{2.8}
\end{equation}
\noindent
$\bullet$ {\it The validity of the mean-field approximation.} Next we
address the question when one is allowed to ignore the fluctuations in
the order parameter field, i.e. to drop $\int{\cal D}\phi$ in Eq.\ 
(\ref{2.6}) and take $F_{mf}$ for $F$. On a microscopic level it
amounts to a replacement of spin-spin interactions by some average
background represented by $\phi\equiv M$. The validity of the mean
field approximation is guaranteed, if the fluctuations
$\delta\phi\equiv\phi(x)-\phi_0$ in the order parameter field are
small compared to the order parameter itself, i.e.
\begin{equation}
  \left<\left(\delta\phi\left(x\right)\right)^2\right>
  \ll {\phi_0}^2 ,
  \label{2.9}
\end{equation}
where the average $\left<\ldots\right>$ should be taken over all cells
of the coarse grained lattice. Upon using the fluctuation dissipation
theorem, Eq.\ (\ref{2.9}) translates to (see e.g. Binder, 1987)
\begin{equation}
  1\ll R^d (1-T/T_c)^{(4-d)/2} ,
  \label{2.10}
\end{equation}
where $R$ is the interaction range. Mean-field theory becomes exact,
if either the dimensionality, the range of interactions or the number
of interacting neighbors (large $N$ approximation, where $N$ is the
number of order parameter components) becomes infinite.

Mean-field theory often gives correct qualitative predictions for
phase diagrams of three-dimensional systems. Three-dimensional systems
play some role as effective models for high temperature QCD. As the
dimensionality increases, mean-field theory improves, while numerical
calculations get harder. The critical dimension depends on the form of
Landau's free energy expansion.  Symmetry considerations are essential
for constructing Landau's free energy. As both finite temperature
transitions of QCD are supposed to be driven by symmetry breaking, it
is natural to construct effective actions for QCD by using Landau's
concept as a guide-line.

If the condition (\ref{2.10}) is violated, one should use the
renormalization group approach of (Wegner, 1972; Wilson and Kogut, 1974;
Fisher, 1974; Ma, 1976) to describe critical phenomena.

\paragraph*{The renormalization group approach in the infinite
  volume limit}
In this section we recall the main steps of the
renormalization group approach in the {\it infinite volume limit} as a
tool to describe critical phenomena for second order transitions. The
approach leads to predictions of critical indices and scaling
relations between them.

In the renormalization group approach one attempts to solve the path
integral (\ref{2.8}) in iterated steps. (It is useful to visualize
these steps as block spin transformations in a spin system.) One step
refers to one application of a renormalization group transformation
$R_b$ in coordinate space.  The transformation increases the scale by
a factor $b>1$, $b$ is called the scale factor. A series of
Hamiltonians ${\cal H}^{(n)}$ is generated along with the iterated
application of $R_b$, ${\cal H}^{(n)}$ depends on the fields
$\phi^{(n)}$ and couplings $g^{(n)}$ after $n$ steps. For simplicity
let us denote the set of couplings after $n$ steps $g$ and after
$(n+1)$ steps $g^{(1)}$. The relation between ${\cal H}^{(n+1)}$ and
${\cal H}^{(n)}$ implies for the coarse grained free energy density
$f(g)$ after one application of $R_b$
\begin{equation}
  f(g) = G(g)+b^{-d} f(g^{(1)})
  \label{2.11}
\end{equation}
with the same functional dependence $f$ on both sides of the equality
sign. The rescaling factor $b^{-d}$ is necessary to account for the
reduction of the effective degrees of freedom after one step by
$N^{(1)}=b^{-d}N$.  To keep the free energy $F=\int d^dx f(x)$
constant under the operation $R_b$, one has to compensate the reduced
volume on the block lattice by a factor $b^{-d}<1$. The term $G(g)$
denotes the contribution to $f$ which is regular in $g$.

Similarly the {\it correlation length} on the `block lattice' is
reduced by a factor $b^{-1}$ in one step
\begin{equation}
  \xi(g^{(1)})=b^{-1}\xi\left(g\right) .
  \label{2.12}
\end{equation}

At criticality ($T=T_c$), ${\cal H}^{(n\to\infty)}$ is supposed to
converge to a nontrivial fixed point Hamiltonian ${\cal H}(g^*)$,
which is invariant under the transformation $R$. The set of fixed
point couplings $g^*$ which is invariant under $R$ implies
\begin{equation}
  \xi(g^*)=b^{-1}\xi(g^*) .
  \label{2.13}
\end{equation}
Eq.\ (\ref{2.13}) is only compatible with a vanishing or a diverging
correlation length, since $b>1$ by assumption.

Singularities of thermodynamic functions arise as the critical point
is approached. Their power law divergences are characterized by
critical exponents.  Critical exponents can be derived in the
renormalization group approach, when the transformation $R$ acting in
the parameter space of couplings is linearized in the vicinity of the
fixed point. Let ${\cal H}(g)$ denote a Hamiltonian ``close'' to the
fixed point Hamiltonian, close in the sense that $g\approx g^*$, such
that we can write
\begin{equation}
  {\cal H}(g)={\cal H}(g^*)+ \sum_i q_i O_i +
  o\left(\left(\sum_i q_i^2\right)^{1/2}\right) .
  \label{2.14}
\end{equation}
The scaling operators $O_i$ are taken as eigenfunctions of $R$, the
coefficients $q_i$ are the {\it scaling fields}, measuring the
distance in coupling parameter space between $g$ and $g^*$. In a
typical second order transition we have
\begin{eqnarray}
  q_1&&=k_1 t\nonumber\\
  \label{2.15}
  q_2&&= k_2 h\\
  t&&=(T-T_{c})/T_{c}\nonumber ,
\end{eqnarray}
where $k_1$ and $k_2$ are constants, $t$ is the reduced temperature,
$h$ is an external field. The eigenvalues $\Lambda_i$ of the scaling
operators $O_i$ may be written as $b^{\lambda_i}$ due to an imposed
semigroup-property of $R$.  Thus we have
\begin{equation}
  R_b {\cal H}(g) = {\cal H}(g^*) + \sum_i q_i b^{\lambda_i} O_i
  \label{2.16}
\end{equation}
under the action of $R$. Scaling fields for which $\lambda_i>0$ are
called {\it relevant}, those with $\lambda_i=0$ {\it marginal} and
with $\lambda_i<0$ {\it irrelevant}. In the linearized version Eq.\ 
(\ref{2.16}) the relevant and marginal scaling fields (or couplings)
have to vanish (e.g. $T\to T_c$ and $h\to 0$) for ${\cal H}(g)$
converging to ${\cal H}(g^*)$.

The free energy density $f$ and the correlation length $\xi$ can be
expressed as functions of the scaling fields by using Eqs.\ 
(\ref{2.11})-(\ref{2.12}). As we are interested in a temperature
driven second order phase transition in the presence of an external
field $h$ we choose $q_1$ and $q_2$ as the only relevant scaling
fields according to Eq.\ (\ref{2.15}). A frequently used notation is
$\lambda_1\equiv y_t$ for the thermal exponent and $\lambda_2\equiv
y_h$ for the magnetic exponent.  Here we only indicate the subsequent
steps. For a concrete application one has to specify the relevant
scaling fields. If one is interested in temperature driven transitions
in the presence of an external field (later we are interested in this
case in applications to QCD, cf. section \ref{sec:III.C.3}), it is
sufficient to choose two fields $q_{1}$ and $q_{2}$ according to Eq.\ 
(\ref{2.15}).  A frequently used notation is $\lambda_{1}\equiv y_{t}$
for the thermal exponent and $\lambda_{2}\equiv y_{h}$ for the
magnetic exponent. Further fields $q_{\alpha}$ refer to marginal and
irrelevant scaling fields. The singular parts of $f$ and $\xi$ are
written in terms of the reduced temperature times a {\it scaling
  amplitude}. The scaling amplitudes are assumed to admit an analytic
expansion in their arguments. This implies that in the vicinity of
$T_{c}$ ($|t|\ll 1$) contributions of irrelevant scaling fields may be
neglected. We explicitly mention this assumption entering the final
formulas, because the condition $|t|\ll 1$ is easily violated in
numerical applications.

The asymptotic behavior of $f$ is then given as
\begin{equation}
  f(t,h=0) \approx G(0) + A_f^{(\pm)} \left|t\right|^{d/\lambda_1} ,
  \label{2.17}
\end{equation}
where $A_{f}^{\pm}$ denotes the scaling amplitude for $f$, the signs
$(\pm)$ refer to the approach of $T_{c}$ from above or below. This
behavior implies the scaling relation
\begin{equation}
  2-\alpha=d/\lambda_{1} ,
  \label{2.18}
\end{equation}
$\alpha$ is the critical exponent characterizing the scaling behavior
of the specific heat close to $T_{c}$. For the correlation length one
obtains
\begin{equation}
  \xi(t,h=0) \approx A_\xi^{(\pm)} 
  \left|t\right|^{-1/{\lambda_1}}~, \qquad
  |t|\ll 1 .
  \label{2.19}
\end{equation}
From Eq.\ (\ref{2.19}) it is seen that the critical exponent $\nu$
characterizing the divergence of the correlation length may be
identified as
\begin{equation}
  \nu=1/\lambda_{1} .
  \label{2.20}
\end{equation}
Finally we denote the relations for the magnetization $M$ and the
susceptibility $\chi$ in the zero field limit ($h=0$)
\begin{mathletters}
  \label{2.21}
  \begin{equation}
    M(t, 0) =
    {\left.{\partial f\over\partial h}\right|}_{h=0} \approx
    M_0 + A_M^{(-)}\left(-t\right)^{d/{\lambda_1}
      -\lambda_2/\lambda_{1}},~
    t<0
    \label{2.21a}
  \end{equation}
  \begin{equation}
    \chi(t, 0)=
    {\left.{\partial^2 f\over \partial h^2}\right|}_{h=0}
    \approx \chi_0 +
    A_\chi^{(\pm)}\left|t\right|^{d/{\lambda_1}-2\lambda_2/\lambda_{1}},~
    |t|\ll 1
    \label{2.21b}
  \end{equation}
\end{mathletters}
with an analogous splitting to Eq.\ (\ref{2.17}) in singular parts and
additive analytic contributions $M_{0}$ and $\chi_{0}$, originating in
derivatives of the analytic part $G$. The exponent $\lambda_{1}$ is
distinguished as long as we consider temperature driven transitions.

Note that only the leading singular terms in the expansion of
$A_f^{\pm}|_{h=0}$ and $A^{\pm}_{\xi}|_{h=0}$ in powers of
($q_{\alpha}|t|^{-\lambda_{\alpha}/\lambda_{1}}$) with
$\lambda_{\alpha}/\lambda_{1}<0$ have been kept in Eqs.\ (\ref{2.21}),
(\ref{2.19}), (\ref{2.17}). As emphasized above this is justified as
long as $|t|\ll 1$. Due to critical slowing down it is inherently
difficult in a Monte Carlo simulation to satisfy the condition $|t|\ll
1$. If one observes deviations in measured critical indices from
theoretical expectations, one should keep in mind that contributions
from irrelevant terms are a possible explanation among others.

%
\subsubsection{Finite size scaling analysis}
\label{sec:II.A.2}
Monte Carlo simulations are necessarily performed in a finite volume.
Effects of the finite size are certainly important, if either the
volume is small compared to physical length scales or the physical
scales are given by long-wavelength fluctuations or massless modes.
Massless modes feel the finite volume however large the size of the
system.

It is well known that finite size effects render it difficult to infer
the order of a phase transition. Even qualitative conclusions may be
misleading. Thus a truly second order transition may look like first
order showing a double peak structure in the probability distribution
of the internal energy or magnetization. The double peak structure is
a typical signal for the coexistence of phases at a first order
transition. In this case, however, it is a finite size artifact. It
would eventually vanish, if the volume is further increased.

Vice versa large correlation lengths in a finite volume suggest a
second order transition. One possible reason is that the correlation
length is truly large, but finite. The finite value would be visible,
if the volume is further increased. A further increase may involve a
practical problem of computer time. A second reason for a large
correlation length may be a particle-like excitation in the spectrum
which gets its mass from the volume and decouples in the infinite
volume limit. In such a case the associated correlation length does
not correspond to a bulk correlation length. Moreover one has to be
careful about the order of the critical ($T\to T_c$) and the
thermodynamic ($V\to\infty$) limits. If $T\to T_c$ is taken first, the
correlation length diverges also in case of a (truly) first order
transition. A misinterpretation of data for a correlation length was
at the origin of the controversy about the order of the deconfinement
transition in the pure $SU(3)$ gauge theory (cf. section
\ref{sec:III.B.1}).

Fortunately there are characteristic signatures in a finite volume,
which anticipate the behavior in the thermodynamic limit. A careful
analysis of these signatures allows one to determine the order of the
transition. The measurements should be performed as function of a
varying volume. A measurement for a single size is not conclusive in
general.

In the following we discuss the finite size scaling behavior of
singularities in thermodynamic functions. We consider the case of
second order transitions first.

\bigskip {\it Finite size scaling analysis for second order
  transitions.} The renormalization group approach is an appropriate
framework to discuss finite size effects (see e.g. Barber, 1983).  It
puts the size of the system on an equal footing with other relevant
scaling fields like temperature, masses, magnetic fields.

The central conjecture of a finite size scaling analysis refers to a
thermodynamic function $P$, which admits a critical singularity in the
infinite volume limit, characterized by an index $\mu$
\begin{equation}
P_\infty(t)\sim A_P^\pm \left|t \right|^{-\mu}~,
\label{2.22}
\end{equation}
where the reduced temperature $t$ has been defined in Eq.\ 
(\ref{2.15}), and $A^\pm$ are the amplitudes obtained in the limit
$T\to T_c$ with $t>0(+)$ or $t<0(-)$. Consider a system, which is
finite at least in one dimension with linear size $L$. The conjecture
about the finite size behavior of $P$ is the following
\begin{equation}
P_L(t) = \left|t\right|^{-\mu} Q_P^{(\pm)} \left(L/\xi_\infty(t)\right)~,
\label{2.23}
\end{equation}
where $L$ is assumed to be large and $T$ close to $T_c$, $L\gg 1$,
$\left|t\right|\ll 1$, $\xi_\infty$ denotes the correlation length in
the infinite volume limit.  The derivation of Eq.\ (\ref{2.23}) in a
renormalization group approach is essentially the same as in the
infinite volume limit. The additional ingredients are arguments, why
$L$ can be considered as a scaling field. There may be a conflict in
principle. Already in a single renormalization transformation
non-local terms are generated in the Hamiltonian. Thus the system size
should be sufficiently large to account for all non-local couplings
which one wants to keep.

The assumption is that close to the critical point one needs to keep
only a few couplings. Once this set of couplings has been fixed, $L$
can be chosen sufficiently large compared to the typical interaction
range. In further distinction to the infinite volume limit the
renormalization group transformations can be iterated only a finite
number of times, as the scale increases by a factor $b>1$ in each
step. Here the allowed number of steps should be sufficiently large.
It is also taken for granted that the stability of a set of fixed
point couplings $g^*=R_b g^*$ and the semigroup property of $R$ are
preserved in a finite volume. Under these assumptions $L^{-1}$ may be
treated as an additional scaling field.

The ansatz for the free energy density in terms of scaling fields and
eigenvalues of $R$ takes then a similar form as in the infinite volume
analysis. It is given as
\begin{eqnarray}
  f(k_1t,&&k_2h, q_{\alpha},L)=G(k_1t,k_2h,q_{\alpha}) \nonumber\\
  &&+|t|^{d/\lambda_1}\cdot
  f^{(\pm)}\left(h|t|^{-\varphi_2},q_{\alpha}|t|^{-\varphi_2},
    L|t|^{1/\lambda_1}\right) ,
  \label{2.24}
\end{eqnarray}
where $b$ has been chosen large enough that $b^{\lambda_1}|t|\sim 1$,
$\phi_{2}$ stands for $\lambda_{2}/\lambda_{1}$.  Eq.\ (\ref{2.24})
implies the finite size scaling behavior of various derived
quantities. For the zero-field susceptibility we find
\begin{eqnarray}
  \chi_L(t) &&= \chi(t,h=0,L)
  ={\partial^2f\over \partial h^2}\bigg|_{h=0}
  \nonumber\\
  &&=  \chi_0 + A^{(\pm)}_{\chi}(L\, |t|^{1/\lambda_1})
  \cdot |t|^{d/\lambda_1-
    2\varphi_2} .
  \label{2.25}
\end{eqnarray}

Irrelevant terms have been dropped on the right hand side of Eq.\ 
(\ref{2.25}). The analytic part $\chi_0$ stays finite as $T\to T_c$.
It may be dropped for large volumes, if the second term is singular as
$T\to T_c$. If the argument of $A_\chi^{(\pm)}$ is expressed in terms
of $\chi_\infty(t)$, it is easily shown that the scaling behavior of
$\chi$ takes the form of Eq.\ (\ref{2.23}) with $P\equiv\chi$ and
$\mu\equiv\gamma$.

Let us see, what Eqs.\ (\ref{2.22}) and (\ref{2.23}) imply for the
explicit $L$-dependence of $P$. In order to guarantee that
\begin{equation}
  \lim_{L\rightarrow\infty} P_L(t)=P_{\infty}(t) \, ,
  \label{2.26}
\end{equation}
the amplitude $Q^{(\pm)}_P$ has to satisfy
\begin{equation}
  \lim_{x\rightarrow\infty} Q^{(\pm)}_P(x)=A_P^{(\pm)}\, ,
  \label{2.27}
\end{equation}
where $x\equiv L/\xi_{\infty}$. The limit $x\rightarrow 0$ is
realized, if $T\rightarrow T_c$ for fixed $L<\infty$. This is the
limit we are interested in. In this case $P_L(t=0)$ should be finite.
Hence the singularity of $P_{\infty}(t)$ has to be compensated by the
scaling function $Q^{(\pm)}_P(x)$ according to
\begin{equation}
  Q^{(\pm)}_P(x)\sim x^{\mu/\nu}\quad \quad {\hbox {for}} \quad
  x\rightarrow 0 \, ,
  \label{2.28}
\end{equation}
where we have used $\xi_{\infty}(t)\sim |t|^{-\nu}$ for $t\rightarrow 0$.
It follows from Eqs.\ (\ref{2.23}) and (\ref{2.28}) that
\begin{equation}
  P_L(T_c)=P_L(t=0)\sim L^{\mu/\nu}\, .
  \label{2.29}
\end{equation}
The last relation (\ref{2.29}) predicts for the susceptibility
\begin{equation}
  \chi_L(T_c)\sim L^{\gamma/\nu}
  \label{2.30}
\end{equation}
and for the specific heat
\begin{equation}
  c_L(T_c)\sim L^{\alpha/\nu}\, .
  \label{2.31}
\end{equation}

Finally we state the scaling behavior of the correlation length in a
finite volume
\begin{equation}
  \xi_L(t)=\xi_{\infty}(t)Q^{(\pm)}_{\xi}(L/\xi_{\infty}(t))\, .
  \label{2.32}
\end{equation}

\vskip10pt {\it Refined criteria} in Monte Carlo calculations for QCD
account for the finite lattice size according to formulas
(\ref{2.30}), (\ref{2.31}), (\ref{2.32}), when they test on a second
order transition. Note that these formulas have been derived under the
conditions $L\gg 1$ and $|t|\ll 1$. One should make sure that these
conditions are satisfied in the actual simulations, otherwise
subleading corrections must be kept.

\vskip10pt An important feature of the finite size scaling analysis in
the renormalization group approach is the similar treatment of the
system size and other scaling fields. The treatment on an equal
footing is manifest in Eq.\ (\ref{2.24}). The inverse extension $1/L$
appears as a relevant scaling field with eigenvalue $\lambda_i=1$,
$i\equiv1/L$. To obtain the critical limit, all scaling fields which
are relevant with respect to the fixed point set of couplings have to
vanish: $t\rightarrow 0 \, (T\rightarrow T_c)$, $h\rightarrow 0$ and
$1/L \rightarrow 0$. As long as $\xi_{\infty}(t)\ll L$, finite size
effects are negligible. If $|t|$ is so small that
$(\xi_{\infty}(t)/L)\sim 0(1)$, or, if $L$ is not sufficiently large,
finite size corrections may drastically alter the thermodynamic
behavior.

Thus the formal treatment of $1/L$ is the same as the treatment of an
external field. The external field may be a mass. From physical
applications it is well known that the finite volume can act like a
mass. An obvious manifestation is the finite correlation length as
result of a finite volume or of a nonvanishing mass. The analogy goes
even further. Consider an Ising model in four dimensions in a finite
volume. In the broken phase there is always a finite probability in a
finite volume for tunneling between degenerate states with
magnetization $\pm 1$. Below $T_c$ the Ising model can be approximated
by the effective potential of an anharmonic oscillator, where the mass
in the effective action is proportional to the volume. The
corresponding ``particle'' decouples in the infinite volume limit. The
kink describing the tunneling between states of opposite magnetization
is associated with a so called vacuum tunneling energy. This energy
goes to infinity along with the volume (spin flips of the entire
system are rather unlikely in a very large volume) (see
e.g.~Meyer-Ortmanns, 1989).

\bigskip {\it Finite size scaling analysis for first order
  transitions.} The finite size scaling analysis for first order
transitions is inherently more difficult than for second order
transitions. The correlation length stays finite even in the infinite
volume limit as $T\to T_c$ after $L\to\infty$, $\xi_\infty/L$ is no
longer a sensible scaling variable as in Eq.\ (\ref{2.28}). In the
second order case it is the thermal eigenvalue $1/\nu$ of the
renormalization group transformation that controls rounding and
shifting of the algebraic singularities, both being of the order of
$L^{-1/\nu}$.

The goal of the finite size scaling analysis for first order
transitions is to predict the rounding and shifting of
$\delta$-function singularities in the second derivatives of a
thermodynamic potential due to the finite volume (e.g.  in the
specific heat due to a latent heat, or in the susceptibility due to a
jump in the order parameter).

The volume has to be finite at least in one direction. This time the
rounding of the singularity, the width of the sharp crossover region,
and the shift in the transition temperature are predicted to be of the
order $L^{-d}$, where $d$ is the dimension of the system.

Before we enter the subtleties of refined criteria, which are based on
the precise form of the rounding and shifting as function of $L$, we
summarize some qualitative signatures of a first order transition in a
typical Monte Carlo simulation.

Qualitative signatures are essentially the large or infinite volume
signatures, which we call `naive' criteria for inferring the order of
the transition.

\vskip12pt
{\it Naive criteria}

\begin{itemize}
\item Some thermodynamic quantities (e.g. the internal energy) are
  almost discontinuous at the transition.
\item A starting configuration half-ordered, half-disordered relaxes
  to very different equilibrium states on both sides of $T_c$ (rather
  than frequently tunneling between both sides). In a second order
  transition the system relaxes to an equilibrium configuration
  independently of the initial condition.
\item At infinite volume tunneling between both phases is completely
  suppressed, at small volumes tunneling may be mixed up with
  fluctuations of statistical origin. For large volumes {\it tunneling
    events} are clearly visible in a Monte Carlo simulation. The
  system is in the ordered phase with a probability
  $\exp\{-L^d\,F_0(\beta)\}$, and in the disordered phase with
  $\exp\{-L^d\,F_d(\beta)\}$, where $F_0$ and $F_d$ denote the free
  energies in the ordered and disordered phases, respectively. If the
  time history is followed over a number of Monte Carlo iterations, it
  shows flip-flops between states of different ``magnetizations''. The
  frequency of flip-flops decreases with increasing volume.
\item Another manifestation of tunneling shows up in the probability
  distribution $P_E$ for the internal energy $E(\beta)$. For large
  volumes it is sharply peaked at the energy values of the ordered
  ($E_0$) and disordered ($E_d$) phases. The deep valley between these
  peaks reflects the rare number of tunneling events. If the initial
  condition is an ordered start, the probability is large to find
  $E_0$ for $E$, for a disordered start it is large to find $E_d$.
  This is nothing but a sign for {\it metastability}.
\item {\it Hysteresis effects} are observed even if one is less close
  to the transition temperature.
\end{itemize}

\vskip4pt\noindent The qualitative signature of a pronounced double
peak structure can be made more quantitative, when it is analyzed as a
function of the lattice size. This way we are led to refined criteria.

\bigskip {\it Refined criteria}.  Different approaches have been
studied to determine the precise form of the finite size scaling. One
possibility is the incorporation of first order transitions in the
renormalization group approach. First order transitions are then
associated with a discontinuity fixed point at zero temperature
(Nienhuis and Nauenberg, 1975; van Leeuwen, 1975). In this approach
one would derive the finite size scaling as special cases of Eq.\ 
(\ref{2.24}).  For example Eq.\ (\ref{2.30}) for the susceptibility
$\chi_L\sim L^{(\gamma/\nu)}=L^{(2\lambda_2-d)}$ gives a scaling
proportional to the volume $\chi_L\sim L^d$ with $\lambda_2=d$.
Similarly Eq.\ (\ref{2.31}) for the specific heat $c_L\sim
L^{\alpha/\nu}= L^{2\lambda_1-d}$ gives $c_L\sim L^d$ with
$\lambda_1=d$.  The dimensionality $d$ is the only eigenvalue of the
discontinuity fixed point.

Scaling behavior at a first order transition can be discussed also
independently of the existence of such a discontinuity fixed point. A
phenomenological approach has been developed by (Imry, 1980; Fisher
and Berker, 1982; Privman and Fisher, 1983; Binder and Landau, 1984;
Challa et al., 1986).

More recently exact results have been obtained by Borgs and Kotecky
(1990) and Borgs et al.  (1991) for models which can be represented by
a contour expansion with small activities like the $q$-state Potts
model for large $q$. The partition function $Z(\beta, L)$ in a finite
lattice of volume $L^d$ with periodic boundary conditions is expanded
according to
\begin{eqnarray}
  Z(\beta,L)=&&e^{-L^{d}\beta f_d(\beta)}+q\cdot e^{-L^d\beta f_o(\beta)}
  \nonumber\\
  &&+{\cal O}(e^{-bL})e^{-\beta f(\beta)L^d} \quad  b>0  \, .
  \label{2.33}
\end{eqnarray}
The free energy is the minimum of $f_d$ and $f_o$, where the indices
$o$ and $d$ stand for ordered and disordered, $b$ is a constant larger
than zero. In what follows we discuss predictions of the
phenomenological two Gaussian peak model for the energy probability
distribution $P_L(E)$ which has been introduced by Binder and Landau
(1984) and Challa et al. (1986). This model is related to the rigorous
expansion (\ref{2.33}) by an inverse Laplace transform.  Therefore
results of the phenomenological model may be compared to exact results
and improved in a more systematic finite size scaling analysis
(Billoire et~al., 1990 a and b; Billoire, 1990). We present part of
their results in the following.

\bigskip {\it The two Gaussian peak model}.  The basic quantity for
temperature (energy $E$) driven transitions in the first order case is
the probability distribution $P_L(E)$ of the internal energy.  (For
field driven transitions $E$ is replaced by the magnetization $s$,
(Binder and Landau, 1984).) For large volumes and away from the
transition point $P_L(E)$ is given by a simple Gaussian. This is seen,
when the exponent is expanded around its maximum as function of the
internal energy. The double peak structure of $P_L(E)$ in the vicinity
of a first order transition is then easily understood by noticing that
the exponent is extremal for two values $\widetilde{E_o}(\beta)$ and
$\widetilde{E_d}(\beta)$, one for each of the coexisting phases. As a
phenomenological ansatz the energy probability density $P_L(E)$ is now
replaced by a {\it sum of two Gaussian distributions}. It is plausible
that their width should be determined by the distance from $T_c$.
Expanding $\widetilde{E}(\beta)$ around $T_c$ leads to
\begin{eqnarray}
  P_L(E) = A\bigg[ && {a_o \over \sqrt{c_o}} e
  {-L^d\beta^2_c(E-E_o-c_o\delta T)^2
    \over 2 c_o} \nonumber\\
  + && {a_d\over\sqrt{c_d}}e {-L^d\beta^2_c(E-E_d-c_d\delta T)^2
    \over 2 c_d} \bigg] \, .
\label{2.34}
\end{eqnarray}
Here $E_{o,d}$ and $c_{o,d}$ are the infinite volume energy and
specific heat in the pure ordered and disordered phases, respectively,
$A$ is a normalization factor, $\delta T$ is defined as $\delta
T=(\beta_c-\beta)/\beta^2_c$, and $a_o$, $a_d$ are weight factors for
each phase
\begin{equation}
  a_o=q e^{-\Delta}\, , \quad \quad a_d=e^{\Delta}\, .
  \label{2.35}
\end{equation}
The factor $q$ corresponds to $q$ ways of realizations of the broken
phase in the $q$-state Potts model, $q=2$ for the ($\pm$)
magnetizations in the Ising model. Finally $\Delta$ is given as
\begin{eqnarray}
  \Delta=&&{L^d\over 2}\beta\left(F_o(\beta)-F_d(\beta)\right) 
  \nonumber\\
  \sim && {L^d\over 2}(\beta-\beta_c)
  (E_o-E_d+{1\over 2}(c_o-c_d)\delta T)\, .
\label{2.36}
\end{eqnarray}
Eqs.\ (\ref{2.34})-(\ref{2.36}) predict energy moments correctly up to
order $L^{1/d}$.

A specific feature for a first order transition is the sharpening of
the double peak structure as the volume increases.  The tunneling
between both phases is less likely the larger the volume is. The spin
configurations corresponding to mixed states lead to an energy $E$,
which lies between the locations of the peaks at $E_o$ and $E_d$. (The
distribution of these ``intermediate'' energies is not correctly
represented by the superposition of Gaussians according to Eq.\ 
(\ref{2.34}).) Contributions of mixed states vanish in the infinite
volume limit, as long as they are equilibrium configurations.  Nothing
is said about the phase conversion itself, which may proceed as an
off-equilibrium process with ``droplet'' formation.

The double peak structure and the criteria we present in the following
are essentially based on the double valued internal energy $E_o$ and
$E_d$ at the transition point. The free energies are degenerate in
this case.

Let us first consider the rounding in the specific heat and the
shifting of the transition point. The specific heat in the volume
$L^d$ is as usual obtained as
\begin{equation}
  c_L=L^d\beta^2(\langle E^2\rangle_L-\langle E\rangle^2_L)~,
  \label{2.37}
\end{equation}
where
\[
  \langle E\rangle_L =\int_{-\infty}^{+\infty}P_L(E)\, E\, dE
\]
with $P_L(E)$ given by Eqs.\ (\ref{2.34})-(\ref{2.36}). Its maximum
$c_{L max}$ occurs for an inverse temperature
\begin{equation}
  \beta(c_{L max})=\beta_c-{\ln q\over E_d-E_0}{1\over L^d}+0(1/L^{2d})
  \, .
  \label{2.38}
\end{equation}
Eq.\ (\ref{2.38}) shows that the shift of the critical temperature in
the finite volume compared to the infinite volume value is of the
order of $1/L^d$.  The height of the maximum in the specific heat
$c_{L max}$ is derived to be proportional to the volume
\begin{equation}
  c_{L max}=L^d{\beta^2_c\over 4}(E-E_d)^2+{\hbox{terms of order}}\, (1)\, .
  \label{2.39}
\end{equation}
Thus Eqs.\ (\ref{2.38})-(\ref{2.39}) display the anticipated volume
dependence of shifting and rounding in a first order temperature
driven transition .  Subleading corrections of $0(1/L^{2d})$ in Eq.\ 
(\ref{2.38}) and of $0(1)$ in Eq.\ (\ref{2.39}) have been determined
by Lee and Kosterlitz (1990).

\bigskip {\it Binder's cumulant.} Rather than calculating the $P_L(E)$
distribution itself, certain moments and cumulants of the energy
probability distribution have been proven to be suitable indicators of
the order of the transition in a finite volume. An important example
is Binder's cumulant (Challa et al., 1986) defined as
\begin{equation}
  B={1\over 3}\left(1-
    {\langle E^4\rangle\over\langle E^2\rangle^2}\right)\, .
  \label{2.40}
\end{equation}
It vanishes in the infinite volume limit for all temperatures apart
from the transition, where $B$ gets a finite value from tunneling
between coexisting phases of a first order transition. It also
vanishes in the case of a second order transition. Calculating the
moments $\langle E^4\rangle$ and $\langle E^2\rangle$ from Eqs.\ 
(\ref{2.34})-(\ref{2.36}) with use of Eq.\ (\ref{2.37}), $B$ takes its
minimum $B_{min}$ at (Billoire et al., 1990a and b)
\begin{equation}
  B_{min}={-(E_o^2-E^2_d)^2\over 12(E_o E_d)^2}+O(1/L^d)
  \label{2.41}
\end{equation}
at an inverse temperature
\begin{equation}
  \beta(B_{min})=\beta_c
  -{\ln(q(E_o/E_d)^2)\over E_d-E_o}\cdot {1\over L^d}
  +O({1\over L^{2d}})\, .
  \label{2.42}
\end{equation}
The shifting in the critical temperature in Eq.\ (\ref{2.42}) is again
of order $1/L^d$. A nonvanishing value of $B_{min}$ signals a first
order transition . Analogous formulas can be derived for the spin
probability distribution $P_L(s)$, implying that the maximum of the
susceptibility grows proportional to the volume.

Binder's cumulant depends on the choice of an arbitrary additive
constant which may be added to the energy. This has led to consider
the quantity $U4$ (Billoire et al., 1992)
\begin{equation}
  U4\equiv {\langle (E-\langle E\rangle )^4\rangle\over 
    \langle (E-\langle E
    \rangle)^2\rangle^2}\, .
\label{2.43}
\end{equation}
This quantity can be shown to be larger than 1 apart from a first
order transition point $U4$ can be shown to be larger than 1. Its
minimum is given as
\begin{equation}
  U4_{min}=1+{8(C_o+C_d)\over L^d\beta^2_t(E_o-E_d)^2}+O(1/L^{2d})\, .
  \label{2.44}
\end{equation}
The power law corrections to $C_{L max}$, $B_{min}$ and $U4_{min}$
should be contrasted with exponential corrections to bulk quantities
like the average internal energy $E$ which are directly obtained as
derivatives of $Z$. According to Eq.\ (\ref{2.33}) $Z$ itself has only
exponential corrections from the finite volume.


In Table~\ref{tab:1} we summarize criteria to distinguish first and
second order transitions in a finite volume. The indicated volume
dependence should be understood as leading term in a large volume
expansion. Similarly the numbers $0$ or $1$ indicate limiting values
which are approached as $T\rightarrow T_c$. Recall that $\alpha /\nu$
and $\gamma /\nu\leq d$ for a second order transition. Thus a typical
test in a Monte Carlo simulation could be a calculation of $c_{L
  max}/L^d$. If this ratio as function of $L^d$ goes to a
nonvanishing constant for large values of $L$, a first order
transition is signalled, if it approaches zero, the transition must be
of second or higher order.

\bigskip The two-phase coexistence at $T_c$ refers to the Ising-model.
The analytic results for leading and subleading finite size scaling
behavior have been derived by Billoire et al. (1992) for the more
general case of $d$-dimensional q-state Potts models which include the
familiar Ising model for $q=2$. Potts models have served as testing
ground for the above criteria.

\bigskip
\noindent
{\it $q$-state Potts models.} The Hamiltonian of a $d$-dim $q$-state
Potts model is given as
\begin{equation}
  H=-\sum_{\langle ij\rangle}\delta_{s_i,s_j}\, .
  \label{2.45}
\end{equation}
The spin variables $s_i$ are associated to sites of a $d$-dim
hypercubic lattice. They can take $q$ different integer values. The
sum extends over nearest neighbor pairs $\langle ij\rangle$. The
symmetry group leaving $H$ invariant is the permutation group of $q$
elements. The $q=3$-Potts model in three dimensions plays a
distinguished role in view of QCD. It shares the global
$Z(3)$-symmetry with the pure SU(3)-gauge theory.  The restoration of
the spontaneously broken $Z(3)$-symmetry at finite temperature is
assumed to drive the deconfinement transition in the pure gauge
theory. \par Beyond this application to QCD, Potts models provide a
suitable testing ground for numerical methods. Exact results are
available for comparison in certain special cases of $q$ and $d$.  The
strength of the first order varies as a function of $q$. The
$q=5$-model has a weakly first order transition accompanied by a tiny
latent heat of $\Delta Q=E_d-E_o=0.05292$ in lattice units and a large
correlation length of 2000 in lattice units at the transition point
(Billoire, 1991).  The transition in the $q=10$-model is strongly
first order, $\Delta Q=0.69605$ and $\xi\sim 6$, both quantities
measured in lattice units (Billoire, 1990). Obviously the transition
in the $q=5$-Potts model is easily misinterpreted as being of second
order.  The linear lattice size has to exceed the large value of
$\xi\sim 2000$.

Simulations of the $2d$-$q$=10-Potts model have been performed by
Billoire et al. (1992) to check the limiting behavior of $(c_{L
  max}/L^d)$, $B_{min}$, $U4_{min}$ and $c(\beta_t)$ ($\beta_t$ being
the infinite volume transition point). According to the above
predictions these quantities should scale as (const1 + const$2/L^d$).
Deviations have been observed, which do not behave in a simple way as
function of $L$. Several possible reasons for the discrepancies are
mentioned by the authors: $q=10$ may not be in the large $q$-limit,
for which Eq.\ (\ref{2.33}) has been derived, $L$ may not be
sufficiently large to be in the asymptotic regime.

Therefore a $2d$-$q$=20-Potts model has been simulated in addition
(Billoire et al., 1992). The first order transition there is yet
stronger than in the $q=10$-case. Thus the asymptotic behavior in $L$
should earlier set in than for $q=10$, 'earlier' means for smaller
volumes. (The ratio which counts is $\xi/L$.) Here the agreement with
the theoretical predictions is reasonable.

Let us summarize so far. Computer simulations (as well as real
experiments) are always performed in a finite volume, whereas the
order of a phase transition is usually formulated in the infinite
volume limit. Although this formulation is very convenient, it is not
the only appropriate way to proceed. The good message comes from
finite size scaling analysis. In a finite volume it is prevailed in a
specific way, whether the transition is going to be of first or second
order in the thermodynamic limit. (Likewise a small mass can play the
role of a scaling field. A finite mass scaling analysis is suited for
an extrapolation to the zero mass limit.)
\noindent
From a practical point of view a finite size scaling analysis may be
less useful, since it is derived for finite but large volumes.  The
original hope was to disentangle first and second order signatures
already for moderate lattice sizes by using refined criteria. The
results for the $2d$-$q$=10-Potts model involve a somewhat discouraging
message. One has to go to so large volumes to verify the predictions
of a finite size scaling analysis that the `naive' criteria are
equally well applicable for inferring the order of the transition.
This message refers to weakly first order transitions.  Otherwise the
asymptotic region of large volumes is easily realized as $\xi/L\ll 1$
for moderate values of $L$ when $\xi$ is sufficiently small.

Typical volumes in Monte Carlo calculations for QCD are marginal in
large finite size.  For QCD transitions it seems to be rather
difficult to ever get into the asymptotic regime of large volumes.
Right from the beginning one is restricted to much smaller volumes
when simulating full QCD rather than a spin model. The final
efficiency of refined criteria is, however, a question of size, which
has to be answered in the concrete model of interest. The value of the
correlation length in a first order transition depends on details of
the dynamics. For QCD the largest correlation length is likely not
small compared to the typical lattice size.

\bigskip
This concludes our section about concepts and results of
statistical mechanics.  As we turn now (and for the remainder of this
review) to applications in QCD, we give a dictionary between
thermodynamic quantities in statistical physics and their pendants in
QCD in Table~\ref{tab:2}.

The second and third columns refer to transitions in liquid/gas or
ferromagnet systems. Two magnet-fluid analogies are seen, the first
one between the set $(T,p,V)$ of a fluid and $(T,H,-M)$ in a magnet,
and the second one between $(T,\varrho,\mu)$ in a fluid and $(T,M,H)$
in a magnet (Stanley, 1971). The external field $H$ is the
thermodynamic variable conjugate to the order parameter $M$. Likewise
the chemical potential is conjugate to the density $\varrho$,
$\varrho$ being the order parameter for a fluid/gas transition. A
substitution of $V$ by $(-M)$ and $p$ by $H$ transforms almost all
equations for a fluid/gas system in the corresponding equations for a
magnet.

In the two last rows various response functions are listed, describing
a response of the system to a stimulus in the temperature or in an
external field. They are second derivatives of the thermodynamic
potential $\Omega$, while the order parameter is obtained as a first
derivative of $\Omega$ with respect to the conjugate field. The last
column shows the associated critical exponents, characterizing the
singular behavior of thermodynamic functions in the case of the second
order transition.

The QCD transitions refer to limiting cases of vanishing quark masses
(chiral symmetry) or infinite quark masses (Z(3) symmetry). A further
column is devoted to the chiral transition described on an effective
level (here in the linear sigma model), where the quark- and gluonic
substructures of mesons are disregarded. If one is interested in the
evolution of a QCD-plasma ``fluid'', (T, p, V) is an appropriate set
of thermodynamic variables entering the equation of state.

Formally the current quark masses $m_q$ will be shown to play the same
role as an external magnetic field $H$ in a ferromagnet. On an
effective level this analogy is more manifest, as nonvanishing quark
masses will be described by external fields $\varepsilon_0,
\varepsilon_8$, see section \ref{sec:IV.A.4}. The analogy is
extensively utilized, when effects of finite quark masses on the order
of the QCD transitions are conjectured from results in the massless or
pure gauge limits.

The most popular order parameters in the QCD case are the quark
condensate $\langle \bar{q} q\rangle$ (or $\langle \bar{q} q\rangle$
and $\langle \bar{s} s\rangle$), or mesonic condensates $\langle
\sigma_0\rangle$, $\langle \sigma_8\rangle$ on the mesonic level for
the chiral transition, and the expectation value of the Wilson line
$\langle L\rangle$ for the deconfinement transition. The possibility
to predict the sigma- and pion-meson mass from a critical equation of
state (section \ref{sec:IV.A.1}) sounds less surprising, if one keeps
this dictionary in mind.

The indices $b$ and $s$ at $T_c$ refer to the approach of $T_c$ from
the symmetric or the broken phase. For the deconfinement transition
the phase of broken Z(3)-symmetry is realized {\it above} $T_c$, thus
the commonly used ($\pm$) signs to indicate the approach from above or
below $T_c$ would be misleading in this case. Note that the singular
behavior of the ``magnetization'', i.e. the order parameter, is
characterized by a thermal exponent $\beta$ as $T\rightarrow T_c^b$
and by a ``magnetic'' exponent $1/\delta$ as ``H''$\rightarrow 0$ {\it
  at} $T_c$. Further explanations to the QCD-part will be given in the
succeeding sections.

%
\subsection{Phase transitions in QCD}
\label{sec:II.B}
In a continuum notation the QCD-Lagrangian is given as
\[
L=L_g+L_m
\]
\vskip-6pt\noindent
where
\vskip-12pt
\begin{eqnarray}
  L_g=&-{1\over 4}F^a_{\mu\nu}(x)F^{a\mu\nu}(x) \nonumber\\
  L_m=&\bar{\psi} (x)(i\gamma^{\mu} D_{\mu}-M)\psi(x)& \nonumber
\end{eqnarray}
\vskip-6pt\noindent
and
\vskip-12pt
\begin{eqnarray}
  F^a_{\mu\nu}(x)\equiv &&\partial_{\mu}A^a_{\nu}(x)
  -\partial_{\nu}A^a_{\mu}(x)+
  g\cdot f^{a b c }A^b_{\mu}(x)A_{\nu}^c(x) \nonumber\\
  D_{\mu}\equiv &&\partial_{\mu} 
  - i\cdot g {\lambda_a\over 2}A^a_{\mu}(x)\, .
  \label{2.46}
\end{eqnarray}
\noindent
Here $A^a_{\mu}$ are the gauge fields, $\psi_{\alpha ,f,c}$,
$\bar{\psi}_{ \alpha, f, c}$ denote the quark fields, where $\alpha$
is a Dirac index, $f=1,\ldots, N_f$ labels the flavors, $c=1,\ldots
,N_c$ labels the colors ($N_f$ is the number of flavors and $N_c$ the
number of colors).  The gauge coupling constant is denoted as $g$, the
structure constants as $f_{abc}$, and $\lambda_a$ are the generators
of the fundamental representation of $SU(N_f)$, $a=1\ldots 8$ for
$N_f=3$, where $\lambda_a$ are the Gell-Mann matrices. The quark mass
matrix is denoted as $M$.

As a first step in the investigation of the phase structure one may
consider certain limiting cases of Eq.\ (\ref{2.46}). One limit is the
pure gauge theory, where $L=L_g$. This is the limit of infinitely
heavy quark masses and will be discussed in section \ref{sec:II.B.2}.
The topic of the following section is the limit of massless quarks,
which is called the chiral limit. In the chiral limit the QCD
Lagrangian is invariant under global $U(1)_V\times SU(N_f)_{L}\times
U(1)_A\times SU(N_f)_{R}$ transformations.  The $U(1)_V$-invariance
corresponds to the baryon number conservation. The invariance under
axial $U(1)_A$-transformations is only classically preserved.  On the
quantum level it is broken via the axial anomaly. Even in the presence
of an anomaly there remains axial $Z_A(N_f)$ symmetry (Callan et al.,
1976). At zero temperature the invariance under $Z_A(N_f)\times
SU(N_f)_L\times SU(N_f)_R$ chiral transformations is assumed to be
spontaneously broken by the QCD vacuum to the vector
$SU(N_f)_V$-symmetry. For $N_f=2$ this isospin symmetry is realized in
the hadronic spectrum to a very good approximation, for $N_f=3$ the
realization of the $SU(3)_V$-symmetry is a little more questionable,
nevertheless it is also frequently considered as approximate symmetry
of QCD.

\subsubsection{Renormalization group analysis in the chiral limit}
\label{sec:II.B.1}
Rather than directly studying the QCD Lagrangian (\ref{2.46}) one can
analyze chiral symmetry breaking in an effective Lagrangian which
shares the chiral symmetry properties of QCD. The Lagrangian is
formulated in terms of a self-interacting $N_f\times N_f$ matrix field
$\phi$. Chiral symmetry breaking is parametrized in terms of
$\phi_{ij}$ according to $\phi_{ij}\sim\langle\bar{q}_i
(1+\gamma_5)q_j \rangle$, where $\phi$ transforms under
transformations of $G_f\equiv U_A(1) \times SU(N_f)_{L}\times
SU(N_f)_{R}$ according to
\begin{equation}
  \phi\rightarrow\phi ' =\exp(i\alpha)U_+\phi U_- \, .
  \label{2.47}
\end{equation}
Here $U_+$ and $U_-$ are arbitrary and independent $SU(N_f)$ matrices,
$\alpha$ generates a $U_A(1)$ transformation. The most general
renormalizable Lagrangian in terms of $\phi$ which is consistent with
the chiral symmetry properties of QCD is given as (Pisarski and
Wilczek, 1984)
\begin{eqnarray}
  L= && {1\over 2} \hbox{Tr}~(\partial_{\mu}\phi^+)(\partial_{\mu}\phi)
  -{m^2 \over 2} \hbox{Tr}~(\phi^+\phi)
  - {\pi^2\over3}f_1(\hbox{Tr}~\phi^+\phi)^2
  \nonumber\\
  - && {\pi^2\over 3}f_2 \hbox{Tr}~(\phi^+\phi)^2
  +g\cdot(\det\phi+\det\phi^+)\, .
  \label{2.48}
\end{eqnarray}
\vskip8pt As a necessary condition for stability at large values of
$\phi$, $f_2$ and ($f_1+f_2/N_f$) have to be larger than zero. The
determinant-term accounts for the anomaly. It vanishes in the pure
gauge case ($N_f=0$) and in the limit of infinite colors $N_c=\infty$
(Witten, 1979). At zero temperature the vacuum expectation value
$\langle \phi\rangle$ is $SU(N_f)$-symmetric and different from zero,
$m^2<0$.  The spontaneous symmetry breaking is associated with an
$SU(N_f)$ massless multiplet of Goldstone bosons and a massive flavor
singlet $\eta '$.

In Table~\ref{tab:3} we summarize results of a renormalization group
analysis of Pisarski and Wilczek (1984) about the order of chiral
phase transitions. These transitions are driven by chiral symmetry
restoration, as the temperature is raised.

For each number of flavors three cases are distinguished: a vanishing
anomaly corresponding to a vanishing number of flavors or an infinite
number of colors, $g$=const of order 1, where `const' refers to the
assumed temperature independence of $g$, and $g=g(T)$. Here $g(T)$ is
taken to be approximately equal to the instanton density $d_I$, which
is supposed to vanish at high temperatures. In the following we
summarize, how the conjectures of Table~\ref{tab:3} arise in a
renormalization group analysis.

In the previous section we have mentioned the real space
renormalization group approach as a tool to describe critical
phenomena.  An alternative approach for the construction of
renormalization group equations which is utilized here is the {\it
  $\varepsilon$-expansion} performed in momentum space. Starting from
a Hamiltonian in terms of an order parameter field $\phi$, large
momenta are integrated out (large momenta in contrast to short length
scales in the real space approach). The new Hamiltonian in terms of
the new field variables is arranged to have the same form as the old
one. The new couplings should be understood as renormalized couplings.
They are obtained in a perturbative expansion. The coefficients of the
expansion depend analytically on the dimension $d$. For our
applications the appropriate small expansion parameter is
$\varepsilon=d-4$.

Normally a change in the scale of momenta induces intricate changes in
the action and in derived quantities like correlation functions,
unless there are special instances which we know from the previous
section.  Simple scaling behavior is recovered, if the set of
couplings reaches a value such that any further change in the scale of
momenta does not affect them. This is the point in coupling parameter
space, where the Hamiltonian approaches a fixed point Hamiltonian.

In the present framework fixed points occur as zeros of the
$\beta$-functions.  In the case of one coupling the $\beta$-function
gives the change in the renormalized coupling under a change in
momentum scale. Equivalently the $\beta$-function gives the change of
the dimensionless renormalized coupling $u$ under a change of the
dimensionless bare coupling $u_0$ according to
\begin{equation}
  \beta(u)=  -\epsilon \left({\partial\ln u_0\over \partial u}\right)^{-1}
  = \sum_n a_n(\epsilon)u^n
  \, ,
  \label{2.49}
\end{equation}
where the coefficients $a_n$ depend on $\varepsilon$. Given a set of
bare couplings in parameter space at a certain momentum scale, the
question arises as to whether, and under what conditions the
renormalized couplings $v$ flow into the fixed point $v^*$ (leading
this way to a second order phase transition). The answer is given as a
condition on the occurrence of an IR-stable fixed point $v^*$. The
stability criterion is that the matrix
\begin{equation}
  \omega_{ij}\equiv {\partial\beta_i\over\partial v_j}
  \label{2.50}
\end{equation}
should have real and positive eigenvalues at $v=v^*$. Here $v$ and
$v^*$ stand for sets of couplings, $\beta_i$ are the associated
$\beta$-functions.

The $\beta$-functions for a chiral $SU(N_f)\times SU(N_f)$ linear
sigma model in the absence of an anomaly (case $g=0$ in
Table~\ref{tab:3}) have been derived within an $\varepsilon$-expansion
(Paterson, 1981). When the two functions $\beta_{1}$ and $\beta_{2}$
for $f_{1}$ and $f_{2}$ are inserted in Eq.\ (\ref{2.50}), the
stability criterion may be applied to search for IR-stable fixed
points. The results are summarized in the first column of
Table~\ref{tab:3}. For $0\le N_f < \sqrt{2}$ the IR-stable fixed point
has $f^*_2=0$ with $O(2N_f)$ critical exponents. No infrared stable
fixed point occurs for $N_f>\sqrt{3}$, if $f_1$, $f_2$ are of order
$\varepsilon$.

For some time it has been taken for granted that the absence of an
IR-stable fixed point implies a first order phase transition (Bak et
al., 1976). The very absence or existence of an IR stable fixed point
is, however, less conclusive than it was originally supposed. The
existence of an IR-stable fixed point does not exclude a first order
phase transition (there may be a region in coupling parameter space,
which is not attracted by the fixed point). Vice versa does the
absence of an IR-stable fixed point not exclude a second order phase
transition. Instead of the fixed point criterion precise conditions
for the occurrence of a first order phase transition have been
specified by Iacobson and Amit (1981).  The framework is again the
renormalization group approach, realized in the perturbative tool of
the $\epsilon$-expansion. The prediction of first order phase
transitions applies to all multi-component $\phi^4$-theories with more
than one dimensionless coupling constant. The interaction term of a
multi-component $\phi^4$-theory can be written as
\begin{equation}
  \propto g_{i\,j\,k\,l}\, \, \phi_i \phi_j \phi_k\phi_l\, ,
  \label{2.51}
\end{equation}
where $i,\, j, \, k,\, l \ \in \{1,\ldots ,N\}$. The quartic terms of
the linear sigma model Eq.\ (\ref{2.48}) can be recast in this form.
Explicit expressions can be found e.g. in (Paterson, 1981). As long as
the quartic terms proportional to $f_1$ and $f_2$ in Eq.\ (\ref{2.48})
are independent of each other (as it is the case e.g. for $N_f=3$),
the conditions of the criteria of Iacobson and Amit are satisfied by
the linear sigma model, and a first order chiral transition is
predicted for $N_f\ge 2$.

Note that these predictions are neither of the type of conjectures nor
rigorous statements. They are not only based on the absence of an
IR-stable fixed point, but derived within the perturbative framework
with $\varepsilon \ll 1$. Non-perturbative features are not attainable
in this approach. One may try to extrapolate the results from four-to
three-dimensional models by setting $\varepsilon =1$ (although
$\varepsilon \ll 1$ has been assumed in deriving the
$\varepsilon$-expansion).  Examples are known (Bak et al., 1976),
where the results to leading order in $\varepsilon$ remain a good
guide for $\varepsilon =1$.

The second column of Table~\ref{tab:3} subsums the suggestions for the
order of the chiral transition in the presence of a temperature
independent strength of the anomaly, $g$ is assumed to be of the order
of the other couplings $f_1$ and $f_2$. For two flavors the
$\det$-term in Eq.\ (\ref{2.48}) acts itself like a mass term. Thus it
may change the order of the transition depending on the magnitude of
$g$. For three flavors, the $\det$-term is trilinear in the matrix
elements of $\phi$. A cubic term on the classical level is usually
regarded as sufficient for inducing a first order phase transition
(although we know from the discussion above (section \ref{sec:II.A})
that a weakly first order transition, predicted on the classical
level, may be wiped out by fluctuations; hence the classical cubic
term makes the first order transition likely, but is no guarantee for
its occurrence).

For $N_f=4$ Pisarski and Wilczek (1984) argue again with the absence
of an IR-stable fixed point in favor of a first order transition. (The
$g\ne 0$ case is not covered by the analysis of Iacobson and Amit). As
claimed by Paterson (1981), the fixed point structure of the
$SU(4)\times SU(4)$ linear sigma model is unchanged, when a term
$\propto g(\det \phi +\det \phi^+)$ is included. For $N>4$ the
$\det$-term is an irrelevant operator and should not change the
critical behavior.

The effective symmetry of the linear sigma model Eq.\ (\ref{2.48}) may
change as function of temperature, if the anomaly strength is
temperature dependent and determined by the density of instantons
$d_I(T)$. Since $d_I(T)\rightarrow 0$ as $T\rightarrow \infty$ (Gross
et al., 1981), $g$ could be small at the transition temperature
compared to the $T=0$-value. Thus the axial $U_A(1)$-symmetry would be
partially restored. Predictions accounting for this partial symmetry
restoration as result of decreasing $g$ are listed in the third column
of Table~\ref{tab:3}. For further details about consequences of an
approximate $U_A(1)$-restoration we refer to the original references.

The effective Lagrangian (\ref{2.48}) can be extended to include
nonzero bare (meson) masses. The most simple ansatz has the form (tr M
$\phi$), it is linear in the mass matrix $M$. Formally it acts as a
background magnetic field.  The formal analogy is evident, if we
recall that the linear sigma model can be rewritten as a
multi-component $\phi^4$-theory with an additional (multi-component)
$\phi^3$-term (the $\det$-term) and a symmetry breaking mass term,
which is linear in $\phi$. This way it takes the form of Landau's free
energy functional Eq.\ (\ref{2.2}) for a multicomponent order
parameter field $\phi$ in an external field.

For later comparison we notice that the magnetic field term is
proportional to the mass and vanishes in the chiral limit. In the
other extreme case of infinitely heavy quark masses, the effective
magnetic field of pure gauge theories will be shown to be proportional
to ($e^{-m}$), thus vanishing in the infinite mass limit
($m\rightarrow \infty$). We come back to this point at the end of the
next section.

To summarize so far, conjectures about the order of the chiral
transition as function of $N_f$ are based on a perturbative
renormalization group analysis in momentum space. The analysis has
been performed for an effective $SU(N_f)\times SU(N_f)$ linear sigma
model sharing the chiral symmetry properties with QCD. None of the
conjectures is rigorously proven. Some of them ($g=0$) are verified
within an $\varepsilon$-expansion, others ($g\neq 0$) are based on the
absence of an IR-stable fixed point, when a first order transition is
predicted, a criterion which should be taken with care.

From statistical physics it is known that a first order transition
remains first order when a background field is introduced and the
field is sufficiently weak. Otherwise the transition may be washed out
completely.  The relative size of the latent heat compared to the
strength of the external field, i.e. the values of the quark masses,
decides whether the $N_f=3$ chiral transition is preserved under
realistic QCD conditions or not. This question cannot be answered
within a renormalization group analysis, but only by detailed
calculations. The renormalization group approach is nevertheless a
good guideline to start with.

%
\subsubsection{The limit of a pure $SU(N_c)$ gauge theory}
\label{sec:II.B.2}
In this section we deal with the {\it quenched limit} of QCD. The
quenched limit is obtained as the number of flavors goes to zero or
the quark masses are sent to infinity. The gluonic vacuum in a
background of infinitely heavy quarks may be probed by test quarks. In
perturbation theory this means that virtual quark loops are
suppressed.

In the quenched limit the theory has an extra global symmetry, which
results from the periodicity of the gauge fields in the temperature
direction. The partition function may be represented as functional
integral over gauge fields, which are periodic in Euclidean time with
period $\beta=1/T$. The periodicity condition arises as a consequence
of the trace in the definition of the thermodynamical potential
\begin{equation}
  Z=\intop_{A_\mu(\beta,\vec{x})=A_\mu(0,\vec{x})} \hskip-3em{\cal D}A\;
  \exp\left\{
    -{1\over g^2}\int_0^\beta dt\int d^3x\; 
    {1\over4} \hbox{Tr} F^2_{\mu\nu} \right\}~.
  \label{2.52}
\end{equation}
(We still use a continuum notation for the path integral over all
gauge fields $A_\mu$ with Yang Mills field strength $F_{\mu\nu}$.)
Gauge transformations which are compatible with the periodicity
condition need only be periodic up to an element $c_N$ of the center
$Z(N)$ of the gauge group $SU(N)$. Thus a gauge transformation must
obey
\begin{equation}
  V(\vec{x},0) = c_N V(\vec{x},\beta) \hbox{\hskip2ex for all
    }\vec{x}
  \label{2.53}
\end{equation}
with $c_N\in Z(N)$. The $n$-th element $c_N^n$ of $Z(N)$ is given as
$\exp(2\pi i n/N)$. It is easily checked that the generalized gauge
transformation (\ref{2.53}) leaves all topologically trivial Wilson
loops invariant (in particular those appearing in the action) and thus
the action itself.

This means that the extra symmetry is generated by the action of local
gauge transformations, which are periodic up to an arbitrary element
of the center modulo strictly periodic local gauge transformations.
The essential ingredient, which ensures the invariance of
topologically trivial loops under the additional symmetry, is the
property of center elements to commute with all elements of the group.

The issue of the transition from a confinement phase at low
temperatures to the deconfinement phase at high temperatures can be
related to the issue of whether the pure glue vacuum of QCD is
$Z(N)$-invariant like the action or not. As it turns out, the
transition from the confinement- to the deconfinement phase may be
explained as a spontaneous breaking of the extra $Z(N)$-symmetry at
finite temperature.

Qualitatively the quark gluon plasma can be probed by a heavy test
quark. The free energy of this test quark should be infinite in the
confinement phase, but finite in the deconfinement phase. It may be
computed as the expectation value of the Wilson line (synonymously for
Polyakov loop or thermal Wilson loop). The Wilson line $L$ is defined
as the spatially local operator
\begin{equation}
  L(\vec{x})\equiv\hbox{ Tr } P \exp\left(\intop_0^\beta
    dt\,A_0(t,\vec{x})\right)\;, 
  \label{2.54}
\end{equation}
where $P$ stands for the path ordered product. The Wilson line is a
topologically nontrivial loop. It is closed due to the periodic
boundary conditions and transforms nontrivially under the center
transformations
\begin{equation}
  L(\vec{x})\longrightarrow c_NL(\vec{x})\;. 
  \label{2.55}
\end{equation}
Thus $L$ can have a vanishing expectation value. Vice versa a
nonvanishing expectation value can be taken as signal of a
spontaneous breakdown of the global $Z(N)$-symmetry of the action. We
have
\begin{equation}
  \left\langle L(\vec{x})\right\rangle=e^{-\beta
    F(\vec{x})}=\left\{{0\atop\hbox{finite}}\right.\hskip1ex
  {\hbox{confinement}\hfill\atop\hbox{deconfinement}}\;.
  \label{2.56}
\end{equation}
Here $F(\vec{x})$ denotes the free energy of an isolated test quark.
This behavior qualifies the Wilson line expectation value as an order
parameter for the confinement/deconfinement phases of QCD in the
absence of dynamical quarks.

It is natural to look for an effective action of the $SU(N)$ gauge
theory in terms of the order parameter field, this means in term of
$L$. Such an effective action could simplify the investigation of the
phase structure of pure QCD, if universality arguments may be used.
Svetitsky and Yaffe (1982a, 1982b) have given plausible arguments that
a suitable candidate for such an effective action is a $Z(N)$-spin
theory.

The result of an integration over spatial gauge fields is an effective
action in terms of Wilson lines. The Wilson lines (originally given in
terms of elements of the $SU(N)$-Lie algebra), are represented by
$Z(N)$ variables. The action is argued to be short ranged. The high
temperature behavior of the $3+1$-dimensional theory is determined by
the dynamics of the three-dimensional $SU(N)$ gauge theory. Three
dimensional $SU(N)$ gauge theories show an area-law for Wilson loops
(Feynman, 1981) with short range correlations. The range of
interactions is supposed to stay finite over the entire temperature
range.

The effective action is invariant under $Z(N)$ transformations. The
$Z(N)$ symmetry can be spontaneously broken at high temperatures, but
is restored at low temperatures and ensures confinement.

The path integral representation of the original $SU(N)$ gauge theory
is then replaced by
\begin{equation}
  e^{-\beta V\Omega} = \sum_{\{s_i\}} e^{-{1\over g^2(T)}H[\{s_i\}]}\;.
  \label{2.57}
\end{equation}
Here $s_i\in Z(N)$, the sum extends over all $Z(N)$-spin
configurations, $\Omega$ is the thermodynamic potential.

\paragraph*{Effective vs physical temperatures}
In the original action of Eq.\ (\ref{2.52}) $g$ parametrizes the
interaction strength.  When the time dependence of the field is
dropped, the $\int_0^\beta dt$ can be performed to yield a prefactor
of the remaining 3-dim action $(-\beta/g^2)$.  The $\beta$-factor can
be absorbed by a rescaling of the fields such that the action in three
dimensions takes the same form as the original action in four
dimensions. Now the prefactor plays the role of an inverse temperature
of the 3-dim model. Thus $g^2$ is identified with $T_{eff}$, the
effective temperature of a classical spin system described by the
partition function of Eq.\ (\ref{2.57}). For $T_{eff}$ small, the
`spin system' is in the ordered phase, we have $\left<L\right>\neq 0$
corresponding to deconfinement.  Deconfinement is realized at high
physical temperatures $T$, for which $g$ is small due to asymptotic
freedom. A small value of $g$ means a small value of $T_{eff}$, which
is consistent with the initial assumption. Thus a low effective
temperature $T_{eff}$ corresponds to a high physical temperature $T$
and vice versa. This explains, why the order parameter vanishes in the
low $T$-phase and signals ``order'' in the high $T$-phase.

The coupling $g$ becomes manifestly temperature dependent, when high
frequency contributions are integrated out in passing from the $3+1$-
to the 3-dimensional theory.

\paragraph*{The phase structure of $Z(N)$-spin models}
The phase structure of $Z(N)$-spin models has been studied in
statistical physics for various values of $N$ and space dimension
$d-1$. If a second order transition is predicted in the spin model,
one could attempt to locate the renormalization group fixed point and
consider the simpler spin model as fixed point theory in the
universality class of the $SU(N)$-theory.

A table for various $Z(N)$-, $SU(N)$- and $U(1)$-models in dimensions
$d=2,3$ and $\geq 4$ can be found in (Svetitsky and Yaffe, 1982b).
Here we mention the cases of two and three colors.

The spin model, which is associated with a $(3+1)$-dimensional $SU(2)$
gauge theory at high temperatures, is the Ising model in three
dimensions. The 3-d-Ising model is known to have a second order phase
transition.

The case of $N_c=3$ in $d=3$ is particular. In the space of
$3d$-$Z(3)$- symmetric theories no IR-stable renormalization group
fixed point is known. A specific realization of a $Z(3)$-symmetric
spin theory is the 3-state Potts model (cf. section \ref{sec:II.A}),
which is known to have a first order transition. The famous cubic term
on the classical level, driving the transition to first order, is
allowed by the $Z(3)$-symmetry. The potential of a $Z(3)$-symmetric
theory of a single complex scalar field $L(\vec{x})$ may be written as
a sum of a $U(1)$-symmetric term depending on $|L|^2$ and a term
depending on $\hbox{Re }L^N$. In $d=3$ a $U(1)$ gauge theory has a
second order transition (see e.g. Pfeuty and Toulouse, 1977). A term
$\hbox{Re }[L(\vec{x})^3]$ explicitly breaks the $U(1)$-symmetry down
to $Z(3)$. In the renormalization group sense it is relevant enough to
affect the critical behavior of the $U(1)$-theory.

Thus the conjecture of Svetitsky and Yaffe is that in the absence of
an IR stable fixed point the original (3+1)-dimensional $SU(3)$ gauge
theory has a first order finite temperature phase transition as well.
The prediction has been confirmed in early Monte Carlo calculations
(see e.g. Celik et al., 1983a, 1983b), was doubted in the meantime
(Bacilieri et al., 1988, see also the introduction and section
\ref{sec:III.B} below) and has been reestablished in more recent
calculations by Fukugita et al.~(1989). Before we report on this
controversy in section \ref{sec:III.B.1}, let us consider the
influence of dynamical quarks on the deconfinement transition.

\paragraph*{Inclusion of dynamical quarks}
From spin systems we know that a first order
transition stays first order, if the magnetic field is sufficiently
small. The gaps in thermodynamic quantities are only continuously
deformed for a perturbatively small external field. The magnetization,
however, ceases to be a good order parameter, as it is always finite
due to the presence of an external background field.

Similarly the Wilson line may be expected to fail as an order
parameter when dynamical quarks are included. The physical reason is
easily understood. The free energy of an isolated test quark does no
longer diverge ($\left<L\right>\neq 0$ for all temperatures).  When
the flux tube between two test quarks is sufficiently stretched, a
$q\bar{q}$-pair is popped out of the vacuum. The test quarks may
always form finite energy bound states with dynamical quarks.

Since the deconfinement transition was supposed to be driven by the
spontaneous symmetry breaking of the global $Z(3)$-symmetry, we
consider the invariance of the QCD action under $Z(3)$-transformations
in the presence of dynamical quarks. The generalized gauge
transformations Eq. (\ref{2.53}) are periodic only up to an element of
the center of the gauge group. While the periodic boundary conditions
on the gauge field are preserved under the generalized
transformations, the antiperiodic boundary conditions on the quark
fields are not. Imposing
\begin{equation}
  \psi(\vec{x},\beta)=-\psi(\vec{x},0)
  \label{2.58}
\end{equation}
on the quark fields as before, Eq.\ (\ref{2.58}) transforms under a
generalized gauge transformation with
$U(\vec{x},\beta)=c_N^iU(\vec{x},0)$ ($i$ labeling the center
elements) according to
\begin{eqnarray}
  \psi(\vec{x},\beta)\to U(\vec{x},\beta)\psi(\vec{x},\beta)
  =&&c^i_NU(\vec{x},0)\psi(\vec{x},\beta)
  \nonumber\\
  =&&-c^i_NU(\vec{x},0)\psi(\vec{x},0)\;.
  \label{2.59}
\end{eqnarray}
Eq.\ (\ref{2.59}) shows that the gauge transformed fields $U^.\psi$
(under the generalized gauge transformation) do no longer satisfy
antiperiodic boundary conditions. If we compare the path integral over
all gauge field configurations in the absence and presence of
dynamical quarks satisfying Eq.\ (\ref{2.58}), configurations
differing by generalized gauge transformations get the same Boltzmann
weight in the absence of quarks, but a different weight in their
presence. Hence the effect of dynamical quarks is an explicit symmetry
breaking of the $Z(N)$-symmetry. Its strength is determined by the
values of the quark masses.

\paragraph*{Finite quark masses and external fields}
The analogy between an external field in a spin system and dynamical
quarks in QCD becomes manifest, when the quarks are integrated out.
The integration over quark degrees of freedom induces an external
field on the effective level of spin models (Banks and Ukawa, 1983;
DeGrand and DeTar, 1983). We refer to the work by DeGrand and DeTar
(1983) and anticipate results from lattice gauge theory.

Consider a 4-d-$SU(3)$-lattice gauge theory with fermions at high
temperatures. The original action consists of a gauge field part $S_g$
and a fermionic part $S_F$. The fermionic part may be written as
$S_F(U)=\sum\bar{\psi}M(U)\psi$ (dropping all indices, $U$ is the
$SU(3)$ gauge field and $M$ the fermion matrix, explicit expressions
will be given later). When the Grassmann variables are integrated out,
the effective action takes the form
\begin{equation}
  S_{eff}[U]=S_g[U]+\hbox{Tr }\ln M[U]\;.
  \label{2.60}
\end{equation}
In the high temperature- and strong coupling limit of the
$SU(3)$ gauge theory it can be shown that the gauge part $S_g$
transforms into the action of a 3d- 3-state-Potts model. (For the
Potts model cf. section \ref{sec:II.A}.) The fermionic part of Eq.\ 
(\ref{2.60}) simplifies, when it is treated in a hopping parameter
expansion.  The hopping parameter $\kappa$ may be related to the bare
quark mass $m$ of the original action according to
\begin{equation}
  \kappa\sim{1\over2}e^{-m\tau}\;,
  \label{2.61}
\end{equation}
where $m^.\tau$ is the bare quark mass in lattice units, $\tau$ being
the lattice spacing in time direction. (Relation (\ref{2.61}) only
holds at strong coupling and for small values of $\kappa$). The result
for the fermionic term together with the simplified gauge part leads
to the following effective action (for zero chemical potential)
\begin{equation}
  S_{eff}=-\left[
    {\beta a\over \tau} \sum_{n,\mu} \hbox{Re } z_n^* z_{n+\hat\mu} 
    + h(\kappa) \sum_n \hbox{Re } z_n
  \right]\;.
  \label{2.62}
\end{equation}
Here $a$ is the lattice spacing in spatial directions. The sums go
over all sites $n$ and nearest neighbor pairs ($n$,$n+\hat\mu$),
$\hat\mu$ is the unit vector in $\mu$-th direction, $z_n$ are elements
of $Z(3)$. The very coupling $h$ is the external field, which is a
remnant of the dynamical fermions. The $\kappa$-dependence is
explicitly known and approximately given as
\begin{equation}
  h(\kappa)\sim24\kappa\;.
  \label{2.63}
\end{equation}
Together with Eq.\ (\ref{2.61}) we see that $h$ vanishes with $\kappa$
for infinitely heavy quark masses, in the limit of a pure gauge
theory. The strength of the external field grows with decreasing quark
mass. Hence sufficiently light quark masses will completely wash out
the first order transition in the 3-d-3-state-Potts model. DeGrand and
DeTar (1983) obtain for the critical strength of the magnetic field
$h_{cr}={2\over3}(\ln2-{2\over3})$.  The result has been obtained in a
mean-field analysis. It holds for a cubic lattice in three dimensions.
At the critical field (mass) the line of first order transitions
terminates at a second order critical point and disappears for larger
fields (smaller masses).

A particular virtue of the lattice ``derivation'' of the effective
action Eq.\ (\ref{2.62}) is that it allows a calculation of the
effective coupling in terms of parameters of the original action (here
$h=h(\kappa)$). In this aspect the effective Potts model differs from
an effective Lagrangian of chiral perturbation theory (see section
\ref{sec:IV.A.2}), where the Lagrangian parameters have to be fixed
from an experimental input. The price one has to pay for such a
derivation on the lattice are the involved approximations (strong
coupling and high temperature expansions). Strong coupling results on
the lattice must be extrapolated to continuum results over a long way
in coupling parameter space.  It remains to be shown that the
qualitative predictions of the effective $Z(N)$-Potts models survive
the continuum limit.

\bigskip\bigskip In summary of section \ref{sec:II.B}, the
renormalization group analysis of Pisarski and Wilczek, and of
Svetitsky and Yaffe refer to the idealized limits of vanishing or
infinite quark masses, respectively.  They have been performed in the
$SU(N_{f})\times SU(N_{f})$ linear sigma model and in
$Z(N)$-spin-models rather than directly in QCD.  Studies in these
limiting cases reveal a dependence of the order of QCD transitions on
the number of flavors and the number of colors. The case of physical
interest is included for three colors and two or three light flavors.
The outcome of the renormalization group analysis is the message that
one has to do hard work in the following sense.  Model calculations in
terms of a scalar field theory with an $N$-component order parameter
field would suffice to model QCD, if the transition were conjectured
to be of second order also in case of three light or, more
realistically, two light and one heavier flavor.  One would be free to
choose as simple a model as possible within the conjectured
universality class.  Since the deconfinement and the chiral
transitions are expected to be of first order (for three colors and
three massless flavors) (see e.g.  Karsch, 1990), the transition
depends on details of the dynamics. Here this is the dynamics of full
QCD.

Compared to the scale of $T_{c}$, two quarks (up, down) are light,
three are heavy (charm, bottom, top), but the strange quark happens to
be just of the order of $T_{c}$. In the thermodynamics of QCD the
quark masses play the same role as external magnetic fields in
temperature driven transitions of ferromagnets. From this analogy one
must expect that the effect of finite masses depend on their actual
values. Results in the idealized limit cases may even qualitatively
change under their influence. Thus a challenge for further
investigations of the phase structure of QCD is to find out, which of
the alternatives displayed in Fig. \ref{fig:2}(a)-(b) is realized.
Partly conjectural diagrams are shown in the $(m,T)$-plane, where $m$
stands for a generic current quark mass, and $N_{f}=3=N_{c}$ is
assumed. For comparison a $(H,T)$-diagram for a ferromagnet and a
$(p,T)$-diagram for a liquid/gas system are shown in Fig.s
\ref{fig:2}(c)-(d). For $m=0$, a first order chiral transition is
predicted at some temperature $T_{ch}$. The chiral transition
continues to be of first order, but in a weakened form, as long as the
mass is smaller than a critical value (corresponding to
$h_{ch}<h_{cr}$). For $m=\infty$ or $\kappa=0$ the first order
deconfinement transition occurs at some temperature $T_{d}$. It will
persist as long as $h(\kappa)\equiv h_{d}$ is smaller than some
$\tilde{h}_{cr}$ or $m>m_{cr}$. Both transitions may ``meet'' and
coincide for intermediate mass values (Fig. \ref{fig:2}(a)) or not
(Fig. \ref{fig:2}(b)). In the latter case the discontinuities
disappear completely. The dynamical quark masses are then too large
for the chiral transition and too small for the deconfinement
transition to persist. The chiral symmetric deconfining high
temperature region is then smoothly connected with the chiral symmetry
broken, confined, low temperature world of daily life. To date lattice
results suggest that Fig. \ref{fig:2}(b) gives a more realistic
description.  Lattice calculations are the topic of the next chapter.

%
%
\section{THE LATTICE APPROACH TO THE QCD TRANSITION}
\label{sec:III}
\subsection{A primer to lattice gauge theory}
\label{sec:III.A}
This section serves mainly to fix the notation of lattice gauge
theory. We list the basic definitions, summarize some tools which are
relevant to the phase transitions, and point out sources of errors for
misleading results. For pedagogical introductions we refer to the
literature (see e.g. (Kogut, 1983; Creutz et al., 1983) or more
recently (Rothe, 1992; Montvay and M\"unster, 1994)).

The partition function of an $SU(N)$ gauge theory interacting with
matter fields can be written in the Euclidean path integral
formulation as
\begin{equation}
  Z = \int{\cal D}A_\mu{\cal D}\psi{\cal D}\bar\psi
  \exp\left[-S(A_\mu,\bar\psi,\psi;g,m_i)\right] ,
  \label{3.1}
\end{equation}
where the action depends on the gauge fields $A_\mu$, quark fields
$\bar\psi$, $\psi$, the gauge coupling $g$ and the quark masses $m_i$.
The action is given by
\begin{equation}
  S = \int_0^\beta dt \int d^3x~
  {\cal L}(A_\mu,\bar\psi,\psi;g,m_i) ,
  \label{3.2}
\end{equation}
where ${\cal L}$ is the QCD Lagrangian (Eq.\ (\ref{2.46})) for $N_f$
flavors and $N_c$ colors. The chemical potential is set to zero. At
finite temperature the gauge- and matter fields have to satisfy the
following boundary conditions
\begin{eqnarray}
  A_\mu(\vec x, 0) &&= A_\mu(\vec x, \beta) \nonumber\\
  \label{3.3}
  \psi(\vec x, 0) &&= -\psi(\vec x, \beta)
  \qquad \rm{for~all}~\vec x,\mu \\
  \bar\psi(\vec x,0) &&= - \bar\psi(\vec x,\beta) \nonumber
\end{eqnarray}
One usually has periodic boundary conditions in the spatial directions
in computer simulations.  Note that the finite temperature enters only
via the boundary conditions, $\beta$ denotes the inverse temperature.

In finite temperature physics a typical task is to evaluate the
thermal expectation value of a physical observable $O$. The thermal
expectation value, defined as
\begin{equation}
  <O>~= {1 \over Z}~{\rm Tr}~e^{-\beta H} O
  \label{3.4}
\end{equation}
takes the following form in the path integral formulation
\begin{equation}
  <O>~=
  {{ \int_{pbc} {\cal D} A_\mu {\cal D}\psi
      {\cal D}{\bar\psi} O e^{-S} } \over
    { \int_{pbc} {\cal D}~A_\mu {\cal D}\psi 
      {\cal D}{\bar\psi} e^{-S} } }, 
  \label{3.5}
\end{equation}
where $pbc$ stands for the periodic (antiperiodic) boundary conditions
as specified in Eq.\ (\ref{3.3}). The functional integral (\ref{3.5})
is by itself not well defined. Many schemes may be used to perform
the functional integration in an approximate way.

One possibility, which is particularly suited in view of the
nonperturbative nature of the phase transition is the lattice
formulation. The (3+1)-dimensional spacetime continuum is discretized
commonly on a hypercubic lattice. The lattice provides a gauge
invariant regularization scheme. Lattice results may be extrapolated
to the continuum limit in the very end.  Lattice artifacts can be
controlled via renormalization group equations and should vanish in
the continuum limit. In practice it may be quite subtle and intricate
to disentangle lattice artifacts from continuum physics.

A gauge invariant lattice action is conveniently defined in terms of
link variables $U(x, \mu)\equiv U_x^\mu$, associated with a link
$(x,\mu)$ leaving site $x$ in direction $\mu$, and matter variables
$\psi(x)\equiv\psi_x$, $\bar\psi(x)\equiv\bar\psi_x$, associated with
the sites of the lattice. We keep the continuum notation for the
discrete sites $(\vec x, x_0)$. The link variables are elements of the
gauge group $SU(N)$ and replace the continuum gauge fields $A_\mu$. In
the continuum limit these are related as
\begin{equation}
  U_x^\mu =
  \exp\left(i~g~a \sum_{i=1}^8 \lambda^i A^i_\mu(x)\right) .
  \label{3.6}
\end{equation}
If the matter fields are fermions, as in the QCD action, $\psi$ and
$\bar\psi$ are Grassmann variables satisfying the usual
anti-commutation relations
\begin{equation}
  \left\{\psi(x),\psi(y)\right\}=0, \qquad
  \left\{\bar\psi(x),\bar\psi(y)\right\}=0 .
  \label{3.7}
\end{equation}
The matter fields carry spin, color and flavor indices (as indicated
in Eq.\ (\ref{2.46})).  The lattice discretized version of the
partition function then takes the form
\begin{eqnarray}
  Z = \int_{pbc} && \prod_{x,\mu} d U_x^\mu ~
  \prod_x d\psi(x) d\bar\psi(x) \nonumber\\
  \cdot\; && \exp\left[-S(U_x^\mu,\bar\psi_x,\psi_x;~
    g^2,m_i,a_\sigma/a_\tau)\right] .
  \label{3.8}
\end{eqnarray}
In contrast to the symbolic notation in Eq.\ (\ref{3.5}) the measure
${\cal D}U~{\cal D}\psi~{\cal D}\bar\psi \equiv \prod dU \prod d\psi
d\bar\psi$ has a well defined meaning in Eq.\ (\ref{3.8}). The
products extend over all link- and site variables of the lattice,
$dU_x^\mu$ refers to the (gauge invariant) Haar measure on the
$SU(N)$ gauge group, $d\psi~d\bar\psi$ is the usual measure over
Grassmann variables for each site $x$. The action depends on the gauge
coupling $g^2$ and the quark masses $m_i$ ($i=1\ldots N_f$) as before,
but has an additional dependence on the lattice spacings in spatial
($a_\sigma$) and temporal ($a_\tau$) directions.  The lattice spacing,
or equivalently, the bare coupling may be chosen independently in the
spatial and temporal directions. This is sometimes utilized in finite
temperature calculations of thermodynamical quantities.

\paragraph*{The lattice action}
Two requirements have to be satisfied by a lattice action.
The first one is the reproduction of the classical continuum limit. In
the limit of $a_{\sigma}=a_{\tau}\to 0$, the action must approach the
continuum form of Eq.\ (\ref{3.2}). The second one is local gauge
invariance, i.e. an invariance under local gauge transformations
$V(x)$, acting according to
\begin{eqnarray}
  U_x^\mu \to (U_x^\mu)'\; && = V(x)~U_x^\mu~V(x+\hat\mu) \nonumber\\
  \label{3.9}
  \\
  \psi(x) \to \psi'(x)\; && = V(x)~\psi(x)\;. \nonumber
\end{eqnarray}
(The second requirement is commonly used, but actually not necessary.
Actions with non-compact gauge groups recover local gauge
invariance only in the continuum limit.) These requirements are not
very restrictive. They leave us with a variety of possible lattice
formulations. Here we state only the most popular choices, which are
sufficient for what follows.

For the gauge part it is the Wilson form (Wilson, 1974), for the
fermionic part two choices are {\it Wilson fermions} (Wilson, 1974)
and {\it staggered or Kogut-Susskind fermions} (Kogut and Susskind,
1975; Susskind, 1977). We distinguish explicitly between spacelike and
timelike quantities, to make the appearance of temperature explicit.

The Wilson form for the gauge part is defined as
\[
S_g = {2 N\cdot a_\tau \over g^2_\sigma\cdot a_\sigma}
\sum_{\scriptstyle x \atop \mu<\nu<4} P_x^{\mu\nu} +
{2 N\cdot a_\sigma \over g^2_\tau\cdot a_\tau}
\sum_{\scriptstyle x \atop \mu<4} P_x^{\mu 4} ,
\]
\vskip-12pt
\noindent where
\vskip-12pt
\begin{equation}
  P_x^{\mu\nu} = 
  1-{1 \over N}~{\rm Re}~{\rm Tr}~U_x^\mu~
  U_{x+\hat\mu}^\nu~U_{x+\hat\mu}^{\mu+}~U_x^{\nu+}
  \label{3.10}
\end{equation}
contains a product of gauge field variables along the boundary of an
elementary plaquette of the lattice, and $N=3$ for $SU(3)$, ${\rm Tr}$
denotes the trace in color space.  \goodbreak The first term of the
action is a sum over plaquettes, which contain only spacelike links,
the second term involves also timelike links. It can be easily checked
that Wilson's formulation (\ref{3.10}) satisfies both requirements for
a lattice action.

To find an appropriate form for the fermionic part of the action turns
out to be more difficult. A naive translation of the continuum Dirac
action leads to a lattice action which actually describes $2^d(=16)$
species of fermions in the naive continuum limit rather than the
intended one species of the original continuum action.  ``No-Go''
theorems by Nielsen and Ninomiya (1981a, 1981b) explain the failure of
naive transcription and tell us that no lattice formulation of the
fermionic action exists that is satisfactory in all aspects. If one
insists on a local action, either continuous chiral symmetry is
completely lost on the lattice or one ends up with too many flavors.

We consider two popular choices, {\it Wilson fermions} and {\it
  staggered fermions}. The former choice gets rid of the species
doubling at the expense of breaking all continuous chiral symmetries
explicitly in the limit of $M\to 0$. The latter choice keeps a
$U(1)\times U(1)$ chiral symmetry for all lattice couplings, a welcome
feature in view of an investigation of chiral symmetry restoration.
The price is too many flavors in the continuum limit (although the
number is reduced with respect to the naive formulation) and a broken
flavor symmetry on the lattice. For derivations and details of lattice
fermions we refer to textbooks (e.g. Rothe, 1992; Montvay and
M\"unster, 1994) or reviews (Kogut, 1983; Creutz et al., 1983). Here
we summarize only the results.

\paragraph*{The Wilson action and hopping parameters}
In the Wilson action the bare masses $m_{i}$ are hidden in the hopping
parameters in spacelike ($\kappa_{\sigma}^{(i)}$) and timelike
($\kappa_{\tau}^{(i)}$) directions, one for each flavor $i$, where $i$
labels the $N_{f}$ species of flavors in the continuum limit.  For the
free gauge theory ($U_x^\mu=1$) and $a_\sigma=a_\tau$,
$\kappa_\sigma^{(i)}=\kappa_\tau^{(i)}=\kappa^{(i)}$, the relation to
the bare quark masses is
\begin{equation}
  {1 \over 2} (1/\kappa^{(i)}-1/\kappa_c) = \exp(m_i a)-1
  \label{3.11} 
\end{equation}
with $\kappa_c=1/8$. For massless quarks $\kappa^{(i)}=\kappa_c$. In
the interacting theory, the dependence of $\kappa_c$ on $g^2$ is not
known. This is an undesirable feature in Monte Carlo calculations,
when hopping parameter values should be translated to quark masses to
check the relevance of the results for realistic mass values.

While the flavor symmetry is well defined and conserved for all
lattice spacings by the Wilson action, all chiral symmetries are
explicitly broken even in the massless case ($m_i=0$ or
$\kappa^{(i)}=\kappa_c$). `All' chiral symmetries should be contrasted
with remnants of the full invariance under certain subgroups. Recall
our original intention to study the phase structure within the lattice
approach. A chiral symmetry restoration at finite temperature in the
massless limit should be signalled by a melting of the condensate
$\left\langle \bar\psi\psi\right\rangle$. Such a `melting' is
prejudiced right from the beginning, if an explicit chiral symmetry
breaking is involved as it is in the Wilson formulation.

\paragraph*{Staggered fermions and flavor symmetries}
This choice of lattice fermions is frequently used in studies of the
chiral phase transition. A careful inspection of the origin of the
species doubling in the naive formulation suggests the possibility of
eliminating the unwanted fermions by doubling the effective lattice
spacing. This amounts to a distribution of the fermionic degrees of
freedom over the original lattice in such a way that the effective
lattice spacing for each type of Grassmann variables is twice the
fundamental lattice spacing. It turns out that in $d$ dimensions
$2^{d/2}$ fermion fields are necessary to place a different fermionic
degree of freedom at each site of an elementary hypercube on the
lattice. This indicates the origin of the integer multiple of four
flavor degrees in the continuum for four dimensions, which is
attainable by a description with staggered fermions.

A realization of these ideas is rather involved. The sites of the
hypercube will be occupied by single-component spinors $\chi_f$,
$\bar\chi_f$, which may be multi-component in flavor space,
$f=1,\ldots,n$.  (This number $n$ should be distinguished from the
desired number $N_f$ of continuum flavors.)  The $\chi$'s and
$\bar\chi$'s are certain linear combinations of the original fields
$\psi$, $\bar\psi$.  In the end of a simulation in terms of $(\chi,
\bar\chi)$-fields, the results for the ($\psi, \bar\psi$)-fields have
to be reconstructed.

The action for staggered fermions in terms of $(\chi,
\bar\chi)$-fields is defined by
\[
  S_F \equiv \sum_{x,x'} \sum_{f=1}^n \bar\chi_f(x)~
  {\cal Q}_{x,x'}^f~\chi_f(x')
\]
\vskip-12pt
\noindent with
\vskip-12pt
\[
  {\cal Q}_{x,x'}^f = 
  {1 \over a_\sigma} \sum_{i=1}^3 D_{x,x'}^{(i)} +
  {1 \over a_\beta} D_{x,x'}^{(4)} + m_f~\delta_{x,x'}
\]
\vskip-12pt
\noindent and
\vskip-12pt
\begin{eqnarray}
  D_{x,x'}^{(\mu)} = {1\over 2}~\Gamma_\mu(x)
  \left[U_x^\mu\cdot\delta_{x,x'-\hat\mu} -
    U_{x'}^{\mu+}\cdot\delta_{x,x'+\hat\mu}\right]
  \label{3.12}
  \\
  \Gamma_\mu(x) = (-1)^{x_1+x_2+\ldots+x_{\mu-1}}\;. \nonumber
\end{eqnarray}
The only remnant of the Dirac structure is hidden in the phases
$\Gamma_\mu$.  The staggered fermion action depends on the quark
masses $ m_f$. The index $f$ labels the species of 'staggered'
flavors. Their number $n$ is often set to 1, since $n$ staggered
species on the lattice correspond to $N_f=4\cdot n$ flavors in the
continuum. The case $n=1$ comes closest to $N_f=3$ flavors, for which
QCD is approximately chiral invariant. More precisely, for $n=1$ and
$d=4$ dimensions one has 16 $\chi$'s in each elementary hypercube.  It
can be shown that these 16 $\chi$'s can be combined to define a quark
field with four flavors in the continuum, more generally with
$N_f=4\cdot n$ flavors. A non-zero lattice spacing breaks the flavor
symmetry between the $4n$ flavors, and it breaks the chiral
$U(4n)\times U(4n)$-symmetry of the continuum limit of the staggered
fermion action for $m_f=0$ down to $U(n)\times U(n)$.

It is the $U(1)\times U(1)$ remnant of the full chiral symmetry (that
is left for $n=1$ and arbitrary lattice spacing), which is the desired
feature for investigating chiral symmetry restoration.

Wilson and staggered fermions appear quadratically in the action. In
general, fermion fields can only occur quadratically in a
renormalizable four-dimensional field theory. Thus fermions are
usually integrated out. This seems to be the only tractable way in
numerical and analytical calculations, since the fermion fields are
Grassmann variables. The integrals are performed using the well-known
formulas of Berezin integration (Berezin, 1966). The result is
\begin{equation}
  Z = \int_{pbc} \prod_{x,\mu} dU_x^{\mu}~e^{-S_g}~
  {\det}^l({\cal Q}) ,
  \label{3.13}
\end{equation}
where ${\cal Q}$ is given by the action for Wilson fermions and by
Eq.\ (\ref{3.12}) for staggered fermions. The power $l$ equals to $N_f$,
the number of continuum flavors, for Wilson fermions, and $l$ equals
$n$ for staggered fermions (recall that $n=N_f/4$). For a positive
definite ${\cal Q}$ one may use that $\det {\cal Q}=\exp({\rm Tr}\ln
{\cal Q})$, and is led to
\[
  Z = \int_{pbc} \prod_{x,\mu} dU_x^{\mu}~e^{-S_{\rm eff}}
\]
\vskip-12pt
\noindent with
\vskip-12pt
\begin{equation}
  S_{\rm eff}(\{U_x^{\mu}\}) = S_g(\{U_x^{\mu}\})-l \cdot
  {\rm Tr} \ln \left[{\cal Q}(\{U_x^\mu\})\right] ,
  \label{3.14}
\end{equation}
where the trace extends over spin, color and flavor indices. Thus a
simulation of Eq.\ (\ref{3.14}) with $l=1$ describes $N_f=4-{\rm
  flavors}$ in the continuum limit, if staggered fermions are used for
${\cal Q}$. For Wilson fermions $\det {\cal Q}(U)$ is real and positive
for values of the hopping parameter less than $1/8$ (Seiler, 1982).

\paragraph*{Sources of errors} 
Also for staggered fermions (and vanishing chemical potential) $\det
{\cal Q}$ is positive, but its eigenvalues are not always positive.
This is an annoying feature for a representation in terms of
pseudofermions.  Therefore in actual simulations ($-l\cdot{\rm
  Tr}~{\cal Q}(_x^\mu)$) in the effective action of Eq.\ (\ref{3.14})
is replaced by ($-l\cdot{\rm Tr}\ln [\hat{\cal Q}\hat{\cal Q}^+(U)]$).
Consider first $\hat{\cal Q}={\cal Q}$. The term ${\rm Tr}\ln{\cal
  Q}{\cal Q}^+$ is proportional to $(\det{\cal Q})^2$, thus the
replacement induces a further doubling of flavor degrees of freedom
from $n$ to $2n$. To compensate for this doubling, the number of
degrees of freedom in ${\cal Q}$ is reduced by a factor of 2 by a
further doubling of the effective lattice spacing.  Let us call the
corresponding fermion operator $\hat{\cal Q}$. Due to an even-odd
symmetry of the determinant one has $\det{\cal Q} = \det\hat{\cal
  Q}_{\rm even} \cdot \det\hat{\cal Q}_{\rm odd}$. Thus ${\rm
  Tr}\ln{\cal Q}$ is actually simulated as ${\rm Tr}\ln\hat{\cal
  Q}\hat{\cal Q}^+$. This induces an error, which is not well under
control. Another uncontrolled error entering the formulation for
staggered fermions is due to flavor exchanging currents. The flavor
symmetry is violated for a finite lattice spacing. By claiming that
their contribution vanishes in the continuum limit a similar type of
error may be induced as when one says that the irrelevant additive
terms in the Wilson action vanish in the continuum limit. It is
well-known that just the irrelevant terms of the Wilson action are
essential for reproducing the right axial anomaly in the continuum
limit.

\paragraph*{Translation from lattice units to physical units}
A basic step in understanding the lattice approach is the translation
of lattice results into physical units. For illustration let us
consider the measurement of a mass.  Such a measurement on the lattice
typically yields a dimensionless number of the order of 1. The
dimensionless lattice mass $m_{\rm{latt}}$ is related to the physical
mass $m$ via the lattice spacing $a$. From simple dimensional
arguments we have
\begin{equation}
  m=m_{\rm{latt}}\;a^{-1}
  \label{3.15}
\end{equation}
in units where $\hbar\cdot c=1$. (In these units $100~[\rm{MeV}]\sim
3.2~[\rm{fm}^{-1}]$.) In other words, the lattice mass $m_{\rm{latt}}$
is measured in units of $a^{-1}$. Other variables are obtained
similarly, energy densities in units of $a^{-4}$, etc.

When a lattice mass is interpreted as $m\cdot a$ and $m$~[MeV] is
known from experiments, Eq.\ (\ref{3.15}) gives ``the'' lattice
spacing in physical units. More precisely it gives
$a(g)~[\rm{MeV}^{-1}]$ at coupling $g$, if $g$ stands for the bare
input parameter(s) of the lattice Lagrangian, which have been used in
the measurement of $m_{\rm{latt}}$. In a pure gauge theory $g$ is the
bare gauge coupling. The lattice spacing which is obtained this way is
only unique, i.e. independent of the choice of the physical input
$m$~[MeV], if one is in (or close to) the continuum limit.

Let us assume we have determined $a~[\rm{MeV}^{-1}]$ or $a [\rm{fm}]$
from a first mass measurement as indicated above and measure a second
mass $\tilde{m}_{\rm{latt}}$ in lattice units. The physical value
$\tilde{m}$~[MeV] is then predicted from the lattice simulation and
can be compared with the experimental value. Furthermore, once $a(g)$
is known as function of $g$ in units of $\rm{MeV}^{-1}$ or fm, it
makes sense to associate strong couplings ($g\gg 1$) with coarse
grained lattices (say $a>1$ fm), and weak couplings with a fine
grain size close to the continuum description. The very existence of
such a universal mapping between bare couplings and lattice spacings
in physical units is based on the renormalization group equation. In
case of an $SU(N_{c})$ gauge theory with $N_{c}$ colors and $N_{f}$
massless flavors the renormalization group equation relates $g^2$ and
$a$ in the continuum limit (for perturbatively small couplings $g^2\to
0$) according to
\begin{eqnarray}
  a~\Lambda_L && = (b_0g^2)^{-b_1/2b_0^2}~e^{-1/2b_0g^2} , \nonumber\\
  b_0 && = {1\over 16\pi^2} 
  \left[11{N_c\over 3}-{2\over 3} N_f \right]
  \label{3.16}
  \\
  b_1 && = \left({1\over 16\pi^2}\right)^2 
  \left[{34\over 3} N_c^2 - \left({10\over 3}N_c+
      {N_c^2-1 \over N_c}\right)N_f\right] . \nonumber
\end{eqnarray}
If lattice calculations are performed at sufficiently small values of
$g$, an observable in lattice units should scale as a function of $g$
in a way, which is determined by $a(g)$ according to Eq.\ 
(\ref{3.16}). For a mass in lattice units this implies
\begin{equation}
  (m\cdot a)(g) = m\cdot(a(g))\;.
  \label{3.17}
\end{equation}
If such a scaling behavior is observed, the asymptotic scaling regime
has been reached.  The lattice spacing can then be replaced by the
scale parameter $\Lambda_{\rm latt}$, which represents the only
parameter in QCD with $N_{f}$ massless flavors, and has to be fixed
from experiment.

The physical volume $V$ and temperature $T$ are given in terms of the
lattice spacing as
\begin{equation}
  V = N_\sigma^3a_\sigma^3 , \quad
  {1\over \beta} \equiv T = {1\over N_\tau a_\tau} .
  \label{3.18}
\end{equation}
For $a_\tau=a_\sigma=a$, the volume and temperature may be converted
in units of fm, if $a(g)$ is taken from Eq.\ (\ref{3.16}).

For $a_\tau\neq a_\sigma$, the relations $a_\tau(g_\sigma,g_\tau)$ and
$a_\sigma(g_\sigma,g_\tau)$ can be determined in the case of a pure
gauge theory (Eq.\ (\ref{3.10})) as follows. We use the notations
\begin{equation}
  {2N \over g_\sigma} \equiv \beta_\sigma ,~
  {2N \over g_\tau} \equiv \beta_\tau ,~
  \xi \equiv {a_\sigma\over a_\tau} \equiv 
  {a\over a_\tau},~
  \beta_g \equiv {2N \over g^2} ,
  \label{3.19}
\end{equation}
where $g_\sigma$, $g_\tau$ have been introduced in Eq.\ (\ref{3.10})
and $\beta_g$ denotes the bare coupling on an isotropic Euclidean
lattice.  Here $\beta_g$ is large when $g$ is small like the physical
temperature (cf. the remarks in section \ref{sec:II.B.2}). Thus one
may also think of $\beta_{g}$ as some kind of {\it effective
  temperature} (rather than effective {\it inverse} temperature as the
notation suggests). The two relations which replace $g(a)$ of the
isotropic case are written as
\begin{eqnarray}
  \beta_\sigma &= \xi^{-1}\beta~[1+c_\sigma(\xi)g^2+O(g^4)]
  \nonumber\\
  \beta_\tau &= \xi\beta~[1+c_\tau(\xi)g^2+O(g^4)]\;.
  \label{3.20}
\end{eqnarray}
The coefficients $c_\sigma$ and $c_\tau$ have been perturbatively
determined by Karsch (1982).  With $\xi=a/a_\tau$ and $g=g(a)$ from
Eq.\ (\ref{3.16}), Eq.\ (\ref{3.20}) can be solved for $a=a_\sigma$
and $a_\tau$ in terms of $\beta_\sigma$ and $\beta_\tau$, the
couplings of the pure gauge action. The relations
$a_\sigma(\beta_\sigma,\beta_\tau)$ and
$a_\tau(\beta_\sigma,\beta_\tau)$ replace $a(g)$ on an anisotropic
lattice in thermodynamic calculations.

\paragraph*{The critical temperature $T_c$}
In QCD the critical temperature $T_{c}$ depends on the number of
colors ($N_{c}$), the number of flavors ($N_{f}$), the current quarks
masses $m_{q}$, and the volume $V$. The finite volume induces a shift
of $T_{c}$ of the order of $1/V$, if $V$ denotes the $d$-dimensional
volume of the system. The finite light quark masses give an effect of
a few percent compared to the value of $T_{c}$ in the chiral limit.
The strongest dependence comes from the number of flavors. $T_{c}$
varies about 100~MeV between $T_{c}\sim 150$~MeV for two light flavors
and $T_{c} (N_{f}=0)\sim 260$~MeV for a pure gauge theory (DeTar,
1995). This is easily understood in a percolation picture. The
transition is assumed to occur when a critical hadron density is
reached where the hadrons start overlapping which we have briefly
mentioned in the introduction. A much lower temperature is then needed
for creating a critical hadron density out of light pions with a mass
of the order of $T_{c}$ than out of heavy glueball states. The
lightest states in a pure gauge theory are glueballs with a mass of
the order of $5 T_{c}$.

For a smooth crossover phenomenon $T_{c}$ is no longer defined.
However, if there is a narrow temperature interval with rapid changes
in thermodynamic quantities, it makes still some sense to associate a
`{\it pseudocritical}' or {\it `crossover' temperature} $T_{c}$ with
the rapid crossover region. More precisely, the crossover temperature
is defined as the temperature of maximum change in an observable (e.g.
the chiral condensate). Alternatively it is the temperature of the
peak in some susceptibility. Both pseudocritical values for $T_{c}$
must only agree, if the susceptibility is the derivative of the
particular observable whose maximum change defines $T_{c}$.

\paragraph*{A test of asymptotic scaling}
Usually one measures the critical coupling $g_c$ rather than the
critical temperature $T_c$. Order parameters are plotted as functions
of $g$ (or $\beta=1/(6g^2)$) to show their behavior as a function of
$T$. For simplicity let us consider an isotropic lattice, i.e.
$a_\sigma=a_\tau=a$, and the limit of a pure gauge theory with bare
coupling $g$. For a given number $N_\tau$ of lattices sites in the
timelike direction the temperature may be varied by tuning $a$ via $g$
according to Eq.\ (\ref{3.18}). From a physical point of view it is
not surprising that the temperature may be implicitly varied via the
coupling, as both are related through asymptotic freedom. Starting in
the strong coupling region $g\gg 1$ and lowering $g$, the transition
from the confinement to the deconfinement phase will be encountered at
a certain coupling $g_c$. Instead $g$ can be kept fixed, but $N_\tau$
varied from smaller to larger values (say from $N_\tau=2$ to
$N_\tau=8$). This way the transition region is passed from the
deconfinement to the confinement phase, if $g$ and $N_\tau$ are in an
appropriate range.

One would like to simulate the lattice system for large values of
$N_\tau$ to keep the finite size effects small. For large $N_\tau$ a
rather small coupling is necessary to reach the high temperature
phase. The larger $N_\tau$, the smaller ist the critical coupling
$g_c$, which corresponds to the transition temperature $T_c$. In
practice, the computer time rises rapidly, if $g$ is small. This
explains why typical extensions in time direction are limited to
$N_\tau=$ 6, 8 or 12, when fermions are included.

The effect of fermions is a further reduction of the effective lattice
spacing $a_\tau$ at the same coupling $g$ compared to the pure gauge
case. Even larger values of $N_\tau$ are needed to reach the
transition region.  Small values of $N_\tau$ require stronger
couplings $g$ in the presence of fermions. One should keep in mind
that one is far outside the asymptotic scaling region, if dynamical
fermions are simulated at small $N_\tau$.

\vskip12pt A {\it test of asymptotic scaling} is the very first check,
whether the measured critical coupling $g_c$ has some relevance for a
critical temperature in the continuum limit. When the temporal size
$N_\tau$ is increased, the transition should occur at a smaller
lattice spacing $a(g_c)$ such that $T_c=1/(N_\tau{}a(g_c))$ stays
constant. For small couplings $g$, the relation $g_c(N_\tau)$ should
scale as
\begin{equation}
  T_c/\Lambda_L = N_\tau^{-1}(b_0 g_c^2)^{b_1/2 b_0^2}
  e^{1/2 b_0 g_c^2}\;.
  \label{3.21}
\end{equation}
Eq.\ (\ref{3.21}) follows from Eqs.\ (\ref{3.16}) and (\ref{3.18}).
Calculations for an $SU(2)$- and $SU(3)$ gauge theory have been
performed up to $N_\tau=16$. A measurement of $g_c$ shows strong
violations of asymptotic scaling according to Eq.\ (\ref{3.21}) (see
e.g.  Karsch, 1992a).

Asymptotic scaling violations have been also observed in other
observables. They seem to be universal. Thus it should be possible to
absorb the scaling violations in a `renormalization' of the bare
coupling to an effective coupling which accounts for rapid
fluctuations in the action in an intermediate coupling range. This
idea turned out to be successful (Altmeyer et al., 1992). As a
function of an effective coupling $T_{c}/\Lambda_{\overline{MS}}$
depends much weaker on $N_{\tau}$. An extrapolation to the continuum
limit seems to be justified leading to (Karsch, 1992a)
\begin{equation}
  {T_{c} \over \Lambda_{\overline{MS}}} =
  \left\{
    \matrix{
      1.23\pm 0.11,~SU(2)~\rm{gauge~theory} \cr
      1.03\pm 0.19,~SU(3)~\rm{gauge~theory}
      }
  \right.
  \label{3.22}
\end{equation}

\paragraph*{Translation to physical units}
A second check whether the critical temperature on the lattice has
some relevance for the continuum limit is a translation to physical
units. The outcome should be independent of the choice of the
experimental input. In the pure $SU(3)$ gauge theory $T_{c}$ is
approximately independent of the physical input. It is 239(13)~MeV
from the string tension $\sqrt{\sigma}=420\pm 20$~MeV, 239(23)~MeV
from the $\rho$-mass and 225(30)~MeV from the nucleon mass (Karsch,
1992a).

For QCD with two light flavors in the staggered fermion formulation
the {\it crossover temperature} is estimated as 140-160~MeV (DeTar,
1995). The experimental input comes from the $\rho$-mass with 770~MeV.
The critical temperature is then determined as
\begin{equation}
  \label{3.23}
  T_{c}~[\rm{MeV}]={770~[\rm{MeV}] \over
    (m_{\rho}\cdot a)(6/g_{c}^{2},m_{q}a)\cdot N_{\tau}} .
\end{equation}
The $\rho$-mass in lattice units $m_{\rho}\cdot a$ (depending on the
bare gauge coupling $g_{c}$ and the bare quark masses $m_{q}$) is
calculated as a function of $g_{c}$ and $m_{q}\cdot a$ from a fit to
several zero temperature simulations (see DeTar, 1995, and references 
therein). Note that in Eq.\ (\ref{3.23})
the $\rho$-mass [MeV] is kept fixed at its physical value of 770~MeV,
while the bare quark masses $m_{q}\cdot a$ are allowed to vary to
unphysical values leading to unphysical $\rho$-masses as well. As
emphasized by DeTar (1995), one should further keep in mind that the
nucleon to rho mass ratio is approximately 20 \% above its physical
value over the range of lattice parameters, where $m_{\rho}\cdot a$ is
measured. Also the $m_{\pi}/m_{\rho}$ ratio comes out twice to three
times the experimental value, because the bare quark masses are still
too large. Thus the $\rho$-mass is not (yet) a perfect candidate for
the conversion from lattice units into physical units.

In earlier lattice simulations thermodynamic quantities have been
plotted as function of $6/g^{2}$. Nowadays one often finds plots vs
$T$~[MeV]. The translation should be taken with care, as the mapping
between $6/g^{2}$ and $T$ is not (yet) unique.

\paragraph*{Numerical simulations}
So far we have specified the measure, boundary conditions
and lattice action for the path integral of Eq.\ (\ref{3.8}). We have
sketched, how numbers from the lattice can be translated into physical
units, and what a criterion on the relevance of lattice results for
continuum physics looks like. Numerical and analytical methods can be
utilized to attack the functional integrations of Eq.\ (\ref{3.8}).
Analytical methods are usually applicable in limits where small
expansion parameters are available. The {\it hopping parameter}
expansion is an expansion in small values of $\kappa$, it is an
expansion around the limit of infinite quark masses. The {\it strong
  coupling} expansion applies for large values of the bare gauge
coupling $g$.  As mentioned above, it is intrinsically difficult to
find a small expansion parameter in the transition region, where $g$
is neither weak nor strong. Thus it is not surprising that many
results on the phase structure in the vicinity of $T_c$ are based on
{\it numerical} simulations.  The most important approach is the Monte
Carlo method.

\vskip12pt\noindent
{\it Monte Carlo on the lattice.} In a Monte Carlo procedure a set of
field configurations $\{U^\mu_x\}^{(1)}\ldots\{U^\mu_x\}^{(N)}$ is
generated such that the Boltzmann factor is absorbed in the selection
of configurations. For a positive definite and real action
$S[\{U^\mu_x\}]$ the expectation value of an observable $O$ defined as
\begin{equation}
  \langle O\rangle=
  {\displaystyle\int\prod_{x,\mu}dU^\mu_x\,e^{-\beta S[\{U^\mu_x\}]}
    O(\{U^\mu_x\})
    \over
    \displaystyle\int\prod_{x,\mu}dU^\mu_x\,e^{-\beta S[\{U^\mu_x\}]}}
  \label{3.24}
\end{equation}
(a lattice version of Eq.\ (\ref{3.5}) for a pure gauge theory) equals
to
\begin{equation}
  \langle O\rangle
  =\lim_{N\to\infty}{1\over N}\sum^N_{i=1}O(\{U^\mu_x\}^i) ,
  \label{3.25}
\end{equation}
and is approximated by the arithmetic average over a finite number of
configurations
\begin{equation}
  \langle O\rangle_N\approx{1\over N}\sum^N_{i=1}O(\{U^\mu_x\}^i)\;. 
  \label{3.26}
\end{equation}
The observables in Eq.\ (\ref{3.26}) are evaluated on an ensemble of
configurations which are representative of the coupling (temperature)
$\beta^{-1}$. The configurations are selected according to the
probability distribution
\begin{equation}
  P(U){\cal D}(U)\equiv{e^{-\beta S[U]}\over Z}{\cal D}U\;,
  \label{3.27}
\end{equation}
where ${\cal D}U$ and $Z$ are short hand notations of the expressions
in Eq.\ (\ref{3.24}). The fundamental idea of the Monte Carlo method
is stated by the law of large numbers, which tells us that under very
general conditions on the probability measure
$\lim_{N\to\infty}\langle O\rangle_N=\langle O\rangle$.

The sequence of configurations is generated with a Markov process. An
easy way of constructing a Markov process, which has a particular
distribution $P$ as its fixed point, is to choose the transition
probabilities $Q:(\{U^\mu_x\}^{(i)}\longrightarrow\{U^\mu_x\}^{(f)})$
from an initial configuration $\{U^\mu_x\}^i$ to a final configuration
$\{U^\mu_x\}^f$ to satisfy the detailed balance condition
\begin{eqnarray}
  P[U^{(i)}]&&Q\left(U^{(i)}\longrightarrow U^{(f)}\right){\cal D}U
  \nonumber\\
  &&= P[U^{(f)}]Q\left(U^{(f)}\longrightarrow U^{(i)}\right){\cal D}U\;.
  \label{3.28}
\end{eqnarray}
The fixed point distribution is in our application the Boltzmann
equilibrium distribution. A particular way of realizing the detailed
balance condition is the Metropolis algorithm (Metropolis et al.,
1953).

The Metropolis algorithm is an example of a local updating procedure.
One starts from an initial configuration, which may be chosen ordered
($U^\mu_x=1\hskip-0.6ex{\rm l}$ for all links (cold start)) or
disordered ($U^\mu_x\in SU(N)$ random for all links (hot start)) or
mixed. A single new link variable ${U^\mu_x}'$ is then chosen randomly
and always accepted as a replacement for the old variable, if it
lowers the action, i.e. $\Delta S=S(\{U\}')-S(\{U\})<0$. Otherwise the
change is accepted with a conditional probability. For $\Delta S>0$ a
random number $r$ with $0\leq r\leq1$ is selected.  If $r<e^{-\Delta
  S}$, the new variable $U'$ is still accepted for replacing the old
one, otherwise not, one goes back to $\{U^\mu_x\}$ and repeats the
steps as indicated above.

This way all links of the lattice are changed, either randomly or
successively. Such a sweep through the entire lattice is counted as
one Monte Carlo iteration. Usually hundreds or thousands of such
iterations are necessary, before the equilibrium distribution is
reached. The expectation value of the observable is then obtained
according to Eq.\ (\ref{3.26}), where the sum extends only over the
last $N$ equilibrated (or thermalized) configurations.

When fermions are included, one comes across a technical problem due
to the nonlocality of the effective action Eq.\ (\ref{3.14}). The
nonlocality refers to the det-term or equivalently to the
Tr~ln~$Q$-term. For staggered fermions the matrix $Q$ has
$(3N_\sigma^3{}^.N_\beta)^2$ complex elements. In the local Metropolis
algorithm this matrix should be calculated for every link in every
iteration, which renders it impracticable.

\vskip12pt {\it The hybrid Monte Carlo algorithm.} In the last
decade much effort has been invested in improving algorithms for QCD
with dynamical quarks (for a review see e.g. the lectures by Toussaint
(1988) or Herrmann and Karsch, 1991).  The integration over the
fermions had led to the factor ${\det}^l({\cal Q})$ in Eq.\ 
(\ref{3.13}). While ${\cal Q}$ has been a local operator, coupling
only nearest neighbors on the lattice, $\det{\cal Q}$ is nonlocal. A
calculation of $\det{\cal Q}$ with $\exp\{-S_{g}\}$ as Boltzmann
factor should be avoided. 
The idea is to bosonize $\det{\cal Q}$ and
to ``lift'' ${\cal Q}$ in the exponent.  The bosonization is performed
with the {\it pseudofermion method} of Petcher and Weingarten (1981).

{\it The pseudofermion method} (Petcher and Weingarten, 1981) is based
on the following formula for the determinant
\begin{equation}
  \hbox{det }Q={1\over\hbox{det }Q^{-1}}=
  \int{\cal D}\lambda^*\,{\cal D}\lambda\,e^{-\sum_{x,y}\lambda^*_xQ^{-
      1}_{xy}\lambda_y}\;.
  \label{3.29}
\end{equation}
Here $\lambda,\lambda^*$ are complex bosonic (pseudofermionic)
variables. Note that $Q$ has to be a positive definite matrix for the
Gaussian integral in Eq.\ (\ref{3.29}) to converge. If we identify $Q$
with the fermionic matrix of Eq.\ (\ref{3.13}), $Q$ has negative
eigenvalues. This is the reason, why $Q$ is replaced by $Q^+Q$ at the
cost of doubling the number of fermionic flavors or (for compensating
this doubling) by $\hat Q^+\hat Q$, where $\chi$-fields live on even
and $\bar\chi$-fields on odd sites. (Another reason for the
replacement is that it is convenient for easily generating the
pseudofermion fields from a heat-bath.) Via the doubling ($Q\to QQ^+$)
configurations leading to negative eigenvalues of $Q$ are mapped on
the same contribution to the effective action as those with positive
eigenvalues, although $\ln\;\hbox{det }Q$ in the original effective
action is not given as Tr~$\ln Q$ for negative eigenvalues. Thus the
error entering this way is difficult to control.

Like $\det {\cal Q}$ the action in Eq.\ (\ref{3.29}) is nonlocal as
well. Hence we see, when the bosonic part
(${-\sum_{x,y}\lambda^*_xQ^{-1}_{xy}\lambda_y}$) is added to the gauge
part of the QCD action to account for fermions, we need a nonlocal
updating procedure. A practicable method, which is free of
uncontrolled errors and nowadays often used in QCD applications, is
the {\it hybrid Monte Carlo algorithm} (Duane et al., 1987). It has to
be combined with the pseudofermion representation to yield a feasible
simulation of QCD. The hybrid Monte Carlo algorithm is a stochastic
hybrid of a Langevin and a microcanonical algorithm, where an
additional Monte Carlo acceptance test has been built in to remove the
systematic errors of the stochastic hybrid. The stochastic hybrid
algorithm combines the virtues of the Langevin and the microcanonical
(or molecular dynamics) algorithms.  Both algorithms use equations of
motion to select new variables.

In the {\it Langevin algorithm} (Parisi and Wu, 1981; Fukugita and
Ukawa, 1985; Batrouni et al., 1985) the equation of motion is a
partial differential equation, which is of first order in time. The
finite step size in time entering the discretized version has to be
extrapolated to zero by varying its size. This is time consuming and
an unwanted feature.

{\it The microcanonical algorithm} (Callaway and Rahman, 1982 and
1983) is based on the observation that the Euclidean path integral of
a quantum field theory in four dimensions can be written as a
partition function for a system of classical statistical mechanics in
4 spatial dimensions with a canonical Hamiltonian that governs the
dynamics in a fifth new `time' variable. This time $\tau$ may be
identified with the simulation time. The algorithm determines the new
variables $\Phi_x(\tau)$ and their canonically conjugate momenta
$\pi_x(\tau)$ in a fully deterministic way. Here one moves faster
through phase space $(\phi,\pi)\to (\phi',\pi')\to\ldots$, but on the
price that ergodicity is lost.

{\it The stochastic hybrid algorithm} (Duane, 1985) interrupts the
integration of Hamilton's equation of motion along a single trajectory
with fixed energy. The interruption is made once in a while for a
refreshment of momenta. The new parameter, which enters and should be
optimized is the frequency of momenta refreshment.  If it is low, the
algorithm is as slow as the Langevin algorithm, if it is too high,
ergodicity will be violated. Yet there is a systematic error in the
hybrid algorithm introduced by the finite time step in the integration
procedure, which leads to a violation of energy conservation.  

The idea of the {\it hybrid Monte Carlo algorithm} is to absorb this
energy violation in a `superimposed' Monte Carlo procedure. 
The phase space configurations at the end of every
molecular dynamics chain are subjected a Metropolis acceptance test.
If the time step is too large, the energy violation too strong, the
$(\Phi,\pi)$ configuration is likely to be rejected. In the case the
energy is conserved, the configuration is always accepted. This way
the systematic error due to finite integration steps is eliminated.

\vskip12pt Above we have sketched a number of intermediate steps in
developing the hybrid Monte Carlo algorithm. It may have indicated,
how complex the problem of finding efficient algorithms is. Although
the hybrid Monte Carlo algorithm moves the system fast through
configuration space, it is slow due to a matrix inversion,
which enters the equations of motion. The matrix inversion is usually
performed with the conjugate gradient method. The number of conjugent
gradient steps is proportional to $(m_f^{-1})$ and takes more than
90\% of the CPU-time. The hybrid Monte Carlo acceptance rate is
proportional to $(m_f^{-3/4})$ and the autocorrelation time $\propto$
$m_f^{-1}$ (Karsch, 1992). For example, one lattice update on a
$16^3\times8$ lattice with $m_f=0.01$ takes about 20 min on a
200~MFlop-machine. Here we see, why it is so difficult to work with
small bare quark masses. Quark masses should be as small as possible
to avoid an uncontrolled extrapolation to the chiral limit, but some
extrapolation is unavoidable. 

More generally we see, why Monte Carlo simulations of lattice QCD are
so time consuming, when dynamical fermions are included. It is
extremely hard to satisfy the following requirements simultaneously:
\begin{itemize}
\item The masses are small enough to guarantee a safe extrapolation to
  the chiral limit.
  
\item The volume is large enough to avoid finite size effects.
  
\item The bare coupling is small enough to be in the asymptotic
  scaling regime.
\end{itemize}

\subsection{Pure gauge theory}
\label{sec:III.B}
\subsubsection{The order of the $SU(3)$-deconfinement transition}
\label{sec:III.B.1}
The deconfinement transition in a pure $SU(3)$ gauge theory is
nowadays believed to be of first order. Over several years there has
been a controversial discussion in the literature. It was due to the
subsequent use of different criteria to determine the order from a
numerical analysis. In the background from statistical physics
(section \ref{sec:II}) we have distinguished between `naive' and
`refined' criteria. In this section we will illustrate their
application to the deconfinement transition.

\vskip12pt Early calculations of Kogut et al. (1983) and Celik et al.
(1983) found a strong first order deconfinement transition. The
conclusions were based on hysteresis effects, signals for coexisting
states, and abrupt quantitative changes in bulk thermodynamic
quantities like the internal energy density. One finds in fact an
explicit suggestion that hydrodynamic models of the quark gluon plasma
and models of hadronization should incorporate a ``hard, first-order
transition'' (Kogut et al., 1983).

These results have been questioned by the APE-group 
(Bacilieri et al., 1988) claiming to see strong 
indications for a second order deconfinement transition. 
The claim has been based on finding a correlation length 
that grows proportional to the lattice  size, as the size 
is increased. The APE-results led to a careful 
reinvestigation of the deconfinement transition in 
different aspects.

From section \ref{sec:II.B} we recall the renormalization group 
analysis of Svetitsky and Yaffe (1982 a and b). There it 
has been argued that an effective theory for an order 
parameter of a (3+1)-dimensional $SU(N)$ gauge theory 
shares the global symmetry with 3-dimensional $Z(N)$-spin 
models. In particular knowledge about $Z(3)$-spin models 
was exploited to predict a first order phase transition for 
the case of an $SU(3)$ gauge theory. The knowledge refers 
to the absence of an IR-stable fixed point in $Z(3)$-spin 
systems.

A possible second order phase transition in an 
$SU(3)$ gauge theory would cast some doubts on the 
universality arguments. While it was not difficult to 
`derive' the $Z(3)$-symmetry of the effective action in 
terms of the order parameter field, it was harder to 
argue in favor of the locality of the spin model. Thus 
the question of non-local interaction terms in the 
effective action has been addressed by Fukugita et al. 
(1989 and 1990a) and by Gavai et al. (1989) and Gavai and 
Karsch (1989) in subsequent papers.

In general the order of the transition does depend on the 
type of effective couplings. These couplings are 
generated, when the original degrees of freedom are 
partially integrated out. Non-local couplings could cause 
the $SU(3)$ transition to be of second order. This 
suspicion was supported by earlier results of Fucito and 
Vulpiani (1982) stating that already small 
anti-ferromagnetic next-to-nearest-neighbor (nnn-)
couplings in addition to the leading ferromagnetic 
nearest neighbor (nn-)coupling induce a second order 
transition in a 3d-$Z(3)$-Potts model. It was suggested 
that likewise a small anti-ferromagnetic nnn-coupling could 
induce a second order transition in the 
$SU(3)$-deconfinement transition.

Both transitions in variants of the $Z(3)$-Potts model and in the
original $SU(3)$ gauge theory have then been studied by applying
refined criteria compared to earlier calculations. Here we report in
some more detail on results by Gavai et al. (1989) on the $Z(3)$-Potts
model and by Fukugita et al. (1989) on the $SU(3)$ gauge theory. In
both models the first order nature of the transition has been verified
with a variety of criteria, and the results of the APE-group are
understood most likely as a misinterpretation of the measured data.

Let us recall from section \ref{sec:II} the distinction between naive
and refined criteria. Naive criteria are signs of metastabilities,
hysteresis effects, two-state-signals or jumps in thermodynamic
quantities. Typically all these signatures occur in a truly first
order transition. If we call them `naive', it refers to their
incautious use when the signatures are taken for granted from
measurements for a single lattice size. Early lattice simulations were
`naive' in this sense. Compared to nowadays standard they were also
performed on rather small lattices.

Refined criteria refer to a finite size scaling analysis of bulk and
`special' quantities. Examples for bulk quantities are the internal
energy density $\varepsilon$, the specific heat $c$ or the order
parameter susceptibility $\chi$. Finite size corrections to the
average internal energy are {\it exponentially} suppressed. The
shifting and rounding effects in the peaks of $c$ and $\chi$ are of
the order of $1/V$ in first order transitions (see section
\ref{sec:II}).

With `special quantities' we mean distribution functions of the order
parameter $P(O)$, and derived moments or cumulants like Binder's
cumulant, see Table~\ref{tab:2}. In general they have power-law finite
size corrections and are practically applicable for very large
volumes, where these power-law corrections are suppressed.

Similarly correlation lengths show {\it power-law} corrections. Even
their very definition in a finite volume is rather intricate compared
to bulk quantities.  Their practical utility for a finite size scaling
analysis turns out to be questionable, as we are going to show below.

\vskip12pt We turn now to concrete applications of both kinds of
criteria. Kogut et al. (1983) have measured large gaps in the internal
energies of gluons (and fermions in the quenched approximation). The
lattice size was $2\times 8^3$ and $4\times 8^3$. Celik et al. (1983)
have recorded two-state signals from hot and cold starts, indicating
two coexisting phases with different expectation values of the Wilson
line at the critical coupling. On lattices with $8^3$ and $10^3\times
2$ (or $\times 3$, 4) sites they have also found a large latent heat
as difference in the gluonic internal energies
($\Delta\varepsilon/T_c^4=3.75\pm 0.25$). Lattice expressions for the
latent heat and other thermodynamic quantities will be given in
section \ref{sec:III.B.2}.

Brown et al. (1988) have performed a concurrent study to Bacilieri et
al. (1988). They still find a first order deconfining transition, but
it is less strong with signs of an increasing correlation length. The
lattice volumes are enlarged to $16^3\times 4$ and~$16^3\times 6$ and
$24^3\times 4$ and~$24^3\times 6$. The values for the latent heat
(obtained from the gap in ($\varepsilon+p$)) are
$\Delta\varepsilon/T_c^4=2.54\pm 0.12$ for $N_\beta=4$ and
$\Delta\varepsilon/T_c^4=2.48\pm 0.24$ for $N_\beta=6$.  This is
$60\%$ and $25\%$ smaller than the earlier values on smaller lattices,
respectively.

Correlation lengths are determined from a fit of Wilson line
correlation functions (cf.~Eq.\ (\ref{3.40}) and (\ref{3.41}) below
for the definition) with a string potential ansatz in the confinement
phase and a screened Yukawa potential in the deconfinement phase. The
results are increasing correlation lengths by a factor of 2 to 3 on
both sides of the transition. The relatively large correlation length
can be traced back to the 3d-$Z(3)$-Potts model (cf.  our discussion
in section \ref{sec:II.A}).

Bacilieri et al. (1988) focus their study on Wilson line correlation
functions. Large correlation lengths are found, increasing
proportional to the lattice size. This increase is interpreted as
suggestion of a second order transition. Only a careful finite size
scaling analysis of correlation lengths can show that this
interpretation is most likely not adequate. Such an analysis is rather
involved. Therefore we devote the following section to it.

\paragraph*{Correlation lengths, mass gaps and tunneling events}
We use a notation, which is independent of the application in a spin-
or gauge model. We will distinguish {\it physical} and {\it tunneling
  correlation lengths} $\xi^{(p)}$ and $\xi^{(t)}$ with associated
mass scales $m^{(p)}$ and $m^{(t)}$ both in the finite and the
infinite volume. A physical or bulk correlation length in the infinite
volume limit can be defined via the decay of the connected
pair-correlation $C_c$ as function of the distance $z$ according to
\begin{mathletters}
  \label{3.30}
  \begin{eqnarray}
    \label{3.30a}
    C_c(z) &&= C(z) - B 
    \mathop{\smash\longrightarrow}\limits_{z\to\infty} A\cdot 
    e^{-m^{(p)}z} \\
    \label{3.30b}
    \xi^{(p)} &&\equiv 1/m^{(p)}\;.
  \end{eqnarray}
\end{mathletters}
Here $C$ is the unsubtracted, disconnected pair correlation and $B$
stands for the constant disconnected part, which is different from
zero in the phase of spontaneously broken symmetry. The inverse
correlation length $\xi^{(p)-1} = m^{(p)}$ is called {\it the physical
  mass gap}.

In the conventional lore it is the infinite volume correlation length,
which diverges as $\beta\to\beta_c$ for a second order transition and
stays finite in the first order case. Here we would like to stress the
order of limits.

\begin{eqnarray}
  \lim_{\beta\to\beta_c^-}~&&\lim_{L\to\infty}
  \xi_L^{(p)}(\beta,L)
  \nonumber\\
  =&& \lim_{\beta\to\beta_c^-} \xi^{(p)}(\beta)
  \left\{\matrix{
    \to\infty & \rm{2nd~order} \cr
    \to \hbox{const} <\infty & \rm{1st~order}.
    }\right.
  \label{3.31}
\end{eqnarray}
In the opposite order
\begin{eqnarray}
  \lim_{L\to\infty}~&&\lim_{\beta\to\beta_c^-}
  \xi_L^{(p)}(\beta,L)
  \nonumber\\
  =&& \lim_{L\to\infty} \xi_L^{(p)}(\beta_c,L)
  \left\{\matrix{
    \to\infty & \rm{2nd~order} \cr
    \to\infty & \rm{1st~order}.
    }\right.
  \label{3.32}
\end{eqnarray}
The `critical' correlation length $\xi_L^{(p)}(\beta_c,L)$ diverges at
$\beta_c$ in the infinite volume limit also in the case of a first
order transition (Bl\"ote and Nightingale, 1982; Privman and Fisher,
1983). This behavior is familiar, if we recall from section
\ref{sec:II.A} that singularities in second derivatives of
thermodynamical potentials occur for first order transitions as well.
The only difference to the second order case is the type of
divergence: $\delta$- function singularities may occur in the specific
heat and the magnetic susceptibility, when the infinite volume limit
is taken {\it at} criticality.

Thus a diverging correlation length $\xi_L^{(p)}(\beta_c,L)$ for
$L\to\infty$ is not conclusive for a second order transition.

In the following discussion we consider infinite volume correlation
lengths according to Eq.\ (\ref{3.31}). In Table~\ref{tab:4} we
compare the behavior of the physical and tunneling correlation lengths
and their associated mass scales in the infinite volume limit.

In the first row of Table~\ref{tab:4} we have listed the behavior of
the physical mass gap and its associated bulk correlation length, when
the critical coupling $\beta_c$ is approached from the symmetric phase
(i.e. $\beta_c$ from below) denoted as $\beta\to\beta_c^-$. In the
first order case $\xi^{(p)}$ stays finite, while it diverges as
$\beta\to\beta_c^-$ for a second order transition. In the broken phase
the physical mass gap stays finite in both cases (second row).

In the lower two rows we introduce an `unphysical' mass gap
$m^{\rm(unp)}$ and its corresponding tunneling correlation length
$\xi^{(t)}$. In the symmetric phase both quantities coincide with
$m^{(p)}$ and $\xi^{(p)}$, respectively, while $\xi^{(t)}$ is always
infinite in the broken phase in the infinite volume limit. Here
$m^{\rm(unp)}$ may be defined via an analogous relation to Eq.\ 
(\ref{3.30a}), but this time via the decay of the disconnected
correlation function
\begin{equation}
  C(z) 
  \mathop{\smash\longrightarrow}\limits_{z\to\infty} e^{-
    m^{\rm(unp)}z}~.
  \label{3.33}
\end{equation}
The vanishing of $m^{\rm(unp)}$ defined via Eq.\ (\ref{3.33}) just
reflects the non-vanishing constant $B$ in the broken phase, which
prevents an exponential decay of the correlation due to spontaneous
symmetry breaking. The relation to the associated tunneling
correlation length is not as simple as in Eq.\ (\ref{3.30b}).

From an analysis of the transfer matrix one expects in general for the
(disconnected pair) correlation function that it is a coherent sum of
its eigenvalues. If we consider only the two smallest eigenvalues
above the ground state, the correlation $C(z)$ may be written as
\begin{equation}
  C(z) \sim a_1~e^{-z/\xi^{(p)}} + 
  a_2~e^{-z/\xi^{(t)}} + \ldots~,
  \label{3.34}
\end{equation}
which should be compared with Eq.\ (\ref{3.33}). Eq.\ (\ref{3.34})
defines the tunneling correlation length $\xi^{(t)}$, or more
precisely, $\xi^{(t)}$ arises in the spectrum of the transfer matrix
from the level splitting associated with tunneling between different
degenerate vacua. (We have anticipated this effect in section
\ref{sec:II.A}.)

This tunneling has a finite probability as long as the volume is
finite, resulting in a finite $\xi_L^{(t)}$, but the potential barrier
between the degenerate ground states becomes infinitely high in the
infinite volume limit, leading to
\begin{equation}
  \lim_{L\to\infty} \xi_L^{(t)} = \infty~.
  \label{3.35}
\end{equation}
From Eq.\ (\ref{3.34}) we recognize that as $L\to\infty$ the diverging
$\xi^{(t)}$ gives rise to the constant disconnected term $B$ in the
correlation function Eq.\ (\ref{3.30a}).

In a {\it finite volume} both length scales $\xi^{(p)}$ and
$\xi^{(t)}$ are necessarily finite and (when considered at
criticality, cf.~Eq.\ (\ref{3.32})) both increase with increasing
volume at a first order transition. Thus there is a two-fold risk of
misinterpreting an increasing correlation length: the order of limits
has not been properly arranged, and $\xi^{(t)}$ is intertwined with
$\xi^{(p)}$.

{\it Concrete realizations} of the correlation function $C$
are spin-spin- or Wilson line-pair correlations, if we deal with
applications in the $SU(3)$-deconfinement transition. We give the
infinite volume definitions, written for a large, but finite volume
$V=L^3$.

In the 3-d three-state Potts model considered by Gavai et al. (1989),
the spin-spin correlations are defined as
\begin{equation}
  C_1(z) = {1\over 6V}~\left\langle \sum_i 
    s_i~s_j^+\right\rangle\;,
  \label{3.36}
\end{equation}
where $z=|i-j|$ and the corresponding zero-momentum 
projection reads
\begin{equation}
  C_0(z) = {1\over 6L}~\left\langle \sum_{i=1}^L \bar 
    s_i~\bar s_j^+\right\rangle\;,
  \label{3.37}
\end{equation}
where $s_i$ is the spin at site $i$ and $\bar s_i$ the average spin
over a plane $i$, here $z$ denotes the distance between planes $i$ and
$j$ along one of the principal axis of the lattice. The connected part
\begin{equation}
  C_{i,c}(z) = C_i(z) - \lim_{z\to\infty} C_i(z),
  \quad i=0,1
  \label{3.38}
\end{equation}
defines the physical correlation length $\xi^{(p)}$ as
\begin{equation}
  1/\xi^{(p)} = - \lim_{z\to\infty}
  {1\over z} \ln C_{i,c}(z)
  \label{3.39}
\end{equation}
in accordance with Eq.\ (\ref{3.30a}).

The analogous quantity $C_0(z)$ for an $SU(3)$-lattice gauge theory
has been considered by Fukugita et al.  (1989) and is the
zero-momentum correlation function of Wilson lines
\begin{equation}
  C_0(z) = {1\over 3L} \sum_{i=1}^3 \sum_{z'=1}^L~
  < \bar\Omega_i(z')~\bar\Omega_i^+(z'+z) >~,
  \label{3.40}
\end{equation}
where $i$ labels the planes, $\bar\Omega_i(z)$ is the 
average of Wilson lines $\Omega_n$ over the $i$th-plane 
with $n_i=z$, where
\begin{equation}
  \Omega_n = {1\over 3}~\hbox{Tr}~
  \left( \prod_{\tau=1}^{N_\tau}~U_{(n,\tau)}^{(4)} \right)~,
  \label{3.41}
\end{equation}
i.e. the trace of the product over gauge field variables
$U_{(n,\tau)}^{(4)}\in SU(3)$ associated to timelike links leaving the
site $(n,\tau)$ in direction 4, $n$ denote the spatial coordinates.
The additive constant $B$ of Eq.\ (\ref{3.30a}), that can be split off
in $C_0(z)$, is given by $|<\Omega>|^2$.

Bacilieri et al. (1988) have calculated Wilson line pair correlations
essentially with the same formulas as (\ref{3.40}) and (\ref{3.41}).
They use a source method, where all link variables with z-coordinate 1
are fixed to the identity matrix $1\kern-0.7ex1$. The correlation
reads
\begin{equation}
  C_1(z) =~< {1\over L_x L_y}~\hbox{Re} \sum_{x,y}
  \Omega_{(x,y,z)} \cdot 1 >~,
  \label{3.42}
\end{equation}
where $\Omega_{(x,y,z)}\equiv \Omega_n$ from Eq.\ (\ref{3.41}).

{\it Adaption to the finite volume.} The above
definitions for the various correlation lengths are applicable in the
large volume limit, i.e. if $\lim_{L\to\infty}$in Eq.\ (\ref{3.31}) is
approximately realized. As subleading corrections from the finite
volume are power-like, the large distance behavior of correlations
($z\to\infty$) is not accessible on lattices of moderate size.

For first order transitions it is rather hard to predict the finite
size dependence from first principles. (The approach of Borgs and
Kotecky (1990) comes closest to a rigorous derivation.) In general one
has to make at some place an ansatz. The double peak Gaussian ansatz
for the distribution function $P_L(E)$ has led to definite predictions
of the finite size behavior.

Likewise it requires an {\it ansatz} for the functional form of the
correlation function $C_L(Z)$ in a finite volume. A procedure, which
has been adopted by Gavai et al. (1989) and Bacilieri et al. (1988)
is to form ratios of correlation functions
\begin{equation}
  R_i(z) = {C_i(z) \over C_i(z+1)}~.
  \label{3.43}
\end{equation}
A finite volume ansatz for $C_i(z)$, which is inspired by the infinite
volume decay and adapted to the choice of periodic boundary conditions
on a cubic lattice is
\begin{equation}
  C_i(z) = A_i~[\exp(-z/\xi)/z^i + 
  \exp(-[L-z]/\xi)/(L-z)^i]~,
  \label{3.44}
\end{equation}
where $\xi$ stands for a generic correlation length, and $i=0$, 1.
Eq.\ (\ref{3.44}) includes only the first image of a source in
$z$-direction, the one which is a distance $L$ apart. Multi-mirror
images (accounting for images a distance $n\cdot L$ apart) should be
also included, if the correlation can become large compared to the
considered volume.

The careful analysis of Gavai et al. for the 3d-$Z(3)$-Potts model
shows that different methods of extracting the physical correlation
length $\xi^{(p)}$ from data on finite lattices do not agree in the
critical region, although they do agree away from $T_c$. Depending on
the ansatz for $R_i$, $\xi$ seems to grow with $L$, or to stay finite
($\xi\sim10$ in lattice units) in the vicinity of $T_c$. The former
behavior is compatible with $\xi=\xi^{(t)}$ as $L\to\infty$, the
latter with $\xi=\xi^{(p)}$, i.e. a physical mass gap which stays
finite at $T_c$ and indicates the first order of the transition.

\vskip12pt\noindent {$\bullet$} To summarize, it is the very
definition of a correlation length in a finite volume, the order of
limits ($L\to\infty$ and $\beta\to\beta_c$) and the occurrence of the
tunneling correlation length, which impede the finite size analysis of
correlation functions.

\vskip12pt After this digression we proceed with concrete applications
in the $SU(3)$-deconfinement transition.  Bacilieri et al. (1988)
considered ratios of finite volume correlations according to Eq.\ 
(\ref{3.44}) with $C_1$ given by Eq.\ (\ref{3.42}), very similar to
the observables, which were later considered by Gavai et al. (1989).
Rather than measuring the correlations $C_i(z)$, they applied a so
called smearing procedure (Albanese et al., 1987) to generate
operators weakly coupled to the high-energy fluctuations. We do not go
into technical details, although the smearing technique is quite
important for following the decay of the correlation over a large
distance up to 15-$z$-slices on a lattice of cylinder geometry
$12^2\times 48\times 4$.

Alternatively ``the'' correlation length $\xi$, defined via Eqs.\ 
(\ref{3.30a}) and (\ref{3.42}) has been measured from a
three-parameter fit with $C(z)=A\cdot\cosh~[m {1\over 2} L_z - z]$.
The statements of the paper by Bacilieri et al. about the increasing
correlation length refer to this length scale $\xi=1/m$. When $m$
coincides with the physical mass gap $m^{(p)}$, it is related to the
string tension $\sigma$ via $m^{(p)}=\sigma\cdot N_\beta$. This
relation holds in a pure gauge theory. Thus the statement about the
physical correlation length $\xi^{(p)}$ tells us the behavior of the
string tension in the critical region of the $SU(3)$- gauge theory.

The resolution of the contradicting statements about the order of the
phase transition seems to lie in this ansatz and its interpretation
\begin{equation}
  C(z) \to A~e^{-m z} + B~.
  \label{3.45}
\end{equation}
In the critical region it should be probably replaced by an ansatz
like Eq.\ (\ref{3.34}) to account for both correlation lengths
$\xi^{(p)}$ and $\xi^{(t)}$, or, if $\xi^{(t)}$ dominates, $m$ of Eq.\ 
(\ref{3.45}) reflects the increase of $\xi^{(t)}$ rather than
$\xi^{(p)}$.

\vskip12pt A systematic finite size study of the 3d-3-state Potts
model has been performed by Gavai et al. (1989). The Hamiltonian of
the Potts model has been specified in section \ref{sec:II}. It is
equivalent to a $Z(3)$-spin model. The volume dependence has been
studied for bulk quantities like the average action $E$ and the order
parameter $S$.  When $S$ and $E$ are plotted as function of $\beta$,
the crossover region becomes sharper with increasing volume.  Also a
typical manifestation of coexisting phases has been verified, the
function $\beta(L,E)$ is multi-valued.  One coupling (temperature) is
compatible with several values of the free energy at the transition
point.

Flip-flop signals in the order parameter are observed as function of
the Monte Carlo time. They are reflected in a double peak structure of
the order parameter distribution $P_L(S)$. Fig.~\ref{fig:3} verifies
what we expect from the general considerations of section
\ref{sec:II}. With increasing volume the double peak structure gets
more pronounced. (In a truly second order transition it fades away at
larger volumes.)

In the critical region the rounding and shifting of thermodynamic
quantities is governed by the thermal exponent $\lambda_1\equiv y_T =
1/\nu = d/(2-\alpha)$, where $y_T=d$ for a first order transition (cf.
section \ref{sec:II}). In particular the critical coupling should be
shifted according to
\begin{equation}
  \beta_{c,L} - \beta_c \propto L^{-y_T}~,
  \label{3.46}
\end{equation}
the width of the critical region $\sigma_L$ should shrink as
\begin{equation}
  \sigma_L \propto L^{-y_T}~.
  \label{3.47}
\end{equation}
The finite volume critical coupling $\beta_{c,L}$ and the width
$\sigma_L$ have been extracted from a Gaussian fit for the relative
population density of both phases in the critical region, and,
alternatively, by the midpoint and length of the coupling interval
($0.367\le\beta\le 0.36703$), where flip-flops have been observed on
an $48^3$ lattice. Results for both quantities are compatible with
$y_T=d$, i.e. with a first order finite size scaling behavior.

The careful analysis of finite volume correlation lengths has been
anticipated in the previous subsection. An estimate for the inverse
physical mass gap leads to $\xi^{(p)}(\beta_c)\sim 10$ in lattice
units. Thus it stays finite at $T_c$ as it should for a first order
transition.

The inclusion of an antiferromagnetic next-to-nearest-neighbor
coupling with a relative coupling strength of $\gamma=-0.2$ does not
change the qualitative result of a first order transition (Gavai and
Karsch, 1989). The discontinuity in the order parameter is of the same
magnitude as for $\gamma=0.0$, but the latent heat is found to be
smaller by a factor of 2.

This result is interesting, as it corrects the naive expectation 
that a large gap in the order parameter goes along with a
large latent heat. It raises the possibility that the deconfinement
transition in the $SU(3)$ gauge theory is associated with a small
latent heat due to slightly nonlocal effective couplings, even if the
jump in the order parameter is not small. For the actual size of the
latent heat in the $SU(3)$ gauge theory cf. section \ref{sec:III.B.2}.

\vskip12pt Finally we come to the $SU(3)$ gauge theory as considered
by Fukugita et al. (1989). The lattice size ranges over $8^3\times 4$
to $36^3\times 4$ sites. Two-state signals have been observed for the
size $N_\sigma=8- 28$. The distinction between the two-states becomes
conspicuous with increasing volume . Likewise the double peak
structure in the order parameter distribution, here $P(|\Omega|)$
($\Omega$ being the Wilson line operator of Eq.\ (\ref{3.41}) is more
pronounced for a larger lattice size.

The susceptibility of the Wilson line is given as
\begin{equation}
  \chi = V~[~<(\hbox{Re}~\Omega)^2>~-~
  <\hbox{Re}~\Omega>^2~]~.
  \label{3.48}
\end{equation}
Here $\Omega$ has been defined in Eq.\ (\ref{3.41}), where
$\hbox{Re}~\Omega$ is taken as the projection of $(\sum_n \Omega_n/V)$
onto the nearest $Z(3)$-axis on the complex $\Omega$-plane. The
results for the maximum in the susceptibility lead to a scaling of
$\chi_{\rm max}\propto V^{\gamma/(d\cdot\nu)}$ with
$\gamma/(d\cdot\nu)=0.99(6)$ and a width $\Delta\beta\propto
V^{-1/(d\nu)}$ with $1/(d\nu)=0.95(5)$ in reasonable agreement with
the expected values of 1.

The `unphysical' mass gap $m^{\rm(unp)}$ defined in Eq.\ (\ref{3.33})
is obtained from a fit of
$C_0(z)=A\cdot\cosh~[m^{\rm(unp)}(z-N_\sigma/2)]$. The expected
behavior in the infinite volume limit has been shown in
Table~\ref{tab:4}. For a first order transition in a {\it finite}
volume $m^{\rm(unp)}$ should sharper decrease as $V\to\infty$ than for
a second order transition. In fact, a crossing of the family of
functions $m^{(unp)}(L,\beta)$ is observed, which is qualitatively in
agreement with the expectations, although a detailed quantitative
understanding is missing.

The physical mass gap $m^{(p)}$ is read off from an ansatz as Eq.\ 
(\ref{3.30a}) with $B=\left|<\Omega>\right|^2$, where the exponential
is replaced by a $\cosh$ to account for the periodic boundary
conditions. It is small, but stays finite at $\beta_c$
(cf.~Table~\ref{tab:4}).

Fukugita et al. (1989 and 1990a) have also addressed the question of
nonlocal couplings on an effective level of the $SU(3)$ gauge theory.
An answer requires an explicit construction of the effective $Z(3)$
action from the original $SU(3)$ gauge theory.  The $Z(3)$ variables
are not introduced ad hoc (i.e. by assumption), but expressed in terms
of the original field variables. An `effective' $Z(3)$-spin $\omega_n$
is defined as the projection of $\Omega_n$ to the nearest $Z(3)$ axis
(normalized to unit length). The generic form of an effective action
in terms of $\omega_n$-variables is written as
\begin{equation}
  S_{\rm eff} = \sum_\alpha~\beta_\alpha~
  O_\alpha  (\{\omega_n\})~.
  \label{3.49}
\end{equation}
Here $O_\alpha$ are $Z(3)$-invariant monomials in $\omega_n$,
$\beta_\alpha$ are the associated couplings, local as well as
non-local, e.g. nnn-couplings. First correlation functions of the
monomials $O_\alpha$ are calculated from the ensemble of $\omega_n$'s,
generated in the original $SU(3)$-Monte Carlo runs. The number of
terms labeled by $\alpha$ is truncated by hand. Next the couplings
$\beta_\alpha$ are determined with the aid of Schwinger-Dyson
equations. This method is known from Monte Carlo renormalization group
calculations (Gonzalez-Arroyo and Okawa, 1987).

The results are the following. Effective two-spin couplings decrease
to values less than $10^{-3}$ for distances larger than 3 between the
two spins (on a $24^3\times 4$ lattice with $\beta=5.6925$). Three-
and four-spin couplings are smaller than $(1-2)\times 10^{-3}$. The
effective couplings decrease exponentially with the distance as
anticipated by Svetitsky and Yaffe from the phase structure of the
associated {\it three}-dimensional $SU(3)$ gauge theory.

The two-spin couplings (in particular the nnn-couplings), which are
derived quantities in this case, turn out to be all positive. There is
no sign of ``antiferromagnetism''.

\vskip12pt Like other numerical investigations this derivation is not
rigorous. In particular it involves a truncation in the effective
action, which is, however, unavoidable. In this sense none of the
Monte Carlo results does actually {\it prove} the first order of the
deconfinement transition, but different derivations give strong hints
pointing in the same direction: the deconfinement transition in the
pure $SU(3)$ gauge theory is of first order. The controversy about
this topic has settled. The situation becomes considerably more
involved, when fermions are included.

\subsubsection{Thermodynamics on the lattice}
\label{sec:III.B.2}
An important contribution of lattice QCD to phenomenological
applications of the QCD transition is a prediction of QCD's equation
of state. The behavior of the pressure $p$, the internal energy
density $\epsilon$ and the entropy density $s$ are of most interest
for seeking observable effects in heavy-ion collisions. In this
section we summarize the lattice expressions of thermodynamic
quantities and outline an approach due to Engels et al. (1990b) to
derive the equation of state in an entirely nonperturbative framework.
The method overcomes former inconsistencies observed for the pressure.
Here we consider its application to the pure $SU(N)$ gauge theory
followed by a discussion of finite size effects in the regions of
critical and high temperature.

From section \ref{sec:II}.A we recall the continuum expressions for
the energy density $\epsilon$ and the pressure $p$ as
$\epsilon=(T^2/V)^.{\partial\over\partial T}\ln Z$ and
$p=T{\partial\over\partial V}\ln Z$. One way of calculating these
quantities on the lattice is to transcribe the derivatives which
respect to $T$ and $V$ on the lattice.  This has led to introducing
different lattice spacings for the temporal ($a_\beta$) and spatial
($a_\sigma$) directions, and is the reason, why we have specified the
lattice action directly on an asymmetric lattice. The derivatives are
replaced according to
\begin{eqnarray}
  {\partial\over\partial T}~ && \longrightarrow~
  {1\over N_\tau}{\partial\over\partial a_\tau} \nonumber\\
  \label{3.50}
  \\
  {\partial\over\partial V}~ && \longrightarrow~
  {1\over 3a_\sigma^2N_\sigma^3}{\partial\over\partial 
    a_\sigma}\;. \nonumber
\end{eqnarray}
After performing the appropriate lattice derivatives of ln~$Z$, the
lattice spacings are set equal again, $a_\sigma=a_\beta=a$.

Let us consider the first derivatives of the partition function. The
order parameter for the pure gauge theory is the thermal Wilson line.
Its lattice expression has been given in Eqs.\ 
(\ref{3.41})-(\ref{3.42}). The order parameter for the chiral
condensate will be given in Eqs.(\ref{3.112}a-b) in section
\ref{sec:III.C}.

The lattice expression for the internal energy density gets a
contribution from the gluonic part and the fermionic part
\begin{equation}
  \epsilon=\epsilon_G+\epsilon_F\;.
  \label{3.51}
\end{equation}
For completeness we give also the formula for the fermionic part. The 
continuum expression
\begin{equation}
  \epsilon=-{1\over V}{\partial(\ln Z)\over\partial(1/T)}={1\over V} 
  \left\langle{\partial S\over\partial(1/T)}\right\rangle
  \label{3.52}
\end{equation}
is transcribed to
\begin{equation}
  \epsilon={1\over N_\sigma^3a_\sigma^3N_\tau}\left\langle{\partial 
      S\over\partial a_\tau}\right\rangle
  \label{3.53}
\end{equation}
with $S$ given by the Wilson action or the action for staggered
fermions, cf. section \ref{sec:III.A}.

The energy density contains a contribution from the vacuum, which is a
$\beta$-independent infinite constant, cf. Bernard (1974). When the
vacuum contributions from the gluons $\epsilon_{G,vac}$ and the
fermions $\epsilon_{F,vac}$ at $T=0$ are subtracted, the resulting
lattice expressions are
\begin{eqnarray}
  \label{3.54}
  \epsilon_G&&=2N(N_\sigma^3N_\tau a^4)\Biggl[g^{-
    2}\left\langle\sum_{\scriptstyle x\atop\mu<\nu<4}P_x^{\mu\nu}-
    \sum_{\scriptstyle x\atop\mu<4}P_x^{\mu4}\right\rangle 
  \nonumber\\
  && -c_\sigma'\left\langle\sum_{\scriptstyle 
      x\atop\mu<\nu<4}P_x^{\mu\nu}\right\rangle-
  c_\tau'\left\langle\sum_{\scriptstyle 
      x\atop\mu<4}P_x^{\mu4}\right\rangle\Biggr]-\epsilon_{G,vac}
\end{eqnarray}
and
\[
  \epsilon_F=N_f(4N_\sigma^3N_\tau a^4)^{-1}\left\langle\hbox{Tr 
      }\left\{D^{(4)}(D+m)^{-1}\right\}\right\rangle-\epsilon_{F,vac}\;,
\]
where $m_f$ has been set to $m$ for all flavors $f$ in Eq.\ 
(\ref{3.54}) and
\begin{eqnarray}
  a=a_\sigma=a_\tau\;, \quad && D=\sum_{\mu=1}^4D^{(\mu)}\;,
  \nonumber\\
  c_\sigma'=-a^.\left.{\partial g_\sigma^{-2} / \partial 
      a_\tau}\right|_{a_\tau=a_\sigma} , \quad && 
  c_\beta'=-a^.\left.{\partial g_\tau^{-2} / \partial 
      a_\tau}\right|_{a_\tau=a_\sigma} , \nonumber\\
  \label{3.55}
  \epsilon_{G,vac}=N(N_\sigma^4 a^4)^{-1} && [-c_\sigma'-
  c_\tau']^.\sum_{\scriptstyle x\atop\mu<\nu}P_x^{\mu\nu}\;, \nonumber\\
  \epsilon_{F,vac}= && 3N_f(16a^4)^{-1} . 
\end{eqnarray}

The notation has been introduced in section \ref{sec:III.A}. Note that
the energy densities scale with $(1/a^4)$. An uncertainty in $a(g)$ in
units of [MeV] is amplified to the fourth power in the results for
$\epsilon$.  Further we see that $\epsilon$ is expressed in certain
plaquette expectation values, which are easily accessible in a Monte
Carlo simulation.  In principle the coefficients $c'_{\sigma}$ and
$c'_{\tau}$ can be determined nonperturbatively from lattice
simulations (for first attempts see (Blum et al., 1995a)), but in the 
past the
perturbative values (Karsch, 1982; Trinchero, 1983) have been used in
the otherwise nonperturbative expressions.  The use of perturbative
values is questionable from the beginning. For the $SU(3)$ gauge
theory large deviations from the perturbative $\beta$-function are
known to occur for $g\geq1$, and the $\beta$-function $a^.dg/da$ is
related to the coefficients $c_\sigma'$ and $c_\tau'$ via
\begin{equation}
  a{dg\over da}=g^3\left({\partial g_\sigma^{-2}\over\partial\xi}
    +{\partial g_\tau^{-2}\over\partial\xi}\right)_{\xi=1}
  \label{3.56}
\end{equation}
with $\xi\equiv a_\sigma/a_\tau$, cf. section \ref{sec:III.A}.
Therefore significant nonperturbative contributions must be expected
for $c_\sigma'$ and $c_\tau'$ as well.

The perturbative ingredients have actually led to 
inconsistencies in measurements of the pressure. The 
pressure turned out to be negative and discontinuous 
across the transition point. In consequence the values 
for the latent heat were not consistent. For a continuous 
pressure the latent heat $\Delta L$ should be the same, 
whether it is calculated as
\begin{equation}
  \Delta(\epsilon+p)/T_c^4\equiv\Delta L_1
  \label{3.57}
\end{equation}
or
\[
  \Delta(\epsilon-3p)/T_c^4\equiv\Delta L_2\;.
\]
Both Eqs.\ (\ref{3.57}) are simple expressions in terms of plaquette
expectation values. Results of Fukugita et al.  (1990a) for an $SU(3)$
gauge theory gave
\begin{equation}
  \Delta L_1=2.54(12)\hbox{\hskip2em and\hskip2em} \Delta 
  L_2=3.78(20)\;.
  \label{3.58}
\end{equation}
The discrepancy between $\Delta L_1$ and $\Delta L_2$ is clearly
outside the statistical error bars.  In an independent simulation Deng
(1989) found
\begin{equation}
  \Delta p/T_c^4=-0.29(15)
  \label{3.59}
\end{equation}
for the discontinuity in the pressure.

These difficulties have been overcome by a different 
approach to lattice thermodynamics, which can be entirely 
performed within a nonperturbative framework. The 
additional quantity now is the $\beta$-function 
$a^.dg/da$ from QCD, which is known in a nonperturbative 
region of couplings $g$ from Monte Carlo renormalization 
group calculations. Otherwise only plaquette expectation 
values have to be calculated as before. This approach has 
been proposed by Engels et al. (1990b). The steps are as 
follows.

\begin{itemize}
\item The free energy density is calculated from an integration over
  its derivative with respect to $\beta$, since the logarithm of the
  partition function is not directly accessible within the Monte Carlo
  approach. With $f=-{T\over V}\ln Z$ one takes the derivative with
  respect to the gauge coupling $\beta=2N/g^2$
  \begin{equation}
    -{\partial\ln 
      Z\over\partial\beta}=\left\langle 
      S_G\right\rangle=6N_\sigma^3N_\tau P_T\;,
    \label{3.60}
  \end{equation}
  where $S_G$ is the gauge part of the action (e.g. given by Eq.\ 
  (\ref{3.10})) and $P_T$ is a short notation of the plaquette
  expectation value at temperature $T$, calculated on a lattice of
  size $N_\sigma^3N_\tau$. If $P_0$ denotes the corresponding zero
  temperature expectation value, evaluated on a lattice of size
  $N_\sigma^4$, the difference of free energy densities $f$ at
  coupling $\beta$ and $\beta_1$ is obtained as
  \begin{equation}
    \left.{f\over T^4}\right|_{\beta_1}^\beta=-
    6N_\tau^4\intop_{\beta_1}^\beta d\beta'\;[P_0-
    P_T].
    \label{3.61}
  \end{equation}
  
\item Now an additional assumption enters, which is
  strictly satisfied only for an infinitely large lattice.  It is the
  relation $\ln Z=V\partial\ln Z/\partial V$, which is valid for
  homogeneous systems in large volumes. From
  $p=T{\partial\over\partial V}\ln Z$ and Eq.\ (\ref{3.61}) one
  obtains
  \begin{equation}
    p(\beta)=-[f(\beta)-f(\beta_1)]\;.
    \label{3.62}
  \end{equation}
  In Eq.\ (\ref{3.62}) $\beta_1$ has been chosen small enough that
  $p(\beta_1)\sim0$.
  
\item A second quantity, which has appeared in Eq.\ 
  (\ref{3.57}) and is easily calculable on the lattice, is the
  interaction measure $\Delta$
  \begin{equation}
    \Delta={\epsilon-3p\over T^4}=-12NN_\tau^4a{dg^{-2}\over da}[P_0-
    P_T]\;.
    \label{3.63}
  \end{equation}
  As the name suggests, $\Delta$ vanishes for an ideal gas (where
  $\epsilon=3p$, see section \ref{sec:V}).
  
\item The advantage of calculating first $\Delta$ rather than
  directly $\epsilon$ is that $\Delta$ depends no longer on
  $c_\sigma'$ and $c_\tau'$, but on the QCD-$\beta$-function $B(g)$
  \begin{equation}
    B(g)=-a{dg^{-2}\over da}\;.
    \label{3.64}
  \end{equation}
  
  When $\Delta$ is to be evaluated in a nonperturbative coupling
  regime, it requires a knowledge of the $\beta$- function for
  nonperturbative couplings as well. This function has been inferred
  from Monte Carlo renormalization group studies (Hoek, 1990; Akemi et
  al., 1993). The functional form suggested in (Hoek, 1990) has been
  used by Laermann et al. (1995) in fits to the more recent data from
  Akemi et al. (1993). The remaining terms in $\Delta$ and $f$ are
  calculated in a Monte Carlo simulation. Once we have $f$ and
  $\Delta$, we know $\epsilon$ and the entropy density $s$ (with
  $f=\epsilon-Ts$).  Using this approach, the previously mentioned
  problems with the pressure have disappeared, $p$ stays positive and
  behaves continuously at $T_c$.
\end{itemize}

A fully nonperturbative approach alone does not guarantee that the
equation of state on the lattice is relevant for continuum physics.
Earlier simulations have been performed for a time extension of four
lattice spacings, although $N_{\tau}\sim 10$ is needed for the
continuum limit.

The $N_{\tau}$-dependence has been studied in case of the
$SU(3)$-deconfinement transition by Boyd et al. (1995) (see also
Karsch and Laermann, 1995). The temporal lattice size has been varied
from ($N_{\tau}=4$ with $N_{\sigma}=16$) to ($N_{\tau}=6$ with
$N_{\sigma}=32$), and ($N_{\tau}=8$ with $N_{\sigma}=32$).  The
dominant type of finite size effects depends on the temperature.  At
high temperatures there is an intimate relation between finite size
(IR-) and finite cut-off (UV-)artifacts, although the distinction
between them may naively suggest their decoupling. The relation
between IR- and UV-artifacts is seen as follows. A reduction of
$a_{\tau,\sigma}(g)$~[fm] as function of $g$ goes along with an
increase in temperature $T$~[MeV], if $N_{\tau}$ is kept fixed. This
way continuum behavior is mixed with high temperature behavior. For a
small lattice spacing just the high momentum modes are cut off which
give the main contribution to the energy density and pressure at high
temperatures. For a larger lattice spacing at lower temperatures also
lower momentum modes are cut off, but the energy density and pressure
grow approximately with $T^{4}$, thus the finite cut-off effects,
induced by the finite number of time slices, are largest for high
temperatures.

The $N_{\tau}=1/(a\cdot T)$-dependence of the gluonic part of the
energy density of an ideal gluon gas is estimated as (Karsch, 1995)
\begin{equation}
  \label{3.65}
  \epsilon_{\rm{gluonic}}(N_{\tau})=(N_{c}^{2}-1)\cdot
  \left[{\pi^{2}\over 15}+{2\pi^{4}\over 63}{1\over (N_{\tau})^{2}}
    +O({1\over N_{\tau}^{4}})
  \right] .
\end{equation}
For $N_{\tau}=4$, $\epsilon_{\rm{gluonic}}(4)$ is 50 \% larger than
the corresponding Stephan-Boltzmann value in the continuum. The
leading $1/N_{\tau}^{2}$-dependence has been used by Karsch et al. to
extrapolate the $N_{\tau}=6,8$-data for $\epsilon$ and $p$ to the
continuum limit. Fig.~\ref{fig:4}(a-b) shows the energy density and
pressure for the various lattice sizes. It prevails the expected
increase of the finite size/finite cut-off effects with temperature.
There are strong deviations by about 20~\% from ideal gas behavior
(indicated by the horizontal lines) for temperatures at least up to
$3T_{c}$ in the continuum limit. It should be further noticed that the
pressure $p$ increases more slowly with $T$ than the energy density
$\epsilon$. This leads to a strong deviation of the velocity of sound
from the ideal gas value. The velocity of sound can be an input from
first principles to applications in heavy-ion collisions, cf. section
\ref{sec:V}.

The latent heat is indicated by the vertical bar in Fig.~\ref{fig:4}(a),
the value is taken from Iwasaki et al. (1992). It is further reduced
compared to earlier results (e.g. of Brown et al., 1988). The result
is
\begin{equation}
  \label{3.66}
  {\Delta\epsilon \over T_{c}^{4}} = 
  \left\{
    \matrix{
      2.44 \pm 0.24 & N_{\tau}&=4 \cr
      1.80 \pm 0.18 & N_{\tau}&=6 .
      }
  \right.
\end{equation}
The latent heat is only 30~\% of the energy of an ideal gas at
$T_{c}$.

The interaction measure $\Delta$ is shown in Fig.~\ref{fig:5} for
three lattices. The deviation from ideal gas behavior is largest in
the transition region, it goes logarithmically to zero with increasing
temperature. The interaction measure is obviously less sensitive to
finite size/finite cut-off effects than $\epsilon$ and $p$ separately.
The deviations from the ideal gas behavior are assumed to be due to
the nonperturbative infrared structure of QCD, which is less sensitive
to UV-effects. The results of Fig.~\ref{fig:5} support this assumption
and indicate that QCD's IR-structure plays an important role even at
high temperatures.  As mentioned in the introduction (section
\ref{sec:I}), it is too naive to argue from asymptotic freedom that
QCD behaves fully perturbatively at high temperatures.

In physical units the result for the interaction measure is given by
(Karsch, 1995)
\begin{equation}
  \label{3.67}
  \epsilon-3p\approx 2.4~\rm{GeV/fm}^{3}\quad\rm{at}~T=1.1T_{c} .
\end{equation}
In Eq.\ (\ref{3.67}) the string tension $\sigma$ with
$\sqrt{\sigma}\sim 420$~MeV has been used as physical input. Thus the
main contribution to the interaction measure comes from the zero
temperature gluon condensate
$\left<G^{2}\right>_{0}=2.0~\rm{GeV/fm}^{3}$, noting that the
trace anomaly ($\epsilon-3p$) is related to the gluon condensates at
zero and finite temperature according to
\begin{equation}
  \label{3.68}
  \epsilon-3p=\left<G^{2}\right>_{0}-\left<G^{2}\right>_{T} .
\end{equation}

\vskip12pt As argued above, UV-artifacts induced by the finite
temporal extension play a dominant role in the high temperature
region, but a minor role in the transition region. The main
contributions to the finite size effects in the transition region come
from the low momentum modes, in particular, if a correlation length
becomes large. In case of the pure $SU(2)$ gauge theory the standard
finite size scaling analysis for second order phase transitions can be
applied to the nonanalytic behavior of the free energy density. The
finite size effects are controlled by the ratio
$N_{\sigma}/N_{\tau}=TV^{1/3}$ rather than by $1/N_{\tau}$ as in the
high temperature region. In particular the normalized critical energy
density $\epsilon^{\rm{crit}}/T_{c}^{4}$ should scale as function of
$T\cdot V^{1/3}$ with critical indices of the three-dimensional Ising
model (cf. section \ref{sec:II.B.2}). This scaling behavior has been
verified by Engels et al. (1995). An extrapolation to the infinite
volume limit leads to
\begin{equation}
  \label{3.69}
  \epsilon^{\rm{crit}}/T_{c}^{4}= 0.256(23)
\end{equation}
for a pure $SU(2)$ gauge theory. This result may be considered as free
of lattice artifacts. It is a continuum result although not continuum
physics because of two colors and the absence of dynamical fermions.

\vskip12pt We summarize. Reliable data for QCD's equation of state
exist for the limiting cases of a pure $SU(2)$ and $SU(3)$ gauge
theory. Earlier discrepancies for results on the pressure have been
removed by using fully nonperturbative ingredients in the derivation.
To test the relevance of the lattice results for continuum physics, a
careful finite size scaling analysis has been performed. The finite
size effects are under control, as the predicted dependence on
$N_{\tau}$ and $N_{\sigma}$ has been confirmed in the simulations.
Therefore an extrapolation to the continuum limit
($N_{\tau}\to\infty$, $a\cdot T\to 0$, $T$~[MeV] fixed) is safe. Such
tests should become standard in future lattice simulations.

The fully perturbative equation of state can be also applied when
dynamical fermions are included. We report on first attempts in this
direction in section \ref{sec:III.C.6}. However, a finite size scaling
analysis gets much more involved in the presence of dynamical
fermions. For sufficiently large quark masses the singularities of the
first order $SU(3)$ or the second order $SU(2)$ deconfinement
transitions will be rounded anyway. One then has to disentangle, which
part of the rounding comes from the finite volume and from the finite
masses (see section \ref{sec:III.C}.2 and 3).

\subsubsection{Interface tensions in QCD}
\label{sec:III.B.3}
In recent years the interface (or surface) tension has been frequently
calculated for models of QCD. Like the latent heat the surface tension
is an important measure for the strength of a first order transition.
The interface occurs between phases, which coexist at the critical
temperature.  A large surface tension leads to a strong supercooling
effect. The onset of the phase conversion is delayed, as the critical
temperature is reached. One possible scenario for the phase conversion
is droplet formation. Once the system has sufficiently supercooled
below the critical temperature, the gain in energy from the conversion
to the new phase can compensate for the energy loss in interface free
energy, and the phase conversion sets in.

In applications to QCD different kinds of interfaces must be
distinguished. In a first order chiral transition we have interfaces
between phases with broken and restored chiral symmetry. Interfaces
between the deconfinement and the confinement phases may occur in a
first order deconfinement transition. They will be considered below
for calculations of interface tensions. Furthermore different
realizations of the plasma phase are separated by interfaces. Such
interfaces do not only occur in the region around $T_{c}$, but
throughout the phase of broken $Z(3)$-symmetry above $T_{c}$. One
specific phase realization corresponds to a spatial domain, which is
characterized by a certain expectation value of the Wilson line
operator. The associated interface free energy density is called an
{\it ordered/ordered interface tension} $\sigma_{oo}$, whereas an
interface between the plasma and the hadronic phase leads to an {\it
  ordered/disordered interface tension} $\sigma_{od}$. We write
`plasma phase' as a common name for the deconfinement and/or chiral
symmetric phase and `hadronic phase' for the confinement and/or chiral
symmetry broken phase. Via the relative magnitude of both quantities
at $T_{c}$ one may gain some insight about the degree of wetting.

{\it Wetting} is an alternative or competing mechanism to droplet
formation.  One phase may spread along a 2-dimensional front into the
other.  Stated in a liquid/gas picture a droplet of liquid may wet an
interface between a gas and a solid when these phases coexist. Whether
the wetting is complete or not depends on the relative size of
$\sigma_{oo}$ and $\sigma_{od}$.

In QCD the issue of wetting has been raised by Frei and Patk\'os
(1989). It has been studied in systems which share the global
Z(3)-symmetry, in the three-state Potts model and in an effective
model of Polyakov loops. Later the question of wetting at the
deconfinement transition has been investigated in an effective
three-dimensional $Z(3)$- symmetric $\Phi^4$-model for the Polyakov
loop (Trappenberg and Wiese, 1992). In this model the confinement
phase completely wets the different realizations of the deconfinement
phase at $T_{c}$.

So far these results are not applicable to QCD transitions under
realistic conditions, since fermions have been excluded. Recall that
the inclusion of fermion masses may completely wash out the first
order nature of the chiral and the deconfinement transition. A smooth
crossover phenomenon prevents the occurrence of different {\it
  coexisting} phases and their associated interfaces. If the hints of
today will be confirmed in the future that both transitions fade away
for physical fermion masses, the various interface tensions have no
impact on applications in the early universe or relativistic heavy-ion
collisions.

\paragraph*{How to measure an interface tension}
Consider a system, where a domain with volume $V_H$ in the (hadronic)
phase $H$ is embedded in a second phase $Q$ (the quark-gluon-plasma
phase) with volume $V_Q$. If we denote by $f_h$ and $f_Q$ the free
energy densities of both phases, the free energy $F$ of the total
system is given as
\begin{equation}
  F=F_s + \left(V_H f_H + V_Q f_Q\right)\;,
  \label{3.70}
\end{equation}
$F$ differs from the sum of the domain free energies just by an amount
$F_s$, $F_s$ equals the free energy which is associated with the
interface separating both domains.

As long as $V_H$ and $V_Q$ are finite, and an external field $\beta$
is applied, which is homogeneous, $F_s$ will be actually zero. Thermal
equilibrium forbids the coexistence of phases as stable configurations
in a finite volume. The system will tunnel from one phase into the
other. A transient coexistence can only result from metastability
effects. If one wants to measure the surface tension directly as the
excess free energy due to an interface in a finite system, one has to
stabilize the interface by an external field gradient $\Delta\beta$.
Again the order of limits is essential
\begin{equation}
  \lim_{\Delta\beta\to0} \;\; \lim_{V_H, V_Q\to\infty} \;\; 
  F_s/A=\sigma\;,
  \label{3.71}
\end{equation}
where $\sigma$ is the surface free energy $F_s$ per unit interfacial
area $A$, $\Delta\beta$ stands for a generic field gradient.

For the temperature-driven transitions in QCD an appropriate
`field'-gradient is provided by the temperature. To have a preferred
direction, we consider a spacetime lattice with spatial cylinder
geometry, i.e.  $L_x \times L_y \times L_z \times N_\tau$ with
$L_x=L_y\ll L_z$. Choosing half of the lattice links $(z=1,2,\dots
L_z/2)$ at a temperature larger than $T_c$ (corresponding to a
coupling $\beta\equiv 6/g^2>\beta_c$) and the other half
$(z=L_z/2+1,\dots,L_z)$ below $T_c$ $(\beta<\beta_c)$, an interface
between a deconfinement and a confinement phase has been enforced onto
the system. It will be located somewhere `between' $z=L_z/2$ and
$L_z/2+1$. Early calculations of the surface tension in an $SU(3)$
gauge theory have been performed along these lines (Kajantie et al.,
1990). The procedure involves the $\Delta\beta\to0$-extrapolation in
the very end. The result was $\sigma/T_c^3=0.24\pm0.06$. It has been
obtained for a time extension of $N_\tau$=2. The value is probably too
large by an order of magnitude (cf.  below). One reason are possible
finite size effects due to the small extension in time direction. The
other reason is that the extrapolation to $\Delta\beta\to0$ may be not
safe. An external field can lead to a suppression of fluctuations of
the interface, if it is strong enough. This way the interface looks
more rigid than it would be in the $\Delta\beta\to0$ limit.

Most of the work for interface calculations in an $SU(3)$ gauge theory
has been concentrating on time extensions of $N_\tau=2$, as the
measurements are rather time consuming (Kajantie et al., 1991; Potvin
and Rebbi, 1989; Huang et al. 1990 and 1991; Grossmann et al., 1992;
Janke, 1992). Results for $N_\tau=4$ can be found in (Brower et al.,
1992; Potvin and Rebbi, 1991; Grossmann and Laursen, 1993). In the
following we report on work by Iwasaki et al. (1994), who extend the
number of time slices up to $N_\tau=6$.

Iwasaki et al. apply the {\it histogram method}, which has been introduced
by Binder (1981 and 1982) for Ising-type-systems, to an $SU(3)$ gauge
theory. The histogram method is based on the analysis of probability
distributions $P(\Omega)$ of order parameters $\Omega$ in the vicinity
of the phase transition. As we have seen in section \ref{sec:II.A},
the probability distribution $P(\Omega)$ develops a characteristic
double peak structure close to $T_c$. The structure gets more
pronounced as the volume is increased. This behavior had been
identified as a characteristic signature for a first order transition
in a finite volume, cf. Fig.~\ref{fig:3}. The valley between both
peaks corresponds to configurations with interfaces, which are the
more suppressed the larger the volume.

As the suppression comes from the extra costs in interface free
energy, it is plausible that the value of $\sigma$ may be inferred from
the position of the maxima and minimum of $P(\Omega)$ as a function of
volume. In the formulas below cylinder geometry $(N_x,N_y\ll N_z)$ and
periodic boundary conditions are assumed. Thus two-interfaces will be
preferably be performed in the $(x,y)$-plane with total area
$2A=2N_xN_y\cdot a^2$, $a$ being the lattice constant.  The order
parameter $\Omega$ will be identified with the Polyakov loop.

The ansatz for the probability distribution $P(\Omega)$ which has been
chosen by Iwasaki et al. (1994) is then given as
\[
  P(\Omega) = P_1(\Omega) + P_2(\Omega) + P_m(\Omega)
\]
with
\vskip-12pt
\begin{eqnarray}
  P_i(\Omega) &&= c_i\exp \left( - f_i V/T \right) \cdot \exp 
  \left[ - \left(\Omega - \Omega_i \right)^2 / d_i^2 
  \right] \nonumber\\
  && \qquad\qquad (i=1,2)
  \label{3.72}
  \\
  P_m(\Omega) &&= c_m \exp \left[ - \left( f_1 V_1 + f_2 V_2 
    \right) / T - \sigma2A/T\right] , \nonumber
\end{eqnarray}
where the following notations are involved.  The order parameters
$\Omega_1$ and $\Omega_2$ denote the values in the confinement and
deconfinement phases, $f_1$ and $f_2$ the corresponding free energy
densities, $V_1$ and $V_2$ are the associated volumes occupied by each
of the coexisting phases. The coefficients $c_i$ and $d_i$ depend on
the volume. In Eq.\ (\ref{3.72}) for $P(\Omega)$ we recognize the
superposition of the Gaussians around the characteristic peaks for
each phase. The third term $P_m(\Omega)$ gives the probability for
finding the mixed phase. Here is the place where the interface tension
enters. The probability $P_m(\Omega)$ depends on $\Omega$ via the
volumes $V_1$ and $V_2$ which are occupied by both phases such that
$\Omega V = \Omega_1 V_1 + \Omega_2 V_2$.

The weight factors $\exp( - f_i V/T) (i=1,2)$ are proportional to the
probabilities for the system to reside in the confinement or
deconfinement phase. While $f_1$ and $f_2$ are degenerate at $T_c$ in
the infinite volume limit, they are in general different in a finite
volume. To project on $\sigma$ in a finite volume $V$, i.e.  on
$\sigma_V$ or $\hat\sigma_V\equiv \sigma_V/T_c^3$, one has to define
$\hat \sigma_V$ in a way that the leading $V_i$-dependence in the
exponent cancels out. The cancellation is achieved, if $\hat \sigma_V$
is defined according to
\begin{equation}
  \hat\sigma_V \equiv { - N_\tau^2 \over 2 N_x N_y} \cdot 
  \log { p_{min} \over \left( p_{max,1}\right)^{\gamma_1} 
    \left( p_{max,2} \right)^{\gamma_2} }\;.
  \label{3.73}
\end{equation}
Here $p_{max,1}$ and $p_{max,2}$ are the two maxima of $P(\Omega)$,
while $p_{min}$ denotes the minimum between the two peaks. The powers
$\gamma_1$ and $\gamma_2$ denote the weights of the contributions of
$\Omega_1$ and $\Omega_2$ to $\Omega$ at the minimum, i.e.
$\Omega=\gamma_1\Omega_1+\gamma_2\Omega_2$. The infinite volume limit
of $\hat \sigma_V$ is the value for the interface tension in units of
$T_c$
\begin{equation}
  \hat \sigma\equiv \sigma/T_c^3 = \lim_{V\to\infty} \hat 
  \sigma_V\;.
  \label{3.74}
\end{equation}

The actual measurement of $\hat\sigma_V$ is more subtle. 
The subtleties concern the choice of the order parameter, 
the critical coupling in a finite volume and the 
determination of $p_{max}$ and $p_{min}$. Candidates for 
order parameters are the action density or the Polyakov 
loop. The Polyakov loop is a complex-valued observable. The 
distribution $P(\Omega)$ develops 4 peaks near $T_c$ 
corresponding to the confinement phase with $\Omega_i=0$ 
and 3 realizations of the deconfinement phase in the 
directions $\exp (i 2 \pi n/3),n=0,1,2$. To reduce the 
numerical effort for obtaining high quality data, Iwasaki 
et al. have projected $\Omega$ on the real axis by taking 
the absolute value $\Omega_{abs}$ or by rotating it with 
$\exp(i 2 \pi n/3)$ such that $-\pi/3 <\; {\rm arg}\;\Omega 
\leq \pi/3$ and taking the real part, leading to 
$\Omega_{rot}$. The dependence of the results on this 
choice has been checked.

The choice of the critical coupling in a finite volume is a 
nontrivial issue. It should guarantee the correct infinite 
volume limit of $\beta_c$. Here $\beta_c$ has been adjusted 
such that the peaks of the histograms for $\Omega_{rot}$ 
had equal height.

The actual minimum and maxima of the histograms may be 
contaminated due to statistical fluctuations. Therefore the 
extrema were read off from 3rd order polynomial fits to the 
histograms in the vicinity of the extrema.

Only the leading volume dependence drops out of $\hat \sigma_V$, when
it is calculated according to Eq.\ (\ref{3.73}).  Subleading
corrections in $1/V$ arise e.g. from fluctuations of $\Omega$ in the
bulk phases, capillary wave fluctuations of the interfaces and zero
modes corresponding to the translation of interfaces in the direction
perpendicular to the interfaces. If the geometry of the lattice
deviates from an idealized cylinder in $z$- direction, interfaces are
no longer restricted to the $xy$-plane, but sweep out in other
directions as well. All of these finite volume effects have been taken
into account by making an appropriate ansatz for the volume dependent
prefactors in the formula for $p_{min}/p_{max}$. For further details
we refer to Iwasaki et al. (1993).

An important point about the actual measurement of $\sigma$ 
concerns the generation of histograms. In case of the 
$SU(3)$ gauge theory high statistics histograms were 
available from the QCDPAX-collaboration (Iwasaki et al., 
1991 and 1992a). They were used as input in the $\sigma$-
measurements of Iwasaki et al. (1994). If such histograms 
are not available for the quantities one is interested in, 
one can dispose of a more advanced method to generate such 
histograms by means of so-called multicanonical algorithms.

\paragraph*{Multicanonical algorithms}
We briefly digress from the $\sigma$-measurement to sketch 
the idea of a multicanonical updating procedure. It has 
been developed by Berg and Neuhaus (1991) and first applied 
to 2-dimensional $q$-state-Potts models. For large volumes 
it is notoriously difficult to generate configurations 
corresponding to the valley between the two peaks. These 
configurations are suppressed by the large amount of 
surface free energy, which is just the quantity of 
interest. The reason is that usual local upgrading 
procedures like the Metropolis- or heat bath algorithms are 
designed for single-peaked distributions $P(\Omega)$. 
Configurations are selected with an importance sampling 
according to their Boltzmann weight $\exp \{-\beta S\}$. A 
precise calculation of the maxima and minimum of the 
double-peak structure of $P(\Omega)$ requires frequent 
tunneling through the minimum, implying exponentially large 
autocorrelation times $\tau\propto\exp\{2\sigma L^{d-1}\}$.

The idea of the so-called {\it multicanonical} approach is to generate
an auxiliary distribution $P'(\Omega)$ (see Fig.~\ref{fig:6}) which
does not suffer from the original problems, but can be related to the
canonical distribution in a controllable way. Fig.~\ref{fig:6} gives
an example for the auxiliary distribution as function of the action
density $s\equiv S/V$, which is approximately flat between the two
maxima. We have
\begin{equation}
  P_L'(S) \propto n_L(S) P_B'(S)\;.
  \label{3.75}
\end{equation}
The index $L$ stands for the linear size $L$, $n_L(S)$ 
denotes the spectral density and $P_B'(S)$ has the form of 
a Boltzmann weight factor. In practice one starts with a 
guess for the parameters in the exponent of $P_B'(S)$ such 
that $P_L'(S)$ is approximately flat. Configurations with 
the multicanonical probability $P_B'$ are generated with a 
generalized Metropolis or heat bath updating procedure. The 
original distribution $P_L(S)$ (or $P_L(\Omega)$) is then 
obtained from $P_L'(S)$ by reweighting (Baumann and Berg, 
1985; Ferrenberg and Swendsen, 1988 and 1989).

The efficiency of the multicanonical algorithm has been 
well established (see e.g. Berg and Neuhaus, 1992; Janke, 
1992). The algorithm has been generalized to an $SU(3)$ 
gauge theory by Grossmann et al. (1992), where the time 
extent was restricted to $N_\tau=2$. For $N_\tau=4$ 
Grossmann and Laursen (1993) have used the histograms of 
the QCDPAX collaboration (Iwasaki et al., 1991 and 1992a) 
and Fukugita et al. (1989a and b). Their $N_\tau=4$-
results for $\sigma$ are compatible with the results of 
Iwasaki et al. (1994) which we discuss in the following.

The results of Iwasaki et al. (1994) are based on the 
histograms of the QCDPAX-collaboration for lattices of 
temporal extent $N_\tau=4$ and 6 and spatial sizes ranging 
from $N_xN_yN_z=12^2\times24, 24^2\times 36, 20^3, 24^3$ and 
$36^2\times48$. An extrapolation to the infinite volume 
limit leads to
\begin{equation}
  \sigma/T_c^3 = \left\{
    \matrix{
      0.0292(22) & {\rm for}~N_\tau=4&\cr
      0.0218(33) & {\rm for}~N_\tau=6&.\cr
      }\right .
  \label{3.76}
\end{equation}
The results do not depend on the actual choice for the order parameter
($\Omega_{rot}$ or $\Omega_{abs}$). To assess the relevance of these
numbers for the continuum limit, one has to check the scaling
property, i.e. sending $N_\tau\to\infty (a\to0)$ while keeping $T_c$
in physical units fixed. It is obvious from Eq.\ (\ref{3.76}) that
$\sigma/T_c^3$ violates scaling. The scaling violation is even more
obvious, if the $N_\tau=2$ value is included
($\sigma/T_c^3=0.103(7)$). A similar violation of scaling has been
observed for the latent heat $\Delta L$, when it is measured from the
same data of the QCDPAX-collaboration (Iwasaki et al., 1991 and
1992a). Thus it is not surprising that the scaling violation
approximately drops out, if certain ratios of $\Delta L$ and $\sigma$
are considered.  Such a ratio is given by
\begin{equation}
  \alpha^2 = 16 \pi \sigma^3 /\left( 3 \Delta L^2 \cdot T 
  \right)\;,
  \label{3.77}
\end{equation}
which enters the amount of supercooling and the average 
distance between nucleated bubbles in the early universe. 
Let us see, what the results for $\sigma/T_c^3$ of Iwasaki 
et al. (1994) imply for possible relics of the QCD 
transition in the early universe.

\subsubsection{Phenomenological implications for the early universe}
\label{sec:III.B.4}
The QCD transition happened at $\sim10^{-6}$sec after the 
big bang, when the universe cooled down to a temperature of 
$\sim(150-200)$~MeV. In the following we describe one 
possible scenario for the succeeding evolution which has 
been frequently discussed in the literature (see e.g. 
Kajantie and Kurki-Suonio, 1986; Fuller et al., 1987; Meyer 
et al., 1991; Applegate et al., 1987; Applegate, 1991; 
Schramm et al, 1992).

If the QCD transition is of first order and proceeds via 
bubble formation, the nucleation will not set in exactly at 
$T_c$, but start at some lower temperature $T_i$ at time 
$t_i$. The plasma phase supercools, until the gain in free 
energy due to the phase conversion can compensate the costs 
in interface free energy between the coexisting phases. The 
degree of supercooling cannot be too large, as the cooling 
of the universe goes adiabatically slowly as viewed from the 
scale of QCD.

In the vicinity of the QCD transition two very different time scales
enter the evolution: the QCD time scale of the order of
$1/T_c\sim1~\rm{fm/c}\sim 0.33\cdot 10^{-23}~\rm{s}$ and the Hubble
time of the order of $1/\chi=0.36\cdot 10^{-4}~\rm{s}\sim 10^{19}/T_{c}$
with $\chi=\sqrt{8 \pi G B/3}$, where $G$ is the gravitational
constant and $B$ the vacuum energy density represented by the bag
constant.  The time dependence of temperature $T(t)$ follows from
Einstein's equations combined with QCD's equation of state (e.g. the
bag model equation of state (Kajantie and Kurki-Suonio, 1986)). It is
given by
\begin{equation}
  {- dT \over T dt} = {1 \over 2 t} \propto {T^2 \over M_{Pl}}\;
  \label{3.78}
\end{equation}
with $M_{Pl}$ denoting the Planck mass,
i.e. the expansion rate is $10^{-19}$ times more slowly than typical
time scales in QCD.

At time $t_i$ and temperature $T_i$ nucleation sets in, 
bubbles of critical radius are formed. Their growth rate is 
slow compared to the shock waves which they emit. The shock 
waves propagate with the velocity of sound 
$v_{sh}=1/\sqrt3$. They reheat the plasma to some higher 
temperature $T_i$. The release of latent heat prevents 
further nucleation in regions of space which have been 
affected by shock waves. In such regions only bubbles will 
grow which have been created before. The creation and 
growth of bubbles continues until the shock waves 
collide, which are emitted from various nucleation centers.

At this time ($t=t_{PT}$), when the entire universe has 
been affected by shock waves, nucleation shuts off. A new 
scale becomes important. It is the {\it average distance R} 
between hadronic bubbles at the time when the shock waves 
collide. This scale is a measure for the inhomogeneities in 
the baryon number density. During the coexistence of both 
phases most of the baryon number resides in the plasma 
phase. Since the hadronic bubbles are nucleated through 
random thermal or quantum processes, they contain on 
average no net baryon number. In thermal and chemical 
equilibrium the chemical potentials $\mu_Q(B)$, $\mu_h(B)$, 
associated with the elementary baryon charge, are equal in 
both phases (h and q). The ratio of the baryon number 
density in the quark ($q$) and hadron ($h$) phases 
$n_B(q)/n_B(h)$ is approximately given by 
$\exp(m_Nc^2/T_c)$, where $m_N$ is the nucleon mass, since 
the light current quark masses are almost zero.

A competing scale to the scale of baryon number density 
fluctuations is the {\it neutron diffusion length}. The 
diffusion length of neutrons depends on the age of the 
universe. In the period where the age of the universe is 
$20~\mu\rm{s}-1~\rm{s}$, the weak interactions are fast enough 
to ensure a uniform neutron to proton ratio. The ratio of 
neutrons to protons is given by $(\exp(m_n-m_p)/k_BT)$. 
Protons and neutrons diffuse along with each other. 
Neutrons really diffuse, protons diffuse by converting into 
a neutron through the weak interaction, diffusing, and 
converting back. At the age of $\sim$1sec, when the 
temperature falls to $\sim 1$~MeV, the weak and 
electromagnetic interactions decouple. The weak interaction 
drops out of equilibrium. The neutrons stay neutrons, the 
protons stay protons. The proton diffusion length is very 
short, because the proton is electrically charged. The 
diffusion length of neutrons is larger by several orders of 
magnitude. If the baryon number density fluctuations 
survive until the decoupling of weak interactions the 
neutrons will diffuse out of regions which are rich in 
baryon number. This way baryon rich regions are 
transformed into neutron poor (or proton rich) regions, 
vice versa, baryon poor regions become rich in the number 
of neutrons. The ratio $\exp((m_n-m_p)/k_BT)$ becomes 
spatially inhomogeneous.

Neutron rich and proton rich regions constitute different 
initial conditions for the primordial nucleosynthesis. An 
inhomogeneous universe uses neutrons less ``efficiently'' 
than a homogeneous one. (In proton rich regions 2 neutrons 
makes 1 alpha, while 4 neutrons are needed in a neutron 
rich part ($4n\to2p+2n\to1\alpha$). It has been shown by 
Applegate et al. (1987) that primordial nucleosynthesis in 
an inhomogeneous universe leads to a distinctive pattern in 
the light element abundances. In particular the production 
of $^7$Li is increased as compared to the standard scenario 
without baryon number inhomogeneities.

Thus the alluring prediction of this scenario is that the 
observation of the light element abundances nowadays 
provides a snapshot of the universe at its infancy (the 
time period of the QCD transition). Clearly it is a 
question of relative size of the competing scales, as to 
whether the inhomogeneities in the baryon number densities 
characterized by the mean interbubble separation $R$ 
transform into variations in the $n/p$-ratio determined by 
the neutron diffusion length. If $R$ is too large, 
neutron diffusion will not leave an effect. If $R$ is too 
small, the baryon number inhomogeneities are washed out, 
as both neutrons and protons diffuse over these scales. 
In addition back diffusion of neutrons into proton rich 
regions must be taken into account. They further reduce 
the range of length scales, over which the QCD 
inhomogeneities can effect the primordial nucleosynthesis.

\vskip12pt
Let us finally indicate, how the average interbubble 
distance $R$ depends on the surface tension $\sigma$. 
(Notice that in this context $\sigma$ belongs to one of the 
few quantities which can be calculated from first 
principles and do not depend on the ``choice'' of the 
scenario.) The average separation $R$ defined at the time 
when the shock waves collide is given by (Kajantie and 
Kurki-Suonio, 1986)
\begin{equation}
  R = t_{growth} \cdot v_{sh}\;,
  \label{3.79}
\end{equation}
where $v_{sh}=1/\sqrt3$ is the velocity of the shock waves 
(as introduced above), $t_{growth}$ denotes the length of 
the time interval between the onset of nucleation and its 
completion in the sense that shock waves start colliding. 
The duration $t_{growth}$ is known, once the fraction $f$ 
of the universe which has been affected by shock wave 
fronts goes to 1. This fraction is determined by the 
nucleation rate $\Gamma$ per unit time and volume 
according to
\begin{equation}
  f(t) = \int\limits_{t_i}^t dt' \Gamma (t') {4\pi \over 3} 
  \left[v_{sh} \cdot (t'-t) \right]^3 \;.
  \label{3.80}
\end{equation}

In classical nucleation theory the rate $\Gamma$ is 
proportional to $T^4 \cdot \exp (-F_b/T)$, where $F_b$ is 
the free energy of the bubble of critical size or the 
minimal work for nucleating a bubble. Before the nucleation 
the free energy $F_{before}=-p(V+V')$, where $V$ and $V'$ 
are the volumes of the coexisting phases. After the 
nucleation we have $F_{after}=-pV-pV'+\sigma\cdot A$. For 
one spherical bubble it follows $F_{after}-F_{before}=-(p'-
p){4\pi\over3}r^3+\sigma\cdot4\pi r^2$. A bubble of 
critical radius satisfies $\partial(F_{after}-
F_{before})/\partial r=0$, leading to
\begin{equation}
  F_b = {16 \pi \over 3} {\sigma^3 \over (p_h - p_q)^2}\;.
  \label{3.81}
\end{equation}
It remains to express the difference in pressures of both 
phases via the latent heat $\Delta L$. We have $\Delta L = 
T_c {\partial \over \partial T} (p_q - p_h)$ such that in 
the vicinity of $T_c$ it follows $(p_h - p_q) = \Delta L 
\cdot (T_c - T)/T_c$ or
\begin{equation}
  F_b=\alpha^2 T_c \cdot \left( {T_c - T \over T_c}\right)^{-2}
  \label{3.82}
\end{equation}
with $\alpha^2 = (16 \pi \sigma^3 / (3 \Delta L^2 \cdot 
T_c))$. Notice that the surface tension and the latent heat 
enter the nucleation rate only in the combination of 
$\alpha$. The {\it degree of supercooling} is determined as 
$(T_c - T_{PT}) / T_c$, $T_{PT}$ is defined as the 
temperature, where the nucleation shuts off or the fraction 
$f(t)$ approaches 1. From $\Gamma(T)$ and $f(T)$ the degree 
of supercooling is finally obtained as (Iwasaki et al., 
1994)
\begin{equation}
  (T_c - T_{PT}) / T_c \propto \alpha \cdot \tilde\chi^{-1/2}
  \label{3.83}
\end{equation}
and the average separation of nucleation centers as
\[
  R \approx v_{sh} \pi^{1/3} e^{\tilde\chi / 4} \cdot 
  \tilde\chi^{-3/2} T_c^{-1} \cdot \alpha
\]
with
\begin{eqnarray}
  \label{3.84}
  \tilde\chi &&\equiv 4 \ln (T_c^2 / {\cal C}) \\
  {\cal C} &&= - \partial T / \partial t\;. \nonumber
\end{eqnarray}
For the QCD transition in the early universe 
$\tilde\chi\approx173$, $T_c\approx150$~MeV, 
$v_{sh}\approx1/\sqrt3$. Now we are ready to insert the 
value for $\alpha$ which has been obtained by Iwasaki et 
al. (1994) from the measurement of the surface tension and 
the value $\Delta L$ taken from the QCDPAX collaboration 
(Iwasaki et al., 1991 and 1992a). The degree of supercooling 
turns out to be as small as
\begin{equation}
  (T_c - T_{PT}) / T_c = 5.6 (1.4) \times 10^{-4}~,
  \label{3.85}
\end{equation}
and the average distance between hadronic bubbles
\begin{equation}
  R = 22(5)~{\rm mm}\;.
  \label{3.86}
\end{equation}
The value of $R$ is clearly too small to lead to an observable relic
in the present day light element abundances (see e.g. Meyer et al.,
1991).

One should keep in mind that the calculation of $\sigma$ in 
the pure $SU(3)$ gauge theory does not directly apply to 
the QCD transition in the early universe, although the main 
conclusions probably would remain unchanged. Dynamical 
matter fields will not increase the degree of supercooling. 
In general matter fields weaken (or even completely wash 
out) a first order transition.

\subsection{Including dynamical fermions}
\label{sec:III.C}
The ambiguity in studying the order of the deconfinement transition in
the $SU(3)$ gauge theory has been caused by the finite volume. A
finite size scaling analysis has resolved the controversy. It is
natural to attempt a finite size scaling analysis also in the presence
of dynamical fermions. Other artifacts can superimpose on finite size
effects. In part 1 of the following section we summarize some typical
pitfalls in lattice calculations. In the second part we report on the
{\it finite size scaling analysis} of the chiral transition by
Fukugita et al.  (1990b). The results are not yet conclusive. Possible
effects of finite quark masses are disregarded, although finite masses
amount to an ordering effect, which may well compete with the finite
volume. Therefore we turn in part 3 to a {\it finite mass scaling
  analysis}. It serves to discriminate possible UV-artifacts in the
chiral transition. {\it Bulk transitions} are a further manifestation
of UV-effects. With the example of $N_f=8$- flavor QCD we illustrate
bulk transitions in section 4.  Results for the more realistic cases
of 2 and 3 flavors are summarized in section 5.
The equation of state for two-flavor QCD is the topic of section 6.
In section 7 we
summarize some progress which has been achieved with {\it Wilson
  fermions}.

\subsubsection{Pitfalls on the lattice}
\label{sec:III.C.1}
\begin{itemize}
\item {\it Finite size effects.} Finite size effects are specific for
  the Monte Carlo approach, which is frequently used in lattice
  calculations. They contaminate any numerical calculation which is
  performed in a finite volume, if the largest correlation length of
  the system is not small compared to the smallest linear size. In the
  case of first order transitions we have presented a
  phenomenological approach (cf. section \ref{sec:II}). The formulas
  of section \ref{sec:II} for the scaling of the specific heat or the
  susceptibility as function of the linear size $L$ hold in the {\it
    zero external field limit $h=0$}.
  
  As we have seen in section \ref{sec:III.A}, Monte Carlo calculations
  of QCD with dynamical fermions are necessarily performed at finite
  bare quark masses. The updating time is estimated to increase
  proportional to $m_f^{-11/4}$. The chiral limit $m_f\to0$ must be
  extrapolated. Thus a second pitfall is a contamination of data
  through finite mass effects.
  
\item {\it Finite mass effects}. In section \ref{sec:II} we have
  stressed the similarity between the finite volume $L^3$ and an
  external field $h$, and between an external field and finite quark
  masses. In the vicinity of $T_c$ one may consider a {\it finite mass
    scaling analysis} at vanishing $1/L$ rather than a {\it finite
    size scaling analysis} at vanishing $h$. For a moment let us
  assume that the condition $1/L\to0$ is satisfied to a sufficient
  accuracy.  A mass scaling analysis then provides a tool to control
  the rounding of algebraic singularities in correlation functions due
  to finite quark masses. The precondition is a second order phase
  transition.  (An analogous analysis may be performed for a first
  order transition and rounding effects due to finite masses as well).
  The need of extrapolation to zero masses is specific for exploring
  the chiral limit, when dynamical fermions are included.
  
  A third class of pitfalls which is known from the pure $SU(N)$ gauge
  theory and is common to any latticized version of a continuum model
  are artifacts due to the UV-cut-off.
  
\item {\it Artifacts due to the UV-cut-off.} Here we distinguish three
  manifestations. The most obvious one occurs in the interpretation of
  any physical observable, when the lattice units are translated to
  physical units. In section \ref{sec:III.A} we have already argued
  about the ambiguity in predicting $T_c$~[MeV]. Such an ambiguity
  occurs for the QCD transitions in the presence of dynamical quarks,
  when the quark mass in lattice units, e.g. $ma=0.025$, shall be
  translated in units of MeV, although the lattice is rather coarse
  grained and a splitting of a mass according to $m^.a$ is not unique.
  
  A less obvious and more subtle UV-artifact is a change in the
  effective symmetry group of the lattice action as a function of the
  bare coupling. This artifact is lately a consequence of the
  `No-Go'-theorems referring to lattice regularized actions for
  fermions. As we have mentioned in section \ref{sec:III.A}, the
  lattice action with massless fermions has a global $U(n)\times U(n)$
  symmetry for $n$ species of staggered fermions. Only in the
  continuum limit the full $SU(N_f)\times SU(N_f)$ flavor symmetry
  will be restored with $N_f=4^.n$. Far outside the continuum region
  the lattice action has only the global $U(1)\times U(1)$ symmetry
  for $n=1$.
  
  A global $U(1)\times U(1)$ symmetry may trigger a second order phase
  transition with $O(2)$- critical exponents, whereas the spontaneous
  breaking of the restored $SU(4)\times SU(4)$ symmetry in the
  corresponding continuum limit is supposed to induce a first order
  transition (Pisarski and Wilczek, 1984). Thus the order of the
  chiral transition may change when passing from strong to weak
  couplings. Such a possibility has been discussed by Boyd et al.
  (1991) and will be the topic of section 3.
  
  If one expects an $SU(2)\times SU(2)$ flavor symmetry for the
  continuum limit, the continuum phase transition would be also of
  second order. In that case only the critical exponents would change
  in passing from strong to weak couplings.
  
  If one wants to describe an odd number of flavors in the continuum
  limit within the staggered fermion formulation, one usually
  ``corrects'' for the desired number of continuum flavors on the
  level of determinants.  The correction is performed after
  integrating over the fermionic degrees of freedom, whereas the
  fermionic operator $Q$ is that of the original lattice action. The
  flavor symmetry in the original action is broken and describes at
  least four flavors in the continuum limit. The representation of the
  prefactor of the determinant-term in Eq.\ (\ref{3.14}) as $N_f/4$
  cannot be derived. It should be considered as a {\it prescription}
  when $N_f$ is not an integer multiple of 4. In this case it is even
  more difficult to infer the symmetry that drives the phase
  transition and leads to a vanishing condensate above $T_c$.
  
  So far the UV-effects are supposed to modify the critical indices or
  the order of the chiral transition.  A third manifestation of the
  UV-cut-off are {\it bulk transitions}. Here the very transition
  itself is an artifact of the lattice. Bulk transitions are not
  restricted to QCD with dynamical fermions. They occur in pure
  $SU(N)$ gauge theories as well. As an illustration we consider the
  example of $N_f=8$-flavor-QCD in section \ref{sec:III.C.4}. For 8
  flavors the bulk transition is either dominant or superimposed on
  the usual finite temperature transition of QCD.
\end{itemize}

\subsubsection{Finite size scaling analysis}
\label{sec:III.C.2}
We report on the work by Fukugita et al. (1990b) on a finite size
scaling analysis of the chiral transition in the presence of fermions.
The fermionic action is given by Eqs.\ (\ref{3.12}). The quark mass
$m_f$ is chosen as $0.025$ for $N_f=2$ flavors and as $0.025$ and
$0.0125$ for $N_f=4$ flavors (with $l=N_f/4$ in Eq.\ (\ref{3.14})).
The chiral condensate $\left\langle\bar\chi\chi\right\rangle$ and the
Wilson line $\Omega$ are take as order parameters, although they
strictly have this meaning only in the limiting cases $m_f\to0$ and
$m_f\to\infty$, respectively.  The expectation value of the chiral
condensate is calculated according to Eq.\ (\ref{3.112}) below. The
Wilson line expectation value is defined as
\begin{eqnarray}
  \left<\rm{Re}~\Omega\right>&&={\left<\sum_x\hbox{Re 
        }\Omega(x)\right>\over N_\sigma^3} \nonumber\\
  \label{3.87}
  \\
  \rm{Re}~\Omega(x)&&=N^{-1}\hbox{ Re Tr 
    }\prod_{\tau=1}^{N_\tau}U_{x,\tau}^{(4)} \nonumber
\end{eqnarray}
with $N=N_\sigma^3{}^.N_\tau$. Otherwise the notations have been
introduced in section \ref{sec:III}. The expectation value
$\left<\ldots\right>$ is calculated with the measure and the
effective action of Eqs.\ (\ref{3.14}). The temporal extent is fixed
to $N_\tau=4$, while the spatial size is varied between
$N_\sigma^3=4^3$ to $12^3$. The adopted algorithm is the R-algorithm
of Gottlieb et al. (1987a) with a time step size of $\Delta\tau=0.02$.
The accuracy of this algorithm has been checked against the exact
hybrid Monte Carlo algorithm.

Fig.~\ref{fig:7} shows the time history of the Wilson line Re~$\Omega$
for $N_f=2$ and $m_f=0.025$ on an $N_\sigma^3\times4$ lattice for
various spatial lattice extents. It illustrates that the `naive'
criteria for identifying a first order chiral transition are not
unique on small lattices.

The signatures of interest are signs of metastabilities and associated
two-state signals.

The time evolution should be compared with Fig.~\ref{fig:8}, where the
time history of Re~$\Omega$ is displayed for the 4-flavor case at
$m_f=0.025$ on an $N_\sigma\times4$ lattice for several values of
$N_\sigma$. In the $N_f=4$ case distinct flip-flops are seen for
$N_\sigma=$6 and 8, and a single event for $N_\sigma=12$. (The
tunneling rate goes to zero in the infinite volume limit.) For
$N_\sigma=10$ no flip-flop is visible, but a metastability is observed
as a two-state separation. When transitions between both phases occur
via tunneling and tunneling events are rare, the phases are metastable
over a long time. The system remains in the ordered phase over
thousands of iterations starting from an ordered start and in the
disordered phase for a random start.

A crucial ambiguity is hidden in what is called `long time'. A
comparison of Figs.~\ref{fig:7} and \ref{fig:8} shows a significant
difference in time scales. The fluctuations in Fig.~\ref{fig:7} look
more irregular on a large scale. On a short time scale the
interpretation may change. Consider Fig.~\ref{fig:7} for the
$8^3\times4$ lattice. If the runs would have been stopped after 2000
iterations, the fluctuations (probably of statistical origin) could have 
been misinterpreted as a two-state signal. In the longer run of 10000
iterations they look like a statistical fluctuation.

Fig.~\ref{fig:8} indicates for the $10^3\times4$ lattice, how
important a check of the dependence on the initial conditions is
(whether one uses a hot or a cold start).  From a single run one can
never exclude that a flip-flop will occur, if one waits for a long
enough time. For large volumes the metastabilities can be so
pronounced that the system is in one phase over the entire simulation.
The phase is interpreted as a truly stable phase and a first order
transition is easily overlooked. Thus the choice of the volume has to
be optimized. It may not be too small to see distinct flip-flops, but
not too large to see them at all.

Further indications for the first order transition are 
taken from the finite size scaling behavior of various 
susceptibilites. Susceptibilities are `magnetic' response 
functions which can be expressed in terms of an order 
parameter $O$ according to
\begin{equation}
  \chi=V\left(\left\langle O^2\right\rangle-
    \left\langle O\right\rangle^2\right)\;.
  \label{3.88}
\end{equation}
Fukugita et al. have investigated the finite size scaling 
of $\chi$ for different choices of the order parameter 
field: the Wilson line Re~$\Omega$, the chiral condensate 
$\bar\chi\chi$ (not to be confused with the susceptibility 
itself), and the average plaquette $P\equiv\hbox{Tr 
}U(\partial P)$ ($U(\partial P)$ denotes the product of 
$U$'s along the boundary of a plaquette).

From section \ref{sec:II} and Table~\ref{tab:2} we recall that the
peak in the susceptibility $\chi_{max}$ should scale with the volume
$V$ according to $\chi_{max}=const+a^.V^p$, where $p=1$ for a first
order transition and $p<1$ for a second order transition. The constant
accounts for a contribution from the regular part of the free energy,
which cannot be neglected as long as the volume is relatively small.

Runs with $N_\sigma\geq10$ showing no flip-flops are of no 
practical use for the finite size scaling test with $\chi$, 
since the formulas are based on a phase coexistence in the 
sample with a non-vanishing weight. A two-parameter fit for 
$a$ and $p$ of the ansatz for $\chi_{max}$ with various 
values of $const\;(\geq0)$ leads to an estimate of 
$0.78<p<1.1$ for the Wilson line-susceptibility in the 
case of 4 flavors. For comparison we mention the result 
for the pure $SU(3)$ gauge theory with a first order 
transition (Fukugita et al., 1989 and 1990a), where the 
exponent $p$ has been estimated as $0.86<p<1.0$. The 
authors conclude from the estimate in the 4-flavor case 
that it is consistent with the first order nature of the 
transition.

Results for the other susceptibilites lead to even larger 
bounds on $p$, in particular for the 2-flavor case they are 
not conclusive.

For a further criterion to test on the order of the phase transition
we recall the definition of Binder's cumulant Eq.\ (\ref{2.40}). The
minimum of Binder's cumulant vanishes for a second order transition,
but goes to a finite value $(\geq0)$ in the case of a first order
transition. Fukugita et al. have used a slightly different
normalization
\begin{equation}
  V_L=1-{1\over 3}{\left\langle P^4\right\rangle\over 
    \left\langle P^2\right\rangle^2}\;.
  \label{3.89}
\end{equation}
Here $P$ denotes the average plaquette. The minimum 
$V_{L,min}$ should approach 2/3 for a second order 
transition, as the volume increases. (For the explicit 
volume dependence cf. Eq.\ (\ref{2.41})). When $V_{L,min}$ is 
plotted as a function of $1/V$, the trend is obvious that 
$V_{L,min}$ does not approach 2/3 for 4-flavor QCD. This 
is a further indication for a 1st-order transition. For 2 
flavors the deviation from 2/3 is smaller than for 4 
flavors, although larger than in the pure gauge limit 
$(N_f=0)$. The difference of $(V_{L,min}-2/3)$ goes 
almost to zero for $N_f=0$. This is rather surprising, as 
the transition is known to be of first order for zero 
flavors. Thus the analysis of Binder's cumulant turns out 
to be not conclusive as well.

It has been already suggested by Fukugita et al. (1990b) 
that the relatively large quark mass of $m_f=0.025$ may 
considerably interfere with the finite size of the system. 
Therefore the 2-flavor simulations have been repeated for a 
smaller mass value of $m_f=0.0125$ (Fukugita et al., 1991). 
The characteristic features are very similar to those for 
$m_f=0.025$, and the order of the 2-flavor transition could 
not be determined there either. The favored possibilities 
are `no transition for finite $m_f$' or `a second order 
transition in the $m_f\to0$-limit', but a first order 
transition is not excluded.

This concludes our discussion of IR-artifacts. We turn to 
UV-artifacts in the next two sections.

\subsubsection{Finite mass scaling analysis}
\label{sec:III.C.3}
In this section we discuss the UV-artifacts on phase 
transitions in some more detail. We refer to the work by 
Boyd et al. (1991) dealing with possible changes in 
critical indices and a change in the order of the chiral 
transition as function of the bare coupling.

In the strong coupling limit the staggered fermion action of QCD is
given by Eqs.\ (\ref{3.12}) with one difference. An additional
coupling $\gamma$ is introduced in the action, which is related to the
anisotropy in the spatial and temporal lattice spacing in the weak
coupling limit. It should not be mixed up with the critical index
$\gamma$ which is also considered later. The notation should be clear
from the context. The fermion operator $Q$ is written as
\begin{equation} 
  Q_{xx'}=\sum_{\mu=1}^3D_{xx';\mu} + \gamma D_{x,x';\sigma} + 
  m^.\delta_{xx'}
  \label{3.90}
\end{equation}
with $D$ as in Eqs.\ (\ref{3.12}). The temperature is varied by
varying the coupling $\gamma$ on lattices with fixed number of time
slices $N_\tau$. The chiral symmetry restoring transition is observed
at some value $\gamma(N_\tau)$ with $\gamma (N_\tau)\to\infty$ for
$N_\tau\to\infty$. The role of $\gamma$ is analogous but not identical
to the anisotropy parameter $\xi$ (Eq.\ (\ref{3.19})) at weak
couplings which is directly related to the lattice spacings $a_\beta$
and $a_\sigma$ and this way also to the temperature
$T=1/(N_\tau{}^.a_\beta)$. The particular role of the strong cooling
limit can be guessed from the special case where $\xi=1$ and
$a_\sigma=a_\beta$. An attempt to reach the high temperature region in
the strong coupling $(\beta=0)$-limit requires $N_\tau<1$.

For vanishing quark masses the action (\ref{3.12}) for one species of
staggered fermions ($n=1$) is invariant under global $U(1)\times U(1)$
transformations acting on even lattices sites according to
\[
  \chi'=e^{i\alpha}\chi\hskip1cm\bar\chi'=e^{-
    i\beta}\bar\chi
\]
and on odd sites
\vskip-12pt
\begin{equation}
  \chi'=e^{i\beta}\chi\hskip1cm\bar\chi' = e^{-
    i\alpha}\bar\chi\;.
  \label{3.91}
\end{equation}
For finite mass $m$ in Eq.\ (\ref{3.90}) this symmetry is explicitly
broken to $U(1)$.

Mean field calculations of Damgaard et al. (1986) and 
F\"aldt and Petersson (1986) and numerical calculations 
of Klaetke and M\"utter (1990) for $SU(2)$ suggest that 
the chiral QCD transition is of second order in the 
strong coupling limit. Second order transitions are 
immediately washed out in the presence of an external 
field, as we know from statistical physics. The 
singularities in thermodynamical functions will be 
rounded, when the infinite volume limit is taken at fixed, 
non-vanishing quark mass.

The critical index characterizing the `finite mass scaling' at
criticality is $1/\delta$ (cf. Table~\ref{tab:2}). It is defined as
\begin{mathletters}
  \label{3.92}
  \begin{equation}
    \lim_{N_\sigma\to\infty}\left. 
      \langle \bar\psi \psi \rangle(m,t,N_\sigma) \right|_{t=0} 
    \propto m^{1/\delta}\;,
    \label{3.92a}
  \end{equation}
  where the reduced temperature $t=(T-T_c)/T_c$ is now replaced by
  \begin{equation}
    t={\gamma-\gamma_0\over\gamma_0}\;,
    \label{3.92b}
  \end{equation}
\end{mathletters}
and the order parameter
\begin{equation}
  \langle\bar\psi\psi\rangle={1\over N_\sigma^3N_\tau}
  {\partial\over\partial m}\ln Z(m,\gamma)
  \label{3.93}
\end{equation}
is the chiral condensate of strong coupling QCD. Here we have renamed
the staggered fermion fields $\chi$, $\bar\chi$ as $\psi$, $\bar\psi$
to avoid confusion with the susceptibility $\chi$ later on. It can be
expressed in terms of expectation values of monomers, which are easily
calculable on the lattice (see e.g. Karsch and M\"utter, 1989).

Further recall the critical index $\gamma$ (cf. Table~\ref{tab:2})
which specifies the singular behavior of the susceptibility $\chi$,
when $T$ approaches $T_c$ at zero field (here zero mass)
\begin{eqnarray}
  \chi(t)&&=\lim_{m\to0} \lim_{N_\sigma\to\infty} 
  \chi(t,m,N_\sigma) \propto t^{-
    \gamma} \nonumber\\
  \chi(t,m,N_\sigma)&&={\partial^2\ln Z\over\partial m^2}\;.
  \label{3.94}
\end{eqnarray}
Note the order of limits. To measure $\gamma$, first the infinite
volume limit has to be taken and next the zero mass limit. (In the
opposite order the order parameter would vanish also in the broken
phase due to tunneling events in the finite volume). Eq.\ (\ref{3.94})
is still a zero mass limit.

A finite {\it mass} scaling analysis proceeds in complete 
analogy to a finite {\it size} scaling analysis. The 
rounding and shifting effects on singularities are derived 
from a scaling ansatz for the free energy density. The non-
analytic part of the free energy density of a generic 
statistical ensemble is the presence of an external field 
(the finite quark mass) is written as
\begin{equation}
  f(t,m)=b^{-1}f\left(b^{y_t}{}^.t,b^{y_h}{}^.m\right)\;,
  \label{3.95}
\end{equation}
where $b$ is an arbitrary scale factor as in section \ref{sec:II},
$y_t$ and $y_h$ are the thermal and magnetic critical exponents
($y_t\equiv\lambda_1,y_h\equiv\lambda_2$ in our notation of section
II).  Choosing the scale factor
\begin{equation}
  b=m^{-{1\over y_h}}\;,
  \label{3.96}
\end{equation}
the free energy density transforms to
\begin{equation}
  f(t,m)=m^{1\over y_h} f\left(tm^{-{y_t\over y_h}},1\right)\;.
  \label{3.97}
\end{equation}
Eq.\ (\ref{3.97}) implies the finite mass scaling behavior of the
order parameter $\langle\bar\psi\psi\rangle(t,m)$ and the
susceptibility $\chi(t,m)$ in the vicinity of $\gamma_0$ (the critical
coupling)
\begin{equation}
  \langle\bar\psi\psi\rangle(t,m)= m^{1 \over \delta} F 
  \left( tm^{ - {y_t\over y_h }}\right)
\label{3.98}
\end{equation}
and
\begin{equation}
  \chi(t,m)=
  {1\over\delta}m^{{1\over\delta}-1}
  \left[
    F\left(tm^{-{y_t\over y_h}}\right) -
    {y_t\over1-y_h}tm^{-{y_t\over y_h}}
    F'\left(tm^{-{y_t\over y_h}}\right)
  \right]
  \;,
  \label{3.99}
\end{equation}
where $F$ and $F'$ are scaling functions. It follows that 
the peak of the finite mass susceptibility occurs at
\begin{equation}
  t_m=c^.m^{y_t\over y_h}
  \label{3.100}
\end{equation}
and scales according to
\begin{equation}
  \chi(t_0=0,m)\propto m^{{1\over\delta}-1}\;.
  \label{3.101}
\end{equation}
Eq.\ (\ref{3.101}) should be compared with Eq.\ (\ref{2.30}) for the
finite {\it size} scaling of the peak in the susceptibility. Eq.\ 
(\ref{3.100}) gives the shift in the critical coupling $\gamma_0$ or
in the reduced critical coupling $t_0\to t_m$ due to the finite quark
mass.

With $\delta=y_h/(1-y_h)$ it is seen that the exponent $\delta$ can be
measured either from the shift in $t_m$ according to Eq.\ 
(\ref{3.100}) or from Eq.\ (\ref{3.99}) for the susceptibility or the
mass dependence of the order parameter Eq.\ (\ref{3.98}) at
$t=t_0\equiv0$.

In case of a second order transition the exponent $\delta$ 
is characteristic for the universality class of the action. 
Thus a measurement of $\delta$ is of much interest for 
verifying that the restoration of the $U(1)\times U(1)$ 
symmetry drives the phase transition at $\gamma_0$.

Boyd et al. (1991) have proposed a related quantity, the {\it chiral
  cumulant}, to measure $\delta$ and $t_0$ from finite mass
calculations.  It has analogous properties to Binder's cumulant (cf.
Eq.\ (\ref{2.40})) and is defined as
\begin{equation}
  \Delta(t,m)={m\chi\over\langle\bar\psi\psi\rangle}
  ={1\over\delta}-{y_{t}x F'(x) \over y_{h} F(x)}\;.
  \label{3.102}
\end{equation}
with $x\equiv t\cdot m^{-y_{t}/y_{h}}$.  It follows from Eqs.\ 
(\ref{3.98}) and (\ref{3.99}) that $\Delta(0,m)$ gives $1/\delta$. The
slope of $\Delta$ increases with decreasing $m$
\begin{equation}
  \Delta'(m)=\left.{\partial\Delta \over \partial 
      t}\right|_{t=0} \propto m^{-{y_t\over y_h}}
  \label{3.103}
\end{equation}
in a way that the ratio $\Delta$ itself has a $\Theta$-
function shape for vanishing $m$
\begin{equation}
  \lim_{m\to0}\Delta(t,m)=\left\{
    \matrix{1&t>0\cr1/\delta&t=0\cr0&t<0\;.\cr}
  \right.
  \label{3.104}
\end{equation}
Eq.\ (\ref{3.104}) follows from the definition of $\Delta$ and Eqs.\ 
(\ref{3.98}) and (\ref{3.99}). When the ratios of $\Delta(t,m)$ are
plotted as a function of $t$ for various values of $m$, the curves
cross at the $m=0$ critical point $t_0$. The crossing comes from the
fact that $\Delta$ increases with decreasing $m$ for $\gamma>\gamma_0$
and decreases with decreasing $m$ for $\gamma<\gamma_0$. This behavior
is analogous to Binder's cumulant $(\langle O^4\rangle/\langle
O^2\rangle^2)$ where $O$ stands either for the order parameter or the
internal energy $E$. This way one may extrapolate the zero mass
critical point $t_0$ from a series of finite mass measurements, cf.
also (Binder, 1981).

Such a determination of the zero mass critical coupling is in
principle free of an uncontrolled extrapolation.  However, corrections
originate in irrelevant terms, which may not be sufficiently
suppressed in the vicinity of $\gamma_0$, and in contributions coming
from the regular part of the free energy (cf. section \ref{sec:II}). A
further source which leads to interfering effects with the finite mass
scaling behavior is the finite lattice volume used in the Monte Carlo
simulation. As we have seen above, the formulas (\ref{3.98}),
(\ref{3.99}), (\ref{3.100}), (\ref{3.104}) hold in the infinite volume
limit. In general finite mass- and finite volume effects are competing
ordering effects. It is a question of relative size of both effects
whether one is allowed to neglect one with respect to the other.

Let us compare the mass scaling behavior of the peak in the
susceptibility in the infinite volume (Eq.\ (\ref{3.99}))
\begin{equation}
  \chi_{peak}\propto m^{{1\over\delta}-1}\hbox{\hskip2ex for 
    }L\to\infty
  \label{3.105}
\end{equation}
with the finite size scaling behavior in the zero mass limit (Eq.\ 
(\ref{2.30}))
\begin{equation}
  \chi_{peak}\propto L^{\gamma\over\nu}\hbox{\hskip2ex for 
    }m\to0\;.
  \label{3.106}
\end{equation}
The correction coming from $L<\infty$ as $m\to0$ is small compared to
the rounding due to $m>0$ as $L\to\infty$, if
\begin{eqnarray}
  m&>const^.L^{-b} \nonumber\\
  b&={\gamma\delta\over\nu(\delta-1)}\;,
  \label{3.107}
\end{eqnarray}
following from Eqs.\ (\ref{3.105}) and (\ref{3.106}). On the other
hand the quark mass has to be chosen sufficiently small in order to
keep the contributions from the regular part of the free energy
density small. To keep the corrections small, the quark mass must be
reduced proportional to $1/\sqrt{N_\tau}$, when $N_\tau$ is increased
in the contributions from the regular part.

\vskip12pt The chiral condensate, the susceptibility and $\Delta$ have
been also measured in a Monte Carlo simulation for strong coupling QCD
using the monomer-dimer-algorithm (see e.g.  Karsch and M\"utter,
1989). The lattice size $N_\sigma^3\times N_\tau$ is chosen as
$N_\sigma=$4, 8, 16 and $N_\tau=4$. The values for the bare fermion
masses are $m=$0.005, 0.01, 0.02, 0.04 and 0.1. The results are in
agreement with the expectations: the chiral condensate vanishes in the
symmetric phase, when the mass approaches zero. In the critical region
$\gamma=2.3$-$2.4$ the finite size effects are strong. For a smaller
lattice a larger mass is necessary to keep the finite size effects
negligible in accordance with the condition (\ref{3.107}). In Eq.
(\ref{3.107}) the critical indices $\gamma$, $\delta$, $\nu$ are taken
from 3-dimensional $Z(2)$ or $O(N)$-models, for which $b\sim O(2)$.
Hence the finite size effects should be small for the $16^3\times4$
lattice over the entire mass range.

The chiral condensate $\langle\bar\psi\psi\rangle$ smoothly decreases
in the transition region $\gamma\sim2.3-2.4$ as function of $\gamma$
without any signal of a first order phase transition. The distribution
$P_m(\langle\bar\psi\psi\rangle)$ of the order parameter does not show
any sign of a double peak structure (cf.  section \ref{sec:II}). We
recall that the same lattice action as Eqs.\ (\ref{3.12}) with
(\ref{3.90}) in the weak coupling limit has led to a first order
transition (Gavai et al., 1990), where the continuum symmetry group
$SU(4)\times SU(4)$ of 4-flavor QCD seems to be sufficiently restored
to induce the first order transition.

The height of the peak in the susceptibility increases with decreasing
$m$, but the statistics was not sufficient to measure $\delta$
according to the formulas (\ref{3.98}), (\ref{3.99}).  Instead
$\delta$ and the critical coupling $\gamma_0$ are estimated from a
simulation of $\Delta$ with the result
\begin{equation}
  \matrix{2.35<\gamma_c<2.4&\cr
    0.18<{1\over\delta}<0.25&.}
  \label{3.108}
\end{equation}
The curves $\Delta(\gamma)$ for different quark masses 
cross in the vicinity of $\gamma=2.35$. The result for 
$\delta$ is in agreement with $\delta$ for the three-
dimensional $O(2)$ spin model ($\delta=4.755(6)$). It 
justifies the hypothesis that strong coupling QCD in the 
staggered fermion formulation is in the same universality 
class as the 3-d $O(2)$ spin model. The reason is the 
$U(1)\times U(1)$ remnant of the $SU(N_f)\times SU(N_f)$ 
continuum symmetry for 1 species of staggered fermions.

\vskip12pt A more subtle case is that of $N_f=2$-flavors in the
continuum, where the renormalization group analysis of Pisarski and
Wilczek suggests a second order transition with $O(4)$ critical
exponents.  This hypothesis has been recently questioned by Koci\'c
and Kogut (1995) in an analysis of the three-dimensional Gross-Neveu
model. Koci\'c and Kogut find mean-field scaling behavior rather than
the expected universality of the two-dimensional Ising model. This
sheds doubts on the conventional lore that actions for {\it composed}
scalars (i.e. scalar mesons composed of quark bilinears) share the
universality class with sigma models for {\it fundamental} scalars. In
particular the chiral transition in two-flavor QCD might have
mean-field rather than $O(4)$-critical exponents.

In Monte Carlo simulations with staggered fermions $l=n=N_f/4$ is set
to 1/2 in the effective action Eq.\ (\ref{3.14}). No sign for a first
order transition has been found by Brown et al. (1990a and b) for the
2-flavor case. Thus a second order transition is favored in the
massless limit. The question now arises as to whether this continuous
behavior is a remnant of the strong coupling $U(1)\times U(1)$
symmetry restoration or a restoration of the desired $SU(2)\times
SU(2)$ continuum symmetry with either $O(4)$- or mean-field exponents.
Karsch (1994) and Karsch and Laermann (1994) performed a finite mass
scaling analysis of critical exponents in 2-flavor QCD to decide which
of the three possibilities is realized: $O(2)$, $O(4)$ or mean-field
exponents. The exponent $1/(\beta\cdot\delta)=y_{t}/y_{h}$ was read
off from Eq.\ (\ref{3.100}), expressed in terms of the critical and
pseudocritical couplings $g_{c}$ and $g_{pc}$, respectively
\begin{equation}
  \label{3.109}
  6/g^{2}_{pc}(m_{q}\cdot a)=6/g_{c}^{2}(0)
  +(m_{q}a N_{\tau})^{1/\beta\delta} .
\end{equation}
The exponent $1/\delta$ was determined from the chiral cumulant Eq.\ 
(\ref{3.102}) evaluated at $x=t=0$. The results are summarized in
Table~\ref{tab:5} and compared with $O(4)$, $O(2)$ and mean-field
values. The QCD result for $1/(\beta\delta)$ is consistent with the
$O(2)$ and mean-field results, but slightly outside the error bars of
the $O(4)$ value. For $1/\delta$ the QCD result is less consistent
with mean-field, but in good agreement with the $O(4)$ and $O(2)$
values.

DeTar(1995) has analyzed the scaling behavior of
$\left<\bar\psi\psi\right>$ as function of a scaled temperature (cf.
Eq.\ (\ref{3.98})) with $O(4)$ critical exponents over a wide range of
available data. The agreement is good apart from $N_{\tau}=12$ data.
The origin of this discrepancy has still to be clarified. Possible
explanations are a shift of the crossover temperature as function of
$N_{\tau}$, erroneous extrapolations of lattice mass measurements or
the very scaling hypothesis with $O(4)$ exponents.  A clear
identification of mean-field values in two-flavor QCD would certainly
initiate much effort to get a deeper understanding of critical
behavior in field theories with fermions.

The caveat to see unwanted features of the strong coupling symmetries
also applies to the case of three flavors in the staggered fermion
formulation. In the ``worst'' case, a crossover phenomenon might be a
result of a second order transition in the strong coupling regime,
that is immediately washed out, when finite masses are included.  The
crossover phenomenon then would be an UV-artifact indicating that the
bare coupling $g$ was yet too strong.

\vskip12pt Once a second order phase transition has been identified, a
measurement of critical indices is not a minor detail for its further
characterization. As we have seen, at finite temperatures and fixed
$N_{\tau}$ the continuum limit coincides with the high temperature
limit. Thus one has to increase $N_{\tau}$ to shift the critical
coupling towards smaller values. The large volume limit is
necessary for the continuum limit at low temperatures. In Monte Carlo
simulations {\it extrapolations} to zero mass, zero lattice spacing
and infinite volume are unavoidable. In the vicinity of a second order
phase transition the critical exponents enter the extrapolation
formulas. Thus their correct identification and knowledge of their
precise values are needed for taking the right limits.

If finite mass- or finite size scaling analysis are not practicable or
not applicable , an alternative fermion formulation (Wilson fermions)
should be explored to reestablish the type of phase transition which
has been observed in the staggered formulation. In particular, when
odd numbers of flavors are described in the staggered fermion
formulation, it is difficult to control, how the trick of doubling and
reducing the flavor degrees of freedom affects the effective symmetry
of the lattice action which triggers the phase transition.

\subsubsection{Bulk transitions}
\label{sec:III.C.4}
Bulk transitions are phase transitions at zero temperature. 
Their very occurrence is a lattice artifact. Latticized 
systems are systems of statistical mechanics with their own 
dynamics. They do not care about a well defined continuum 
limit. Generic phase transitions can occur at some critical 
coupling, while the temperature is zero. Their physical 
meaning depends on the context. They have no physical 
relevance for continuum QCD, if they do not ``survive'' the 
continuum limit. Thus one would like to ignore them 
completely, but one is not allowed to do so.

The phenomenon of bulk transitions in lattice gauge 
theory is known from the pure gauge sector. According to 
the conventional lore on four-dimensional lattice gauge 
theories there should be no zero-temperature phase 
transition for non-abelian $SU(N)$ gauge groups, which 
separates the strong coupling ($g^2\gg 1$) from the weak 
coupling ($g^2\ll 1$) region. This is a desired feature, as 
it should guarantee that the continuum limit of lattice 
gauge theory includes both the confinement properties 
(proven on the lattice for strong couplings) and the 
asymptotic freedom in the weak coupling regime. A bulk 
transition may in principle destroy the confinement 
properties of the strong coupling regime.

Monte Carlo calculations for the $SU(2)$- and $SU(3)$- gauge groups
with fields in the fundamental representation have verified the
conventional lore. For $N\geq4$ first order bulk transitions have been
found for the $SU(4)$ lattice gauge theory by Moriarty (1981) and for
the $SU(5)$ theory by Creutz (1981). Creutz argues, however, that the
confinement property is not lost at weak couplings in spite of the
transition.

In $SU(2)$- and $SU(3)$ gauge theories one observes a rapid 
crossover phenomenon between the strong and weak coupling 
regime instead. This is explained by a nearby critical 
point in the $(\beta, \beta_A)$ plane, where $\beta\equiv 
2N/g^2$ as above and $\beta_A$ denotes the coupling of a 
${\rm Tr}_A U(\partial P)$ term in the adjoint 
representation of $SU(N)$.
For two colors we have a mixed $SU(2)$-$SO(3)$ lattice 
action. A small $\beta_A$ leads to a bulk transition 
(Bhanot and Creutz, 1981), which is absent for $\beta_A=0$.

The interplay of bulk and thermal transitions in $SU(2)$ and $SU(3)$
gauge theories has been recently reinvestigated by Gavai et al.
(1994), Gavai (1995), and Blum et al. (1995b). A bulk transition is
signalled, if the location of the phase boundary stays fixed in
coupling parameter space, independently of the number of time slices.
In contrast, continuum universality for the thermal transition
requires a shift of the critical couplings towards smaller values, as
$N_{\tau}$ is increased, cf. section \ref{sec:III.A}. For the mixed
fundamental/adjoint $SU(2)$-action Marthur and Gavai found a shift
towards weaker couplings of both the thermal and the bulk transition
boundaries at $N_{\tau}=6$. This result sheds some doubts on the
interpretation of the first order phase boundary as a bulk transition.

For the $SU(3)$ mixed fundamental/adjoint action the bulk and thermal
phase boundaries coalesce for $N_{\tau}=4$, but split into two lines
for small enough couplings and larger values of $N_{\tau}$
($N_{\tau}=6$, 8) (Heller et al., 1995). The shift of the thermal
transition line towards weaker couplings for increasing $N_{\tau}$
supports the hypothesis of continuum universality.

The peculiar behavior, which is sometimes found for the phase
structure of QCD with dynamical fermions, has been addressed to bulk
transitions in mixed fundamental/adjoint actions that are induced as
effective actions from the integration over fermions. It is still an
open question, whether the inclusion of dynamical {\it Wilson} fermion
leads to a strong enough adjoint term in the $SU(3)$-mixed action that
could explain the phase structure for Wilson fermions at large
$\kappa$ and $N_{\tau}=6$ (Rummukainen et al., 1995).

Similarly the integration over the eight dynamical flavors in the
staggered fermion scheme has been conjectured to induce an adjoint
term of the $SU(2)-SO(3)$ mixed action in the effective action. This
provides a possible explanation of the bulk transition, which is seen
in $N_f=8$-flavor QCD (for $N_\tau\geq8$-timeslices) (Brown et al.,
1992). We discuss the 8-flavor case now in more detail.

\vskip12pt The action for two species of staggered fermions is given
by Eqs.\ (\ref{3.12}). Integration over the fermionic variables leads
to
\begin{equation}
  S = -{1 \over 3}~\beta~\sum_p~{\rm Re}~{\rm Tr}~
  U(\partial P) - 
  {N_f\over 4} \ln\det ({\cal Q}{\cal Q}^+)
  \label{3.110}
\end{equation}
\vskip-6pt
\noindent
with
\[
  {\cal Q} = D + m\cdot a\;,
\]
and $D$ is the Dirac operator acting on an $SU(3)$-triplet field
$\phi$ according to
\begin{equation}
  (D\phi)_x = {1 \over 2}~\sum_\mu~\Gamma_\mu(x)~
  [U_x^{\mu+} \phi_{x+\hat\mu}-
  U_{x-\hat\mu}^\mu \phi_{x-\hat\mu} ]~.
  \label{3.111}
\end{equation}
Here the number of continuum flavors $N_f$ is chosen as 
8, corresponding to $l=N_f/4=2$ species of staggered 
fermions. The dynamical quark mass $m\cdot a$ is chosen 
flavor independently as 0.015 throughout all simulations. 
The gauge coupling $\beta=6/g^2$ is varied between 4.5 
and 5.0. The lattice size is $16^3\times N_\tau$ with 
$N_\tau=4$, 8, 16 and 32.

One may wonder why the special case of 8 light flavors is 
of any interest at all, as only 2 (or 3) quark flavors 
are approximately massless in nature. Nevertheless the 
reason for us is a physical one. Usually the strength of 
the chiral transition is argued to grow with an 
increasing number of flavors. Earlier work on $8^3\times 4$ 
(Kogut et al., 1985; Fukugita et al., 1988), $6^4$ (Kogut 
and Sinclair, 1988a), $8^4$ (Kogut and Sinclair, 1988a; 
Fukugita et al., 1988),
and $16^3\times 4$ and $\times 6$ (Ohta and Kim, 
1991) lattices shows a strong first order transition. A 
comparison of the transition for 2, 3, 4 and 8 flavors 
reveals a strengthening of the transition as the number 
of flavors increases (Ohta and Kim, 1991; see e.g. also 
Gottlieb, 1991), where the range of time slices $N_\tau$ 
lies between 4 and 8.

Usually this tendency is interpreted as reflection of a 
physically plausible effect: The chiral transition gets 
more pronounced the higher the number of flavors is which 
drive the transition, whatever the `driving dynamics' in 
detail may be. Results about the $N_f=8$-transition 
expose this tendency as a possible lattice artifact and 
give rise to further studies in effective models about 
the supposed flavor dependence of the chiral transition.

\vskip12pt\noindent Let us first give a qualitative description of the
diagram displayed in Fig.~\ref{fig:9}. It is partly conjectural and
summarizes the results for the $N_f=8$-transition of Brown et al.
(1992), see also (Christ, 1992b; Dong and Christ, 1992).  

The solid line locates the bulk transition separating the 
parameter space in a weak coupling 
($\beta\ge\beta_c=6/g_c^2$) and a strong coupling 
($\beta\le\beta_c$) phase. For a temporal extent 
$N_\tau\ge 8$ the transition becomes $N_\tau$-independent 
and occurs at the same critical coupling 
$\beta_c=4.73(1)$ for $N_\tau=8$ and 16 or at 
$\beta_c=4.64(1)$ for $N_\tau=8$ and $\beta_c=4.62(1)$, 
when the finite time step interval $\Delta\tau$ in the 
integration step is varied (see below). This behavior 
signals the bulk feature of the transition.

\vskip12pt {\it Identification of the lattice artifact.} As mentioned
above, in a usual finite temperature transition with a correspondence
in continuum field theory, the critical coupling (temperature) should
scale with $N_\tau$ such that the physical transition temperature
$T_c=(N_\tau a(g))^{-1}$ remains constant (cf.  section
\ref{sec:III.A}.). Thus it is the absence of a shift in $\beta_c$ when
$N_\tau$ is increased, which suggests the possible lattice artifact.
The persistence of the transition on a symmetric ($T=0$) $16^4$
lattice further supports the interpretation as a bulk transition.

While the strong coupling phase shares features with the 
usual chirally broken phase at $N_\tau=4$, 6, the 
structure of the weak coupling phase (on the right hand 
side of the bulk transition) is more complex and less 
obvious in its physical meaning. The weak coupling phase 
is crossed by a hypothetical finite temperature 
transition, which should be relevant for the continuum 
limit. It separates regions with chiral symmetry broken 
($\left<\bar\chi \chi\right>~\neq 0$) and chiral symmetry restored
($\left<\bar\chi \chi\right>~=0$) features. The conventional finite 
temperature transition may be recovered on very large 
lattices ($N_\tau>32$).

The solid squares refer to parameters of actual 
simulations, the open squares locate the critical 
couplings.

In what follows we sketch the criteria which have led to 
the conjectures of Fig.~\ref{fig:9}.
\vskip12pt\noindent
{\it Basic observables}. Basic observables are the chiral 
condensate and hadron masses. Of particular interest is 
their dependence on the bare quark masses. The chiral 
condensate of staggered fermion fields, defined as 
\begin{mathletters}
  \label{3.112}
  \begin{equation}
    \left\langle \bar\chi \chi \right\rangle = {1\over 
      3}~{1\over N_\sigma^3 N_\tau}
    \sum_x~\left\langle \bar\chi_x \chi_x \right\rangle
    \label{3.112a}
  \end{equation}
  with a sum over all lattice sites $x$ is estimated by
  \begin{eqnarray}
    \left\langle \bar\chi \chi \right\rangle&&= {1\over 
      3}~{1\over N_\sigma^3 N_\tau}
    \left\langle \left\langle \sum_{x x'}~h_x(D+m)_{x,x'}^{-
          1}~h_{x'} \right\rangle\right\rangle \nonumber\\
    =&& {1\over 3}~{1\over N_\sigma^3 N_\tau}
    \left\langle \left\langle \sum_{x x'}~h_x(m(D D^++m^2)^{-
          1})_{x x'}~h_{x'} \right\rangle\right\rangle .
    \nonumber\\
    \label{3.112b}
  \end{eqnarray}
\end{mathletters}
For each site $x$, $h_x$ is an independent, complex three-vector of
Gaussian random numbers. This representation of $\left<\bar\chi
  \chi\right>$ follows from the effective action in terms of
pseudofermionic fields $h_x$, cf. section \ref{sec:III.A}. Eq.\ 
(\ref{3.112b}) shows the nonlocality of the fermionic condensate. The
expectation value $<<\ldots>>$ denotes an average over gauge fields
and random three-vectors $h_x$.

The mass $m$ entering Eqs.\ (\ref{3.112}) should be identical with $m$
of the effective action, as it enters the determinant in the path
integral. Strictly speaking the full calculations should be repeated
for several quark mass values, if the quark condensate and the hadron
propagators are to be checked on their quark mass dependence. The
observables are then measured on new sets of equilibrated gauge field
configurations depending on the quark mass via the fermionic
determinant. This way of proceeding costs an enormous amount of
computing time.

In the actual measurements Brown et al. have calculated the
observables on the same set of configurations, generated for an action
with $m\cdot a=0.015$, but with varying {\it valence quark} masses:
$m_{\rm val}=0.004$, 0.01, 0.025, and 0.05. Valence quark masses enter
the observables via the propagator, whereas {\it sea quark} masses are
used for the determinant in simulations with dynamical fermions. The
mass entering Eqs.\ (\ref{3.112}) should be identified with a valence
quark mass.

A conclusive criterion for studying the chiral transition 
is the scaling behavior of the chiral condensate 
$\left\langle\bar\chi\chi\right\rangle$ and the pion mass 
$m_\pi$ as a function of the quark mass $m$ in the limit 
of $m\to0$. Spontaneous chiral symmetry breaking is 
indicated, if $\left\langle\bar\chi\chi\right\rangle$ 
stays finite for $m\to0$, while $m_\pi(m)$ should vanish 
in the same limit. From chiral perturbation theory and 
PCAC-relations it is expected that $m_\pi^2$ vanishes 
linearly in the quark masses. As it is argued by Brown et 
al., the limit $m_{val}\to0$ for fixed $m_{sea}$ may be 
conclusive as well.

{\it The simulation method} is the R-algorithm of 
Gottlieb et al. (1987a). The R-algorithm contains finite 
time-step errors of the order $(\Delta\tau)^2$, but 
requires half the number of Dirac propagator inversions per 
unit Monte Carlo time compared to the exact hybrid Monte 
Carlo method of Duane et al. (1987), when it used for the 8-
flavor case.

\vskip12pt\noindent
{\it Determination of the critical coupling.} An accurate 
determination of the critical coupling is essential for 
an identification of the bulk transition. Here one has to 
face the usual dilemma, which is encountered for {\it 
strong first order} transitions. If the volume is chosen 
too large, strong metastabilities make both phases stable 
within a large range of couplings. This requires long Monte 
Carlo runs on large lattices. Metastabilities are less 
pronounced when the volume is small, but then the data are 
contaminated with strong finite size effects.

This problem is solved in the simulations of Brown et al. 
by starting with a mixed phase configuration. Small 
changes in the coupling cause the system to rapidly 
evolve into one of the phases, which by itself is a 
typical signal for a first order transition. Starting 
from a mixed phase configuration, the evaluation of 
$\left<\bar\chi \chi\right>$ is followed for several values of 
$\beta$ to get upper and lower bounds on $\beta_c$.

\noindent
{\it The first order of the bulk transition} is concluded from
Figs.~\ref{fig:10} and \ref{fig:11}. Fig.~\ref{fig:10} shows the
evolution of $\left<\bar\chi \chi\right>$ starting from hot and cold
starts for $\beta=4.65$ on a $16^3\times 16$ lattice. The first order
is signalled by the persistence of two phases over a time scale, which
is considerably larger than the equilibration time. This criterion has
been mentioned in section \ref{sec:II} as one of the naive criteria
which are applicable for strongly first order transitions. A further
evidence comes from the jump in the order parameter seen in
Fig.~\ref{fig:11}, which is interpreted as a tunneling event between
coexisting phases in the transition region (cf. section \ref{sec:II}).
The evolution of the order parameter for the $16^3\times 32$ lattice
at $\beta=4.6$ is shown for a cold start. A cold start corresponds to
an `ordered' configuration typical for the weak coupling phase (for a
finite temperature transition typical for the deconfinement phase).
Since the critical coupling $\beta_c=4.73$, the value $\beta=4.6$
belongs to the strong coupling regime. The tunneling event that occurs
at $\tau\sim 250$ time units suggests that the enforced weak coupling
phase (via the initial configuration) becomes unstable at the strong
coupling value $\beta=4.6$.

\vskip12pt\noindent {\it Indications for chiral symmetry restoration}.
For $N_\tau=4$, 6 and 8 time slices the weak coupling phase (on the
right hand side of the solid line in Fig.~\ref{fig:9}) shows the
typical features of chiral symmetry restoration. The condensate
$\left<\bar\chi \chi\right>$ extrapolates linearly to zero as $m_{\rm
  val}\to 0$.  For $N_\tau=16$ and/or 32 the strong coupling phase has
the typical features of a chiral symmetry broken phase.

Thus one would like to observe indications of chiral symmetry
restoration in the weak coupling phase. The value of $\left<\bar\chi
  \chi\right>$ becomes smaller when going from strong to weak
couplings, but less small than expected from naive scaling arguments.
It extrapolates to zero as $m_{\rm val}\to 0$, but in a non-linear
way, while $m_\pi^2$ extrapolates to a non-vanishing value. It is
speculated that for $N_\tau=16$ and $\beta=4.65$ the weak coupling
phase is itself near a finite temperature transition region.

For a larger value of $\beta$ ($\beta=5.0$) chiral 
symmetric behavior is manifest: hadronic screening 
lengths show parity doubling, $\left<\bar\chi \chi\right>$ 
extrapolates linearly to zero as $m_{\rm val}\to 0$ and 
$m_\pi\neq 0$ for $m_{\rm val}=0$, varying little with 
$m_{\rm val}$. For further (technical) details we refer to 
the original reference (Brown et al, 1992).

\vskip12pt
Our discussion may have demonstrated that the bulk 
transition does not merely replace the finite temperature 
transition and can be ignored as a lattice artifact. It 
seems to be superimposed on the structure of a finite 
temperature transition. For a smaller number of flavors 
($N_f=2$, 3, 4) no bulk transition has been observed. There 
it may be even more difficult to discriminate precursors of 
the bulk transition at $N_f=8$ from continuum behavior.

Brown et al. (1992) attempt an explanation for the bulk 
transition as an outgrowth of the rapid crossover region 
seen in the pure $SU(3)$ gauge theory for $\beta=5.6$. 
When passing from strong to weak couplings in this 
region, a strong deviation from the scaling behavior 
predicted by the perturbative $\beta$-function is seen 
(Kennedy et al., 1985). Adding light dynamical quarks to 
the pure gauge action, the crossover region narrows with 
increasing number of flavors. The sharper the crossover, 
the stronger the violation of perturbative scaling and 
the larger the increase in slope of $N_\tau$ vs
$\beta=6/g^2$. A plot of $N_\tau$ vs $\beta_c$ for 
$N_f=0$, 2, 4, and 8 flavors supports this view.

\vskip12pt\noindent
In concluding we summarize the phase structure, which is 
expected for various temporal extents. For small values of 
$N_\tau$ ($N_\tau\le 4$) one has to work with a coarse 
grained lattice to reach the transition region.The scale is 
controlled by the lattice spacing. A single finite 
temperature transition is observed, but its relevance to a 
finite temperature transition in the continuum is not 
obvious.

At intermediate values of $N_\tau$ ($4\le N_\tau\le 16$) 
the corresponding critical couplings fall in the 
crossover region. A change occurs in what is called the 
relevant scale. At the end of the crossover region the 
scale becomes loosely related to the lattice spacing, but 
controlled by the continuum behavior. This rapid change 
of relevant scales is manifest in a bulk transition 
preventing any smooth change of $N_\tau$ vs $\beta$, 
both quantities are apparently unrelated.

For larger values of $N_\tau$ ($\ge 32$) (and also a 
simultaneous extension of the spatial lattice size to 
mimic a finite temperature box) the transition region 
is expected to lie in a coupling regime, where the lattice 
is rather fine grained. The bulk transition has 
disappeared, a finite temperature transition recurs. This 
time it will be related to the chiral transition of 
continuum QCD (taking for granted that it does recur).

Thus the $N_f=8$-simulations -- although far from modelling 
realistic QCD systems -- allow an identification of lattice 
artifacts, which may also influence more realistic lattice 
simulations ($N_f=2$, 3) in a weakened form.

\subsubsection{Results for two and three flavors}
\label{sec:III.C.5}
The limiting cases we have discussed in the previous 
sections are not close to the realistic conditions of QCD. 
The pure gauge theory, the limit of four and eight nearly 
massless flavors and the strong coupling approximation 
may be regarded as tools to gain some insight in the 
dynamical origin of the QCD transitions, when it is viewed 
from certain limiting cases. They can further give some 
hints about the stability of QCD results with respect to 
variations of input parameters. Of particular interest is 
the role of the quark masses. For example a crossover 
phenomenon for experimental quark masses may be 
understood as a result of mass values, which are too small 
to sustain a first order deconfinement transition and too 
large to sustain the chiral transition. Such a conclusion 
can be drawn, if unrealistic mass values have been 
studied before.

Next we turn to the cases of two and three flavors, which 
come closest to the experimental relation of two light 
($m_u\sim m_d\sim 5-7$~[MeV]) and one less 
light ($m_s\sim~150-180$~[MeV]) flavor. Here we 
discuss the results by Brown et al. (1990), which are 
still representative for 2 light and 1 heavier flavor.

The simulations are performed in the staggered fermion 
formulation. As we have argued above, the staggered 
formulation represents intrinsically only integer multiples 
of 4 continuum flavors. The projection on 2 or 3 flavors is 
enforced by writing $l$ of the effective action in 
Eq.\ (\ref{3.14}) as $N_f/4$. Although the local fermionic operator 
${\cal Q}$ describes $N_f=4\cdot n$ ($n$~integer) flavors, 
one allows $N_f$ to take the desired continuum value and 
uses as prescription for the effective action
\begin{eqnarray}
  S_{\rm eff} =&& {1 \over 3}~\beta~\sum_p~{\rm Re}~{\rm Tr}~
  U(\partial p)
  - {1 \over 4}~N_{u,d}\ln\det(D+m_{u,d} a) \nonumber\\
  && \qquad - {1 \over 4}~N_s\ln\det(D+m_s a)
  \label{3.113}
\end{eqnarray}
with $\beta=6/g^2$, $U(\partial p)$ denotes the product 
of $U's \in SU(3)$ along the boundary $\partial p$ of a 
plaquette $p$, $N_{u,d}$ is the number of continuum up- 
and down flavors, that is 2, and $N_s$ is the number of 
strange flavors in the continuum, $N_s=1$. The Dirac 
operator $D$ is given by Eq.\ (\ref{3.12}), it describes 4 flavors 
of quarks in the continuum limit. Thus the representation 
of the prefactor $n$ as $N_f/4$ cannot be derived from an 
integration of a local action over fermionic degrees of 
freedom, the local staggered fermion action leads to 
$N_f=4\cdot n$, therefore one should call Eq.\ (\ref{3.113}) a 
`prescription'.

The condensate is given by Eqs.\ (\ref{3.112}). The algorithm 
evolving the gauge fields with respect to the action 
Eq.\ (\ref{3.113}) is the R-algorithm of Gottlieb et al. (1987a) 
with a step size of $\Delta\tau=0.0078$ for $N_f=2$ and 
0.01 for $N_f=3$. The lattice size has been fixed to 
$16^3\times 4$. Table~\ref{tab:6} lists the quark masses which have 
been considered together with the results for the chiral 
transition. The results are based on Figs.~\ref{fig:12}(a)-(d).

No transition for the 2-flavor case is concluded from the time
evolution of the ordered and disordered start, cf.
Fig.~\ref{fig:12}(a). No sign for metastability is seen, as the two
starts mix together without clear tunneling events. In
Fig.~\ref{fig:12}(b) it is the absence of a double peak structure,
which is taken as indication for `no transition'.

Fig.~\ref{fig:12}(c) shows the case of three light degenerate flavors.
Here a two-state signal is visible. Over more than 2000 time units the
system stucks in the ordered (disordered) phase depending on the
starting condition.  For two light and one heavy flavor
(Fig.~\ref{fig:12}(d)) - the case which comes closest to realistic
mass relations - the order parameter evolves similarly to
Fig.~\ref{fig:12}(a). Clear signs of metastability and two-phase
coexistence are absent. This result has been in conflict with earlier
conclusions of a clear transition on an $8^3\times 4$ lattice (Gavai
et al., 1989; Kogut and Sinclair, 1988b) and some evidence for a first
order transition (Kogut and Sinclair, 1989).

At a first glance the absence of first order signals under realistic
quark mass conditions have far reaching consequences for
phenomenological implications in heavy-ion collisions. Many
predictions rely on the first order nature of the chiral transition
for 3 flavors (see section \ref{sec:V}).  From a practical point of
view the alternative between a truly first order transition and a
crossover phenomenon may not be distinguishable, when the volume is
small. The more sensible question to ask is as to whether the
crossover is rapid enough to produce a sufficiently large gap in
entropy densities over a small temperature interval. The preliminary
answer seems to be positive even for the $N_f=2$-flavor case.

The jump in entropy density occurs within a $\beta$-
interval, which corresponds to a temperature interval of 
less than 10~MeV. In section \ref{sec:V.C} we will see that an 
entropy jump over a finite, but small temperature range is 
{\it in principle} sufficient to induce multiplicity fluctuations 
beyond the statistical noise.

Although a translation from lattice units into physical units should
be taken with care for the considered $\beta$-value of 5.171, it is of
interest to estimate bounds on the critical quark masses in units of
MeV. Taking $a~[\rm{MeV}^{-1}]$ from Born et al. (1989), the set of
masses $m_{u}\cdot a=m_{d}\cdot a=0.025$, $m_{s}\cdot a=0.1$
corresponds to $m_{u}=m_{d}\sim 12$~MeV, $m_{s}\sim 50$~MeV. These
mass values give an upper bound on the critical quark masses, for
which the chiral transition just ceases to be of first order and
happens to be of second order.

In the case of three degenerate flavors $N_{f}=3$, two state signals
are observed for $m_{u,d,s}=0.025$ (cf. Table~\ref{tab:6}), while no
clear signals of metastability are seen for $m_{u,d,s}=0.075$ (Gavai
and Karsch, 1985; Gavai et al., 1987), leading to an estimate for the
critical lattice quark masses in physical units 
of $12~\rm{MeV}\leq m_{u,d,s}^{\rm{crit}}\leq
38~\rm{MeV}$. These numbers should be compared with more recent
results on critical quark masses obtained in Wilson's fermion scheme,
see section \ref{sec:III.C.7}. Lattice quark masses in physical units
are related to the current quark masses in units of MeV by an unknown
multiplicative renormalization factor.

Finally we come to the {\it reliability of the results for 2 and 3
  flavors}. All simulations, reported so far, have been performed on a
$16^3\times 4$ lattice. A temporal extent of 4 time slices leads to a
transition region in the coupling range, where the lattice is rather
coarse grained. For a pure gauge theory $N_\tau\ge 10$ is necessary to
reach the continuum region. The effect of fermions is to further lower
the effective lattice spacing, thus an even larger temporal extent
would be necessary to reach the range of asymptotic scaling here.

The ``distance'' from the continuum limit is visible in results of
Brown et al. (1990) for hadron masses, which have been obtained in
separate $T=0$-simulations on a $16^3\times 24$ lattice with
$\beta=5.171$, $m_{u,d} a=0.025$, and $m_s a=0.1$. These calculations
have been performed to test on the very scaling properties in the
considered coupling regime. The masses of two kaons which are
degenerate in their flavor contents in the continuum limit still
differ by a factor of 2. The nucleon over rho mass ratio $m_N/m_\rho$
is obtained as 1.5(1) in contrast to its physical value of 1.22 . The
ratio of $m_K/m_\rho=0.46(1)$ is smaller than its physical value of
0.64 . This suggests that the strange mass entering the lattice
K-meson is smaller than its physical value. Hence the first order
transition for 3 flavors disappears already {\it before} the strange
mass adopts its physical value.  The unphysical masses of the flavor
partners can influence the transition dynamics in a way which is
difficult to control.

Other UV-artifacts due to the coarse grained lattice may be hidden in
the results. From the discussion in section \ref{sec:III.C.3} it
cannot be excluded that the smooth behavior for the 2-flavor case is a
remnant of the strong coupling $U(1)\times U(1)$ symmetry. Going to
larger $N_\tau$-values, a first order transition may recur. The same
warning applies to the 3-flavor case with 2 light and 1 heavy flavors,
as the $U(1)\times U(1)$ symmetry at strong couplings leads to a
second order transition independently of the number of fermionic
flavors.

Concerning possible superimposed features from the bulk 
$N_f=8$-transition, the $N_\tau=4$-extent is probably 
below the dangerous temporal sizes, where the associated 
coupling values fall in the crossover region from strong 
to weak couplings.

From experience with other simulations it is clear that 
the lattice size of $16^3\times 4$ is not large enough to 
be in the asymptotic {\it large volume} region, where 
finite size effects can be excluded.

\vskip12pt {\it More recent results} for the 2-flavor case of
staggered fermions have been obtained for larger $\beta$-values (i.e.
smaller couplings), larger temporal extent $8\times 16^3$, and/or
smaller values of the bare quark masses. The bare quark masses are
$m\cdot a=0.00625$ (HTMCGC collaboration, see Bitar et al., 1993;
Gottlieb et al., 1993) and (Bernard et al., 1995a) , $m\cdot a=0.004$
(Columbia group, see Mawhinney, 1993), and $m\cdot a=0.008$ and 0.016
on $24^{3}\times 12$ lattices (Bernard et al., 1995).  The qualitative
conclusions are the same. There is a rapid crossover for finite quark
masses. As a progress it should be mentioned that the pion mass has
decreased to the order of $T_c$ for the lowest value of the bare quark
mass. In earlier calculations (e.g. Brown et al., 1990) it has been
too large for heavier mesons to decay into pions.

Another good indicator for the crossover phenomenon is the {\it baryon
  number susceptibility}. It is expected to be small in the low
temperature phase, since the baryon number can be only changed by
creating or destroying a baryon, but large in the high temperature
phase, where it is sufficient to create or destroy single quarks
to change the baryon number. In (Bernard et al., 1995) the coupling
dependence of the baryon number susceptibility $\chi$ has been used to
determine the crossover coupling $6/g_{c}^{2}$ for 2-flavor QCD at
given quark masses. The baryon number susceptibility has been also
plotted as function of $T$~[MeV] for various lattice sizes. The
$T$-dependence is obtained from $T/m_{\rho}=1/(N_{\tau}\cdot a\cdot
m_{\rho})$, if the zero temperature spectrum calculations of
$m_{\rho}\cdot a$ are extrapolated to the gauge coupling and the quark
mass values of the thermodynamic simulations. Once the
$N_{\tau}$-dependence is under control, a plot of $\chi(T)$ is of much
interest for phenomenological applications. The baryon number
susceptibility enters QCD's equation of state, hadronization processes
and heavy-ion collisions.

\vskip12pt We conclude with a summary of conjectures and results for 2
and 3 flavors (based on the work by Brown et al., 1990) in
Fig.~\ref{fig:13}. Indicated are the presence or absence of the finite
temperature QCD transitions as a function of the quark masses
$m_s\cdot a$ and $m_{u,d}\cdot a$ in lattice units. The solid line
indicates second order transitions, the shaded areas enclose mass
values leading to first order transitions. The concave shape of the
critical boundary is hypothetical. The solid circles refer to mass
points, where a first order transition is seen, the solid squares,
where no transition is found in the simulations of Brown et al.
(1990). The dashed circle indicates the location of the physical
strange- to up quark mass ratio. Error bars for the location of the
various points relative to the phase boundaries have been left out.
The mass point (0,0) is the chiral limit of 3 flavors with an
$SU(3)\times SU(3)$ symmetry in the continuum limit and a first order
transition.  The pure gauge theory in the upper right corner of the
diagram ($m_{u,d}=\infty$, $m_s=\infty$) has an exact $Z(3)$-symmetry
both in the continuum and on the lattice for 3 colors. The transition
is of first order.

The line ($m_{u,d}=0$, $m_s$) has an $SU(2)\times SU(2)$ continuum
symmetry, and a conjectured second order transition above some
tricritical strange quark mass, where the line of first order
transitions ends. The universality class of QCD may change along the
second order transition line. It has been proposed to be the
universality class of an $O(4)$-model for $m_{u,d}=0$ and sufficiently
large $m_{s}$, a tricritical $\phi^{6}$-theory in the vicinity of the
tricritical point (Wilczek, 1992; Rajagopal and Wilczek, 1993 (cf.
section \ref{sec:IV.A.1})), and Ising like along the concave part with
$m_{u,d}\neq 0$ (Gavin et al., 1994). An identification of the
appropriate universality class of QCD along the critical phase
boundary is still in an exploratory stage. The precise location of the
critical phase boundary is under debate. In particular the location of
the tricritical mass point on the ($m_{u,d}=0$)-axis is yet unknown.
Both fermion schemes agree about the existence of a first order
transition region, but do not agree about its extension. In the
staggered scheme the {\it physical} mass point lies outside, in
Wilson's scheme inside the first order region. Strong coupling
artifacts act in opposite directions in both schemes. As mentioned
above, the smooth behavior in the staggered formulation may be a
result of the $U(1)\times U(1)$-symmetry at strong couplings, while
the first order behavior in Wilson's formulation may be caused by the
light fermion doublers at strong couplings (cf. section
\ref{sec:III.C.7}).  Results for Wilson fermions will be summarized in
section \ref{sec:III.C.7}.

\subsubsection{The equation of state for two-flavor QCD}
\label{sec:III.C.6}
Blum et al. (1995a and b) have studied the energy density and
pressure as a function of temperature for two light flavors in the
staggered fermion formulation of QCD. The approach is based on fully
nonperturbative ingredients along the same lines as in the pure gauge
theory (cf. section \ref{sec:III.B.2}). The generalization comes from
the dependence of observables on the gauge coupling $6/g^{2}$ {\it
  and} the bare light quark mass $m_{q}\cdot a$. Thus the
$\beta$-function has now two components
\begin{equation}
  \label{3.114}
  \beta(6/g^{2},a\cdot m_{q})=
  ({\partial(6/g^{2}) \over \partial \ln a},
  {\partial(a\cdot m_{q}) \over \partial \ln a}) .
\end{equation}
For a number of points in $(a\cdot m_{q},6/g^{2})$-space the
$\beta$-function has been extracted from data for $(m_{\pi}\cdot a)$
and $(m_{\rho}\cdot a)$ measured as functions of $6/g^{2}$ and $a\cdot
m_{q}$. To find the change $\delta(a\cdot m_{q})$ for a given change
$\delta(6/g^{2})$ such that the physics remains unchanged, the mass
ratio $m_{\pi}/m_{\rho}$ has been kept fixed. An alternative way of
finding two equations for the two unknown functions $(a\cdot
m_{q})(a)$ and $(6/g^{2})(a)$ is to fit $m_{\pi}/m_{\rho}$ and
$m_{\rho}\cdot a$ as functions of $m_{q}\cdot a$ and $6/g^{2}$,
leading to $a=a(6/g^{2},m_{q}\cdot a)$. The inverse function then
yields the $\beta$-function corresponding to a symmetric change of
lattice spacings.

The nonperturbative $\beta$-function enters the interaction measure
according to
\begin{eqnarray}
  \label{3.115}
  (\epsilon-3p)\cdot a^{4}
  = &&-2{\partial(6/g^{2}) \over \partial \ln a}(P_{T}-P_{0})
  \nonumber\\
  &&-{\partial(a\cdot m_{q}) \over \partial \ln a}
  (\left<\bar\psi\psi\right>_{T}-\left<\bar\psi\psi\right>_{0}) .
\end{eqnarray}
Here $P_{T,0}$ denote the average plaquette expectation values at
temperature $T$ or zero, respectively. They are calculated as
\begin{equation}
  \label{3.116}
  P_{T,0}={1 \over 2N_{\sigma}^{3}N_{\tau}}
  {\partial \ln Z_{T,0}(a\cdot m_{q}, 6/g^{2}) \over
    \partial(6/g^{2})} ,
\end{equation}
$\left<\bar\psi\psi\right>_{T,0}$ are the corresponding light quark
condensates
\begin{equation}
  \label{3.117}
  \left<\bar\psi\psi\right>_{T,0}=
  {1 \over N_{\sigma}^{3}N_{\tau}}
  {\partial \ln Z_{T,0}(a\cdot m_{q}, 6/g^{2}) \over
    \partial(a\cdot m_{q})} .
\end{equation}
The pressure $p$ is calculated by integrating either of the two
relations
\begin{eqnarray}
  \label{3.118}
  {p \over T^{4}}
  = 2N_{\tau}^{4} \int_{\rm{cold}}^{6/g^{2}} d(6/g'^{2})
  && \left[ P_{N_{\tau}}(6/g'^{2},a\cdot m_{q}) \right.
    \nonumber\\
    &&~\left. -P_{0}(6/g'^{2},a\cdot
    m_{q})\right] ,
\end{eqnarray}
(cf. Eq.\ (\ref{3.61}) in section \ref{sec:III.B.2}) or
\begin{eqnarray}
  \label{3.119}
  {p \over T^{4}}
  = N_{\tau}^{4} \int_{\rm{cold}}^{a\cdot m_{q}} && d(a\cdot m_{q})'
  \left[ \left<\bar\psi\psi(6/g^{2},(a\cdot m_{q})')\right>_{T} \right.
    \nonumber\\
    &&~\left. 
      -\left<\bar\psi\psi(6/g^{2},(a\cdot m_{q})')\right>_{0}\right] .
\end{eqnarray}
The lower limit for the integration should be chosen so that the
contribution of the integrand is negligible, `cold' stands for the
cold symmetric lattice at zero temperature. The energy density is then
obtained from Eqs.\ (\ref{3.118}), (\ref{3.119}) combined with Eq.\
(\ref{3.115}). The results are shown in the two upper curves of
Fig. \ref{fig:14} for two light bare quark masses: The lower curves
denote three times the pressure. Although there is no transition in
the strict sense, there is still a rapid rise in the energy density
over a temperature interval of the order of 10~MeV around a
temperature
of 150~MeV. Also a slower rise in the pressure is seen. A shift of the
transition interval to higher temperatures is observed for lower bare
quark masses. This qualitative feature seems to be at odds with
observations in effective models, where finite light quark masses
delay the melting compared to the chiral limit, cf. section
\ref{sec:IV.A}.2 and 4.

The simulations of Blum et al. (1995a) have been performed for temporal
extensions of $N_{\tau}=4$ or $N_{\tau}=N_{\sigma}$ with
$N_{\sigma}=8$, 12 or 16. From the discussion of finite size effects
in case of the pure $SU(N)$ gauge theories (section III.B.1 and 2) we
expect that the temporal extent of $N_{\tau=4}$ is certainly not large
enough for interpreting the lattice results as continuum physics. An
extrapolation to the high temperature region and a comparison with
perturbative calculations would be a good control which is yet
outstanding. Thus the equation of state for two-flavor QCD is still in
an exploratory stage.

\subsubsection{Simulations with Wilson fermions}
\label{sec:III.C.7}
To confirm the conclusions of the last sections within the lattice
approach, investigations in the Wilson formulation are
indispensable. Staying within the same scheme it would be difficult to
control the approximations of the desired continuum symmetries.

In the staggered fermion formulation the chiral limit can be obtained by
varying the bare mass $m\cdot a$ in the Lagrangian to smaller values
and extrapolating $ma=0$ in the end. In contrast the chiral limit in the
Wilson fermion formulation must be determined as a one-dimensional
submanifold $\kappa_c(\beta)$ in the two-dimensional
$(\beta,\kappa)$-plane, where $\beta=6/g^2$ denotes the inverse gauge
coupling and $\kappa$ the hopping parameters. $\kappa_c(\beta)$ is the
line of critical hopping parameters, which characterize the
chiral limit. The chiral limit can be defined as the vanishing of the
pion mass on zero temperature lattices. Another possibility is to
determine $\kappa_c(\beta)$ by the location of zeros in the fermion
determinant. Both definitions are in general not equivalent, as the
former definition involves an average over many gauge field
configurations, while the latter definition does depend on the
configuration. Other definitions of $\kappa_c$ on {\it finite volume}
lattices have been proposed by Bitar et al. (1991), Bochicchio et al.
(1985) and Iwasaki et al. (1989).

In the following we discuss the QCD transitions in a space of four
parameters: the number of flavors $N_{f}$, the number of time slices
$N_{\tau}$, the coupling $\beta=6/g^{2}$, and the hopping parameter
$\kappa$. The line of finite temperature phase transitions/crossover
phenomena from the confinement/chiral symmetry broken to the
deconfinement/chiral symmetric phase will be denoted by
$\kappa_{T}(\beta)$, while $\kappa_{d}(\beta)$ stands for a line of
bulk deconfinement transitions (further explanations follow below). It
may be useful to visualize the change of parameters by a look at
Fig. \ref{fig:15}. 
Fig. \ref{fig:15} indicates the shift of the thermal transition line
$\kappa_{T}(\beta)$ between the confinement and deconfinement phase as
a function of $N_{\tau}$ and $N_{f}$. Larger $N_{\tau}$ shifts the
transition towards the continuum limit. Note that $\kappa_{T}(\beta)$
hits $\kappa_{c}(\beta)$ only at rather strong couplings. The
qualitative features hold for $N_{f}\leq 6$.

\paragraph*{Obstacles with Wilson fermions}
Fukugita et al. (1986) were the first to point out certain obstacles
that may hamper the investigation of the confinement phase in the chiral
limit with Wilson fermions. For $N_f=4$, $N_\tau=3$ and spatial lattice
sizes of $5^3$ the simulations have been performed for $3.0<\beta<6.0$
and $0<\kappa<0.25$. The transition region of the deconfinement
transition has been defined as the range of $\beta$ over which the
Polyakov loop and the gluon energy density vary rapidly. The result was
that the transition line $\kappa_T(\beta)$ did not meet the chiral line
$\kappa_C(\beta)$ down to $\beta\sim 3.5$. The lines were running almost
parallel to each other to the strong coupling regime without crossing.
The warning of Fukugita et al. was that the line of critical hopping
parameters always stays in the high temperature phase for any finite
lattice size. If the conjecture were proven to be true, it would be
impossible to describe the confinement phase in the chiral limit within
the Wilson formulation. The warning was extended to spectroscopic
calculations with Wilson fermions, which might reflect high temperature
behavior at $\kappa_c$, if the spatial lattice size is not sufficiently
large.

Results of Bitar et al. (1991) for $N_f=2$-Wilson fermions on $4\times
8^3$-lattices pointed in the same direction. The hopping parameter 
was varied between $0.12\leq\kappa\leq 0.19$, the coupling region was
$4.5<\beta<6.0$. The phase transition line $\kappa_T(\beta)$ was
determined in this range, e.g. the transition/crossover coupling
$\beta_T$ lies between 5.12 an 5.13 for $\kappa=0.17$. If the chiral
limit is defined as the vanishing of the pion mass on zero temperature
lattices, the chiral limit is nowhere reached in the low temperature
phase on the $4\times 8^3$ lattice in the above parameter range. The
conclusion is, if the line $\kappa_T(\beta)$ ever reaches the chiral
limit line $\kappa_c(\beta)$, it must be in the strong coupling regime
for these small lattice sizes.

Thermodynamics with Wilson fermions has been reinvestigated by Iwasaki
et al. (1992b and 1993). The number of flavors is varied between 2 and
18. The number of time slices is extended from $N_\tau=4$ to
$N_\tau=6,8$ with spatial lattices of $8^2\times 10$ and to
$N_\tau=18$ with $N_x\cdot N_y\cdot N_z=18^2\times 24$. We distinguish
the cases of $N_f\geq 7$ and $N_f<7$.

\vskip12pt{$N_f<7$}.
In the strong coupling limit ($\beta=0$) quarks are
confined in the chiral limit. For $N_f=6$ the critical hopping parameter
is given as $\kappa_c=0.25$. The line $\kappa_c(\beta)$ is defined as
the values of $\kappa$ where $m_\pi^2$ vanishes in the confinement phase
with a linear extrapolation in terms of $1/\kappa$. The confining
behavior is concluded from the fact that the number of iterations for
the quark matrix inversion $N_{inv}$ exceeds 10 000 for $N_f=6$ with
$N_\tau=4$ at $\kappa_c$, while it is of $O(100)$ for $N_f\geq 7$ under
the same conditions otherwise. The large value of $N_{inv}$ is
attributed to the existence of zero eigenvalues of the quark matrix.

The most interesting question concerns the existence of a crossing point
of the $\kappa_T(\beta)$ and the $\kappa_c(\beta)$ lines at some
$\beta>0$. In the simulations of Iwasaki et al. (1992) the answer is
positive. For $N_f=2$ and $N_\tau=4$ the crossing point $\beta_{cT}$
occurs at $3.9\leq\beta_{cT}\leq 4.0$ with $\kappa_c\sim 0.222$. The
pion mass is consistent with zero at $\beta_{cT}$. The chiral transition
is of second order or a crossover phenomenon. For $N_f=6$ and
$N_\tau=4$, the crossing point lies between $0.2\leq\beta_{cT}\leq 0.3$
with $\kappa_c\sim 0.25$, the chiral transition is of first order, the
value of $m_\pi^2$ depends on the initial configuration.

For $N_f=2$ and $N_\tau=18$ the crossing point $\beta_{cT}$ is only
little shifted towards the continuum region, $\beta_{cT}\sim 4.5-5.0$.
The unpleasant message is that one has to go to temporal extends
$N_\tau>18$ to obtain a confining chiral limit closer to the continuum
limit (i.e.  for $\beta>5.0$).

\vskip12pt{$N_f\geq 7$.}
Here we have to distinguish between a line
$\kappa_d(\beta)$ of bulk transitions which separates the chiral limit
$\kappa_c(\beta)$ from the confinement region, and the line
$\kappa_T(\beta)$ of finite temperature transitions. In the strong
coupling limit ($\beta=0$) the phase of deconfinement and chiral
symmetry restoration is realized for $\kappa>\kappa_d$ ($\beta=0$).
The transition is stable under an increase of $N_\tau$ to $N_\tau=18$.
Therefore the transition is called a bulk transition, cf. the
discussion of bulk transitions for $N_f=8$-Kogut-Susskind fermions in
section \ref{sec:III.C.4}. The $\kappa_d(\beta)$-line of bulk
transitions extends to $\beta>0$. In addition there is the
$\kappa_T(\beta)$-line at $\beta>0$, which reaches the
$\kappa_d(\beta)$-line without crossing the chiral limit. For $N_f=7$
and 12 the quark mass is $O(1)$ in units of $a^{-1}$ at $\kappa_d$. If
this behavior persists in the continuum limit, confinement is lost for
$N_f\geq 7$. Recall that asymptotic freedom is lost in QCD, if
$N_f\geq 17$.

\vskip12pt
Peculiar behavior of the phase structure has been found by the
MILC-collaboration (Blum et al., 1994) for $N_{f}=2$ at $N_{\tau}=4$
or 6. At $N_{\tau}=4$ the transition is smooth for light and heavy
quark masses, but sharp for intermediate masses with $\beta\sim
5.0$. This is at odds with the general expectation that the strength
of the crossover should monotonically decrease with increasing
masses. At $N_{\tau}=6$ a first order phase transition with clear
two-state signals is seen in simulations at $\kappa=0.17$, $0.18$ and
$0.19$. It has been interpreted as a bulk transition located in the
vicinity of the thermal crossover (Blum et al., 1994). Fortunately the
lattice artifacts seem to disappear in a more recent simulation of the
MILC-collaboration (Bernard et al., 1995b) at larger $N_{\tau}$
($N_{\tau}=8$) at $6/g^{2}=5.3$ over a range of $\kappa$-values up to
$\kappa_{c}\sim 0.168$. The thermal crossover is now shifted to a
slightly smaller $\kappa_{T}\sim 0.167$, in qualitative agreement with
Fig. \ref{fig:15}.

\paragraph*{$N_{f}=2+1$ flavors and critical quark masses}
The Tsukuba group (Iwasaki et al., 1995a) performed simulations on
$8^{2}\times 10 \times 4$ and $12^{3}\times 4$ lattices for the
physically interesting case of two light and one heavier flavor. The
lattice parameters are translated to physical quark masses of
$m_{u,d}\sim 0$, $m_{s}\sim 400$~MeV and $m_{u,d}=m_{s}\sim 150$~MeV.
The hopping parameters are chosen as $\kappa^{(u,d)}=\kappa_{c}$ for
the light quarks, while $\kappa^{(s)}$ follows from measurements of
$m_{\rho}\cdot a$, $2m_{q}\cdot a$ as functions of
($1/\kappa-1/\kappa_{c}$) with $\kappa_{c}(\beta)$ referring to the
2-flavor case. The results of $m_{\rho}\cdot a$ as function of
$N_{f}$, $N_{\tau}$, $\beta$ are used to set the scale. If the
physical $m_{\rho}$-mass of 770~MeV is identified with
$(m_{\rho}(\kappa_{c})\cdot a)/a$, it follows a lattice spacing of
$a\sim 0.8~\rm{GeV}^{-1}$ for $\beta\leq 4.7$. Once
$a~[\rm{GeV}^{-1}]$ is known, one can find $m_{q}\cdot a$ and the
associated values of ($1/\kappa^{(s)}-1/\kappa_{c}$) corresponding
to $m_{q}\sim 0$, $m_{s}\sim 400$~MeV and $m_{u,d,s}\sim 150$~MeV.

One should recall that $m_{\rho}\cdot a$, i.e. the $\rho$-mass in
lattice units, has been measured at rather strong couplings
($N_{\tau}=4$) on small volumes with varying quark mass input
$m_{q}\cdot a$. Keeping $m_{q}$ fixed at 770~MeV assumes that
$m_{q}$~[MeV] is independent of $m_{q}$~[MeV] which it is
not. Furthermore the value of $a~[\rm{GeV}^{-1}]$ obtained from
$m_{\rho}\cdot a$ and $m_{\rho}$~[GeV] at such strong couplings is
certainly not universal, i.e. independent of the lattice
observable. Thus the translation from lattice parameters $\beta$,
$\kappa^{(u,d,s)}$ to physical parameters $a~[\rm{GeV}^{-1}]$ and
$m_{u,d,s}$~[MeV] should be taken with care.

The time histories of the plaquette expectation value show two state
signals with states depending on the initial configuration for both
values of $m_{s}$ ($m_{s}=150$ and 400~MeV). Future simulations with
larger temporal extensions at weaker couplings have to show, whether
the first order signals survive the continuum limit or share the fate
with the signals of the $N_{f}=2$-simulations at $N_{\tau}=4$, 6 (Blum
et al., 1994) to fade away on larger lattices. At strong couplings the
fermion doublers in the Wilson formulation are still too light to
decouple. It is known that the strength of the first order transitions
increases with the number of light flavors. Thus the first order at
large $m_{s}$ ($\sim 400$~MeV) might be an UV-artifact of the strong
couplings caused by the small number of time slices.

At present the first order behavior is in disagreement with the
lattice results of the Columbia group (cf. section \ref{sec:III.C.5})
and with results in effective continuum models (Meyer-Ortmanns and
Schaefer, 1995) (see section \ref{sec:IV.A.4}).

It is of interest to compare the estimates for the {\it critical}
quark masses up to which a clear first order signal is seen with
results obtained for staggered fermions and in effective models for
QCD.  For three degenerate flavors it is $m_{u,d,s}^{\rm{crit}}\geq
140$~MeV or $m_{u,d,s}^{\rm{crit}}\cdot a\geq 0.175(2)$ in the Wilson
fermion scheme as compared to 12-38~MeV for staggered fermions
(section \ref{sec:V.C.5}). For nondegenerate flavors it is
$m_{u,d}^{\rm{crit}}\leq 3$~MeV, $m_{s}^{\rm{crit}}\leq 54$~MeV in
the $SU(3)\times SU(3)$ linear sigma model (cf. section
\ref{sec:IV.A.4} below).

We summarize. Although the translation to physical mass units may
suggest continuum physics, it would be prematurate to call the lattice
result of a first order chiral transition for physical quark masses
continuum physics of the `real world'.

\paragraph*{Improved actions}
The experience with Wilson fermions in the original formulation has
shown that the simulations are either inconclusive due to lattice
artifacts at strong couplings, or expensive due to volumes which
should be even larger than for staggered fermions to avoid lattice
artifacts. Thus it is natural to attempt an approach with an {\it
  improved action}. Improved actions are supposed to accelerate the
approach to the continuum limit. One such choice has been adopted by
the Tsukuba group (Iwasaki et al., 1994 and 1995b). So far the
improvement concerns the gauge part of the full QCD-action, while the
fermionic part remains in Wilson's formulation. The form of the
improved pure gauge action was proposed by Iwasaki (1983). Besides the
usual plaquette term it contains a $1\times 2$-loop with coefficients
determined by a block-spin renormalization group analysis. The
preliminary results for full QCD with the renormalization group
improved action look promising. The improved gauge part seems to
reduce (or remove) the lattice artifacts for $N_{f}=2$ and
$N_{\tau}=4$ (6) observed by the MILC-collaboration (mentioned above).

A quantitative measure for the acceleration towards the continuum
limit is obtained from the difference in (inverse) couplings $\beta$,
which lead to approximately the same lattice spacing $a$. The improved
action needs only $\beta=2.0$, while the standard action needs
$\beta\sim 5.0$ for $a^{-1}=1.01$~GeV (if the $\rho$-mass is
calculated on an $8^{3}\times 16$-lattice and used as input to set the
physical scale).

\vskip12pt
It is not surprising that for small lattices without improved actions
the Wilson and staggered fermion schemes give different results. One
manifestation of this discrepancy is given by an estimate of $T_{c}$
in physical units for 2 flavors and 4 time slices. For Wilson fermions
$T_{c}$ is estimated as $221\pm 3$~MeV, whereas $T_{c}=142\pm 6$~MeV
for staggered fermions (Bitar et al., 1991). A more recent estimate of
$T_{c}$ from Wilson fermions leads to $T_{c}\sim 152\matrix{+18\cr
  -14}$~MeV (Bernard et al., 1995) as compared to $140~\rm{MeV}\leq T_{c}
\leq 160$~MeV for staggered fermions (DeTar, 1995). Future
simulations will show whether other quantities show the same tendency
that both fermion formulations approach the same continuum behavior.

\subsection{QCD at finite baryon density}
\label{sec:III.D}
The most convenient way of describing quark or nuclear matter at high
density is to introduce a chemical potential $\mu_{q}$ for
quarks. From asymptotic freedom one would naively expect that chiral
symmetry is restored and quarks are deconfined above a critical value
of $\mu_{q}$, $\mu_{q}^{\rm{crit}}$, and above a critical temperature
$T_{c}$. A critical line ($T_{c}$, $\mu_{q}^{\rm{crit}}$) will
separate the plasma and the hadron phase in a $(T,\mu)$-diagram. The
shape is indicated in Fig. \ref{fig:16}, it is qualitatively
reproduced in a bag model calculation at finite $T$ and finite
$\mu_{q}$ (Cleymans et al., 1986).
Note in particular that chiral symmetry is always restored above the
critical temperature for $\mu_{q}=0$, $T_{c,\mu_{q}=0}$, and above the
critical potential for $T=0$, $\mu_{q,T=0}^{\rm{crit}}$. In the bag
model one considers an ideal gas of nucleons in the hadronic phase,
and of quarks and gluons in the plasma phase. A real-valued
nonnegative energy difference at $T=0$ and the critical chemical
potential $\mu_{b}^{\rm{crit}}$ for baryons is then obtained, if
$\mu_{b}^{\rm{crit}}$ satisfies
\begin{equation}
  \label{3.120}
  M\leq \mu_{b}^{\rm{crit}} \leq 3M/(2\sqrt{2}) ,
\end{equation}
where $M$ denotes the nucleon mass. The chemical potential for quarks
and baryons are related at the transition point via
\begin{equation}
  \label{3.121}
  3 \mu_{q}^{\rm{crit}} = \mu_{b}^{\rm{crit}} .
\end{equation}
Conceptually $\mu_{q,b}^{\rm{crit}}$ should be distinguished from
threshold values $\mu_{q,b}^{t}$. In the bag model the threshold value
for $\mu_{b}$ is given by the nucleon mass $M$
\begin{equation}
  \label{3.122}
  \mu_{b}^{t}=M ,
\end{equation}
because a real-valued, non-vanishing baryon number density is only
obtained for $\mu_{b}>M$. Note that $\mu_{b}^{\rm{crit}}$ within the
bound of Eq.\ (\ref{3.120}) is always close to $\mu_{b}^{t}$ so that
the finite-$\mu$ phase transition at $\mu_{b}^{\rm{crit}}$ may not be
distinguishable from a treshold effect at $\mu_{b}^{t}$ in practical
calculations. Threshold effects are in fact seen on the lattice
(cf. the summary of results below). We turn now to the lattice
approach of finite $\mu$ (and finite $T$) transitions.

The naive translation of the continuum expression on the lattice leads
to quadratic divergences of the internal energy density $\epsilon$ in
the continuum limit. The solution of this earlier difficulty is provided
by introducing a chemical potential for quarks $\mu\cdot a$ into the
fermion matrix as $e^{\mu a}$ multiplying the forward links in time
direction and $e^{-\mu a}$ multiplying the backward links (Hasenfratz
and Karsch, 1983).

For simplicity one may think of performing lattice simulations at finite
density in the quenched approximation. It turns out that the quenched
approximation leads to results which contradict the intuitive
expectations. Physical observables should be independent of $\mu$ as
long as $\mu$ is smaller than some threshold value $\mu_c$. The
threshold is related to the threshold for baryon production. In the
confining phase baryons made of three quarks will be produced which are
bound together with an effective chemical potential $3\mu$. Once $3\mu$
exceeds the value of the baryon mass, a finite baryon density will
populate the system. In the quenched approximation the threshold comes
out as $\mu_c\cdot a=m_\pi^2/2$ even in the case of $SU(3)$, where
$m_\pi$ is the pion mass. The threshold should depend on the baryon
mass, and the baryon mass should stay finite in the chiral limit.

For a finite quark mass the chiral condensate shows some threshold
behavior in the sense that $\left<\bar\psi\psi\right>$ stays constant
for a while and then goes to zero with increasing $\mu$, but
$\left<\bar\psi\psi\right>$ vanishes in the chiral limit as soon as
$\mu>0$. Also this behavior is at variance with the expectation. A
critical density should be exceeded, before chiral symmetry is restored.

Davies and Klepfish (1991) have argued that the strange threshold
behavior $a\cdot\mu_c=m_\pi^2/2$ is really an artifact of the quenched
approximation and not of the finite volume or the strong coupling, as
one may have suspected as well. Their argument is based on the fact that
the gauge loops contributing to the pion propagator are unaffected by
the chemical potential in the quenched approximation. The failure of the
quenched approximation is shown for gauge couplings $\beta=6/g^2$
ranging from 0.0 to 6.2 .

If the full fermion determinant $\det M$ is included in lattice Monte
Carlo simulations, the problem starts for $SU(N)$ gauge theories with
$N\geq 3$. In this case the determinant is a complex number for $\mu\neq
0$. Therefore the factor $\det M \exp(-S_g)$ (where $S_g$ denotes the
pure gauge part of the action) can no longer be used as a probability
for generating configurations in a Monte Carlo simulation as it is
usually possible for $\det M$ being a positive real number. Nevertheless
one can generate gauge field configurations with some probability
$P(U)$, which is frequently chosen as $P(U)=|\det(M)| \exp(-S_g)$. The
expectation value for an observable $O$ is then calculated as
\begin{eqnarray}
  \label{3.123}
  \left<O\right>=&&\left[\int DU O\det(M)e^{-S_g}\right]
  / \left[\int DU\det(M)e^{-S_g}\right] \nonumber\\
  =&&\left<O e^{i\phi}\right>_P / \left<e^{i\phi}\right>_P~,
\end{eqnarray}
where $\left<\ldots\right>_P$ denotes the average over configurations
with respect to the measure $DU P(U)$ and $e^{i\phi}$ is the phase of
the determinant. At $\mu=0$ and $\mu=\infty$
$\left<e^{i\phi}\right>_P=1$. Therefore the simulation according to the
above prescription is feasible for small and large values of $\mu$. For
intermediate values of $\mu$ $\left<e^{i\phi}\right>$ is very small. The
phase $\phi$ fluctuates violently from configuration to configuration. A
report on how far one can push this method from the extreme values
$\mu=0$ and $\mu=\infty$ can be found in Toussaint (1990), see also
(Vladikas, 1988).

One attempt in order to minimize the large fluctuations of $\phi$ is a
detailed study of the grand canonical partition function (Gibbs, 1986;
Barbour and Sabeur, 1990). It can be shown that the characteristic
polynomial for the fermion propagator matrix is an expansion of the
grand canonical partition functions for fixed quark number. Here it is
nontrivial to measure the expansion coefficients $\left<a_n\right>$. Monte
Carlo
measurements of the canonical partition functions within this expansion
have been performed in an $SU(3)$ gauge theory by Barbour et al. (1988),
Barbour and Sabeur (1990), Barbour and Bell (1992) and Hasenfratz and
Toussaint (1992) and are reviewed by Barbour (1992).

\vskip12pt
As an {\it analytical alternative} to the Monte Carlo simulations
lattice calculations have been performed at strong coupling in a
$1/d$- and $1/g^{2}$-expansion (Bili\'c et al, 1992a and b). Here $d$
denotes the spacetime dimension and $g$ is the bare gauge coupling.
The earlier pathological threshold behavior in Monte Carlo simulations
(with a threshold proportional to $m_{\pi}$ and vanishing for
$m_{q}\to 0$) does, fortunately, not occur in the analytical approach.
Compared to the $g=\infty$-limit, at finite coupling the threshold
increases with $1/g^{2}$, while the baryon mass $m_{b}$ decreases. The
difference between the threshold and $m_{b}$ becomes smaller at weaker
couplings.  This is to be expected, as the difference is a measure for
the nuclear binding energy, which is large at strong couplings. Note
that the threshold value is different from the baryon mass in contrast
to Eq.\ (\ref{3.122}), which has been derived in an ideal gas picture
for nucleons.

Above a temperature of the order of 220~MeV chiral symmetry is
restored for any $\mu_{b}\geq 0$ in qualitative agreement with
Fig. \ref{fig:16}. Below this value the chiral transition is of
first order, but occurs at larger critical chemical potentials for
increasing $6/g^{2}$, as one would expect from the considerations
above. For $\mu_{b}<\mu_{b}^{t}$ one is in the chiral symmetry broken
QCD vacuum without nuclear matter (Bili\'c et al., 1992a).

A more detailed study of the threshold effect of {\it zero}
temperature and finite $\mu$ has been performed by Bili\'c et
al. (1992b). The threshold behavior turns out to be flavor-independent
in contrast to the strong flavor dependence of the ($\mu=0$, finite
$T$) transition. Apparently a threshold effect need not be a phase
transition. At strong coupling, however, the threshold effect seems to
coincide with the finite-$\mu$ chiral phase transition. Again the
threshold value at strong coupling lies somewhat below the nucleon
mass. The coincidence of the threshold effect with the finite-$\mu$
phase transition is likely an artifact of strong couplings. At the
threshold value $\mu^{t}$, the baryon number density jumps from zero
incidentally to the maximal occupation of lattice sites.

In the continuum limit at zero temperature one may expect to see a
threshold effect at $\mu^{t}$ separately from the phase transition at
$\mu^{\rm{crit}}>\mu^{t}$. In that case the threshold effect
corresponds to a ``transition'' from the QCD-vacuum to nuclear matter
with broken chiral symmetry, and the phase transition restores chiral
symmetry above the critical $\mu_{q,b}^{\rm{crit}}$.

So far the lattice results confirm some qualitative features of
Fig. \ref{fig:16}, but they are seriously plagued with artifacts of
strong couplings in the analytical approach, and of intermediate
couplings in the Monte Carlo approach.

Mean field calculations of QCD at finite baryon density have been
performed in a number of effective models for QCD in the {\it
continuum}, see e.g. Asakawa and Yazaki (1989), Lutz et al. (1992) or
the review by Hatsuda (1992). Some of these calculations will be also
mentioned in chapter~IV.

%
%
\section{EFFECTIVE MODELS IN THE CONTINUUM}
\label{sec:IV}
\subsection{Models for quark degrees of freedom}
\label{sec:IV.A}
In chapters \ref{sec:II} and \ref{sec:III} we have described effective
models on the lattice, where the $Z(N)$ Potts model was of particular
importance. As a common feature these models could be derived from
lattice QCD in some coupling or temperature limit or in a
renormalization group approach.  They suffer from the same
shortcomings as the lattice approximation to the original model (full
QCD).  Their relevance for the continuum limit has to be first
established. In this section we discuss effective models in the
continuum up to some remarks on dual Ginzburg-Landau models on the lattice
in section \ref{sec:IV.B.2}. Studies of the phase structure of QCD are
frequently performed in a mean-field approach.  The methods which we
have selected for a more detailed discussion in this section go {\it
  beyond} the mean-field level in various ways.

In the first part we illustrate the power of the {\it
  renormalization group} with the work of Wilczek (1992) and Rajagopal
and Wilczek (1993), who consider a three-dimensional $SU(2)\times
SU(2)$-linear sigma model as a model for the chiral phase transition
region. In section \ref{sec:IV.A.2} we outline the approach of {\it
  chiral perturbation theory}, which is well established as
description for low energy and {\it low temperature} QCD. We discuss the
results of Gerber and Leutwyler (1989). The leading term in a series
of actions for chiral perturbation theory is a non-linear $SU(2)\times
SU(2)$ sigma model in four dimensions.  Although chiral perturbation
theory fails to describe the transition region (thus it cannot predict
the order of the phase transition), it allows an estimate of the
latent heat in case of a (hypothetical) first order transition
(Leutwyler, 1992), cf. section \ref{sec:IV.A.3}. In section
\ref{sec:IV.A.4} we discuss the mass sensitivity of the chiral
transition and the equation of state for physical meson masses
(Meyer-Ortmanns and Schaefer, 1995).

%
%
\subsubsection{QCD and dysprosium}
\label{sec:IV.A.1}
In this section we report on the work by Wilczek (1992) and Rajagopal
and Wilczek (1993). It concerns an effective description of QCD which
is supposed to be valid just in the vicinity of the chiral transition.
The approach can be summarized in three hypothesis.

\begin{enumerate}
\item In the limit of 2 massless flavors QCD is in the static
  universality class of an $N=4$-Heisenberg ferromagnet, when all
  other quark masses are sent to infinity and $T\sim T_c$.
  
\item In the limit of 2 massless flavors QCD is in the static
  universality class of the $\phi^6$-Landau-Ginzburg model, when the
  strange quark mass is near a (tricritical) value $m_s^*$, the
  remaining quark masses are sent to infinity and $T\sim T_c$.
  
\item QCD with 2 massless flavors is in the dynamical universality
  class of an $O(4)$-antiferro\-mag\-net.
\end{enumerate}

These conjectures may be regarded as working hypothesis for lattice
simulations of the chiral transition. Before we describe their
predictive power, let us comment on the two basic assumptions.

\begin{itemize}
\item QCD is well approximated by the limit of two massless flavors
  ($m_u=m_d=0$).
  
\item The lightest excitations in the spectrum at zero temperature
  maintain their dominant role up to the transition region ($T\sim
  T_c$). The transition of 2-flavor QCD is expected to be of second
  order.
\end{itemize}

\noindent
Possible objections could be the following.

\begin{description} 
\item{a)} Even the lightest quark flavors are not massless, their
  finite values lead to a pion mass of $m_\pi\sim 135$~MeV. Unless the
  quark masses correspond to critical values (which is not likely),
  a second order transition amounts to an inadequate idealization.
  
\item{b)} From lattice calculations one knows that the chiral
  transition is of first order for 4 flavors, if the masses are not
  too large. For 2 light and 1 heavier flavor the transition is
  replaced by a crossover phenomenon.
  
\item{c)} It is unlikely and not plausible that a theory as intricate
  as QCD can be decimated to a model as simple as a scalar
  $O(4)$-model.
  
\item{d)} Finally the pions loose their dominant role in the
  transition region.
\end{description}

We will argue against the first three objections and postpone
arguments in favor of d) to section \ref{sec:IV.A.2}.

to a). The light quark masses have non-vanishing values, but their
values are small compared to the energy scale of the critical
temperature. Thus it makes sense to consider $m_u$, $m_d$ as
perturbations around the chiral limit, where the transition should be
of 2nd order, if $N_f=2$. The renormalization group approach can
account for finite small mass values. Their ``perturbing'' effect on
the critical behavior in the chiral limit can be parametrized with
critical indices. Thus there is a predictable parametrization of the
deviation from the idealized limit. We recall from chapter~II that the
mass plays a comparable role to the scaling field of the inverse
volume in a renormalization group analysis. Power law singularities of
a second order transition will be rounded due to the finite volume and
due to a finite mass. As we know from chapter~II, the rounding is
specific for the 2nd order, if the volume is sufficiently large. The
deviation from the $L=\infty$-limit can be predicted in a well-known
way.

Similarly there is a good chance that the extrapolation from finite
masses to the chiral limit is under control. It is under control, if
the deviations can be parametrized with the critical indices of a 2nd
order transition. For a check of this assumption it would be
sufficient to measure certain correlations (e.g. the specific heat) at
different small, but finite quark masses and compare the change in the
rounding effects with the predictions of the renormalization group
analysis.

to b). In view of the physical values for the current quark masses
four light flavors are certainly less realistic than two. Typical
signatures for a first order transition (a non-vanishing latent heat,
hysteresis effects and abrupt changes in the order parameters) are
absent for 2 flavors. Support for a 2nd order transition in 2-flavor
QCD is taken from lattice calculations of (Bernard et al., 1992;
Gottlieb, 1991; Gottlieb et al., 1989; Gottlieb et al., 1987b; Brown
et al., 1990a and b). The rise in the specific heat and the critical
slowing down are compatible with a 2nd order transition.  As argued in
chapter~III, lattice results are not yet fully conclusive. Hence they
do not prove the conjecture about 2-flavor QCD.

to c). Note that the hypothesis 1)-3) refer to QCD only in the
critical region ($T\sim T_c$). There it may well be that the
complicated substructure of QCD is not important for the transition
dynamics. In such a case one is free to replace QCD by the most simple
model belonging to the same universality class.  In condensed matter
physics many examples are known, where universality is used in the
same spirit. An example is the critical behavior of a binary mixture
like isobutyric acid plus water, which shares the universality class
of an Ising model in three dimensions.

There remains the question about the right choice of the universality
class. In case of 2-flavor QCD Wilczek's proposal is the universality
class of an $O(4)$-Heisenberg ferromagnet.  This proposal has been
recently questioned by Kocic and Kogut (1995) claiming that QCD with
its quark bilinear composite scalar mesons does not belong to the same
universality class as sigma models with their fundamental scalars. The
warning comes from a study in a three-dimensional Gross-Neveu model
that exhibits mean-field scaling rather than the expected
2-dimensional Ising universality. Let us see, how one is led to the
idea of an $O(4)$-Heisenberg ferromagnet in a natural way, and where
the warnings of Kocic and Kogut would change the line of arguments.

Landau's free energy has been introduced as a first concept to discuss
the phase structure, cf.~section \ref{sec:II.A}, Eq.\~(\ref{2.2}). It
is constructed as a power series in the order parameter field.
Landau's theory corresponds to a mean-field approximation. In general
it leads to wrong predictions of characteristic singularities (e.g.
the correlation length at $T_c$.). The physical reason are large scale
fluctuations in the order parameter field, although the average
magnitude of the order parameter is small.

The theoretical breakthrough came with Wilson's renormalization group
concepts (Wilson and Kogut, 1974). It leads to a systematic way of
constructing an action in terms of the relevant degrees of freedom at
$T_c$ and answers the question, why the resulting action is
representative in some sense. The action is constructed as a limiting
theory (fixed point theory) after a number of renormalization group
transformations. Such a limiting theory will exist, if the theory is
exactly {\it scale invariant} at $T_c$. Scale invariance implies that
models at different scales ($|r|\ll \xi$, i.e. scales much smaller
than the correlation length $\xi$, but larger than the microscopic
scales) share the leading singularity structure.

In case of the chiral transition these renormalization group steps are
not explicitly performed. The limiting theory in terms of pion fields
(and their parity partners) is argued to arise out of such a
procedure. More precisely only the zero modes of these fields are
assumed to survive the iterated renormalization group steps. The pions
are the lightest modes at zero temperature and the zero modes are the
only modes that do not acquire a mass contribution $\propto(2\pi nT)$
from the Matsubara sum. Thus the renormalized mass parameter
$\mu^2(T)$ in the resulting action should be understood as an
effective mass of the zero modes which vanishes as $T$ approaches
$T_c$ from above.

An essential outcome of the renormalization group approach is an
explanation of {\it universality}. Universality defines in which sense
the limiting theory is representative for a whole class of models,
belonging to the same {\it universality class}. Once the order
parameter field is specified, the fixed point theory in terms of these
fields only depends on the dimensionality and the symmetry, which is
assumed to be broken or restored at the transition.  Models with the
same underlying symmetry, order parameter fields and singularity
structure in thermodynamic functions define a universality class.

Thus it is sufficient to find an order parameter field, construct an
action in terms of this field and restrict the allowed terms by the
requirement of chiral symmetry. The order parameter field for the
chiral transition should at least contain the pion multiplet. For two
massless flavors the QCD Lagrangian is invariant under the
$SU(2)_L\times SU(2)_R\times U(1)_B$-symmetry of independent
$SU(2)$-rotations of left- and right-handed fields and the vector
baryon number symmetry. At the phase transition, the symmetry is
assumed to be broken to $SU(2)_{L+R}\times U(1)_{L+R}$. One choice for
an order parameter of the chiral transition is the quark bilinear
\begin{equation}
  M^i_j=\left<\bar{q}^{\,i}_L q_{Rj}\right>
  \label{4.1}
\end{equation}
transforming under  $SU(2)_L \times SU(2)_R$ according to
\begin{equation}
  M\longrightarrow U^+MV~,
  \label{4.2}
\end{equation}
where $U$ and $V$ represent independent unitary transformations of the
left-and right-handed quark fields. In the following the quark
substructure of $M^i_j$ will be disregarded. (This is the main point
which is questionable according to Kocic and Kogut (1995)).

As long as $M$ are general complex $2\times 2$-matrices, the
Lagrangian has too much symmetry ($U(2)\times U(2)$). A restriction of
$M$ to an $SU(2)$-representation removes the additional
$U(1)$-symmetry. A possible choice for the $SU(2)$-representation is
$O(4)\approx SU(2)\times SU(2)$, where $M$ is parametrized in terms of
four real parameters $(\sigma,\vec\pi)=\phi^a$ ($a$=1,...,4)
\begin{equation}
  M=\sigma +i\vec{\pi}\vec{\tau}\, ,
  \label{4.3}
\end{equation}
where $\vec{\tau}$ denote the Pauli matrices. Thus $M$ contains the
pion multiplet along with the scalar meson $\sigma$. The action in
terms of $\phi^a$, which is invariant under $SU(2)\times SU(2)$, is
given as
\begin{equation}
  S=\int d^3x\left\{{1\over 2}\partial^i\phi^a\partial_i\phi_a+
    {\mu^2\over 2}\phi^a\phi_a
    +{\lambda\over 4}(\phi^a\phi_a)^2 \right\}\, .
  \label{4.4}
\end{equation}
The action takes the same form as the Landau-Ginzburg free energy $F$
in section \ref{sec:II.A}. It could be identified with $F$, if the
path integral with $S$ of Eq.\ (\ref{4.4}) is evaluated in the mean
field approximation. Note that $S$ is an action in terms of zero
modes. The Euclidean time dependence of $\phi^a$ has been dropped.
Equivalently the $n\neq 0$-Matsubara modes of the original
4-dimensional theory are neglected. (The treatment of the $SU(2)\times
SU(2)$ sigma-model in the following section \ref{sec:IV.A.2} differs
in this aspect, the time dependence there is kept.)

Here the action Eq.\ (\ref{4.4}) should be understood as an effective
action for the chiral transition. It coincides with the familiar
linear sigma model of Gell-Mann and Levy (1960) up to the absence of
nucleons and the dimension 3.  Nucleon and quark fields are both
omitted, as they refer to ``microscopic substructures'', which are
claimed to be irrelevant for the transition.  Moreover the action
(\ref{4.4}) agrees with the action of an $N=4$-Heisenberg ferromagnet,
which is believed to model a magnetic transition in dysprosium
(Malmstr\"om and Geldart, 1980). This is the reason why Wilczek calls
dysprosium an ``analog computer'' for QCD and explains the title of
this section.

The $O(4)$-model in three dimensions has been extensively studied in
statistical physics (Baker et al., 1978; Lipatov, 1977; Hohenberg et
al., 1976; Bervillier, 1976). It is known to have a second order
transition with $\mu^2(T)\to 0$ as $T\to T_c$ from above. In the
following we summarize the results for critical exponents, taken from
these references and Wilczek (1992).

As usual we distinguish between the response to a stimulus in $T$ and
to a stimulus in $h$, where $h$ is an external field (cf.
Table~\ref{tab:2}). In this context the meaning of the external field
is that of non-vanishing current quark masses.

\vskip12pt\noindent
{\it Response to a stimulus in $T$}

\vskip12pt\noindent The response of the $O(4)$-``magnet'' to a
deviation from $T_c$ is measured by the specific heat $c$ and the
quark pair correlation $\xi$ in the limit, where the quark masses
strictly vanish ($m_\pi=0$). The singular behavior of the specific
heat is characterized by the index $\alpha$, the behavior of the
correlation length by the index $\nu$
\begin{equation}
  c(T)_{~\overrightarrow{T\rightarrow T_c^{\pm}}~} 
  A_{\pm} |T-T_c|^{-\alpha_{\pm}}
  +\hbox{less singular terms}
  \label{4.5}
\end{equation}
with $\alpha=-0.19 \pm 0.06$,
\begin{equation}
  \xi _{~\overrightarrow{T\rightarrow T_c^{\pm}}~} B_{\pm} 
  |T-T_c|^{-\nu_{\pm}} +\hbox{less singular terms}
  \label{4.6}
\end{equation}
with $\nu=0.73 \pm 0.02$, where $\xi$ is determined by
\begin{equation}
  -\ln\left<\bar{q}(x)q(x)\bar{q}(0)q(0)\right> \rightarrow
  {|x|\over \xi}
  \label{4.7}
\end{equation}
at large distances $|x|$. The $\pm$-signs refer to the approach of
$T_c$ from temperatures $T>T_c~(+)$ or $T<T_c~(-)$. The exponents
$\alpha_\pm$ and $\nu_\pm$ do not depend on the approach of $T_c$. The
amplitudes $A_\pm$ and $B_\pm$ are known as well and universal.

Note that $\alpha$ is negative, thus there is only a cusp in the
specific heat, no true singularity in the infinite volume limit. This
is compatible with the definition of a second order transition
(cf.~chapter~II.A), but not so common.  Usually more than one of the
second derivatives of the thermodynamic potential diverge with
power-law singularities.

We recall from Table~\ref{tab:2} that the critical index $\beta$
determines the behavior of the order parameter $\left<|\phi|\right>$
as $T$ approaches $T_c$ in the massless limit. We have
\begin{equation}
  \left<|\phi|\right> \sim \left|{T-T_c \over T_c}\right|^{\beta}
  \quad \hbox{for} \quad T-T_c <0.
  \label{4.8}
\end{equation}
with $\beta = 0.38\pm 0.01$.

The results for the critical indices $\alpha$, $\nu$ and $\beta$
should be checked by lattice simulations of the condensate
correlations Eq.\ (\ref{4.7}) and the specific heat ($T\partial
S/\partial T$). In principle this check can be performed, although an
{\it extrapolation} to the zero mass limit and the struggle with
critical slowing down are unavoidable.

\vskip12pt\noindent
{\it Response to a ``stimulus'' in $m$}

\vskip12pt\noindent The simplest possibility for adding a common mass
term to the action which is left-invariant by the vector
$SU(2)$-symmetry and invariant under discrete $P$- and
$T$-transformations is given by $m\cdot\sigma$, where $m$ denotes the
common quark mass. The dictionary of Table~\ref{tab:2} in chapter~II
tells us how we should translate the compressibility in a liquid/gas
system and the susceptibility in a ferromagnet to the order parameters
for the chiral transition. The corresponding quantity is the variation
of the condensate as a function of $m$
\begin{equation}
  {\partial \left<\bar{q} q\right> \over \partial m} \rightarrow C_{\pm}
  |T-T_c|^{-\gamma\pm}+\hbox{less singular terms}\, .
  \label{4.9}
\end{equation}
The approach to $T_c$ is characterized by the critical exponent
$\gamma$ (with $\gamma_+=\gamma_-$), the amplitudes $C_{\pm}$ are
universal.

The second exponent $\delta$ concerns the response of the condensate
to a change in $m$ at $T_c$
\begin{equation}
  \left<\bar{q} q\right> \rightarrow m^{1/ \delta}
  +\hbox{less singular terms}~,\quad
  T=T_c\, .
  \label{4.10}
\end{equation}
Here the results for the $O(4)\,$-model are $\gamma=1.44 \pm 0.04$ and
$\delta=4.82 \pm 0.05$.

\vskip12pt\noindent
{\it The critical equation of state}.

\vskip12pt\noindent Wilczek's first hypothesis can be further
exploited to predict the temperature dependence of the pion and the
sigma masses in the vicinity of $T_c$. These functions can be
calculated as a function of the underlying quark masses. The important
relation is a critical equation of state, which includes the results
for $\gamma$, $\beta$ and $\delta$ of the previous sections as special
cases. An equation of state in a liquid/gas system can be formulated
as a relation between $p$, $V$ and $T$, $p=p(V,T)$. In a ferromagnet
such a relation translates to $H=H(M,T)$ with the notations of
Table~\ref{tab:2}, in QCD it reads $m=m(\left<\bar q q\right>,T)$, and
$m=m(\left<\sigma\right>=\left<|\phi|\right>,T)$ in the linear sigma
model. In the ``magnetic'' language of Br\'ezin et al. (1973) the
critical equation of state is given as
\begin{equation}
  {H\over M^{\delta}}=\kappa_1\, g(\kappa_2 \, t |M|^{-{1\over \beta}})
  \equiv f({t\over M^{1\over \beta}})\, ,
  \label{4.11}
\end{equation}
where $t={(T-T_c)\over T_c}$, $g$ is a universal function and
$\kappa_1$, $\kappa_2$ are non-universal constants. For the
$O(4)\,$-model the function $f$ has been calculated by Br\'ezin et al.
(1973) to order $\varepsilon^2$ in the $\varepsilon\,$-expansion.
Without derivation we state the expressions for $m^2_{\pi}$ and
$m^2_{\sigma}$ in terms of the function $f$ at small $t$ and $H(\equiv
\left<\sigma\right>)$ (Rajagopal and Wilczek, 1992)
\begin{mathletters}
  \label{4.12}
  \begin{equation}
    m^2_{\pi}= M^{\delta-1} f(x)
    \label{4.12a}
  \end{equation}
  \begin{equation}
    m^2_{\sigma}
    =M^{\delta-1} \left(\delta f(x)-{x\over \beta} f'(x)\right)
    \label{4.12b}
  \end{equation}
\end{mathletters}
with $x\equiv {t/ M^{1/ \beta}}$. Note that the relations hold at
finite temperature, thus we have to specify, which mass it meant by
$m^2_{\pi}$, $m^2_{\sigma}$. The definitions used by Rajagopal and
Wilczek (1992) are the following
\begin{equation}
  m^{-2}_{\sigma} =\int d^3 x G_{00} \qquad
  m^{-2}_{\pi} \delta_{ij} =\int d^3 x G_{ij}
  \label{4.13}
\end{equation}
where $G_{\alpha\beta}(x) \equiv
\left<\phi(x)_{\alpha}\phi(0)_{\beta}\right>
-\left<\phi_{\alpha}\right>\left<\phi_{\beta}\right>$ and
$\vec{\phi_{0}} \equiv \sigma$, $\vec{\phi}_i=\vec{\pi}_i$,
$i=1,2,3\,$.

If the masses are defined according to Eq.\ (\ref{4.13}), they are
related to the spatial correlation functions in the equilibrium
theory. They are sensitive to static screening lengths. ``Static''
correlation lengths should be seen in contrast to dynamic length
scales, which are relevant in the context of the dynamical
universality class of the $O(4)$-Heisenberg antiferromagnet.

In the {\it chiral limit above $T_c$} ($t>0$ and $H\to 0$) the
critical equation of state gives the expected degeneracy between
$m_\pi^2$ and $m_\sigma^2$
\begin{equation}
  m^2_{\sigma}=m^2_{\pi}=c\, t^{\gamma} \,\quad for \quad x\rightarrow
  \infty \, .
  \label{4.14}
\end{equation}
In the {\it chiral limit below $T_c$} ($t<0$ and $H\to 0$) it follows
from Eq.\ (\ref{4.12}) that $m_\pi^2\propto H$, i.e. $m_\pi^2 \propto
m$, the familiar proportionality from zero temperature. On the other
hand, Eq.\ (\ref{4.12a}) follows from $m^2_{\pi}= {H\over M}$, which
reads in a literal translation according to the dictionary of
Table~\ref{tab:2}
\begin{equation}
  m^2_{\pi}={m\over \left<\bar{q} q\right>}\, .
  \label{4.15}
\end{equation}
This is less familiar in view of the $T=0\,$-PCAC-relation
\begin{equation}
  f^2_{\pi} m^2_{\pi}=2 m<\bar{q} q>~.
  \label{4.16}
\end{equation}
Eqs.\ (\ref{4.15}) and Eq.\ (\ref{4.16}) do not contradict, if
$f_{\pi}$ is identified with $f_{\pi}=\left<0|\sigma|0\right>$. The
last relation is compatible with a definition of $f_{\pi}$ via the
axial current of the linear sigma model at zero temperature.  If we
had started with Eq.\ (\ref{4.16}) and used
$f_{\pi}=\left<0|\sigma|0\right>$, the resulting relation at $T \sim
T_c$ may not be considered as derived from the $T=0\,$-relation
(\ref{4.16}). This point has been emphasized by Rajagopal and Wilczek
(1992). The masses $m_\pi^2$ in Eqs.\ (\ref{4.15}) and (\ref{4.16})
cannot be the same as the notation may suggest. In Eq.\ (\ref{4.15})
$m_\pi$ is the mass in the $(3+1)\,$-dim Lorentz invariant
$T=0\,$-theory. In Eq.\ (\ref{4.16}) it resembles a screening mass in
a 3dim-theory.

The result for the sigma-mass at temperature $T<T_c$ in the limit of
vanishing quark mass reads
\begin{equation}
  \left({\beta m^2_{\sigma} \over M^{\delta-1}}\right)^{-1}
  \longrightarrow
  c_1+c_2 \left({H\over M^{\delta}}\right)^{-{\epsilon\over 2}}.
  \label{4.17}
\end{equation}
Here $\beta$ denotes the critical exponent of the ``magnetization'',
$c_1$ and $c_2$ are constants, and $\epsilon$ is the remnant of the
$\epsilon\,$-expansion.

To summarize: The critical equation of state (\ref{4.11}) may serve as
a working hypothesis for lattice simulations. Its various limits are
predictions for the scaling behavior of $m^2_{\pi}$ (Eq.\ 
(\ref{4.15})), $m^2_{\sigma}$ (Eq.\ (\ref{4.17})), or $m^2_{\sigma}=
m^2_{\pi}$ (Eq.\ (\ref{4.17})) in the chiral symmetric phase. In
particular the quantitative predictions for the critical indices
$\beta$, $\gamma$,$\delta$ referring to the condensate as function of
$m$ and $T$ are ready for numerical tests.

\vskip12pt\noindent
{\it The role of the strange quark mass.}

\vskip12pt\noindent We come now to the second hypothesis concerning
the role of the strange quark mass in the chiral transition. So far we
have implicitly assumed that the remaining quark masses $m_s$, $m_c$,
$m_b$ and $m_t$ are infinite. This is certainly justified for the
charm, bottom and top quark masses, which are large compared to the
chiral transition temperature, but the strange quark mass is just of
the order of the transition temperature. Thus it can influence
thermodynamic quantities in a nontrivial way. Such an influence is
already visible in lattice results. If $m_s$ is infinite, the chiral
transition seems to be of second order. We have quoted the references
above. If it is zero, renormalization group arguments predict a first
order transition (Pisarski and Wilczek, 1984). Numerical simulations
for three light flavors verify this conjecture (see e.g.  Gottlieb
1991). Hence a critical value $m_s^*$ should exist, where the second
order changes into a first order transition. Usually such an endpoint,
where the order changes from first into second, is called a {\it
  tricritical point}. The physical value of $m_s$ is unlikely to
coincide with $m_s^*$, but it may be close by. Therefore it is
tempting to describe the realistic mass parameters as a perturbation
around the idealized tricritical limit. If such an ansatz is
justified, the deviations from tricriticality are under control.

According to the second hypothesis a simple model which shares the
universality class of QCD with two massless flavors of quarks, $T\sim
T_c$ and $m_s$ near $m_s^*$ is the $\phi^6$-Landau-Ginzburg model. Its
action reads
\begin{equation}
  S=\int d^3x \left\{{1\over 2}(\nabla\vec{\phi})^2+{\mu^2\over 2}
    \vec{\phi}^2
    +{\lambda\over 4}(\vec{\phi^2})^2+{\kappa\over 6}(\vec{\phi^2})^3
    -H\sigma \right\}\, .
  \label{4.18}
\end{equation}
The field $\vec\phi$ is the same as in Eq.\ (\ref{4.4}). The explicit
symmetry breaking due to $(-H\sigma)$ has been added to account for
finite masses $m_{u,d}$. The $\phi^6$-term arises as follows. The
effect of a finite strange quark mass is to renormalize the mass and
coupling $\mu^2$ and $\lambda$ in Eq.\ (\ref{4.4}). For example one
contribution to the renormalization of $\lambda$ comes from a K-meson
exchange between two pions. The ``amount'' of renormalization depends
on $m_s$. The mere effect of a mass and coupling renormalization is a
shift in $T_c$ as long as $\lambda$ stays positive. If $\lambda$ is
negative the model gets unstable, and a $\phi^6\, $-term is needed for
stabilization. It is easily checked that for $\lambda<0$ and fixed,
the minimum of the free energy jumps discontinuously from zero to a
finite value ${|\lambda|/ (2 \kappa)}$ when $\mu^2= {\lambda^2/
  (4\kappa)}$. (For $\lambda >0$ and fixed, the minimum moves
continuously from zero to positive values, when $\mu^2$ goes through
zero, as one enters the broken phase.) Hence the value of $m^*_s$ can
be defined as the strange quark mass for which the renormalized
coupling $\lambda$ vanishes. At this point the second order of the
$\lambda >0$ region changes into first order ($\lambda <0$ region).

Singularities of thermodynamic functions are universal near
tricritical points.  Thus one may again exploit results from
statistical physics. Tricritical exponents of the $\phi^6$-model have
been calculated by Lawrie and Sarbach (1984). Of particular interest
is the result for $\alpha$, one finds $\alpha={1\over 2}$. Note that
now $\alpha>0$ (in contrast to the 3-dim $O(4)$-model) indicating a
true divergence of the specific heat when $T\to T_c$ and $m_s\to
m_s^*$. In lattice calculations $m_s$ can be tuned to small values.  A
qualitative change in the shape of the cusp in the specific heat would
be a hint on a tricritical point nearby. This illustrates the
predictive power of the second hypothesis of Wilczek and Rajagopal
(1992) and Wilczek (1992).

\vskip12pt The third hypothesis concerns the chiral transition and its
relation to dynamical universality classes. So far we had to deal with
static properties of equilibrium QCD. The third hypothesis may be
relevant for off-equilibrium situations in heavy-ion collisions.
Therefore we postpone the discussion to section \ref{sec:V.D.3}.

\vskip12pt
Lattice simulations in full QCD can measure critical and
tricritical indices in the vicinity of $T_c$ (and $m_s\sim m_s^*$). If
these results would disagree with the predictions of the above
conjectures, it can have different origins.  One possible reason is
the size of the subleading terms in Eqs.\ (\ref{4.6}), (\ref{4.9}),
(\ref{4.10}).  The subleading terms are only small as long as the
parameters $T$ and $m_s$ are in the near vicinity of the critical
values $T_c$ and $m_s^*$. Corrections from subleading terms do not
invalidate the hypothesis. The very ansatz would fail, if the physical
values for $m_u$, $m_d$, $m_s$ perturb the idealized situation of
2-flavor QCD too strongly to reveal the underlying substructure of the
chiral limit. Furthermore $SU(2)\times SU(2)$ is a sensible choice for
the symmetry group, but not unique. If $SU(2)\times SU(2)$ would be
replaced by $SU(3)\times SU(3)$, it is not possible to adapt the
hypothesis to the change in the symmetry group. Along with the second
order of the transition the universality arguments are lost, which
were so welcome for justifying a replacement of QCD by scalar
$O(4)$-models.

Finally it is a question of size whether corrections due to non-zero
modes and heavier mesons are negligible at $T_c$. In the next section
we present an attempt to include heavier mesons (in a four-dimensional
$SU(2)\times SU(2)$ non-linear sigma model). The results shed some
doubts on the very ansatz for the 3-dimensional action in terms of
$\vec\phi\equiv (\sigma,\vec\pi)$.

\subsubsection{The chiral transition in chiral perturbation theory}
\label{sec:IV.A.2}
It is tempting to extrapolate results from low temperatures to the
transition region. Pions as the lightest hadrons are most easily
excited at low temperatures. One may expect that they remain the only
relevant degrees of freedom up to $T_c$. Current quark masses are
small compared to the scale of the transition temperature. Thus the
naive conclusion would be that their effect is negligible at all.
Based on these simplifying assumptions one would study the chiral
limit of an $O(4)$-model as effective description for the chiral
transition region ($T\sim T_c$). Results of Gerber and Leutwyler
(1989) suggest that this way of arguing is too naive. We present their
results in this section.

Chiral perturbation fails to describe the transition region, but leads
to reliable predictions for the condensate and other thermodynamic
quantities at small temperatures. The influence of finite quark masses
on the $T$-dependence of $\left\langle \bar qq\right\rangle$,
$\varepsilon$ and $p$ can be analyzed in the low temperature region.
The effect of heavier mesonic modes may be estimated in a dilute gas
approximation for somewhat higher temperatures. Non-negligible effects
of finite quark masses and heavier mesons show up in the marginal
validity range of chiral perturbation theory. Their extrapolations to
the transition region can be summarized as follows.

\begin{itemize}
\item Below $T=150$~MeV the nonzero quark masses reduce the
  temperature dependence of the condensate by roughly a factor of 2.
  
\item The effect of massive states on the energy density
  $\varepsilon(T)$ is even more significant. In the chiral limit the
  energy stored in the massive states reaches the order of the energy
  stored in the pions when $T\sim130$~MeV. Nevertheless the massive
  states may be diluted at this temperature. The main part of the
  energy is rest energy of the massive states. At $T\sim200$~MeV the
  mean distance between two particles is reduced to $d=0.9$~fm. This
  estimate is based on the density formula for a free gas. The
  approximation of a dilute free gas of massive states is then no
  longer justified. Below $T\sim100$~MeV massive states may be
  neglected.
  
\item At $T\sim200$~MeV the dilute gas approximation for the massive
  modes predicts a melting of the condensate, even if the pions are
  completely ignored.  This sheds some light on the idealization of
  the predominant role of pions in the chiral transition (which we
  have anticipated in the end of the previous section).
  
\item The extrapolated value of $T_c$ decreases from $\sim190$~MeV to
  $\sim170$~MeV, when heavier mesons are included in the chiral limit.
  
\item $T_c$ decreases from $\sim240$~MeV to $\sim190$~MeV, when heavier
  mesons are included at finite quark masses. While heavier mesons
  accelerate the melting of the condensate, non-vanishing quark masses
  delay the melting by $\sim20$~MeV.
\end{itemize}

Let us briefly recall why chiral perturbation theory is a suitable
framework for describing QCD at low temperatures. Chiral perturbation
theory is an expansion in small momenta. Its applicability to QCD is
based on the fact that strong interactions are weak at low energies.
This is a consequence of chiral symmetry. The chiral symmetry of QCD
in the massless limit implies that the interaction strength is
proportional to the square of the energy if the energy is small. At
low temperatures the properties of the hadron gas are determined by
the lightest excitations. The lightest hadrons are the pions.  At low
temperatures the average momenta of the pions are small. Thus the
strong interactions between the pions may be treated {\it
  perturbatively} in the framework of chiral perturbation theory.

To be specific, we consider the $SU(2)_R \times SU(2)_L$ chiral
symmetry. The particle contents is given by the three pion components
$\pi^0$, $\pi^-$ and $\pi^+$. The fourth component corresponding to
the $\sigma\,$-mode is frozen due to the non-linear realization of the
symmetry.  The pion field is described by a matrix field $U(x)\in
SU(2)$. It transforms under global chiral rotations according to
\begin{equation}
  U(x) \longrightarrow V_R U(x) V^+_L\, ,
  \label{4.19}
\end{equation}
where $V_R$, $V_L\in SU(2)$ and U(x) is parametrized as
$U(x)=\exp\left\{i\vec{\tau}{\vec {\varphi} (x) \over f}\right\}$.
Here $\vec{\tau}$ denote the Pauli matrices, $f$ will be later
identified with the pion constant, the components of $\vec{\varphi}$
represent the three components of the pion field.

If one allows only for small four-momenta of the pions, the field
$U(x)$ is slowly varying. It is then convenient to expand the
Lagrangian in powers of derivatives of the fields $\partial_{\mu} U$,
equivalent to a power series in external momenta $p$. The ansatz is
given as
\begin{equation}
  L_{\hbox{eff}} = L^{(0)}+L^{(2)}+L^{(4)}+\ldots
  \label{4.20}
\end{equation}
The upper index counts the number of derivatives $\partial_{\mu}$,
thus $L^{(0)}$ does not depend on U and can be dropped. Lorentz
invariance forbids odd powers in the derivatives. First we consider
the chiral limit with a massless pion field. The form of $L^{(2n)}$ is
then completely determined by the symmetry requirement, i.e.  the
invariance under $SU(2)_R \times SU(2)_L$ chiral transformations. This
leads to
\begin{equation}
  L^{(2)}={f^2 \over 4} Tr\, \partial_{\mu} U^+\partial 
  ^{\mu}U.
  \label{4.21}
\end{equation}
Note that this term is just the familiar non-linear $SU(2)\times
SU(2)$ sigma model. The interaction between the pion fields is
described in a loop expansion with $\vec\varphi$-fields propagating in
the loops. The only place where the finite temperature enters is in
the boundary conditions, which lead to modified propagators.  The
power counting rules for graphs which are associated with the free
energy are modified compared to the rules for chiral perturbation
theory in the loop expansion at zero temperature. The only
modification consists in a replacement of the role of the external
momenta by the temperature. There is a one to one-correspondence
between the order in the low temperature expansion and the loop
expansion of chiral perturbation theory.

The expansion parameter is $T^2\over (8f^2)$. Since an increasing
number in these loops corresponds to a higher order in the low
temperature expansion, $L^{(2n)}$, $n=1,2\dots$ is to be expanded in
powers of $\vec{\varphi}$ to the order which is needed for the desired
accuracy in the low T-expansion.  Up to terms of order $\varphi^6$ we
find for example for $L^{(2)}$
\begin{eqnarray}
  L^{(2)}= {1\over 2}\partial_{\mu}\vec{\varphi}\partial^{\mu}
  \vec{\varphi}
  + && {1\over 6 f^2}(\vec{\varphi}\partial_{\mu}\vec{\varphi}) 
  (\vec{\varphi}\partial^{\mu}\vec{\varphi}) 
  \nonumber\\
  - && {1\over 6 f^2}(\partial_{\mu}\vec{\varphi}
  \partial^{\mu}\vec{\varphi})\vec{\varphi}^2 +
  0(\varphi^6)\;.
  \label{4.22}
\end{eqnarray}

Note that the chiral symmetry fixes $L^{(2)}$ up to one constant $f$
which may be identified with $f_{\pi}$ in the chiral limit. At the
order of $p^4$, two further dimensionless couplings enter the
effective Lagrangian.  They have to be fixed from experiment.

Away from the chiral limit, the mass terms for the pion field induce
an explicit breaking of the chiral symmetry in the effective
Lagrangian. In the chiral limit the guiding principle for constructing
$\cal L$ is an expansion in powers of external momenta. It can be
maintained in the presence of mass terms, if the quark masses are
small compared to the relevant physical scales of the theory.  The
full effective Lagrangian is Taylor expanded in powers of $m$. The
derivate expansion can be superimposed on the Taylor expansion, term
by term. This way additional terms involving powers of $m$ are
generated, which break the chiral symmetry explicitly. The power
counting in external momenta of the chiral limit goes through, if
$m_{\pi}$ is taken to be of the order of $p$. Since the pion mass
square $M^2$ to lowest order in chiral perturbation theory is given as
$M^2=m^2_{\pi}=(m_u+m_d)\cdot B$ (B being the $T=0$-condensate,
$m_\pi$ being the physical pion mass), the quark masses are counted as
quantities of order $p^2$.  Inclusion of mass terms then leads to the
following modifications in $L^{(2)}$
\begin{equation}
  L^{(2)}= {f^2\over 4}\{Tr(\partial_{\mu}U^+\partial_{\mu}U)
  -Tr(M^2(U+U^+))\}\;,
  \label{4.23}
\end{equation}

$L^{(4)}$ depends now on four input parameters $l_1,\dots l_4$ (Gasser
and Leutwyler, 1984). Thus the main effect of finite quark masses is
an increase in the number of input parameters. Once the effective
action is determined and the input parameters are fixed from
experiment, the thermodynamics of the pion gas can be derived in the
usual way.

The partition function is given as the path integral over all pion
field configurations which are periodic in the Euclidean time
direction, $U(\vec x,x_4+\beta)=U(\vec x,x_4)$.
\begin{equation}
  Z=\int{\cal D}U\exp\left\{-\int_{{\rm R\kern-1.7ex 
        I\kern0.75ex}^3\times S^1} d^4 
    x\left(L^{(2)}+\dots\right)\right\}
  \label{4.24}
\end{equation}

where $L^{(2)}$ is given by Eq.\ (\ref{4.23}). The couplings of
$L^{(2n)}$ are treated as temperature independent. The order
parameter, energy density $\varepsilon$, pressure $p$ and entropy
density $s$ follow from the free energy density $\cal F$ in the
standard way (${\cal F}={-1\over \beta V}\ln Z$ in the large volume
limit, where $Z$ is given by Eq.\ (\ref{4.24})).

To obtain a low temperature expansion of $Z$, $L_{eff}$ is expanded in
powers of $\varphi$.

Let us illustrate the power counting in the expansion parameters for
the case where the free energy density is to be determined including
terms of order $T^8$. This is the accuracy, which has been achieved by
Gerber and Leutwyler (1989).

To obtain $\cal{F}$ to the order of $T^8$, L must be expanded to order
$p^8$, contributions come from $L^{(2)}$, $L^{(4)}$, $L^{(6)}$ and
$L^{(8)}$. Tree graphs are generated from all terms $L^{(2)} \ldots
L^{(8)}$, one-loop graphs contain vertices of $L^{(4)}$ and $L^{(6)}$,
two-loop graphs vertices of $L^{(2)}$, $L^{(4)}$, and three-loop
graphs are generated by $L^{(2)}$. Thus $L^{(2)}$ has to be specified
in terms of $\vec{\varphi}$ including powers of the order $\varphi^6$.
Divergent integrals are treated in dimensional regularization.

The leading contribution of the Lagrangian $L^{(2)}$ contains a
free-field term (i.e. a term quadratic in $\vec\varphi$ with mass
$M^2$, the lowest order contribution to the pion mass). Further terms
of the Lagrangian are kept as perturbations depending on their
importance at low energies. This way chiral perturbation theory
provides a systematic expansion also for thermodynamic quantities. It
goes beyond the mean-field level, where one frequently stucks in the
treatment of effective models.

In the {\it chiral limit} the free energy density $\cal F$ takes the
general form
\begin{equation}
  {\cal F}=\sum\limits_{m,n=0,1,\dots}
  c_{mn}{}^.(T^2)^m{}^.(T^2\log T)^n
  +{\cal O}\left(e^{{-M_e\over T}}\right)\;.
  \label{4.25}
\end{equation}

The correction of the order $\exp\{-M_e/T\}$ stands for the
contribution of particles which remain massive in the chiral limit,
$M_e$ denotes the lightest of these masses.

A similar expression can be derived in the presence of finite quark
masses, when a common small expansion parameter $\lambda$ is
introduced such that $T\to\lambda T, m\to\lambda^2m$. The formula for
$\cal F$ then holds in powers of $\lambda$ and $\log\lambda$ with
$\lambda$-independent coefficients $c_{mn}$ which are now nontrivial
functions of $(m_\pi/T)$.

\vskip12pt {\it Free field case.} Let us first ignore the interactions
between the pions and consider $L^{(2)}$. It leads to the following
familiar expressions for the free energy density $\cal F$, the
pressure $p$ and the condensate $\left\langle \bar qq\right\rangle$
for an ideal gas of pions with mass $m_\pi$
\begin{eqnarray}
  {\cal F}&&=\varepsilon_0-p=\varepsilon_0+3T\int 
  {d^3p\over(2\pi)^3}\ln \left\{1-e^{-E/T}\right\}+\dots \nonumber\\
  E&&=\sqrt{{\vec p}^2+m_\pi^2}\;.\
  \label{4.26}
\end{eqnarray}
Here $\varepsilon_0$ is the vacuum energy density at $T=0$.  In the
chiral limit the result for $\cal F$ is
\begin{equation}
  {\cal F} = \varepsilon_0-{\pi^2\over 30}T^4+\ldots 
  \label{4.27}
\end{equation}
with the $T^4$-law for the pressure of a free gas. The condensate is
obtained as
\begin{equation}
  <\bar{q}q>={\partial{\cal F}\over 
    m}={\partial\varepsilon_0\over\partial m}
  +{3\over 2}{\partial m^2_{\pi}\over \partial 
    m}\int{d^3p\over (2\pi)^3}
  {1\over E(e^{E\over T}-1)} \;.
  \label{4.28}
\end{equation}
The first term on the r.h.s. gives the condensate at temperature
$T=0$. Using the PCAC-relation
\begin{equation}
  f^2_{\pi}m^2_{\pi}=-m<0|\bar{q}q|0>\{1+0(m)\}~,
  \label{4.29}
\end{equation}
the condensate in the chiral limit is given by 
\begin{equation}
  <\bar{q}q>=<0|\bar{q}q|0>\cdot\{1-{T^2\over8f^2}+\ldots\}~,
  \label{4.30}
\end{equation}
where $f=f_{\pi}$ up to terms of order $m$ which are counted as order
($T^2$).  Thus the difference shows up in $T^4$-terms. Formula
(\ref{4.30}) displays the melting of the condensate as the temperature
increases. Simultaneously the pion density $n_{\pi}$ grows according
to
\begin{equation}
  n_{\pi}=0.365\cdot T^3
  \label{4.31}
\end{equation}
in the massless limit. Physically the increase in number density can
be made responsible for the melting process.

For $m\ne 0$, the integral in Eq. (\ref{4.28}) becomes a function of
${m_{\pi}\over T}$.

\vskip12pt\noindent {\it Including interactions up to three loop.} Let
us see next, how the results for $p$, $\varepsilon$ and $\left\langle
  \bar qq\right\rangle$ change, when interactions are included. It is
clearly beyond the scope of this review to go into details of the
three-loop calculation. Also we will not explain the way, how the
various constants in the effective Lagrangian and the logarithmic
scales are fixed from experimental observables (the pion mass, the
pion decay constant and various scattering lengths).

\vskip12pt
\noindent
{\it Results in the chiral limit without heavier mesons.} The low
temperature expansion for the pressure is given as
\begin{equation}
  p ={\pi^2\over 30} T^4\left\{1+{T^4\over 36f^4}\ln 
    {\Lambda_{p}
      \over T} +0(T^6)\right\}\;.
  \label{4.32}
\end{equation}
Here $\Lambda_{p}=275\pm65$~MeV, $m_{\pi}$, $f_{\pi}$ and the
I=0, D-wave scattering length have been used as experimental input.

The low temperature expansion for the energy density $\varepsilon
=T({\partial p \over \partial T})-p$ follows as
\begin{equation}
  \varepsilon ={\pi^2\over 10} T^4\left\{1+{T^4\over 
      108f^4}(7\ln {\Lambda_{p}\over T}-1)+0(T^6)\right\}\;.
  \label{4.33}
\end{equation}
The general expression for the condensate can be written as 
\begin{equation}
  \left\langle \bar{q}q\right\rangle = \left\langle 
    0|\bar{q} q|0\right\rangle \left\{1+{c\over f^2}{\partial 
      p\over \partial m^2_{\pi}}\right\}
  \label{4.34}
\end{equation}
with
\[
c=-f^2 { {\left(\partial m^2_{\pi}\over \partial m\right)} 
  \over
  {\left\langle 0|\bar{q}q|0\right\rangle}} \;,
\]
cf.~Eqs.\ (\ref{4.28}), (\ref{4.30}) above. Hence one needs to know
$p=p(m^2_{\pi})$ (Leutwyler, 1988). The temperature independent term
$c$ has to be determined, $c=1$ in the chiral limit. The result for
$\left<\bar{q}q\right>$ including terms of order $T^6$ takes the
following form
\begin{eqnarray}
  \left\langle \bar{q}q\right\rangle&&= \left\langle 
    0|\bar{q}q|0\right\rangle\left\{1-x-{1\over 6}x^2-{16\over 
      9}x^3 \ln ({T\over \Lambda_q})+0(T^8)\right\} \nonumber\\
  x&&={ T^2\over (8f^2)},\qquad \Lambda_q = 470 \pm 110~\hbox{MeV}\;.
  \label{4.35}
\end{eqnarray}
 
The expansion parameter $x$ shows that the temperature scale is
determined by $\sqrt{8}f\sim 250$~MeV. The scale $\Lambda_q$ is fixed
in an analogues way to $\Lambda_{p}$. Both scales $\Lambda_{p}$ and
$\Lambda_{q}$ can be expressed in terms of a single scale
$\Lambda_{b}$ which is fixed from $\pi\pi$-scattering. Thus the
behavior of the pressure $p$ to order $T^8$ and of
$\left<\bar{q}q\right>$ to order $T^6$ needs only two constants
$f_{\pi}$ and $\Lambda_b$. This does not leave much room for
phenomenological fitting and demonstrates the predictive power of
chiral perturbation theory.

The results for $\left\langle \bar{q}q\right\rangle$ and the energy
density $\varepsilon$ are displayed in Fig.~\ref{fig:17}.

It is tempting to read off a transition temperature of
$T_c\sim190$~MeV, if the three-loop curve is extrapolated to a
temperature where the condensate vanishes. This extrapolation suffers
from two error sources. The error bars on the three-loop curve only
refer to the uncertainty in the constants $f$ and $\Lambda_b$, and not
to the finite temperature corrections of Eq.\ (\ref{4.35}).
Contributions of massive states to $\cal F$ have been left out so far.
They are no longer negligible, when $T$ exceeds 150~MeV. Below we will
further discuss the influence of massive states.

The energy density in Fig.~\ref{fig:18} grows with $T$ up to
$T\sim180$~MeV. The decrease of $\varepsilon/T^4$ for higher
temperatures indicates the invalidity of the expansion scheme.

\vskip12pt\noindent {\it Non-vanishing quark-masses.} Non-vanishing
quark-masses change the general expression for the pressure, where
certain constants of the chiral limit now depend on $m_\pi$ also via
$m$. The result for the quark condensate is shown in
Fig.~\ref{fig:19}. The effect of finite quark masses is that it takes
longer for the condensate to melt.  The extrapolated transition
temperature is $\sim240$~MeV (which lies far outside the validity range
of the approximation). The error bars indicate the uncertainties in
four input parameters, which have to be fixed from experiment for
$m>0$.

A comparison of Figs.~\ref{fig:17} and \ref{fig:19} shows that the
effect of finite quark masses is substantial already inside the
validity range of the chiral expansion. The reason why the tiny quark
masses may cause such changes in the final results is that they are
confined to pions. It is the pion mass ($\sim 140$~MeV) that enters the
Boltzmann factors in the thermodynamic formulas for the hadronic
phase.

\vskip12pt\noindent {\it Influence of heavier mesons.} In an expansion
around the chiral limit quark masses must be treated as perturbations
when they are incorporated in an expansion of chiral perturbation
theory. A perturbative treatment of $m_u$ and $m_d$ is certainly
justified, the up- and down-quarks are light compared to the energy
scale of $T_c$. Mesons containing strange quarks are accordingly
heavier than the pions. Frequently they are argued to be negligible in
the thermodynamics, as massive states are suppressed by their
Boltzmann factors $\exp\{-M/T\}$. This argument fails for higher
temperatures, in particular in the vicinity of the transition.

One way of including (part of the) heavier mesons is to extend the
symmetry group from $SU(2)\times SU(2)$ to $SU(3)\times SU(3)$ (cf.
section \ref{sec:IV.A.4}). In the framework of chiral perturbation
theory this means that also $m_s$ should be treated perturbatively.

A further alternative is to treat all massive mesons (others than
pions) on an equal footing and consider their number density and
energy density in a dilute gas approximation (Gerber and Leutwyler,
1989). In a dilute gas approximation the interactions between the
mesons are neglected. This justifies an ansatz for the change in the
free energy density according to
\begin{equation}
  \Delta{\cal 
    F}=-{1\over2}T^4\sum\limits_ih_0\left({T\over M_i}\right)\;.
  \label{4.36}
\end{equation}
The sum extends over all mesons tabulated in the particle data booklet
apart from pions. It is clear that many small contributions may
accumulate to a non-negligible amount.  The change $\Delta\cal F$ is
due to a change in the pressure. The term
${1\over2}T^4h_0\left({T\over M_i}\right)$ gives the pressure of a
free boson gas of particles with mass $M_i$. The function $h_0$ can be
expressed in terms of Bessel functions. The induced change in the
condensate (due to the mass $M_i$) follows from Eq.\ (\ref{4.28}) with
$m_\pi$ replaced by $M_i$. It is given as
\begin{equation}
  \Delta\left\langle\bar 
    qq\right\rangle=T^2\sum\limits_ih_1\left({M_i\over 
      T}\right){}^.\left({\partial M_i\over 
      \partial m^2}\right){}^.M_i
  \label{4.37}
\end{equation}
where
\[
T^2h_1=\int{d^3p\over(2\pi)^3}{1\over E}{1\over e^{E/T}-1}
\]
or equivalently
\begin{equation}
  \Delta \left\langle \bar{q}q\right\rangle=-\left\langle 
    0|\bar{q}q|0\right\rangle{T^2\over f^2}\cdot \sum_i h_1
  ({T\over M_i})\cdot {M_i m\over m^2_{\pi}}\cdot 
  {\partial M_i\over \partial m}\;.
  \label{4.38}
\end{equation}

\noindent
Here we have used the lowest order PCAC-relation for $f^2_{\pi}
m^2_{\pi}$. The derivative ${\partial M_i \over \partial m}$ can be
estimated in the non-relativistic quark model, where it equals $N_i$,
i.e. the number of valence quarks of type $u$ or $d$. Formula
(\ref{4.38}) explains the effect of massive states on the melting of
the condensate.  It accelerates the melting and reduces the transition
temperature. In Fig.~\ref{fig:20} the melting of the chiral condensate
is compared for the three-loop pion contribution and the superposition
of pions plus massive states. The extrapolated transition temperature
is around 190~MeV, smaller than $T_c$ without heavier mesons, but
$\sim 20$~MeV larger than in the chiral limit with heavier mesons. The
shaded area reflects the uncertainty of $\left(m {\partial M_i\over
    \partial m}\right)$ entering the estimate for the contribution of
massive states (Eq.\ (\ref{4.38})). The acceleration of the melting
process due to the massive states is more pronounced for realistic
quark masses than for vanishing ones. In the chiral limit the melting
of the condensate is mainly enforced by the pions, the pions play a
more distinguished role, thus the condensate is less sensitive to
heavier mesons.

\vskip12pt At temperatures $T>150$~MeV pions have lost their dominant
role in melting the condensate. If lattice calculations or other
approximation schemes support a value of $T_c>150$~MeV, it is
challenging to elucidate the way of how the massive modes drive
$\left\langle \bar qq\right\rangle$ to zero. It seems to be very
worthwhile to improve on the dilute gas approximation. A suggestion
due to Gerber and Leutwyler (1989) is to treat the mutual interactions
between the massive states in some bootstrap model, modified and
adapted to QCD (see e.g. Hagedorn, 1985 and 1983).

More recently the temperature dependence of the normalized light quark
condensate has been derived by Bunatian and Wambach (1994) in a
nonperturbative approach. The approach is known from phonon-phonon
interactions in condensed matter physics (see e.g. Luttinger and Ward,
1960; Fulde and Wagner, 1971) and may be more appropriate to the
transition region. In the chiral limit the critical temperature is
identified as the point where the residue of the single-pion
propagator becomes purely imaginary. The pion interactions are
described by the Weinberg Lagrangian (Weinberg, 1966, 1967 and 1979).
The thermodynamic potential for the interacting pion gas is determined
in the Hartree approximation. After using the Gell-Mann-Oakes-Renner
relation, the normalized light quark condensate follows from
$\partial\Omega/\partial m_{\pi}^{2}$, where $\Omega$ denotes the
difference in free energy densities at $T=0$ and $T>0$. The results
confirm chiral perturbation theory up to temperatures $<100$~MeV, but
are clearly different in the transition region, where chiral
perturbation theory is expected to fail.

%
\subsubsection{Clausius-Clapeyron equation for QCD}
\label{sec:IV.A.3}
We recall the Clausius-Clapeyron equation for a liquid/vapor system in
statistical physics. It has been derived in the last century by
Clapeyron (1834) and Clausius (1850). The Clausius-Clapeyron equation
relates the discontinuity in the order parameter to the latent heat
when the phase transition is of first order. For a liquid/gas system
the order parameter is the specific volume or the volume per molecule,
the latent heat equals the entropy gap per molecule. The ratio of the
discontinuities in the specific volume and the entropy is determined
by the dependence of the critical temperature on the pressure during
the transition. The relation is given as
\begin{equation}
  {\partial T_c \over \partial p}={v_{\hbox{gas}}-v_{\hbox{fluid}}
    \over s_{\hbox{gas}}-s_{\hbox{fluid}}} 
  = {\hbox{disc}\, v \over \hbox{disc}\, s}\, ,
  \label{4.39}
\end{equation}
where disc stands for discontinuity.

With the dictionary of Table~\ref{tab:2} Eq.\ (\ref{4.39}) can be
translated to QCD, where it reads
\begin{equation}
  {\partial T_c \over \partial m_q}
  ={T_c~\hbox{disc}\, \left<\bar q q\right>_T\over \hbox{disc}~\epsilon}~.
  \label{4.40}
\end{equation}
The pressure is replaced by the current quark mass $m_q$. The
conjugate variable of $m_q$ is the order parameter $\left<\bar q
  q\right>$ at temperature $T$. The chiral condensate is determined as
response of the partition function to a change in the quark mass
\begin{equation}
  \left<{\bar q}q\right>_T 
  =-{1\over V}{\partial \over \partial m_q} \ln Z \, .
  \label{4.41}
\end{equation}
We sketch the derivation of Eq.\ (\ref{4.40}) and show how it leads to
a bound on the latent heat in a hypothetical first order chiral
transition. The derivation is due to Leutwyler (1992). First note that
the discontinuity in the energy density $\epsilon=(T \partial
p/\partial T-p)$ can be expressed by the gap in the entropy density
\begin{equation}
  {\hbox{disc}}\, \epsilon 
  =T_c\, {\hbox{disc}} {\partial p \over \partial T}\, .
  \label{4.42}
\end{equation}
From Eq.\ (\ref{4.41}) with $\ln Z=V(p-\epsilon_0)$ ($\epsilon_0$
being the vacuum energy density) it follows that
\begin{equation}
  \left<{\bar q}q\right>_T
  ={\partial \epsilon_0 \over \partial m_q}
  -{\partial p \over \partial m_q}
  =\left<0|{\bar q}q|0\right>-{\partial p \over \partial m_q}\, .
  \label{4.43}
\end{equation}
If we expand the pressure in the vicinity of $T_c$ according to
\begin{equation}
  p =p_c + (T-T_c) {\partial p \over \partial T}
  \bigg\vert_{T_c}+\ldots \, ,
  \label{4.44}
\end{equation}
$p$ depends on $m_q$ via $p_c$ and $T_c$
\begin{equation}
  {\partial p \over \partial m_q}={\partial p_c \over \partial m_q}
  -{\partial T_c
    \over \partial m_q}{\partial p \over \partial T}\bigg|_{T=T_c} \, .
  \label{4.45}
\end{equation}
Next we insert Eq.\ (\ref{4.45}) into Eq.\ (\ref{4.43}) and apply
``disc'' on both sides.  The result is
\begin{equation}
  {\hbox{disc}}\, \left<{\bar q}q\right>_T
  ={\partial T_c \over \partial m_q}\cdot{\hbox{disc}}\,
  {\partial p \over \partial
    T}\bigg|_{T=T_c}~.
  \label{4.46}
\end{equation}
While $p$ is continuous at $T_c$ (also in case of a first order
transition), $\partial p \over \partial T$ may jump. From Eqs.\ 
(\ref{4.42}) and (\ref{4.46}) we finally obtain Eq.\ (\ref{4.40}).

Let us see what the Clausius-Clapeyron relation implies for $\Delta
T_c$, i.e.  the change in the critical temperature, when the physical
values for the current quark masses $m_u$ and $m_d$ are replaced by
their chiral limit. As $m_{u,d}$ are small compared to $T_c$, we may
expand
\begin{equation}
  \Delta T_c =m_u{\partial T_c \over \partial m_u}
  +m_d{\partial T_c \over\partial m_d}\, ,
  \label{4.47}
\end{equation}
where $\partial T_c / \partial m_{u,d}$ is given by Eq.\ (\ref{4.40}).
It remains to replace the quark masses in Eq.\ (\ref{4.47}) by means
of the Gell-Mann-Oakes-Renner-relation (Gell-Mann et al., 1968) $m_u
\left<\bar u u\right>_{T=0} + m_d \left<\bar d d\right>_{T=0}\sim
f_\pi^2 m_\pi^2 \sim$~20~MeV/fm$^3$.  The result is
\begin{equation}
  {\Delta T_c \over T_c}
  =\Bigg|{{\hbox{disc}} \left<\bar q q\right>_{T>0} \over \left<\bar q
      q\right>_{T=0}}\Bigg| ~
  \Bigg|{f^2_{\pi}m^2_{\pi}\over {\hbox{disc}}\, \epsilon}\Bigg|\, .
  \label{4.48}
\end{equation}
This relation restricts the size of the latent heat disc $\epsilon$ by
the sensitivity $\Delta T_c / T_c$ of the critical temperature to the
quark masses. An upper bound on disc $\epsilon$ corresponds to an
upper bound on the (hypothetical) gap in the order parameter.

To get an estimate for this bound, take $T_c \sim 140$~MeV as a lower
bound on the chiral transition temperature. The number density $n$ of
the hadron gas grows with temperature. At $T_c \sim 140$~MeV $n$
increases by $\sim 30\%$ in passing from the physical quark masses to
the chiral limit ($m_{u,d} \to 0$) (see Leutwyler, 1992). This
estimate is based on a dilute gas approximation, where the change in
the particle abundances is only induced by the change in their mass as
function of $m_{u,d}$. If the onset of the phase transition is
determined by the average volume per particle, a smaller transition
temperature has to compensate for the increased density to get the
same average volume.  30\% increase in the density $n$ induce a
decrease of 5\% in $T_c$, i.e.  $\Delta T_c/T_c\sim 5\%$ under the
above conditions. For the maximal jump of $\left<\bar q
  q\right>_{T_c}$ one thus finds
\begin{equation}
  {\hbox{disc}}\, \epsilon < 0.4 {{\hbox{GeV}} \over {\rm fm}^3}\, .
  \label{4.49}
\end{equation}
This value is in good agreement with the bound on the latent heat
obtained in a large-$N$-approximation to the $SU(3)\times SU(3)$
linear sigma model (Meyer-Ortmanns and Schaefer, 1995), cf. the next
section. For comparison we mention the gap in the gluonic energy
density in a pure $SU(3)$ gauge theory (Laermann et al., 1995).
Laermann et al. find 
$\hbox{disc}~\epsilon_{\rm{gluonic}}/T_{c}^{4}\sim 1.5$ or 
$\hbox{disc}~\epsilon_{\rm{gluonic}}\sim 0.9~\hbox{GeV/fm}^{3}$
for $T_{c}=260$~MeV.

For a larger value of disc~$\epsilon$, say $\sim 2~\hbox{GeV/fm}^3$,
the Clausius-Clapeyron relation implies ${\Delta T_c/T_c} < 1\%$.
Results of chiral perturbation theory indicate that such a small
change in the critical temperature is rather unlikely. The condensate
and the energy density are rather sensitive to the pion mass $m_\pi$.

Chiral perturbation theory cannot predict the order of the phase
transition, as it looses its validity in the transition region, but it
does provide a useful upper bound on the latent heat. This bound
should be noticed in particular by phenomenologists, who like a large
latent heat for visible effects in heavy-ion collisions. Scenarios
based on such an assumption may be incompatible with the underlying
QCD.

\subsubsection{Mass sensitivity of the chiral transition}
\label{sec:IV.A.4}
In this section we discuss two topics. The first one is the mass
sensitivity of the chiral transition in the $SU(3)\times SU(3)$ linear
sigma model. We give estimates for {\it critical meson/quark masses}
at which the first order chiral transition becomes of second order and
turns into a smooth crossover phenomenon for larger masses. The second
topic is the equation of state for physical meson masses.

\bigskip {\it The aspect of mass sensitivity.} The mass sensitivity of
the order of the chiral transition is essential in view of realistic
predictions for heavy-ion collisions. To estimate the effect of finite
masses on the chiral transition we recall the analogy to a ferromagnet
or a liquid/gas system.Quark masses are analogous to the pressure in a
liquid/gas system or an external magnetic field in a ferromagnet. If
the pressure exceeds a critical value, the first order transition from
a liquid to a gas ceases to occur, it is replaced by a smooth
crossover between the liquid and the gas phases. Similarly a first
order transition in a $Z(3)$-Potts model becomes a second order
transition for a critical value of the external field and disappears
beyond this strength.

When the analogy to a statistical system is translated to an effective
model for QCD, the question concerning the effect of finite masses can
be posed in the following way. Are the physical meson masses too small
for the chiral transition to maintain its first order? In section
\ref{sec:III} we have described an attempt to answer this question
within the lattice approach, cf.  Fig.~\ref{fig:13}. Here
we study the same question on the mesonic level. The bare quark masses
$m_s{}^.a$ and $m_{u,d}{}^.a$ of Fig.~\ref{fig:13} should be replaced
by meson masses with and without strange quark content.

A first attempt to estimate the critical phase boundary between mass
parameters of first order transitions and crossover phenomena has been
made by (Meyer-Ortmanns et al., 1992) in the linear $SU(3)\times
SU(3)$ sigma model. More recently Gavin, Goksch and Pisarski (1994)
addressed a similar question in the same sigma model within a
mean-field approach. In the remainder of this section we will focus on
results of a finite mass analysis of Meyer-Ortmanns and Schaefer
(1995) in a large-$N_{f}$-approximation. As a quantitative measure for
the distance (in mass parameter space) between the physical and
critical meson masses we consider the ratio of the associated critical
to physical light quark masses $m_{u,d}^{\rm{crit}}/m_{u,d}$. If this
ratio turns out to be much smaller than 1, physical masses lie deeply
in the crossover region, and it is difficult to imagine any signatures
specific for the phase conversion from the plasma to the hadron phase.
The most attractive possibility is a ratio of the order of 1. In that
case the physical masses are almost ``critical'', effects due to a
large correlation length should be visible. Non-universal features of
the sigma model are then negligible and the decimation of QCD to an
effective model in the same universality class is an allowed
simplification.

Here we would like to recall that sigma models may not be the right
candidates for sharing a universality class with QCD (Kocic and Kogut,
1995). As long as the deviations from the ideal second order behavior
are perturbatively small, they are theoretically under control (see
e.g. the finite mass scaling analysis of Boyd et al., 1992, Rajagopal
and Wilczek, 1993, or Kocic and Kogut, 1993), and a comparison of
critical indices in the effective model and on the lattice serves as a
check of the conjectured universality class.

\paragraph*{Choice of the model}
In our application we assume that the restoration of the spontaneously
broken $SU(3)\times SU(3)$ symmetry is the driving mechanism for the
chiral phase transition. The deviations in the spectrum from the
idealized octet of pseudoscalar Goldstone bosons are parametrized by
terms which break the $SU(3)\times SU(3)$ symmetry explicitly. The
assumption in our decimation is that only mesons associated with the
$SU(3)\times SU(3)$ multiplets are important for the phase transition.
The criterion is chiral symmetry (rather than the size of the meson
masses, otherwise one should include $\rho$-mesons or others as well).
The reason why we have chosen $SU(3)\times SU(3)$ rather than
$SU(2)\times SU(2)$ is to account for the influence of the strange
quark mass on the thermodynamics. With $SU(3)\times SU(3)$ we also
include some of the heavier mesons. From section \ref{sec:IV.A.2} we
know that heavier mesons are non-negligible in the transition region.

We have chosen the sigma model as an effective model for the {\it low}
temperature phase of QCD ($T\leq T_{c}$). In the low temperature phase
quarks are confined to hadrons, and chiral symmetry is spontaneously
broken. The meson spectrum reflects some remnants of this symmetry
breaking. In the transition region the use of the model becomes
questionable, the model certainly fails above $T_{c}$ as a description 
of the plasma phase.

We use the sigma model in a similar spirit as the $O(4)$-model in
section \ref{sec:II.A.1}. Its action is constructed in terms of QCD's
chiral order parameter field $\phi$, where $\phi$ now is a complex
$3\times 3$-matrix, parametrized as
\begin{equation}
  \phi=1/\sqrt{2}\sum_{l=0}^{8}(\sigma_{l}+i\pi_{l})\lambda_{l} .
  \label{4.50}
\end{equation}
Here $\lambda_{l}$ denote the Gell-Mann matrices, $\pi_{l}$ are the
pseudoscalar mesons, $\sigma_{l}$ are the scalar mesons. The mesonic
order parameter field is a bilinear in the left-handed and
right-handed quark fields
$\phi_{ij}=\left<\bar{q}_{i}^{L}q_{j}^{R}\right>$. In the sigma model
the quark structure is ignored by construction for all temperatures
$T\geq 0$. In terms of $\phi$ the Lagrangian reads
\begin{eqnarray}
  \label{4.51}
  L= && \int d^{4}x\left\{{1\over 2} \rm{Tr}
    (\partial_{\mu}\phi\partial_{\mu}\phi^{+})\right\}
  -{1\over 2}\mu_{0}^{2}\rm{Tr}(\phi\phi^{+})
  \nonumber\\
  + && f_{1}(\rm{Tr}\phi\phi^{+})^{2}
  + f_{2}\rm{Tr}(\phi\phi^{+})^{2}
  +g(\det\phi+\det\phi^{+})
  \nonumber\\
  - && \epsilon_{0}\sigma_{0}-\epsilon_{8}\sigma_{8} .
\end{eqnarray}
Note that there are 2 independent quartic terms with couplings $f_{1}$
and $f_{2}$. The determinant-terms are cubic in the components of
$\phi$, $g$ is the ``instanton'' coupling, it takes care on the right
$\eta$-$\eta'$-mass splitting, $\mu_{0}^{2}$ is the coupling of the
quadratic term. The external field $\epsilon_{0}$ gives a common mass
to the (pseudo)scalar meson octet, while $\epsilon_{8}$ accounts for
the right mass splitting inside the (pseudo)scalar meson octet.

\paragraph*{Tree level parametrization at zero temperature}
The parameters $\mu_0^2$, $f_1$, $f_2$, $g$, $\epsilon_0$,
$\epsilon_8$ of the Lagrangian\ (\ref{4.51}) should be chosen such
that the model reproduces the experimental values of the
(pseudo)scalar meson masses. The parametrization of the sigma model is
not unique (Chan and Haymaker, 1973; Gavin et al., 1994;
Meyer-Ortmanns et al., 1992). Here we are interested in a tuning of
meson masses in terms of a few parameters.  Suitable parameters are
the external fields $\epsilon_{0}$, $\epsilon_{8}$. The induce finite
quark masses according to
\begin{eqnarray}
  \label{4.52}
  -\epsilon_{0}&=&\alpha(2\hat m+m_{s})
  \nonumber\\
  -\epsilon_{8}&=&\beta(\hat m-m_{s})~,
\end{eqnarray}
where $\hat m\equiv(m_u+m_d)/2$, $\alpha$ and $\beta$ are constants.
Eq.~(\ref{4.52}) follows from an identification of terms in the
Lagrangians for quarks and mesons, which transform identically under
$SU(3)\times SU(3)$. The meson masses are determined for given
$\epsilon_{0}$, $\epsilon_{8}$, once the couplings $\mu_{0}^{2}$,
$f_{1}$, $f_{2}$, $g$ are specified and the condensates
$\left<\sigma_{0}\right>_{T=0}$, $\left<\sigma_{8}\right>_{T=0}$ are
calculated for given $\mu_0^2$, $f_{1}$, $f_{2}$, $g$. Thus we vary
the quark and meson masses by varying $\epsilon_{0}$ and
$\epsilon_{8}$.  The chiral limit is obtained for
$\epsilon_{0}=0=\epsilon_{8}$. The couplings $\mu_{0}^{2}$, $f_{1}$,
$f_{2}$, $g$ are then determined from the mass input in the chiral
limit, i.e. $m_{\pi}=m_{K}=m_{\eta}=0$, $m_{\eta'}=850$,
$m_{\sigma_{\eta'}}=800$, $m_{\sigma_{\eta}}=600$, all masses in units
of [MeV], $f_\pi=94$~MeV. Next we keep $\mu_{0}^{2}$, $f_{1}$,
$f_{2}$, $g$ fixed to their values in the chiral limit and change
$\epsilon_{0}$, $\epsilon_{8}$. The choice $\epsilon_{8}=0$,
$\epsilon_{0}\neq 0$ leads to an $SU(3)$-symmetric case with only one
order parameter field $\sigma_{0}$ for which the numerics considerably
simplifies. Meson masses with almost experimental values are induced
for $\epsilon_{0}=0.0265~\rm{GeV}^{3}$,
$\epsilon_{8}=-0.0345~\rm{GeV}^{3}$. 

The results are listed in Table~\ref{tab:7} and compared to the
experimental values.

This way we have constructed a mapping
\begin{equation}
  \label{4.53}
  \{m_{u,d},m_{s}\} \leftrightarrow (\epsilon_{0},\epsilon_{8})
  \leftrightarrow \{m_{\rm{Meson}}^{2}\}
\end{equation}
between quark and meson masses.

It remains to translate the meson condensates at zero temperature to
the light and strange quark condensates. In the same way as we
obtained the relation (\ref{4.52}), we find here
\begin{eqnarray}
  \label{4.54}
  \left<\bar q q\right>&=&\frac{\epsilon_{0}}{2\hat m+m_{s}} 
  \left<\sigma_{0}\right>
  +~\frac{\epsilon_{8}}{2(\hat m-m_{s})}\left<\sigma_{8}\right>
  \nonumber\\
  \left<\bar s s\right>&=&\frac{\epsilon_{0}}{2\hat m+m_{s}}
  \left<\sigma_{0}\right>
  -~\frac{\epsilon_{8}}{\hat m-m_{s}}\left<\sigma_{8}\right>~.
\end{eqnarray}
Eqs.~(\ref{4.54}) are derived at zero temperature. We take these
relations as temperature independent and use them to determine
$\left<\hat q q\right>(T)$, $\left<\hat s s\right>(T)$ from the
measured values for $\left<\sigma_{0}\right>(T)$,
$\left<\sigma_{8}\right>(T)$, see below.

\paragraph*{Critical meson masses in mean-field}
Although the method is crude, it is instructive to get a first
estimate for critical meson/quark masses. Later the results will be
compared with estimates from a large-$N$-approach. In a mean-field
calculation the full effective potential is replaced by the classical
part in terms of two constant background fields $\sigma_{0}$,
$\sigma_{8}$. For simplicity we consider here only the
$SU(3)$-symmetric case, where $\epsilon_{8}=0=\sigma_{8}$,
$\epsilon_{0}\neq 0$, $\sigma_{0}$ denotes a constant background
field. The effect of a finite (high) temperature in a mean-field
calculation is a renormalization of the quadratic term in the
Lagrangian. Thus a finite temperature can be mimiced by tuning
$\mu_0^2$ while keeping the other couplings $f_{1}$, $f_{2}$, $g$
fixed. For a critical field $\epsilon_{0}^{\rm{crit}}$, the first
order transition just disappears, and so does the cubic term in
$U_{\rm{class}}$. At $\epsilon_{0}^{\rm{crit}}$ $U_{\rm{class}}$
starts with a term proportional to
$(\sigma_{0}-\sigma_{0}^{\rm{crit}})^{4}$, where
$\sigma_{0}^{\rm{crit}}$ is the minimum of $U_{\rm{class}}$ for
critical values $\mu_{0}^{2\rm{crit}}$, $\epsilon_{0}^{\rm{crit}}$.
Thus we have
\begin{mathletters}
  \label{4.55}
  \begin{equation}
    \label{4.55a}
    U_{\rm{class}}(\sigma_{0})=-\frac{1}{2}\mu_{0}^{2}\sigma_{0}^{2}
    +\frac{2g}{3\sqrt{3}}\sigma_{0}^{3}+(f_{1}
    +\frac{f_{2}}{3})\sigma_0^4-\epsilon_{0}\sigma_{0} 
  \end{equation}
  \begin{equation}
    \label{4.55b}
    \left.U_{\rm{class}}\right|_{\rm{crit}}(\sigma_{0})
    =\left.\frac{1}{4!}\frac{\partial^{4}U_{\rm{class}}}
      {\partial\sigma_{0}^{4}}\right|_{\rm{crit}}
    (\sigma_{0}-\sigma_{0}^{\rm{crit}})^{4}+o(\sigma_{0}^{5}) .
  \end{equation}
\end{mathletters}
Here $\left.\right|_{\rm{crit}}$ means ``evaluated at critical
parameters''.  Note that $U_{\rm{class}}$ in Eq.~(\ref{4.55a}) takes
the same form as a free energy functional for a liquid/gas system. It
supports the analogy between a liquid/gas system and the chiral
transition in QCD as mentioned in the dictionary of Table~\ref{tab:2}.
The vanishing of the first three derivatives in Eq.~(\ref{4.55b})
determines $\sigma_0^{\rm{crit}}$, $\mu_0^{2\rm{crit}}$ and
$\epsilon_{0}^{\rm{crit}}$ as functions of $f_{1}$, $f_{2}$, $g$. In
the physical case of $\epsilon_{0}=0.0265~[\rm{GeV}^{3}]$,
$\epsilon_{8}=-0.0345~[\hbox{GeV}^{3}]$ we obtain
\begin{equation}
  \label{4.56}
  \frac{m_{u,d}^{\rm{crit}}}{m_{u,d}}\sim 0.03\pm 0.02~.
\end{equation}
Such a small ratio of 3~\% for the critical to physical light quark
masses would mean that the chiral phase transition is easily washed
out by tiny quark masses, and for physical quark masses one is left
with a rather smooth crossover phenomenon.

Our interest in the mean-field result is the order of magnitude of
this ratio. Recall that a first order transition can have different
origins. A first one is a cubic term in the classical part of the
potential. A second one is a $\phi^{6}$-term which may be needed for
stabilization of the free energy when the quartic coupling picks up a
negative sign due to renormalization effects. For two or more
independent relevant couplings a further first order
transition can be a so called {\em fluctuation induced} transition.
Since the linear $SU(3)\times SU(3)$ sigma model contains two such
couplings $f_{1}$ and $f_{2}$, the chiral transition may be mainly
fluctuation induced. This hypothesis has been recently discussed by
Gavin et al. (1994), see also Shen (1993), Paterson (1981),
Jacobsen and Amit (1981).

If the order of magnitude of the ratio~(\ref{4.56}) changes beyond the
mean-field level, it sheds some doubts on the simplified description
of Eq.~(\ref{4.55a}) and favours the hypothesis of a fluctuation
induced transition.

\paragraph*{The large-$N_{f}$-approach}
Ultimately we are interested in the temperature dependence of the
meson condensates $\left<\sigma_{0}\right>(T)$,
$\left<\sigma_{8}\right>(T)$. The condensates are determined as the
minima of the constrained free energy density under the constraint on
the average values
\begin{equation}
  \label{4.57}
  \frac{T}{V}\int_{0}^{1/T}d\tau\int d^{3}x
  \sigma_{0,8}(\vec x,\tau)=\bar\sigma_{0,8}~,
\end{equation}
where $\bar\sigma_{0,8}$ are prescribed values for the background
fields, while the same average value should vanish for
$\sigma_{\ell}$, $\ell=1,\ldots,7$ and $\pi_{\ell}$,
$\ell=1,\ldots,8$. The relation to the full partition function is
given by
\begin{eqnarray}
  Z&=&\int d \sigma_{0} \int d \sigma_{8}
  \exp \{-\beta V U_{\rm{eff}}(\sigma_{0},\sigma_{8})\}=
  \nonumber\\
  &=&\int d \sigma_{0}\int d\sigma_{8} \int {\cal{D}} \phi~
  \hbox{constraint}\cdot e^{-\int d^{4}x {\cal{L}}[\phi]}~,
  \label{4.58}
\end{eqnarray}
with ordinary integrals for the $\sigma_{0,8}$-integrations.

As a next step we quadratize the quartic part of the interaction by
applying a matrix version of the Hubbard--Stratonovich transformation
(Hubbard, 1959; Stratonovich, 1958; Frei and Patk\'os, 1990). It
reads
\begin{mathletters}
\label{4.59}
\begin{eqnarray}
  &&\exp \{-\beta[f_{1}(\rm{Tr}\phi'\phi'^{+})^{2}
  +f_{2}\rm{Tr}(\phi'\phi'^{+})^{2}]\} \nonumber\\
  &&=\hbox{const}\int_{c-i\infty}^{c+i\infty}{\cal{D}}\Sigma(x)
  \exp\{\rm{Tr}\Sigma^{2}+2\epsilon\rm{Tr}(\Sigma\phi'\phi'^{+})
  \nonumber\\
  &&+2\alpha\rm{Tr}(\phi'\phi'^{+})\rm{Tr}\Sigma \}~ ,
  \label{4.59a}
\end{eqnarray}
where $\phi'(x)$ is an $N\times N$-matrix field and
\begin{eqnarray}
  \epsilon^{2}=\beta f_{2} \nonumber\\
  2\epsilon\alpha+3\alpha^{2}=\beta f_{1} .
  \label{4.59b}
\end{eqnarray}
\end{mathletters}
The path integral over the auxiliary field $\Sigma(x)$ is evaluated in
a saddle point approximation. We replace $\Sigma(x)$ by
$\hbox{sad}\cdot\hbox{diag(1,1,1)}$. The saddle point approximation
corresponds to the leading term in a $1/N$-expansion in an
$O(N)$-model (Coleman et al., 1974). The $SU(3)\times SU(3)$ linear
sigma model reduces to an $O(18)$-model for $g=f_{2}=0$. The $N=18$
mesonic modes correspond to $N_{f}=3$ flavors ($N=2N_{f}^{2}$).
Therefore we call our approximation ``large-$N_{f}$''.

The advantage of the large-$N_{f}$-approximation is that we end up
with an effectively free field theory. The only remnant of the
interaction is hidden in the dispersion relation
\begin{equation}
  \omega_{Q}^{2}=p^{2}+\hbox{sad}+\mu_{0}^{2}+m_{Q}^{2} .
  \label{4.60}
\end{equation}
Here $Q=1,\ldots,8$ labels the particle multiplets, $m_Q^{2}$ are the
mass squares defined by the quadratic terms in the fluctuating fields,
sad is the contribution from the auxiliary field.

The final expression for
$U_{\rm{eff}}(\sigma_{0},\sigma_{8},\hbox{sad})$ contains a classical
part $U_{\rm{class}}$, a part coming from the quadratization
$U_{\rm{saddle}}$, a zero point energy part which is dropped, and a
thermal part $U_{\rm{th}}$
\begin{eqnarray}
  &&U_{\rm{eff}}(\sigma_{0},\sigma_{8},\hbox{sad})
  =U_{\rm{class}}+U_{\rm{saddle}}+U_{\rm{th}}
  \nonumber\\
  &&U_{\rm{th}}=\frac{1}{\beta}\cdot \sum_{Q=1}^{8}g(Q)
  \int\frac{d^3p}{(2\pi)^{3}}\ln(1-e^{-\beta\omega_{Q}})~,
  \label{4.61}
\end{eqnarray}
where $g(Q)$ are the multiplicities of the multiplets, e.g. $g(1)=3$
for 3 pions etc. The full expression for $U_{\rm{eff}}$ can be found
in (Meyer-Ortmanns and Schaefer, 1995).

The expression for $U_{\rm{eff}}(\sigma_{0},\sigma_{8},\hbox{sad})$ is
evaluated in a high temperature expansion and---alternatively---fully
numerically. The high-temperature expansion for a free field theory is
standard. Since we are interested in the low temperature phase ($T\leq
T_{c}$), the most we can expect from a high-temperature expansion are
qualitative results. The {\em numerical} evaluation of $U_{\rm{eff}}$
looks quite straightforward, but it is hampered by imaginary parts in
the effective potential. Although the final maximum in sad and the
minima in $\sigma_{0}$, $\sigma_{8}$ turn out to lie in the region of
real valued $U_{\rm{eff}}$, the routines encounter imaginary parts in
intermediate steps. The choice of initial values for sad,
$\sigma_{0}$, $\sigma_{8}$ has to be optimized to make the numerical
evaluation tractable.

\paragraph*{Results for critical quark/meson masses}
In the large-$N_{f}$-approximation the chiral transition is washed out
for an average pseudoscalar octet mass $\geq 203$~[MeV], if the ratio
of $m_{s}/m_{u,d}$ is kept fixed at its realistic value 18.2. This
gives an upper bound on the first order transition region. For the
corresponding {\it critical} quark masses we find
\begin{mathletters}
\label{4.63}
\begin{eqnarray}
  \label{4.63a}
  m_{u,d}^{\rm{crit}}\leq 2.96\pm 0.85~\rm{MeV} \nonumber\\
  \\
  m_{s}^{\rm{crit}}\leq54\pm 15.4~\rm{MeV} , \nonumber
\end{eqnarray}
or a ratio of
\begin{equation}
  \label{4.63b}
  m_{u,d}^{\rm{crit}}/m_{u,d}\sim 0.26\pm 0.08 .
\end{equation}
\end{mathletters}
In the $SU(3)$-symmetric case with three degenerate flavors the common
critical pseudoscalar mass is only $\leq 51$~[MeV], and
$m_{u,d}^{\rm{crit}}\leq 0.9\pm 0.14$~MeV. Thus the critical mass
values depend on the direction in mass parameter space. Our values are
clearly below the estimates for Wilson fermions
($m_{s}^{\rm{crit}}\geq 400$~[MeV] for $m_{u,d}\sim 0$, and
$m_{u,d,s}^{\rm{crit}}\geq 140$~[MeV] in the degenerate case). They
are of the same order as the estimates for staggered fermions
($m_{u,d}^{\rm{crit}}\leq 12$~[MeV], $m_{s}^{\rm{crit}}\leq 50$~[MeV])
for the $N_{f}=2+1$ case, but below the bound for $N_{f}=3$, 12~[MeV]
$\leq m_{u,d,s}^{\rm{crit}}\leq 38$~[MeV], cf. sections
\ref{sec:III.C}.5,7.  Thus our ratio $m_{u,d}^{\rm{crit}}/m_{u,d}\sim
30\%$ lies between the mean-field value of $\sim 3\%$ and the lattice
estimates. The intermediate value for the large-$N_{f}$-result is
plausible, as ``large-N'' goes beyond ``mean-field'' due to the
resummation of a certain subclass of Feynman diagrams (Jain, 1993),
whereas lattice Monte Carlo simulations include all quantum
fluctuations at once. It should be kept in mind, however, that a 
direct comparision between lattice quark masses in physical units
and current quark masses (as we are using here) is questionable 
(cf. the argument in section III.C.5 above).

A ratio of 30\% is certainly not large enough
for predicting visible remnants of a nearby second order chiral
transition. There is some hope that the ratio gets closer to 1, if
further fluctuations are included in the effective model and the true
nature of the chiral transition is fluctuation induced.

We have measured the weakening of the first order transition in the
$SU(3)$-symmetric case ($\epsilon_{8}=0$). The gap in the light quark
condensate above $T=177$~[MeV] slowly decreases for finite meson
masses. It has disappeared for an external field strength of
$\epsilon_{0}=6.6\cdot 10^{-4}~[\rm{GeV}^{3}]$, cf. Fig.~\ref{fig:19X}.
Similarly the barrier height between the coexisting minima in the
effective potential decreases from $1.4\cdot
10^{-4}~[\rm{GeV/fm}^{3}]$ in the chiral limit to $2.1\cdot
10^{-6}~[\rm{GeV/fm}^{3}]$ for $\epsilon_{0}=2\cdot
10^{-4}~[\rm{GeV}^{3}]$ and $\epsilon_{8}=0$.

We want to conclude with an interesting speculation of Gavin et al.
(1994) about the universality class of the $SU(3)\times SU(3)$ sigma
model. It is based on the observation that the mass of
$\sigma_{\eta'}$ vanishes for a specific choice of tree level
parameters $f_{1}$, $f_{2}$, $g$, $\mu_{0}^{2}$. If the
renormalization group flow would drive the couplings towards these
values in the vicinity of $T_{c}$, the light mass could lead to large
correlation volumes. If the $\sigma_{\eta'}$ were the only light mode,
the universality would be Ising like (in contrast to $O(4)$).

\paragraph*{An upper bound on the latent heat}
For physical meson masses we find a sharp crossover phenomenon in the
light quark condensate between $T=181.5$ and 192.5~[MeV], $\left<\bar
  q q\right>_{T}$ decreases to 50~\% of $\left<\bar q q\right>_{T=0}$
over a temperature interval of $\Delta T=10$~[MeV]. The strange quark
condensate $\left<\bar s s\right>_{T}$ stays almost constant up to a
temperature of $\sim 200$~[MeV], cf. Fig.~\ref{fig:20X}.

Similar results are found for the energy density and the entropy
density (Fig.~\ref{fig:21}). Both quantities behave smoothly as a
function of $T$ over the entire temperature range up to large errors
in the transition region. From these errors we obtain an upper bound
on a finite latent heat $\Delta L$ of
\begin{equation}
  \label{4.64}
  \Delta L\leq 0.2~\rm{GeV/fm}^{3} ,
\end{equation}
which is compatible with our data. The value is in agreement
with Leutwyler's bound of $0.4~[\rm{GeV/fm}^{3}]$ (Leutwyler, 1992),
which has been obtained from Clausius--Clapeyron--relations in the
framework of chiral perturbation theory, cf. section \ref{sec:IV.A.3}.
Further note that it is only 10~\% of the value predicted by the
naive bag model equation of state. To our knowledge it is yet unclear,
whether $0.2~\rm{GeV/fm}^{3}$ latent heat are sufficiently large to
induce measurable signatures in heavy-ion experiments.

Furthermore we have calculated the difference
$\epsilon(T_{2})/T_2^{4}-\epsilon(T_{1})/T_{1}^{4}
\equiv\Delta\epsilon/\hbox{``}T_{c}^{4}\hbox{''}$.  Here
$T_{1}=181.5$~[MeV] and $T_{2}=192.6$~[MeV] denote the temperatures,
where the rapid crossover sets in and ends, respectively. In a first
order transition with a finite discontinuity $T_{1}=T_{2}=T_{c}$.
Hence we can compare our value for
$\Delta\epsilon/\hbox{``}T_{c}^{4}\hbox{''}=0.29$ with the ratio for
the gluonic energy density $\Delta\epsilon_{\rm{gluonic}}/T_{c}^{4}$
on the lattice in a pure $SU(3)$ gauge theory. The lattice result is
$\epsilon_{\rm{gluonic}}/T_{c}\sim 1.5$ or disc
$\epsilon_{\rm{gluonic}}\sim 0.9~[\rm{GeV/fm}^{3}]$ for
$T_{c}=260~\rm{MeV}$ as mentioned above. Our bound on the chiral
contribution to the total gap in the energy density over $T_{c}^{4}$
is almost by an order of magnitude smaller than the gluonic
contribution. It should be taken as a warning not to jump to
conclusions from our results in view of relativistic heavy-ion
experiments. So far we have only investigated the
contribution from chiral symmetry to the full equation of state for a
hot hadron gas. The contribution from gluonic degrees of freedom has
been completely left out. It is a question of relative size, which
effects in the full deconfinement/chiral transition are dominant.

\bigskip It seems to us very worthwhile to elaborate alternative
inclusions of heavier mesons in the chiral transition region. From the
work by Gerber and Leutwyler (1989) it has become clear that heavier
mesons are increasingly important when $T$ exceeds $100$~MeV. I regard
our parametrization of the scalar and pseudoscalar meson masses in the
$SU(3)\times SU(3)$ linear sigma model just as one ansatz for
including part of the heavier mesons. This part is determined by the
assumed underlying chiral symmetry. A further assumption is that the
deviations from the broken $SU(3)\times SU(3)$ symmetry can be
parametrized by two external fields, breaking the symmetry explicitly.
The ansatz seems to be justified in the sense that the predictions of
the pseudoscalar meson masses reasonably agree with experimental
values.

An alternative point of view is the following.  $SU(2)\times SU(2)$ is
the true symmetry with only pions as idealized Goldstone bosons and
one external field to account for the finite pion mass, while all
other mesons are treated on an equal footing: scalar, pseudoscalar,
and vector mesons. Gerber and Leutwyler (1989) take this point of
view, when they describe all mesons (apart from the pion) in a dilute
gas approximation to study their influence on the chiral phase
transition. The {\it quantum virial expansion} of thermodynamic
quantities in a hot pion gas provides a further alternative. Heavier
mesons enter the second virial coefficient via their contribution to
experimentally measured phase shifts of $\pi\pi$-scattering (Welke et
al., 1990). A more general framework is the generalized
Beth-Uhlenbeck approach (see e.g. Schmidt et al., 1990), where the
scattering phase shifts of $\pi\pi$-scattering are replaced by
thermodynamic $T$-matrix elements, characterizing the interacting pion
gas at finite temperature and density. When the quantum virial
expansion or the generalized Beth-Uhlenbeck approach are applied to a
pion gas, the underlying chiral symmetry of QCD plays a less
distinguished role in the description as compared to our treatment in
the $SU(3)\times SU(3)$ linear sigma model.

%
\subsection{Models for gluonic degrees of freedom}
\label{sec:IV.B}
\subsubsection{A network of strings}
\label{sec:IV.B.1}
The color flux tube models of Patel (1984a and 1984b) have been
developed to describe the deconfinement phase transition. Quark
degrees of freedom can be included as well. Let us first consider the
case of a pure $SU(2)$- or $SU(3)$ gauge theory. We recall from
section \ref{sec:III} that the spontaneous breaking of the global
$Z(2)$- or $Z(3)$ symmetry is associated with the phase transition
from the confinement to the deconfinement phase. The order parameter
is the expectation value of the Wilson line. It vanishes, when the
free energy to find an isolated test quark in the system grows to
infinity.  It is different from zero in the deconfinement phase.

Effective $Z(N)$-spin models for the deconfinement transition have
been presented above (section \ref{sec:III}). They can be derived from
the $SU(N$) lattice gauge theory in the high temperature limit and
share the essential extra symmetry of QCD. In the limit of infinitely
heavy masses, this is the global $Z(N)$-symmetry. For high
temperatures the gauge fields are almost frozen to unity matrices.
Lowering the temperature increases the disorder. One big cluster of
aligned spins breaks up into several clusters. Below the transition
point the system is completed disordered, if we use a ``magnetic''
language. The formulation in terms of spin systems is natural for
$Z(N)$- models.  Intuitively it is, however, much less clear, how this
picture should be translated to the original $SU(N)$ degrees of
freedom.

Patel's description is complementary to the above model. It starts at
$T=0$ and follows the evolution of the system as the temperature is
increased. The symmetry breaking is no longer the driving force for
the system to undergo a phase transition. It is replaced by entropy
production.

In the first part we present a purely heuristic description. The
picture is based on the flux tube model of the deconfinement
transition. Flux tubes connect quarks and antiquarks in strong
coupling expansions on the lattice.  Thus one may not expect that they
leave some remnant in the continuum limit. On the other hand, they are
also the ingredients of phenomenological continuum descriptions like
string models for hadrons. We add a few remarks about the yo-yo-string
in the end.

Several properties of the flux tubes or strings have to be specified
first.  Strings in Patel's models are characterized by three
parameters: the string tension $\sigma$, the string width $w$ and the
rigidity parameter $a$. We assume that there is a constant energy per
unit length along the string, this is $\sigma$. The string has a
constant width $w$. It has a certain resistance against bending. It
has to go at least a distance $a$ apart, before it can change its
orientation. The distance $a$ is of the order of 1 fm. On the
lattice $a$ naturally coincides with the lattice constant, the strings
bend at right angles there. Here the constant $a$ will not be tuned to
zero in the end. Its role resembles more a lattice constant in models
of condensed matter physics (models in continuum spacetime, where the
constant $a$ is given by the physical lattice constant). In Patel's
flux tube theory the parameters $a$, $\sigma$ and $w$ must be fixed
from experimental inputs.

Two further assumptions about the flux tubes must be specified to get
a well defined model. Flux tubes can terminate only on quarks. Their
interaction occurs at baryonic vertices. The constants $\sigma$, $w$,
$a$ are treated as temperature independent. The only driving force of
the phase transition is the increasing entropy of flux tubes, when the
temperature is turned on.

\paragraph*{The $SU(2)$-case}
Based on these ingredients let us see, how far heuristic arguments can
lead us in a pure $SU(2)$ gauge theory. The only allowed flux tube
structures are closed loops differing only in size and shape, i.e. in
the length of the string. Physically these loops may be interpreted as
glueballs. The partition function is written as
\begin{equation}
  Z=\sum_{loops}N(loops)e^{-E(loop)/T}\;.
  \label{4.65}
\end{equation}
The sum runs over all loops of fixed length $l^.a$ (in physical
units). The energy of such a loop is given as $\sigma^.l^.a$. The
combinatorial prefactor $N$ gives the number of loops of length
$l^.a$. The dependence of $N$ on the rigidity parameter $a$ is
essential as it provides the possibility of a phase transition. (It
depends on temperature, how often strings like to bend.)

Consider random walks of length $l$ (in lattice units) on a lattice
with distance $a$ without backstepping. Backstepping should be
forbidden for a physical string. This number is given as $(2d-1)^l$ in
$d$ dimensions. The constraint that the walks should perform closed
loops leads to a power law correction in $l$ to $N$. It will not be
specified further, because it is irrelevant in the large-$l$-limit.
The partition function is then proportional to
\begin{equation}
  Z\sim\sum_l\exp\left[-l\,\left({\sigma a\over T}-\ln
      5\right)\right]
  \label{4.66}
\end{equation}
in $d=3$ dimensions. The dimension is chosen as 3, since the string
model is constructed to provide an alternative description to the
3-dim $Z(N)$-spin model for the deconfinement transition. As $T$
increases, the average length of the loops becomes larger. A phase
transition is signalled, if $Z$ diverges. This happens for large $l$,
if $(\sigma^.a/T=\ln 5)$, that is at
\begin{equation}
  T_c=\sigma a/\ln 5\;.
  \label{4.67}
\end{equation}
The effective string tension $\sigma_{eff}$ defined via $\exp\{-
la\sigma_{eff}/T\}$ vanishes continuously at $T_c$
\begin{equation}
  \sigma_{eff}=\sigma-{T\ln 5\over a}\;
  {\over {\scriptstyle T\to T_c}}\!\!\!\raise0.3pt\hbox{$\to$}\;
  0\;,
  \label{4.68}
\end{equation}
while the average length of the loops diverges. Both features suggest
a second order phase transition.

To see its relation to the deconfinement transition, we have to probe
the system at $T_c$ with a static $\bar{q}q$-pair. If it costs a
finite amount of energy to isolate both quarks at an infinite
distance, we have reached the deconfinement phase. Equivalently
deconfinement is manifest, if the $q$ and $\bar{q}$ are a finite
distance apart, but only rather loosely correlated.  Color screening
requires that a flux tube connects the quarks of our probe.  With
increasing temperature the flux reorientates itself more and more
often.  It oscillates between its endpoints, see Fig.~\ref{fig:22}.

Since one quark knows about its partner only via the connecting
string, this information is lost, if the string gets infinitely long.
This is just what happens at $T_c$, where it costs no extra free
energy to create an infinitely long flux tube. Thus at $T\ge T_c$ both
quarks ($q$ and $\bar{q})$ are effectively independent of each other
and free. This means deconfinement.

To get a quantitative estimate of the transition temperature, we must
fix the string tension and the rigidity parameter in physical units,
or the product of both in lattice units. In the strong coupling
approximation of the lattice theory, the lowest lying $O^+$-glueball
at $T=0$ is given by a square loop
\begin{equation}
  m_{O^+}=4\,\sigma\, a\;.
  \label{4.69}
\end{equation}
The string parameters are assumed to be temperature independent, hence
one can take this $T=0$-result to fix $T_c$
\begin{equation}
  T_c=m_{O^+}/{4\ln\,5}\;.
  \label{4.70}
\end{equation}
Lattice Monte Carlo calculations give a slightly larger value.

\paragraph*{\it The $SU(3)$-case}
Next let us consider the pure $SU(3)$ gauge theory. Two new features
must be accounted for. Quarks and antiquarks are no longer in
equivalent representations of $SU(3)$. Thus a direction is associated
to a string, indicating whether it terminates in a quark or an
antiquark. (The quarks and antiquarks are test quarks in the pure
gauge theory.) Furthermore, a flux tube representation of a baryon or
antibaryon requires that a string can bifurcate at a vertex $v$, see
Fig.~\ref{fig:23}.  This reflects the previous assumption that flux
tubes only interact at baryonic vertices. The diagrammatic rules for
allowed string structures in the $SU(3)$-case follow from the allowed
vertices, shown in Fig.~\ref{fig:24}.  This excludes closed loops made
up of an odd number of links.

The ansatz for the partition function can be chosen similarly to the
$SU(2)$-case. The sum over loops has to be replaced by a sum over more
complicated topological structures (of the type of nets). A
non-vanishing energy $v$ is associated to each vertex to respect the
additional bifurcation degree of freedom. At zero temperature the
vacuum is filled with closed loops made out of nets of strings, see
Fig.~\ref{fig:25} (It is filled with ``glueballs''.)  

As the temperature is increased, the size and density of these
structures grow. Unlike in the $SU(2)$-case, it is now not the
diverging length of strings that induces a phase transition. A new
qualitative feature enters, the {\it connectivity} of the network. The
phase transition occurs, if an infinite network is generated. It is
infinite in the sense that each string is connected with the entire
volume of the lattice. Between any given pair of flux tube segments
one can find a flux tube path on the lattice that connects them.

The definition of ``critical temperature'' can be made more precise.
This type of phase transitions is well known in condensed matter
physics under the name of percolation transitions. Examples for
percolation transitions are gelation transitions in the context of
polymer chains (Flory, 1941a and b). (We suspect that also jelly is
gelatinous due to polymer networks.) To find a quantitative measure
for the connectivity of the network, one has to distinguish between
relevant and irrelevant links of the network. If relevant links are
removed, the infinite network is destroyed. In the same sense
irrelevant links are not essential for the connectivity. The fraction
$f$ of relevant links depends on the temperature and the specific
underlying dynamics. It can be easily estimated that ${2\over 3} < f <
1$. The lower bound is realized at $T_c$, where the network is
minimally connected. The upper bound means maximal connectivity.

Thus the challenging hypothesis is to look at the {\it deconfinement
  transition as a percolation transition}. Again we have to answer in
a first place, what happens to a static test quark, when it is put
into the network below or above the percolation transition. Below
$T_c$ the free energy of an isolated quark is infinite, because the
flux tube originating from this quark may not terminate on any closed
loop which is a finite distance apart. It is excluded by the
diagrammatically forbidden flux tube structures. Instead the string
attached to the quark has to fuse with the string of an antiquark of
the test probe. The case, where both $q$ and $\bar{q}$ can be
considered as isolated or free, costs an infinite amount of free
energy, since we have assumed a constant energy per unit length of the
string.

Above $T_c$ the distance between the $q$ and $\bar{q}$ need no longer
be infinite for both to loose the correlation (to behave effectively
as free test probes). Already at a finite separation the correlation
is weak due to the presence of an infinite network to which they are
attached. Now the clusters of closed loops are connected, and the
corresponding flux tube structures for a $\bar{q}q$ pair at finite
distance are allowed, cf.~Fig.~\ref{fig:26}.  The costs in energy due
to the long, dense and connected flux tubes are compensated by the
gain in entropy. Thus the free energy of a configuration where the $q$
and $\bar{q}$ test quarks have lost any information about each other
is finite. The $q$ and $\bar{q}$ are effectively free, we have ended
in the deconfinement phase.

For a crude estimate of $T_c$ one has to specify the amount of extra
energy due to bifurcation points.

In the heuristic way of arguing it remains to explain the first order
of the deconfinement phase transition in the $SU(3)$-case. Either one
refers to the percolation transition where the transition is known to
be of first order, or one proves directly for the colored network of
the QCD flux that it costs a finite interface free energy at $T_c$
when both phases coexist. Such a coexistence may be visualized as a
hole in the network. Roughly speaking it is the reduction in entropy
when the loose ends at the boundary of the hole have to be tied up to
the network again, it costs interface energy proportional to the area.

\paragraph*{Inclusion of matter fields}
Flux tube models can also account for the inclusion of dynamical
quarks. By construction the chances are good, since the flux tube
picture does not rely on any symmetry argument, and the extra global
$Z(N)$-symmetry is explicitly broken in the presence of dynamical
quarks. Recall that the limit of a pure gauge theory may be considered
as the $m\rightarrow\infty$-limit of full QCD.  Thus we are interested
in the effect when the quark mass is lowered or a $Z(N)$- symmetry
breaking field is switched on.

Second order phase transitions in ferromagnets are known to disappear,
when the magnets are exposed to an external magnetic field,
independently of its strength. Similarly the second order transition
of the pure $SU(2)$ gauge theory ceases to occur: there is always a
finite probability for an infinitely long string to break up due to
the creation of a $q\bar{q}$-pair out of the vacuum. The probability
is proportional to $l^.e^{-2m/T}$, it is finite even if the quark mass
is very large as long as $l\rightarrow\infty$. (Infinitely long
strings turned out to be responsible for the second order phase
transition in this picture.) The expected change is that the phase
transition from the confinement- to the deconfinement region is
replaced by the crossover from tightly to weakly bound
$q\bar{q}$-pairs.

The $SU(3)$-case is more subtle. Again it is known from statistical
physics that first order transitions are stable with respect to small
perturbations of an external field. Hence we expect for large enough
quark masses still a deconfinement transition, and it is still the
Wilson line expectation value, which would indicate the transition by
a discontinuity at $T_c$. However, $\langle L\rangle$ will fail to
remain a good order parameter, when the quark mass is further lowered.
It was the sensitivity to the $Z(N)$-symmetry which had qualified
$\langle L\rangle$ as an order parameter in the $m\rightarrow\infty$
case. Now, where the symmetry is explicitly broken independently of
the phase, we have to look for new criteria that tell us the phase of
the system at a given temperature. It is the screening property, which
is different in the high- and low temperature phase. At low
temperatures, color charges are screened due to a breakage of strings.
At high temperatures, flux tubes do not break, but get attached to
large networks of strings, at least for very heavy masses. For lighter
masses, quarks can also break up strings at high temperature. (We see
that the clear distinction between low- and high T-phases may get lost
depending on the value of the quark mass.)

The probability for string breakage at high temperature is determined
by the energy costs for popping a $q\bar{q}$-pair out of the vacuum
and the competing gain in entropy by breaking a link. Lighter masses
facilitate breaking of links.  Less surface energy is necessary to
stabilize a hole in an infinitely connected network. Clearly there
should be a critical mass, where the surface energy vanishes at $T_c$
and the number of broken flux tubes exceeds the critical number of
irrelevant links . Once relevant links are broken, by definition, the
network ceases to exist.

In Patel's models one has a heuristic understanding of how the
transition temperature should scale as a function of quark masses. As
the quark mass decreases, string breakage is facilitated. It becomes
more difficult to generate an infinitely connected network. Thus $T_c$
should increase with decreasing $m$. The opposite tendency is observed
in numerical simulations.

From the above line of arguments it is suggested that the transition
will always disappear in the $m\rightarrow 0$-limit. The energy costs
for the creation of a $q\bar{q}$-pair are zero. Recall, however, that
it is just the $m\to 0$-limit, where the chiral symmetry comes into
the game and generates dynamical masses at low temperatures. In the
chiral limit, the scalar mesons are believed to obtain their finite
masses exclusively from chiral symmetry breaking. In the preceding
sections we have investigated just the opposite effect: perturbing
around the chiral limit, turning on an external field by increasing
$m$ from zero towards realistic values.

In the chiral limit the strength of the first order is said to
increase for an increasing number of flavors. On the other hand, in
the same chiral limit, there is a gain in entropy due to easy string
breakage as a remnant of the deconfinement transition. It is difficult
to estimate, whether a critical number of flavors exists where one of
the competing effects will win.

A crossover phenomenon at intermediate masses may result as a
combination of two tendencies. The quark masses are so small that the
deconfinement transition has disappeared, an infinitely connected
network can no longer be formed. They are so large that effects of
chiral symmetry breaking are no longer important, and the chiral
transition is washed out as well. It remains to be seen, whether the
realistic quark masses fall in a window where both transitions are
replaced by crossover phenomena.

As long as chiral symmetry is not implemented in the massless limit,
such competing effects cannot be resolved by Patel's models, even if
matter fields are included.

Our presentation so far might have suggested that the description of
the deconfinement transition in terms of color flux tubes gets stuck
on the heuristic level. This is not the case. A three-dimensional
action can be formulated in terms of occupation number variables
(Patel, 1984b). These variables live on a lattice with lattice
constant $a$ and take on values in the positive integers.  The flux
tube variables are associated with the links, the quarks and vertex
variables with the sites of the lattice. The input parameters for the
action are the string tension $\sigma$, the lattice constant $a$, the
quark mass $m$ and the vertex coupling $v$. If Gauss' law is
implemented, the model is equivalent to a 3-dimensional XY-model with
nearest neighbor coupling in a uniform magnetic field $h$ and a
magnetic field $p$. Mean-field calculations in various limiting cases
confirm the conjectures which are based on the heuristic arguments
(Patel, 1984b). The model can be extended to include several numbers
of flavors and chemical potentials.

\vskip12pt
We summarize the virtues of the color flux tube model. It abandons
symmetry breaking as driving force of the phase transition, at least
at a first place. This is desirable as an alternative way of
understanding the transition dynamics in mass parameter regions, where
the symmetry concept is questionable. Here the deconfinement
transition has been described in analogy to a gas/liquid transition at
finite temperature for different values of the pressure. The pressure
plays the role of the quark masses. The order parameter is the density
or the volume. Its change is discontinuous, when the temperature is
increased as long as the pressure is low enough. In the QCD case,
$\langle L\rangle$ jumps at the transition as long as the quarks are
heavy enough. As in the QCD case with non-zero masses, a symmetry in
the gas/liquid system is absent that could easily distinguish between
both phases. In a $(p,T)$-diagram the liquid gas phases may be
smoothly connected as well. For the analogies we refer to the
dictionary of Table~\ref{tab:2}.

The deconfinement transition here resembles a gelation transition, the
connectivity of the color flux tube network is an essential
characteristic for $SU(3)$. A similar intuitive understanding of the
chiral transition is still missing, when it is described in the
framework of the sigma model. For the chiral transition it is less
clear, what the competing effects to string breakage or network
formation could be.

It seems rather worthwhile to incorporate chiral symmetry properties
in color flux tube models or vice versa to supplement sigma models
with flux tubes for the gluonic degrees of freedom.

We conclude with a few remarks on {\it phenomenological string models}
for hadrons. Experimentalists describe the phenomenology of heavy-ion
collisions with Monte Carlo codes that are based on relativistic
microscopic models. String formation and fragmentation are essential
ingredients. These models share some features with cellular automata.
One tries to find a set of rules for string formation and breakage
(fragmentation) such that observed particle multiplicities and
$\left<p_T\right>$-spectra are reproduced for given initial
conditions. An example for such a string model is the VENUS-code of
Werner (1989) utilizing the concept of yo-yo-strings. Strings are
formed via {\it color exchange}.

One may suspect that yo-yo-strings in VENUS share only the name with
Patel's color flux tube strings. So far it is little more than the
name. The yo-yo-string is elastic, its elasticity is characterized by
the string tension.  Closed loops are forbidden, the flux can only
terminate on colored partons (quarks, diquarks etc.). The endpoints
are characterized by their flavor contents. Yo-yo-strings are allowed
to stretch, break and participate in a collective motion. They cannot
bend or bifurcate. There is no space for thermodynamic concepts, for a
transient plasma in the early stage of the collision, no way to model
a phase transition in the present formulation.

Patel's strings are neither one-dimensional rigid--line tubes
connecting the colored end-points, nor arbitrarily pliable. The
rigidity constant is essential for modulating the entropy and
initiating the phase transition. The range of possible applications of
VENUS-type models may be extended, if the strings are not restricted
to yo-yos, but allowed to close, bend and bifurcate.

\subsubsection{Dual Ginzburg-Landau models}
\label{sec:IV.B.2}
The main idea in dual Ginzburg-Landau models is to explain the
confinement mechanism by a dual Meissner effect. Let us recall the
Meissner effect in a superconductor. The magnetic field is expelled
from the embedding superconducting medium and concentrated in a flux
tube between magnetic monopoles of opposite charge. For large
distances the constant energy per unit flux tube leads to a linear
rise for large distances between the monopole charges. In the
superconducting state of metal, the electrons generate Cooper pairs.
The superconducting medium is characterized by an order parameter
$\psi$, $|\psi|^{2}$ represents the density of Cooper pairs. The order
parameter vanishes in the normal conducting state where Cooper pairs
are absent. The transition to the superconducting phase is triggered
by the condensation of Cooper pairs.

A dual Meissner effect may be an appealing explanation for
confinement. In the dual Meissner effect the role of the
(chromo)electric and (chromo)magnetic fields is reversed. The
chromoelectric field between the chromoelectric charges of quarks and
antiquarks is expelled from the embedding confinement vacuum. The
chromoelectric flux tube leads to a linear rise in the energy between
quarks at far distances. The transition to the confinement phase is
triggered by a condensation of chromomagnetic monopoles.

The interest in models based on such a heuristic explanation has
recently revived. The dual Meissner effect seems to be realized in
compact lattice QED (Polyakov, 1975; Banks et al., 1977; DeGrand and
Toussaint, 1980; Stack and Wensley, 1992). The main task in QCD is to
identify the magnetic monopoles. Due to 't Hooft's idea ('t Hooft,
1981) the identification should be possible after an abelian
projection. An abelian projection of QCD amounts to a partial gauge
fixing. There are many ways of extracting an abelian theory out of
non-abelian QCD. Popular choices are a variety of unitary gauges and
the maximal abelian gauge (Kronfeld et al., 1987; Suzuki and
Yotsuyanagi, 1990 and 1991; Suzuki, 1993). The maximal abelian gauge
turns out to be a particularly useful way of demonstrating the dual
Meissner effect. If the ``non-abelian part'' of the gauge freedom is
fixed, the original $SU(3)$ gauge symmetry is reduced to the maximal
abelian torus group $U(1)\times U(1)$. QCD becomes an effective theory
of abelian ``electric'' charges, ``photons'', and ``magnetic''
monopoles with respect to $U(1)\times U(1)$. The photon fields are the
regular part of the abelian field, the magnetic monopoles the
point-like singular part. The monopole part is isolated from the
$U(1)$-field following a procedure proposed by DeGrand and Toussaint
(1980). 't Hoofts conjecture about the dominant role of monopoles
seems to be realized in the maximal abelian gauge. The conjecture may
be verified in several steps.

\begin{itemize}
\item Typical indicators of confinement are the string tension, the
  Wilson loop, and the Polyakov loop. It is natural to test whether
  their expectation values from full lattice QCD can be reproduced by
  the abelian variables alone. This is indeed what is observed, the
  phenomenon is called {\it abelian dominance} (Suzuki and
  Yotsuyanagi, 1990 and 1991; Hioki et al, 1991; Suzuki, 1993;
  Matsubara et al., 1994).
  
\item More specifically one would like to know which part of the
  abelian residual variables is responsible for the observed abelian
  dominance. Here it turns out that the monopole contributions alone
  determine the string tension and the Polyakov loop in an $SU(2)$ and
  $SU(3)$ pure gauge theory (Shiba and Suzuki, 1994; Kitahara et al.,
  1995; Ejiri et al., 1995; Miyamura, 1995; Matsubara, 1995). The
  similar behavior of the abelian and the conventional Polyakov loop
  expectation value is not only seen in the maximal abelian gauge, but
  also in a number of different $U(1)$-projections.
  
\item Once the relevant degrees of freedom have been identified, the
  next question concerns their dynamics. Which monopoles are important
  in the confinement and deconfinement phases, what drives the phase
  transition ? Here some partial answers are known. In the confinement
  phase one has a long monopole in each configuration, while long
  monopole configurations are absent in the deconfinement region
  (Kitahara et al., 1995). This is easily understood from the main
  difference between low and high temperature QCD. The difference in
  the monopole dynamics comes from the bound on the maximal monopole
  extension due to the large or small number of time slices. The
  driving mechanism for the deconfinement transition is supposed to be
  the balance of the entropy and energy of maximal extended monopoles
  (cf. the preceding section for a similar explanation in the flux
  tube model). Further studies are, however, necessary to clarify the
  calculation of the entropy of monopoles (Kitahara et al., 1995 a and
  b). The difference in the order of the deconfinement transition for
  $SU(2)$ and $SU(3)$ remains to be explained.
  
\item The observation of abelian dominance and the identification of
  long monopoles as the most important field configurations in the
  confinement mechanism suggest a construction of an effective action
  in terms of these monopoles. Effective monopole actions have been
  considered in compact QED and $SU(2)$-QCD (Shiba and Suzuki, 1994
  and 1995). Extended monopoles can be implemented as well by
  performing a block spin transformation on the dual lattice. The
  results obtained so far indicate an interesting scaling behavior,
  giving some hope that the effective action actually describes
  continuum physics.
\end{itemize}

The results in dual Ginzburg-Landau models look rather encouraging.
The dependence on the finite volume deserves further studies on larger
lattices to make sure that the dynamics of abelian $U(1)\times U(1)$
monopoles of lattice QCD survives the continuum limit. The {\it gauge
dependence} of the proposed confinement mechanism is another open
question. So far most of the features of the dual Meissner effect have
been observed in the maximal abelian gauge. Recently Chernodub et
al. (1995) have studied the pure $SU(2)$ gauge theory in different
abelian projections. Their results suggest the conclusion that the
topological configurations, which are relevant for confinement, do
depend on the gauge. {\it Monopole} condensation as mechanism for
confinement appears as specific feature of the maximal abelian
gauge. So called minopoles replace monopoles in the `minimal' abelian
gauge, and string-like topological objects may be important as
well. Such a translation of the confinement mechanism between
different gauges deserves further investigations. Remarks on unitary
gauges can be found in (Matsubara et al., 1995) and (Suzuki and
Yotsuyanagi, 1991).

\subsubsection{Some further approaches}
\label{sec:IV.B.3}
We conclude this chapter about effective models with an outlook for
further directions. Descriptions including quark degrees of freedom
have been left out so far. Also the string networks of the preceding
section cannot be regarded as representative for the variety of
models, which have been proposed for gluonic degrees of freedom.

An important class of models for the gluonic sector are {\it instanton
  descriptions} of the QCD vacuum (Shuryak, 1984). The idea is that
the most important (nonperturbative) contributions to the
QCD-partition function come from instantons and anti-instantons at
short distance scales. Dynakonov and Mirlin (1988) and Kanki (1988)
have considered the instanton vacuum at finite temperature. Within a
variational approach they have shown that the ``evaporation'' of
instantons and anti-instantons out of the vacuum leads to chiral
symmetry restoration as the temperature is increased.

Ilgenfritz and Shuryak (1989) and Novak et al. (1989) have extended
the analysis of instanton models to the inclusion of quark degrees of
freedom. The qualitative conclusion of these authors is a crossover
phenomenon rather than a chiral transition, when finite quark masses
are included.  The results have been obtained within mean-field-type
calculations.

Frequently it is the symmetry of the underlying QCD action, which is
taken as a guiding construction principle for the effective action.
Applied to the gluonic sector one may ask for the analogue of the
chiral condensate and the role it plays in the deconfinement (and
chiral) transition. A natural analogue is the gluon condensate
$\langle F_{\mu\nu} F^{\mu\nu}\rangle$, if $F_{\mu\nu}$ denotes the
QCD field strength tensor.

In view of the gluon condensate we recall the basic symmetry of QCD
related to the condensate. At the classical level the QCD Lagrangian
with massless quarks is invariant under scale transformations
$x^{\mu}\rightarrow e^{\lambda} x^{\mu}$ with some real constant
$\lambda$. The corresponding (classically conserved) current is the
dilatation current $S_{\mu}=T_{\mu\nu} x^{\nu}$, where $T_{\mu\nu}$ is
the energy-momentum tensor. On the quantum level, even massless QCD is
no longer scale invariant. It is broken by the scale- or trace anomaly
of QCD
\begin{equation}
  \partial_{\mu} S^{\mu}=T^{\mu}_{\mu}={\beta(g)\over 
    2g}F_{\mu\nu} F^{\mu\nu}\;.
  \label{4.71}
\end{equation}
Here $\beta(g)$ denotes the beta-function of QCD, g is the coupling.
The perturbative contribution to 1-loop order is given as
$(-g^3/16\pi^2) (11-{2\over3}N_f)$.

Schechter (1980) has proposed an effective action for both mesonic and
gluonic degrees of freedom. In the limit of zero quark masses this
action is invariant under the $SU(N_f)_L\times SU(N_f)_R$ chiral
transformation ($\times U(1)_V$ baryon number), reproduces the axial
$U(1)$ anomaly (the chiral anomaly) and in addition the trace anomaly
of QCD.

For definiteness we state the result for the effective Lagrangian, as
it has been used by Campbell et al.~(1990)
\begin{eqnarray}
  L=&&L_m +L_g \nonumber\\
  L_m =&&  {1\over 16\pi^2} f\pi^2 (\chi/\chi_0)^2 Tr 
  \partial_{\mu}U \partial^{\mu} U^+ 
  \nonumber\\
  &&~-c(\chi/\chi_0)^3Tr\left(m_q(U+U^+)\right)
  -{1\over 2} m^2_0\phi^2_0\cdot (\chi/\chi_0)^4 
  \nonumber\\
  L_g=&&{1\over 2} \partial_{\mu}\chi 
  \partial^{\mu}\chi+B\left[ {1\over 4}\chi^4_0+
    \chi^4 ln (\chi/e^{1\over 4}\chi_0)\right] \,.
  \label{4.72}
\end{eqnarray}
The following notations are involved. Gluonic degrees of freedom are
represented by a scalar gluonium field $\chi$ with scale dimension 1.
It is assumed to take a non-vanishing vacuum expectation value
$\chi_0=\langle 0|\chi|0\rangle$, thus scale invariance is
spontaneously broken on the effective level. The matter fields are
represented by $U$
\begin{equation}
  U(x)= \exp (i\sum_{i=0}^8\lambda_i \phi_i(x)/f_{\pi})\;.
  \label{4.73}
\end{equation}
The fields $\phi_i, i=0,\ldots,8$ denote the nine pseudoscalar meson
fields. The light quark mass matrix is given by $m_q= diag
(m_u,m_d,m_s)$. Up to prefactors depending on $\chi$ and $\chi_0$, the
matter part of the Lagrangian is the familiar non-linear $SU(3)\times
SU(3)$ sigma model (first term of $L_m$), plus a term proportional to
the quark masses breaking chiral symmetry explicitly, plus a third
term ($\propto \phi^2_0$) accounting for the ninth pseudoscalar meson.
All terms are decorated with prefactors correcting for the right scale
dimensions.

The specific form of the potential for the gluonic part $L_g$ has been
proposed by Schechter (1980). It is this term which is designed to
guarantee the trace anomaly according to
\begin{equation}
  \partial_{\mu}S^{\mu}=-B\chi^4\;,
  \label{4.74}
\end{equation}
where $B$ is the vacuum energy density of the pure gauge sector. Eq.\ 
(\ref{4.74}) leads to the identification
\begin{equation}
  \langle 0|(\beta/2g)F_{\mu\nu}F^{\mu\nu}|0\rangle =-B\chi^4_0 
  X \, .
  \label{4.75}
\end{equation}
Campbell et al.~(1990) have used the Lagrangian (\ref{4.72}) as
starting point for thermodynamic considerations.

The matter part of the Lagrangian is easily modified.  Patk\'os (1991)
has used the linear version of the $SU(3)\times SU(3)$ sigma model for
$L_m$ to calculate the surface tension of the first order chiral
transition under the inclusion of gluonic degrees of freedom.

Kusaka and Weise (1992) have chosen the
Nambu-Jona-Lasinio~(NJL)-Lagrangian with $SU(2)_L \times SU(2)_R$
symmetry. An interesting relation can be read off from their
Lagrangian. Due to the prefactors ($\chi/\chi_0)^3$ in Eq.\ 
(\ref{4.72}), the quark condensate is not only determined by the meson
condensate $\langle 0|U+U^+|0\rangle$, but knows about the gluon
condensate as well. Such a relation reads in the effective model of
Kusaka and Weise (1992)
\begin{equation}
  \langle {\bar u} u\rangle =\langle {\bar d}d \rangle = 
  -{1\over N_fG} \left( {\chi_0\over\chi_m} \right)^2 
  \langle\sigma\rangle\;.
  \label{4.76}
\end{equation}
Here $\chi_0=\langle 0|\chi|0\rangle$ denotes the mean-field vacuum
expectation value as above, $\chi_m$ minimizes the potential term of
$L_g$ and $G$ is the four quark coupling.

The interesting feature about this relation arises when it is
extrapolated to temperatures larger than zero. Chiral symmetry may be
either restored ($\langle{\bar u} u\rangle_T\rightarrow 0$), because
of the mesonic part $\langle \sigma\rangle$ has melted or because the
gluon condensate $\chi_0$ vanishes. This leads to an unequality
between the chiral $(T_{ch})$ and the deconfinement $(T_d)$ transition
temperatures
\begin{equation}
  T_{ch}\leq T_d\;.
  \label{4.77}
\end{equation}
The interplay between both condensates can be studied this way. A
question of the type``who drives whom'' in the phase transition may be
answered in these models.

The chiral phase transition has been investigated both as function of
temperature and/or density in a variety of {\it NJL-models} (Klimt et
al., 1990; Lutz et al., 1992; Hatsuda and Kunihiro, 1985 and 1987;
Kunihiro, 1989 and 1991; Bernard and Meissner, 1988). In general the
calculations have been performed within a mean-field approach. The
common conclusion is that under realistic conditions for coupling
strengths (or quark masses) the chiral transition is replaced by a
crossover phenomenon.

An extended NJL-model has been derived by Bijnens et al. (1993) as a
low-energy approximation to QCD. Starting from basic QCD, gluon fields
or alternatively gluon fields and quark fields are integrated out
within a certain approximation in the path integral formulation. The
resulting extended NJL-model includes various low-energy models in
appropriate limit cases, in particular the chiral quark model of
Manohar and Georgi (1984), which extends the linear sigma model of our
previous sections.

Let us finally discuss the {\it chiral quark model} in little more
detail. Originally this model was proposed by Manohar and Georgi as an
effective model for quarks, gluons, and Goldstone bosons to explain
the success of nonrelativistic quark-models. Gocksch (1991) has
considered a version of the chiral quark model, where gluonic degrees
of freedom are dropped, and the Goldstone part is realized as the
linear $SU(2)\times SU(2)$ sigma model.  The Lagrangian is given by
\begin{eqnarray}
  L = && {1\over 
    2}\bigl[ (\partial_{\mu}\sigma)^2 + (\partial_{\mu} 
  {\vec\pi})^2 \bigr] + {\mu^2 \over 2} (\sigma^2 + {\vec 
    \pi}^2) + {\lambda \over 4{\rm !}} (\sigma^2 + {\vec 
    \pi}^2)^2 
  \nonumber\\
  + && c\cdot\sigma + {\bar \Psi} \bigl[ 
  \not\!\partial + g(\sigma -i \gamma_5 {\vec \tau} {\vec 
    \pi}) \bigr] \Psi\;.
  \label{4.78}
\end{eqnarray}
The notation is obvious from former notations in the text.  Note that
the Goldstone bosons ($\vec \pi$) and the $\sigma$-channel are treated
on an equal footing with the quark degrees of freedom. This accounts
for the distinguished role the pions and the sigma-meson play among
the low mass hadrons in the chiral transition. They cannot be
described as states of weakly interacting quarks. (The well-known
price of this treatment is a double counting.  For example there is a
pseudoscalar ${\bar q} q$ bound state as well as the "fundamental"
pion state.)

Gocksch has performed mean-field calculations at finite temperature
based on the Lagrangian (\ref{4.78}). This way it has been possible to
reproduce lattice results for static hadronic screening lengths and
the quark number susceptibility (Gocksch et al., 1988; Gottlieb et
al., 1987c). There are different ways for chiral symmetry restoration
above the transition. One possibility is a parity doubling of the
known hadron spectrum (DeTar and Kunihiro, 1989; McLerran, 1986).
Parity partners should be degenerate in their mass above the chiral
transition, but still in the confinement phase.

The more standard symmetry restoration is realized through massless
quarks. The results of Gocksch indicate that the lattice data for the
$\varrho$-meson and nucleon screening masses can be understood in
terms of nearby massless propagating quarks, while the $\vec \pi$- and
$\sigma$-modes cannot be explained this way. This favors the
realization of chiral symmetry restoration along the "standard" way,
i.e. via massless quarks with degenerate $m_{\pi}$ and $m_{\sigma}$
masses in the chiral symmetric phase.

The chiral quark model (\ref{4.78}) gives a nice example for an
effective model, which allows an explicit comparison with lattice
results and is able to explain them. Such an agreement justifies the
decimation in the very ansatz for the effective Lagrangian. In this
particular case it confirms the distinguished role of the sigma and
pions.  It remains unclear, whether the double counting in the chiral
quark model has any contaminating effect on the results.

\bigskip Interesting directions for future work are combinations of
``chiral'' and ``gluonic'' effective models to study the interplay
between both type of transitions. Dual Ginzburg-Landau-models may be
promising candidates for describing {\it both} aspects, confinement
and chiral symmetry. For first attempts see e.g. Miyamura (1995). It
should be verified that the $U(1)\times U(1)$ abelian monopoles
survive the continuum limit and maintain their dominant role in the
presence of dynamical quarks. An interesting question then is, how the
confinement mechanism in terms of abelian monopole condensation
translates to different conditions for gauge fixing.

Unlike QCD there is no unique favorite action on the effective level.
Therefore one should study the variation of the results under
different decimations of the underlying full QCD. This applies to
generic effective models.

%
%
\section{RELATIVISTIC HEAVY ION COLLISIONS}
\label{sec:V}
\subsection{Scales and observables}
\label{sec:V.A}
The experimental possibilities to test the QCD predictions 
of finite-temperature phase transitions from normal 
hadronic matter to a quark-gluon plasma are limited if we 
face the transition temperature of $\approx10^{12\ 
\circ}K$. Most probably, the QCD transition occurred 
$10^{-6}$~sec after the big bang when the universe 
cooled down to the transition temperature. It is very 
questionable whether any remnants of the transition can be 
seen today. Promising alternatives to reproduce the QCD 
transition in some kind of little bang in laboratory 
experiments are heavy-ion collisions at ultrarelativistic 
energies. In the collision a large amount of the initial 
kinetic energy will be concentrated in a short spacetime 
interval and form a {\it fireball} of matter which could reach 
thermal equilibrium.

In the last decade, it has become technically feasible to 
create matter with energy densities 10-100 times that of
ordinary nuclear matter (see e.g. Jacob and Satz, 1982 or 
Schmidt and Schukraft, 1993). The available energy for 
particle production is specified in terms of $\sqrt s$, 
where $\sqrt s$ is the total center-of-mass energy. In 1992 
the AGS-machine at Brookhaven National Laboratory (BNL) 
has produced very heavy-ion beams of $^{197}$Au with momenta
of 11.4/c~GeV per nucleon 
(Barrette, 1994). At the CERN-SPS machine, 
200~GeV/$n$ are reached for beams up to $^{32}S$, still 
`light' heavy ions. 
Since November 1994, a Pb beam of 160$~GeV/c$ per nucleon is
available at the SPS. Future experiments are planned starting 
in 2006 where a value of $\sqrt s=6300$~GeV/$n$-$n$ 
($n$-$n$ stands for nucleon-nucleon) for
lead should be reached at the CERN-LHC collider. At RHIC of 
BNL $\sqrt s=200$~GeV/$n$-$n$ will be reached in 1999.

Given the value of $\sqrt s$ and the mass number $A$ of 
colliding ions, one would like to estimate the initial 
temperatures that can be reached in such a collision. The 
basic observable is the multiplicity per unit rapidity 
$dN/dy$ of secondary hadrons which are emitted in the 
collision. If one extrapolates the known relation between 
$\sqrt s$ and $(dN/dy)_p$ in proton-proton collisions to 
central nucleus $(A)$-nucleus $(A)$ collisions, the 
relation is given as (Satz, 1990a) 
\begin{equation}
  \left({{dN}\over{dy}}\right)_{AA}=A^\alpha\cdot0.8\cdot\ln\sqrt s 
  \label{5.1}
\end{equation}
with $\alpha\geq1.1$. This leads to multiplicity densities of 
about 480 at midrapidity (including neutrons and neutral pions) 
at the AGS (Barrette et al., 1995) for $Au+Au$, and to an 
estimate of 790 for the maximal $dN/dy$ (when all hadrons are 
included) for $Pb+Pb$ at the SPS (Margetis et al., 1995). At LHC
for $\sqrt s=6300$~GeV/$n$-$n$ multiplicity densities of 1500-2500
are expected according to Eq.$5.1$. The 
multiplicity density of the final state hadrons can be related to 
the initial state energy density $\varepsilon$ either with 
(approximate) energy conservation (free flow) or entropy 
conservation (see section \ref{sec:V.B.1}). For free flow the 
relation is given by
\begin{equation}
  \varepsilon=\{(dN/dy)_Am_T\}/(\pi R^2_A\tau).
  \label{5.2}
\end{equation}
Here $R_A$ is the nuclear radius, $m_T$ the transverse mass 
and $\tau$ the equilibration time. With $\tau\approx1$~fm/c 
and $m_T\approx0.5$~GeV Eq.\~(\ref{5.2}) leads to an average 
initial energy density of 1.5-2.5~GeV/fm$^3$ for central 
$Pb-Pb$ collisions at SPS and 4.6-7.8~GeV/fm$^3$ at LHC. 
For $Si+Pb$-collisions at AGS of BNL the initial energy 
density has been estimated as 0.6~GeV/fm$^3$ (Stachel and 
Young, 1992).

The energy density is translated to a temperature according 
to an ideal gas relation $\varepsilon\propto T^4$, where 
the proportionality constant depends on the number of 
included degrees of freedom. For three massless quark 
flavors, the initial temperature for $Pb-Pb$ collisions at 
SPS ranges from 170-190~MeV (free-flow assumption) or 
160-190~MeV (isentropic expansion). If we compare these 
values to estimates from lattice calculations including 
dynamical fermions, AGS- and SPS-energies could produce 
temperatures slightly above $T_c$. Pronounced signatures of 
a plasma via thermal radiation can be only expected if the 
initial temperature $T_i$ is well above $T_c$. More recent 
estimates of $T_i$ by Shuryak (1992), Kapusta et al. (1992) 
and Geiger (1992a and b) give more optimistic values for 
$T_i$, $T_i\sim500-600$~MeV should be reached (cf. also 
below).

In applying these formulas, we have implicitly made use of the fact
that energy, entropy, temperature are well defined notions in
describing the collision. This assumption will be discussed later on.

At this point, let us pause for a comparison to the 
observation of a phase transition under daily life 
conditions, the boiling of water. Usually one 
does not appreciate all the well defined experimental 
conditions. The fluid container has a fixed volume, the 
fluid is at rest and in thermal equilibrium. Calibrated 
thermometers are at hand, whose Hg-column grows linearly in 
the considered temperature range. The phase transition is 
easily seen as conversion from the liquid to the vapor 
phase while the temperature stays constant. Also heating 
sources are available such that tuning of the temperature 
does not pose a problem either. The temperature may be 
tuned adiabatically or as a quench as one likes.

It is not surprising that at temperatures where hadronic 
matter gets dissolved into its components, the experimental 
conditions drastically change. In collider experiments the 
volume is not fixed in the evolution of the plasma to the 
hadron gas. At high collision energies the nuclei 
interpenetrate each other at the collision and recede as 
Lorentz contracted pancakes, leaving a hot vacuum with 
secondaries of the collisions between them (in the central 
rapidity region). After hadronization this hot area is 
primarily  a pion gas.

At LHC-energies the {\it volume} at the transition is 
 estimated to be 5-8 times larger than the initial volume, 
for Pb-Pb collisions the critical volume is of the order of 
800 - 1200 fm$^3$. After the transition the system 
continues expanding until freeze-out, where interactions 
can be neglected or/and  the mean free path of particles 
gets of the order of the size of the system. (Both 
definitions of `freeze-out' lead to different estimates for 
freeze-out volumes). At LHC freeze-out volumes could be of 
the order of $10^4$ to $10^5$ fm$^3$ (Satz, 1990a).

Next we come to various {\it time scales} which are 
involved in heavy-ion collisions. Note that the total 
duration of a `little bang' is only of the order of 
$10^{-23}$ sec. (The characteristic  length scale of 1 fm 
is a very short distance for light to pass by.) Time scales 
are the equilibration time $\tau_0$, the freeze-out time, 
the delay caused by a possible first order phase 
transition, the conversion rate of one phase into the other 
and the expansion rate. All of them are between say 0.1-100 
fm/$c$. The largest value refers to an extreme delay of 
freeze-out due to strong supercooling, the smallest one 
corresponds to a more recent estimate for the equilibration 
time (Shuryak, 1992). The equilibration time is the time it 
takes until the system reaches a stage of local 
thermodynamic equilibrium, when it has passed an 
intermediate preequilibrium stage after the bang. `Local 
equilibrium' means, that energy, entropy, pressure and 
temperature can be locally defined. In contrast to our 
water boiling experiment, the temperature is -- after it is 
defined at all -- a function of space and time $T(\vec x, 
t)$ in a volume, whose geometry is not even fixed, but 
depends on the impact parameter value $b$.

Volumes of the order of thousands of fm$^3$ and time scales of some
tens of fm/$c$ can be measured with pion-interferometry, especially a
time delay due to a transition (see section \ref{sec:V.C.4}). The
lifetime of the fireball is estimated to be 10-20fm/$c$ for Si+Pb(Au)
collisions at 14.6~GeV/$n$-$n$ from pion-interferometry measurements at
the AGS-machine (see e.g. Stachel, E814 collaboration, 1994).

{\it Temperature scales} which should be distinguished are 
the initial temperature at the onset of local equilibrium 
$T_i$, the transition temperature $T_c$, possible values of 
superheating/supercooling effects and the decoupling 
temperature $T_d$ (of the order of the pion mass).

Candidates for thermometers are thermal photon and dilepton spectra
(Shuryak, 1978; see also Gyulassy, 1984). The differential cross
section of dilepton production as a function of the invariant lepton
pair mass $M$ and the rapidity $y$, i.e. $(d^2\sigma/d M^2 dy)_{y=0}$,
is predicted to scale according to exp $(-M/T)$, $T$ is the
temperature of the emitting system. Unfortunately these thermometers
like to hide in the background of other dileptons. Thermal dileptons
are difficult to be identified in the total dileptonic yield (see
section \ref{sec:V.C.2}). Also pions have been proposed as
thermometers for measuring the freeze-out temperature (Brown et al.,
1991). Pions from the decay of the $\Delta(33)$-resonance have a
characteristic $p_T$-distribution which is very different from that of
primary pions. The ratio of $\Delta(33)$-resonances to nucleons
sensitively depends on the temperature via Boltzmann factors. Thus a
measurement of the $p_T$-distribution allows the identification of
decay pions and a measurement of the temperature dependent ratio of
$\Delta(33)$-resonances to nucleons.

If it is difficult to measure the temperature, let us see, 
how we  can tune it. The initial temperature changes with 
the initial energy density ($\varepsilon \propto T^4$ for 
an ideal gas), and that depends on $\ln\sqrt{s}$. Thus in 
principle one could vary $s$ for a given nucleus $A$. Due 
to the $\ln\sqrt{s}$-dependence this would require a large 
variation in the incident beam energy, accompanied by a 
considerable loss in the luminosity, if $\sqrt{s}$ is 
reduced (Ludlam
and Samios, 1988; Satz, 1990a). For asymmetric 
collisions between small and big nuclei, the energy can be 
increased by going from peripheral to central collisions, 
i.e. by varying the impact parameter. For symmetric 
collisions the realistic possibility which remains is the 
variation of $A$, i.e. the type of nucleus itself. Going 
from $S-S$ to $U-U$-collisions roughly gives a gain  in 
energy density by a factor of $7-8$ at the price that the 
volume changes as well. Going from fixed target to collider 
experiments one expects a factor of $2-4$ in the increase of 
the initial temperature $T_i$. The estimate of $T_i$ at AGS 
and SPS is about 190-200~MeV, at RHIC $T_i$'s will be 
reached between 400-500~MeV, at LHC it may be 600-900~MeV.

Last but not least to complete our comparison to a phase 
transition under 'normal' conditions, we have to find 
observable signatures of the QCD transition. The goal is to 
identify observables that could reflect almost constant 
pressure or temperature over an interval where energy and 
entropy densities ($\varepsilon$ and $s$) rapidly change, 
where the phase conversion takes place. Typical 
observables, which are at our disposal in heavy-ion 
collisions, are multiplicity distributions in rapidity 
space and average transverse momenta $\langle p_T\rangle$. 
It turns out (see section \ref{sec:V.C.1}) that $\langle p_T\rangle$ 
values can be a measure for the initial pressure and 
temperature, while $(dN/dy)$ distributions depend on the 
initial values of $\varepsilon$ and $s$. Thus a 
$T-\varepsilon$ diagram corresponds to a $\langle 
p_T\rangle-dN/dy$ plot.

There are a variety of other signatures which are in 
principle sensitive to a transient plasma and to the 
transition dynamics as well. They are discussed in sections 
\ref{sec:V}.C and D.

\vskip12pt\noindent $\bullet$
We summarize the main complications we have to face in 
heavy-ion experiments. The {\it basic assumptions} that the 
fireball is large enough and long-lived enough to reach 
thermal equilibrium, the application of thermodynamic 
concepts have to be checked and justified. Due to the 
expansion dynamics, different competing time and length 
scales are involved. The spacetime expansion from the  
initial thermalization until freeze-out has to be unfolded 
of the final observables, if one wants to compare the 
signatures with predictions of static, microscopic 
equilibrium quantities. Usually the folding is provided by 
a hydrodynamic description.

\vskip12pt
Hydrodynamics is a useful computational tool in estimating 
bulk features like leptonic or hadronic particle yields, 
multiplicity fluctuations, orders of lifetimes etc. Note 
also that hydrodynamics describes off-equilibrium 
situations although it is based on local equilibrium 
conditions. The system is expanding and cooling and out of 
global equilibrium. In this sense, it is also a conceptual 
framework to treat nonequilibrium situations in a way that 
all the information of equilibrium QCD is not lost, but can 
be built in. This information contains the equation of 
state, elementary cross sections, structure functions and 
other derived quantities of equilibrium thermodynamics 
which we have partly outlined above.

There are a few other concepts to treat off-equilibrium aspects in
heavy-ion collisions. We will discuss the determination of the
nucleation rate of hadronic bubbles in the plasma according to
Langer's approach in condensed matter physics (Langer, 1969; Csernai
and Kapusta, 1992a and b) (section \ref{sec:V.D.1}), an alternative
process for phase conversion (large domain coarsening) in section
\ref{sec:V.D.2} (Borrill and Gleiser, 1995), a calculation of
transport coefficients (based on a combination of linear response
theory with hydrodynamic concepts (section \ref{sec:V.D.3}), and the
concept of dynamical universality applied to the case far
off-equilibrium (Rajagopal and Wilczek, 1992) (section
\ref{sec:V.D.4}).

The effect of a phase transition on the hydrodynamical flow 
depends on the type of phase transition dynamics. If the 
transition is of first order but proceeds smoothly, close 
to equilibrium, the effect is just to slow down the 
expansion. Discontinuities in thermodynamic quantities 
would be reflected in shock-like discontinuities of the 
fluid. If the transition is of first order but involves 
metastable states - a supercooled plasma or a superheated 
hadron gas - deflagrations or detonations may evolve with 
the possible effect of large multiplicity fluctuations 
(section \ref{sec:V.C.5}).

Such explosive processes are one source of entropy production during
the evolution. Other sources are dissipation effects and the
freeze-out transition. All of them are estimated to produce little
extra entropy such that the approximation of entropy conservation
during the evolution seems to be justified (Blaizot and Ollitrault,
1990, and references therein).

Although dissipation effects on entropy production may be small, pure
glue is rather viscous. It is amusing to estimate the viscosity of a gluon
gas at tera degrees in comparison to `normal' gases.
For $T$ close to $T_c$ the shear viscosity is of the order of the
$\Lambda$-QCD scale. Thus it is $10^{16}$ times the viscosity of
classical gases under 'normal' conditions (where it is $10^{-5}$
kg/$m\cdot s$) (Hosoya and Kajantie, 1985).

The main ingredients in a hydrodynamic description are the 
initial conditions (section \ref{sec:V.B.1}) and the equation of 
state (section \ref{sec:V.B.2}). Before we go into detail, let us 
finally quote some numbers from (Cleymans et al., 1986)  to 
show that the basic condition for applying a hydrodynamic 
description is not violated too much: The mean free path of 
a particle in a medium has to be much smaller than the size 
$L$ of the medium. The mean free path of a quark at an 
initial energy density of 2.7~GeV/fm$^3$ for $U-U$ 
collisions is supposed  to be 0.22 fm as compared to a 
diameter of 15 fm of an uranium nucleus. The corresponding 
quark density is 9 quarks/fm$^3$. Groups who are presently 
working on three-dimensional hydrodynamics for relativistic 
heavy-ion collisions are Venugopalan et al. (1994), 
Waldhauser et al. (1992) and Bravina et al. (1993).

We close this section with a warning concerning the 
following subsections. The reader will not find any 
conclusions of the type that there are strong hints for the 
transition to be of either first or second order. We leave 
this question open until the end.

\subsection{The hydrodynamic framework}
\label{sec:V.B}
The derivation of the hydrodynamic equations can be found 
in various textbooks, see e.g. (Landau and Lifshitz, 1959), 
and in their adaption to heavy-ion collisions in the work 
of Cooper and Frye (1974) or in reviews by (Cleymans et 
al., 1986) or (Blaizot and Ollitrault, 1990).  For 
completeness we sketch the main steps in deriving an 
appropriate form for heavy-ion collisions. The adaption to 
heavy-ion collisions amounts to a suitable choice of 
coordinates and a set of initial conditions which is based 
on  experimental observations.

Hydrodynamic equations describe the evolution of a gas (or 
fluid) in space and time. The gas is specified by a local 
temperature, pressure, energy, entropy and velocity. The 
equations result from constraints of energy and momentum 
conservation and other conserved quantities in case, such 
as the baryon number. If we first neglect dissipative 
effects (viscosity, thermal conductivity), the 
energy-momentum tensor of a relativistic perfect fluid in 
motion with velocity $u^\mu$ is obtained by a Lorentz boost 
from its rest frame as
\begin{equation}
  T^{\mu\nu}=(\varepsilon + p)u^\mu u^\nu-
  g^{\mu\nu}p\;.
  \label{5.3}
\end{equation}
The equations for energy-momentum and baryon number 
conservation are
\begin{mathletters}
  \label{5.4}
  \begin{equation}
    \partial_\mu T^{\mu\nu}=0
    \label{5.4a}
  \end{equation}
  and
  \begin{equation}
    \partial_\mu J^\mu=\partial_\mu(n_B u^\mu)=0
    \label{5.4b}
  \end{equation}
\end{mathletters}
respectively, where $n_B(\vec x,t)$ is the local baryon number
density. Using Eq.\ (\ref{5.3}), contraction of Eq.\ (\ref{5.4a}) with
$u_\nu$ leads to
\begin{equation}
  u^\nu\partial_\nu\varepsilon+(\varepsilon+p)\partial_\nu 
  u^\nu=0,
  \label{5.5}
\end{equation}
where $\epsilon$ denotes the energy density, $p$ the 
pressure. An analogous equation can be derived for the 
entropy density $s$, which can be converted to a 
temperature equation in the baryon free case. Contracting  
Eq. (\ref{5.4a}) with $(g_{\nu\rho}-u_\nu u_\rho)$ leads to the
second hydrodynamic equation
\begin{equation}
  (\varepsilon+p) u^\tau\partial_\tau u_\lambda-\partial_\lambda
  p+u_\lambda u^\tau\partial_\tau p=0\;.
  \label{5.6}
\end{equation}
The next step is to choose coordinates adapted to a plasma 
evolution in cylinder geometry, where the $z$-axis is 
commonly identified with the beam axis. Now one can express 
Eqs.\ (\ref{5.5}) and (\ref{5.6}) in coordinates $z$ and $t$, the 
four-velocity $u^\mu$ of the matter is written as
\begin{equation}
  u^\mu={1\over{\sqrt{1-v^2_z-v^2_r}}}(1,v_z,v_r,0)\;.
  \label{5.7}
\end{equation}
For vanishing radical velocity $v_r$ (which is frequently 
used as approximation), the remaining components of $u^\mu$ 
are parametrized according to
\begin{mathletters}
  \label{5.8}
  \begin{equation}
    u^\mu=(\cosh \theta,\sinh\theta, 0,0)
    \label{5.8a}
  \end{equation}
  where $\theta$ is the fluid rapidity, defined via
  \begin{equation}
    \theta=\arctan v_z\;.
    \label{5.8b}
  \end{equation}
\end{mathletters}
A more convenient choice of variables are the space-time 
rapidity
$\eta$ defined as
\begin{equation}
  \eta={1\over2}\ln{{t+z}\over{t-z}}
  \label{5.9}
\end{equation}
and the proper time $\tau$
\[
  \tau=\sqrt{t^2-z^2}
\]
with the inverse transformations
\begin{eqnarray}
  t&=\tau\cosh\eta \nonumber\\
  z&=\tau\sinh\eta\; .
  \label{5.10}
\end{eqnarray}
Note that the fluid rapidity coincides with the space-time 
rapidity in the case of $v_z=z/t$. 

In terms of these new coordinates, the hydrodynamic equations
(\ref{5.5}) and (\ref{5.6}) for a longitudinal motion are
\begin{mathletters}
  \label{5.11}
  \begin{eqnarray}
    \tau{{\partial\varepsilon}\over{\partial\tau}}&&+\tanh(\theta-\eta)
    {{\partial\varepsilon}\over{\partial\eta}}+(\varepsilon+p)
    \left[{{\partial\theta}\over{\partial\eta}}+\tanh(\theta-\eta)\tau
      {{\partial\theta}\over{\partial\tau}}\right]
    \nonumber\\
    &&=0
    \label{5.11a}
  \end{eqnarray}
  and
  \begin{eqnarray}
    {{\partial p}\over{\partial\eta}}&&+\tanh(\theta-\eta)\tau
    {{\partial p}\over{\partial\tau}}+(\varepsilon+p)\left[\tau
      {{\partial\theta}\over{\partial\tau}}
      +\tanh(\theta-\eta){{\partial\theta}
        \over{\partial\eta}}\right]
    \nonumber\\
    &&=0 .
    \label{5.11b}
  \end{eqnarray}
\end{mathletters}
This set is complemented by a third equation which follows 
from baryon number conservation (Kajantie et al., 1983)
\begin{equation}
  \tau{{\partial 
      n_B}\over{\partial\tau}}+n_B{{\partial\theta}\over
    {\partial\eta}}+\tanh(\theta-\eta)
  \left({{\partial n_B}\over{\partial\eta}}
    +n_B\tau{{\partial\theta}\over{\partial\tau}}\right)=0\;.
  \label{5.12}
\end{equation}
Now we have three equations (\ref{5.11})-(\ref{5.12}) for four unknown
functions: the energy density $\varepsilon$, the pressure $p$, the
fluid rapidity $\theta$ and the baryon number density $n_B$, all of
them being functions of $\eta$ and $\tau$ in a longitudinal expansion.
Thus, in order to find solutions of the set (\ref{5.11})-(\ref{5.12}),
we have to supply one additional equation and to specify the initial
conditions.  The additional equation is an {\it equation of state}
relating $\varepsilon$ and $p$ or $T$ and $s$. (Other combinations are
possible as well.) As an example, we will discuss the bag model
equation of state, see section \ref{sec:V.B.2}.  Several proposals
have been made for the initial conditions. Here we sketch only the
Bjorken-Shuryak expansion scenario. We discuss the longitudinal
solutions of Eqs.\ (\ref{5.11})-(\ref{5.12}) for this choice. The
solutions considerably simplify in this special case, which may be one
reason why they are frequently used in the hydrodynamic treatments of
matter evolution in heavy-ion collisions.

Radial solutions of Eqs.\ (\ref{5.11}) are rarefaction waves
propagating from the boundary into the fluid with the velocity of
sound. They differ in an essential way from Bjorken's scaling solution
as they are independent of proper time $\tau$. Radial solutions enter
measurements of enthalpy and pressure, cf. section \ref{sec:V.C.1}.

\subsubsection{Bjorken-Shuryak expansion scenario}
\label{sec:V.B.1}
In the Bjorken-Shuryak scenario (Bjorken, 1983; Shuryak, 
1978), several experimental observations are taken into 
account. The phenomenon of `nuclear transparency' leads to 
a separate treatment of the central rapidity and the 
fragmentation region. Nuclear transparency means the effect 
that a large fraction of the incoming energy is carried 
away by two receding nucleons in a nucleon-nucleon 
collision at high energy. Similarly, in a nucleus-nucleus 
collision the baryon contents of the colliding nuclei 
interpenetrates at the collision and recedes as two Lorentz 
contracted pancakes after the collision. The central 
rapidity region refers to the fluid of quanta contained in 
the region between the receding pancakes. In the hadronic 
phase it mostly consists of pions. Thus it should be a good 
approximation to neglect the baryon number. Setting $n_B=0$ 
leads to a first simplification of the hydrodynamic set of 
equations. The separate treatment of the central and the 
fragmentation regions is justified only if both regions are 
well separated in phase space. Experimental conditions 
should be checked to guarantee this, otherwise an analysis 
in this picture is not adequate (Blaizot and Ollitrault, 
1990). At future colliders (RHIC and LHC) these conditions 
may be satisfied.

Secondly, the pronounced space-time correlations are 
observed in particle production in the sense that particles 
with large longitudinal momenta are produced at a late 
time, those with low momenta promptly, in the 
center-of-mass system. This is nothing but the twin 
paradox. Particles live longer in case their velocities are 
higher. In Bjorken and Shuryak's ansatz the effect of time 
dilatation is incorporated in the boundary conditions. 
Consider an ensemble of particles which are produced at 
$z=0=t$. If it is only the proper time $\tau$ which 
determines the moment of disintegration, all particles 
which measure the same $\tau$ in their rest frame 
constitute an initial condition at $\tau= \tau_0$. That is, 
the initial condition refers to a hyperbola $\sqrt{t^2-z^2} 
=\tau_0$ of constant proper time $\tau_0$. The space-time 
rapidity or light cone variable 
$\eta={1\over2}\ln{{t+z}\over{t-z}}$ specifies the position 
on this hyperbola. Two distinct positions are related via a 
Lorentz boost in $z$-direction. Since particles in the 
fluid element are supposed to move as free particles, their 
velocity component $v_z$ is given by $z/t$. The physics of 
a $z$-slice of a fluid element at time $t$ is equivalent to 
the physics of a $z'$-slice at time $t'=z'/v_z$. This is 
the scaling property in Bjorken's scaling ansatz. For 
$v_z=z/t$, the space-time rapidity $\eta$ equals the 
rapidity $y$
\begin{equation}
  y={1\over2}\ln{{1+v_z}\over{1-
      v_z}}={1\over2}\ln{{E+p_z}\over{E-p_z}}\;,
  \label{5.13}
\end{equation}
if the four-momentum $\underline{p}$ is parametrized as 
$\underline{p}=(E,p_z,p_t)$, $p_t$ being the transverse 
momentum.

A third feature, which is observed in proton-proton 
collisions is the plateau structure of inclusive cross 
sections when they are plotted as functions of $y$. A 
plateau for central values of $y$ is also expected for 
nucleus-nucleus collisions. At least the particle 
multiplicity depends only weakly on $y$ for central 
rapidities (Bjorken, 1983). Accordingly, a further 
simplifying assumption seems to be justified. The local 
thermodynamic quantities like $\varepsilon,p,T,s$ depend 
only on $\tau_0$, but not on $\eta(\tau_0)$, when the 
hydrodynamic expansion commences. Thus the initial 
condition is invariant under Lorentz boosts in 
$z$-direction. The dynamics preserves Lorentz covariance, 
which is most easily seen from the tensor equation (\ref{5.4a}). 
Therefore, we will look for solutions $\varepsilon(\tau), 
s(\tau), T(\tau)$ of Eqs.\ (\ref{5.11}) depending merely on 
$\tau$. Inserting $v_z=z/t$ in the collective four-velocity 
of the fluid leads to
\begin{eqnarray}
  u^\mu&&={1\over{\sqrt{1-v^2_z-v^2_r}}}(1,v_z,v_r,0)
  = (t/\tau,z/\tau,0,0) \nonumber\\
  &&=(\cosh\eta,\sinh\eta,0,0)\;.
  \label{5.14}
\end{eqnarray}
A comparison with Eq.\ (\ref{5.8a}) shows that the fluid rapidity 
can be identified with the space-time rapidity $\eta$, 
which furthermore coincides with the rapidity $y$. With 
$\eta=\theta$, Eqs.\ (\ref{5.11}) simplify to
\begin{eqnarray}
  \tau{{\partial\varepsilon}\over{\partial\tau}}+
  \varepsilon+p&=0 \nonumber\\
  {{\partial p}\over{\partial y}}&=0\;.
\label{5.15}
\end{eqnarray}
The baryon number is set to zero in the following 
considerations. The entropy equation for $s(\tau)$ 
simplifies to
\begin{equation}
  \tau{{\partial s}\over{\partial\tau}}+s=0\;,
  \label{5.16}
\end{equation}
where the relations $u^\eta=0$ and 
$(u^\tau=u^t\cosh\eta-u^z\sinh\eta)$ have been used. The 
solution is
\begin{equation}
  {s\over{s_0}}={{\tau_0}\over\tau}\;.
  \label{5.17}
\end{equation}
As is seen upon integration over $d^3x=\tau dy\ d^2x$, the 
entropy per given rapidity interval remains constant as 
long as the hydrodynamic equations can be applied. This 
need not hold throughout all stages of the expansion, 
especially not close to freeze-out or in the intermediate 
period, where the plasma converts to the hadronic phase in 
one or another way. Let us assume that it approximately 
holds. Then the important feature of Eq.\ (\ref{5.29}) is that it 
allows to infer the entropy density in the initial state 
(more precisely $s_0\tau_0$) from an observation in the 
final state (the pion multiplicity), see section \ref{sec:V.C.1}.
Under the same assumptions as above, the temperature 
equation simplifies in $(\tau,\eta)$-coordinates to
\begin{equation}
  c^2_s+\tau\partial_\tau\ln T=0
  \label{5.18}
\end{equation}
for $\mu=0$, where by definition
\begin{equation}
  c_s^2={\partial p\over \partial\epsilon}\;,
  \label{5.19}
\end{equation}
$c_s$ denotes the velocity of sound. Integration of 
Eq.\ (\ref{5.18}) gives
\begin{equation}
  T=T_0(\tau_0/\tau)^{c^2_s}.
  \label{5.20}
\end{equation}
For a massless free gas, the speed of sound is $1/\sqrt3$ 
in units where $c=1$. Thus the temperature drops more 
slowly than the entropy density. In fact, the predicted 
decrease may be too slow, since transverse expansion has 
been neglected so far.

Finally, we have to solve Eq.\ (\ref{5.15}) for the energy density 
as a function of proper time. One possibility is to use an 
equation of state (in principle, it should be {\it the} 
equation of state of QCD) to eliminate the pressure in 
Eq.\ (\ref{5.15}). The result is that the energy density does not 
merely decrease because of the expanding volume in proper 
time, but also due to the pressure exerted by the gas of 
the covolume. We come back to the equation of state in the 
next section. Note that the velocity in transverse 
direction $v_r$ has been neglected so far. We shall 
sometimes abbreviate the solutions in `Bjorken-Shuryak's 
scenario' as Bjorken's scaling solution.

\subsubsection{The bag model equation of state}
\label{sec:V.B.2}
Although the bag model leads to a crude description of the 
equation of state for QCD, we address a short subsection to 
it, since it is often used in combination with the 
hydrodynamic equations and leads to quantitative 
predictions in the end. In the M.I.T.-bag model, the basic 
features of QCD -- confinement and asymptotic freedom -- 
are effectively incorporated via bags (Chodos et al., 1974; 
Johnson, 1975). In the hadronic phase, quarks and gluons 
are allowed to move freely or with perturbatively small 
interactions inside small volumes of space inside the bags. 
Outside the bags, the quarks are forbidden to move as free 
particles. The vacuum outside the bags is given a constant 
energy density $B$ (the bag constant), which keeps the 
quarks and gluons confined to the bags. During the phase 
transition, latent heat is necessary to liberate the color 
degrees of freedom. It turns out to be proportional to $B$. 
Its original value (0.145~GeV)$^4$ (De Grand et al., 1975) 
was based on fitting the mass spectrum at $T=0$ and low 
density in the M.I.T.-bag model. The effective value for 
the `vacuum pressure` $B$ which should be used in the quark 
gluon plasma phase at a baryonic matter density of 
$n_B=1/$fm$^3$ is 0.5~GeV/fm$^3$ (Shuryak, 1988).

An additive shift $B$ in the energy density of the plasma 
due to the vacuum energy is obtained, if $\ln Z$ of an 
otherwise free gas of quarks and gluons is shifted by 
$-BV/T$, that is
\begin{equation}
  T\ln Z\ \hbox{(plasma phase) = free gas contribution}\ 
  -BV/T\;.
  \label{5.21}
\end{equation}
The free gas contribution follows from the usual expression 
for a free gas of particles and antiparticles with mass 
$m$, chemical potential $\mu$, and degeneracy factor $g$. 
In the large volume limit it is given by
\begin{eqnarray}
  &&\ln Z(T,\mu,V)={{gV}\over{6\pi^2T}}\cdot\int^\infty_0
  dK{{K^4}\over{(K^2+m^2)^{1/2}}} \nonumber\\
  && \cdot~\left[{1\over{\exp\{[(K^2+m^2)^{1/2}-\mu]/T\}\pm1}} \right. 
  \nonumber\\
  &&\qquad \left. +{1\over{\exp\{[(K^2+m^2)^{1/2}+\mu]/T\}\pm1}}\right] ,
  \label{5.22}
\end{eqnarray}
where the ``+''-sign refers to fermions and the ``-''-sign 
to bosons. Adding up the various contributions from bosons 
(gluons), fermions (quarks and antiquarks) and the vacuum, 
one is lead to
\begin{eqnarray}
  T\ln Z=&&{1\over6}N_cN_fV\left({7\over{30}}\pi^2T^4+\mu_q^2T^2+
    {1\over{2\pi^2}}\mu^4_q\right) \nonumber\\
  &&+{{\pi^2}\over{45}}N_gVT^4-BV .
  \label{5.23}
\end{eqnarray}
Here $N_c$ is the number of colors, $N_f$ the number of 
flavors, $N_g$ the number of gluons and $\mu_q$ the 
chemical potential due to quarks. The standard 
thermodynamic relations Eqs.\ (\ref{2.1}) lead to the following 
expressions for the energy density, pressure and entropy 
density in the plasma phase
\begin{mathletters}
  \label{5.24}
  \begin{eqnarray}
    \varepsilon_p= && 111aT^4+B \\
    p_p=&& 37aT^4-B \\
    s_p=&& 148aT^3\;,
  \end{eqnarray}
\end{mathletters}
where $a=\pi^2/90$. These expressions hold for 
$\mu=0,N_c=3,N_f=2,N_g=8$. The general expression for 
$\varepsilon_p$ is
\begin{equation}
  \varepsilon_p={{N_cN_f}\over{\pi^2}}\left({{7\pi^4}\over{60}}T^4\right)
  +{{\pi^2}\over{15}}N_gT^4+B\;.
\label{5.25}
\end{equation}
From Eqs.\ (\ref{5.24}a and b) we can easily read off the bag 
model equation of state in the plasma phase as
\begin{equation}
  p={1\over3}(\varepsilon-4B)\;,
  \label{5.26}
\end{equation}
which remains valid for $\mu\not=0$.

Similarly, expressions for $\varepsilon,p$ and $s$ are obtained in the
hadron phase, when it is described as a free gas of the lightest
mesons and baryons, i.e. pions, nucleons and antinucleons, where the
baryonic contribution is sometimes omitted. For pions analytic
expressions for $\varepsilon,p$ and $s$ can be derived in terms of
modified Bessel functions following from Eq.\ (\ref{5.22}).
Contributions of the nucleon-antinucleon gas can be calculated
numerically. Heavier mass particles are often omitted for moderate
temperatures ($T\leq250~MeV$), although the restriction to pions is
rather questionable above $T_c\geq150$~MeV (cf. section
\ref{sec:IV.A.2}). Here we state the result for the limit of a gas of
{\it massless} pions
\begin{mathletters}
  \begin{eqnarray}
    \varepsilon_h=&&9aT^4 \\
    p_h=&&3aT^4 \\
    s_h=&&12aT^3
  \end{eqnarray}
\end{mathletters}
such that
\begin{equation}
  \varepsilon_h=3p_h
\label{5.28}
\end{equation}
is the bag-model equation of state in the hadron phase. 
The finite $T$ (and finite $\mu$) transition occurs, when 
the following Gibbs criteria are satisfied
\begin{mathletters}
  \label{5.29}
  \begin{eqnarray}
    p_h=&&p_p=p_c \\
    T_h=&&T_p=T_c \\
    \mu=&&3\mu_q=\mu_c .
  \end{eqnarray}
\end{mathletters}
The indices $h,p$ stand for the hadron and plasma phases, 
$c$ for the critical value, $\mu$ is the chemical potential 
associated with nucleons. Pressure balance at $T_c$ relates 
$T_c$ to the bag constant
\begin{equation}
  T_c=(B/34a)^{1/4}\;.
  \label{5.30}
\end{equation}
The latent heat, determined as the gap in the energy 
densities $\varepsilon_p -\varepsilon_h$ at the transition, 
is $4B$ in this model, e.g. for the effective bag constant 
of 0.5~GeV/fm$^3$ the latent heat is 2~GeV/fm$^3$.

Almost by construction the bag model leads to a first-order 
transition at finite $T$ and vanishing $\mu$. There is a 
finite gap in energy and entropy densities, while the 
pressure is continuous. Note that in the mixed phase, the 
velocity of sound $c_s$ vanishes, since 
$c_s=dp/d\varepsilon|_{T_c}=0$. This is true only as long 
as $\mu=0$. An exception is a 2nd order transition at 
$\mu_c\not= 0$, but $T_c=0$ for a particular value of $B$, 
which we will not consider further here (cf. Cleymans et 
al., 1986).

Eqs.(\ref{5.26}) and (\ref{5.28}) can be easily combined to a single 
equation by using $\theta$-functions as projections on the 
distinct phases above and below $T_c$. Similarly it is not 
difficult to formulate an equation of state for a second 
order transition by smoothly interpolating the step-like 
behavior of the bag model equation of state. Such an ansatz 
has been proposed by Blaizot and Ollitrault (1990) for 
$s(T)$. Although the interpolation is ad hoc, it provides a 
useful check, how sensitively phenomenological implications 
depend on the order of the transition. The difficulty is to 
{\it derive} such an equation for a second order transition 
within an effective model. The $O(4)$-model in three 
dimensions allows a second order transition, but is 
supposed to describe only the low-temperature phase of QCD 
($T\leq T_c$) or the immediate vicinity above $T_c$, where 
pions and sigma mesons have not yet dissolved in their 
constituents.

The bag model does not provide an adequate description of 
the transition region. Even at the transition point, the 
plasma and the hadron phases are treated as non-
interacting gases differing only in the degrees of 
freedom and the vacuum energy. An increasing number of 
hints warns of a naive counting of the number of modes 
which are taken to be the same as in the limiting cases 
of high and low temperatures. In particular the number of 
degrees of freedom of a hadron gas is a delicate problem, 
if the change of hadron masses as a function of 
temperature and density is respected. In this case a 
counting of pion and nucleon degrees of freedom is 
certainly insufficient. In the vicinity of $T_c$ the 
confinement/deconfinement properties should be implemented 
in a slightly more sophisticated way than with a single 
parameter $B$. Nevertheless the bag model is frequently 
used for temperatures $T\sim T_c$. The reason is probably 
its very tractable analytic form compared to (preliminary) 
plots of numerical simulations.

At high or low temperatures the bag model is more 
adequate, e.g the gluonic sector may be treated as a gas 
of non-interacting glueballs at low $T$ and of gluons at 
high $T$.

Various improvements on QCD's equation of state have been 
proposed within the framework of the bag model. Finite 
quark masses and perturbative QCD corrections to the 
partons inside the bag can be taken into account (Kapusta, 
1979; Shuryak, 1979) leading to corrections of 
$\epsilon$, $p$ and $s$. Also strange quarks, finite 
masses for pions, and higher mass hadrons can be included.

The ultimate goal is an equation of state from lattice QCD with
dynamical quarks, which is merely based on non- perturbative
ingredients (see section \ref{sec:III.B}). The numerical data should
then be presented in a feasible parametric form to facilitate its
handling.

In the next section we will see, what we can directly 
learn from heavy-ion experiments about QCD's equation of 
state.

\subsection{Signatures sensitive to the nature of the phase transition}
\label{sec:V.C}
Theoretical and experimental tools are described which are 
sensitive to the nature of the transition, in particular to 
its order. The sensitivity holds at least in principle. The 
dependence on the transition dynamics is sometimes hidden 
in the spacetime expansion. The signatures reveal 
characteristic features like time delays, nucleation of 
bubbles, strong correlations or large fluctuations in an 
indirect way.

Let us start with direct experimental tests of the equation 
of state. We sketch the possibilities to measure 
thermodynamic quantities.

\subsubsection{Thermodynamic observables}
\label{sec:V.C.1}
The basic observables which are at our disposal in 
heavy-ion collisions are rapidity distributions of final 
state particles and their transverse momentum 
distributions. An extraction of the equation of state 
requires measurable observables which are related to 
$\varepsilon(T,\mu),\ s(T,\mu)$ or $p(T,\mu)$. The signals 
of a first-order transition in a finite volume may be 
qualitatively very similar to a sharp crossover phenomenon 
consisting in a rapid rise in the effective number of 
degrees of freedom over a small range of temperatures, say 
less than 10~MeV. These numbers are exposed in $s/T^3$ or 
$\varepsilon/T^4$, cf. e.g. Eqs.\ (\ref{5.24}) of the bag model 
equation of state. Thus we have to identify the observables 
which are related to $s,\varepsilon,p$ and $T$. Roughly 
speaking, temperature and pressure are measured by the 
average transverse momentum $\left<p_T\right>$ (as usual under certain 
restrictive conditions), energy and entropy by the particle 
multiplicity distribution in rapidity space.

We consider the baryon-free case, i.e. $\mu=0$. Four 
quantities $s, \varepsilon,p$ and $T$ have to be 
determined. An equation of state is a relation between any 
two of these four variables $(\varepsilon(p),s(T), 
s(\varepsilon)$). It can be obtained from experiments (and 
compared with theoretical predictions) if two relations are 
used as experimental input. For example, if 
$\varepsilon(T)$ and $p(T)$ were known, $s(T)$ follows from 
$dp/dT=s$ and $\varepsilon(p)$ is the equation of state. 
Alternatively, if the initial condition $s_0(\tau_0)$ is 
fixed from a measurement of $dN/dy$, then 
$s(\tau),T(\tau)$, thus  $s(T)$ are known from the 
hydrodynamic equations, and $p(T)$ follows as integral over 
$s$. Suppose that the second relation is provided by a 
measurement of $\left<p_T\right>$ as a function of $s^{1/3}$, where 
$\left<p_T\right>$ is related to $s/\varepsilon$ (see below) and 
$s^{1/3}\propto T$. The knowledge of $(s/\varepsilon)(T)$
yields $\varepsilon(T)$, thus all relations are known
 (Blaizot and Ollitrault, 1990).
 
 In the following subsections we explain relations between $dN/dy$ and
 $s_0$ or $\varepsilon_0,\ <p_T>$ and $\varepsilon/s$ or $T$, as well
 as $\left<p_T\right>$ and $p$ or $(\varepsilon+p)$. We will show why
 $\left<p_T\right>$ vs $\left<dN/dy\right>$ diagrams are roughly
 equivalent to ($T$ vs $s$) or ($T$ vs $\varepsilon$)-diagrams. In
 case of a phase transition a flattening of the
 $\left<p_T\right>$-distribution is expected, which has been
 originally proposed by van Hove (1982) as a possible signature for a
 phase transition.

\paragraph*{Entropy measurements}
The basic step is to identify the entropy density with the 
particle density in the final state. The final state 
particles mostly consist of pions. Pions are nearly 
massless bosons, thus their entropy $S$ is approximately 
proportional to their number $N$. For the densities we have
\begin{equation}
  s=\alpha\cdot n
  \label{5.31}
\end{equation}
with $\alpha=3.6$ for a free gas of massless pions. The 
rapidity distribution is obtained from the particle density 
by an appropriate integration. When the four-volume element 
is expressed in terms of the spacetime rapidity $\eta$ and 
the proper time $\tau$, it follows for $N$
\begin{equation}
  N=\int nd^4x=\int n\tau d\tau d\eta d^2x\;,
  \label{5.32}
\end{equation}
or for the number of particles per unit rapidity at a fixed final proper
time $\tau_f$
\begin{equation}
  {{dN}\over{d\eta}}=\int n(\tau_f,\eta,x)\cdot\tau_fd^2x\;.
  \label{5.33}
\end{equation}
From Eqs.\ (\ref{5.31}) and (\ref{5.17}) we have
\begin{equation}
  {{dN}\over{d\eta}}=\int d^2x\tau_f{1\over\alpha}s(\tau_f,\eta,x)=
  \int d^2x\tau_0{1\over\alpha}s_0(\tau_0,\eta,x)\;.
  \label{5.34}
\end{equation}
If the initial entropy density is taken to be independent 
of the transverse coordinates $x$, we find that the final 
rapidity distribution of
multiplicity is proportional to the initial entropy density
\begin{equation}
  {{dN}\over{d\eta}}=\tau{1\over\alpha}\cdot 
  s_0(\tau_0,\eta)A_T\;,
  \label{5.35}
\end{equation}
where $A_T=\pi R^2$ is the transverse size of the 
nucleus. For example, the multiplicity per unit rapidity 
may be about 150 for a central collision of $^{16}O$ with 
$R\sim3$~fm on a heavy nucleus (Blaizot and Ollitrault 
(1990)). If the initial (= equilibration) time is estimated 
as 1 fm/$c$, it implies a value of $\sim20$~fm$^{-3}$ for 
the initial entropy density. We will see next what is 
further needed to extrapolate the initial temperature and 
the initial energy density.

\paragraph*{Energy measurements}
To derive a similar expression as  Eq.\ (\ref{5.35}) for the 
initial energy density $\epsilon_0$, we need a relation 
between $s_0$ and $\varepsilon_0$ and $\varepsilon_0$ and 
$T_0$ to eliminate $T_0$. From the thermodynamic relation 
($\varepsilon+p=Ts$) and the bag model equation of state 
we find
\begin{equation}
  (\epsilon_0-B)(1+c^2_s)=T_0s_0\;.
  \label{5.36}
\end{equation}
In Eq.\ (\ref{5.36}) the factor 1/3 for a massless free gas has 
been replaced by $c_s^2$. Next we use Eq.\ (\ref{5.24}a) of the 
bag model in the plasma phase (assuming that the initial 
temperature is high enough for the system to be in the 
plasma phase)
\begin{equation}
  \varepsilon_0-B=gT^4_0\;.
  \label{5.37}
\end{equation}
Again the exponent can be generalized to $1+1/c^2_s$. The 
degeneracy factor $g$ in the case of two massless $u$ and 
$d$ quarks, their antiquarks and gluons is given by
\begin{equation}
  g={{\pi^2}\over{30}}(2\times2\times2\times3\times{7\over8}+2\times8)=
  111a\;.
  \label{5.38}
\end{equation}
The different factors correspond to the spin, quark, antiquark, color,
flavor and gluon degrees of freedom, 7/8 arises from the Fermi-Dirac
statistics. When $T_0$ is eliminated in Eq.\ (\ref{5.37}) via
(\ref{5.36}), and (\ref{5.35}) is solved for $s_0$, we obtain
\begin{equation}
  \varepsilon_0=B+g\left[{\alpha\over{A_T(1+c^2_s)g\tau_0}} 
    {{dN}\over{d\eta}}\right]^{1+c^2_s}\;.
  \label{5.39}
\end{equation}
Eq.\ (\ref{5.39}) gives the promised relation between the final 
state rapidity distribution and the initial energy density 
$\epsilon_0$.

\paragraph*{Transverse momentum distributions}
A second important class of observables are average 
transverse momentum distributions. First we consider the 
relation between the average transverse momentum 
$\left\langle p_T\right\rangle$ and $E/S$. For a 
thermalized fluid at rest the momentum distribution is 
isotropic. For an ultrarelativistic fluid the total 
momentum is equal to the energy, such that the average 
transverse momentum is proportional to the energy per 
particle (Blaizot and Ollitrault (1990)
\begin{equation}
  <p_T>={\pi\over4}E/N\;.
  \label{5.40}
\end{equation}
If we replace $E$ and $N$ by the corresponding densities, 
it remains to express the particle density $n$ in terms of 
the entropy density via Eq.\ (\ref{5.31})
\begin{equation}
  <p_T>={{\alpha\pi}\over4}{\varepsilon\over s}\;.
  \label{5.41}
\end{equation}
This is already a relation of the type we are looking for.  Note that
$\varepsilon/s$ scales approximately with $T$ and $s$ with $T^3$.
Measuring $\left<p_T\right>$ as a function of $s^{1/3}_0$ by varying
$dN/dy$ accordingly could in principle provide a check of a
theoretically predicted relation between $\varepsilon/s$ and $T$.

When $\varepsilon/s$ is calculated from the bag model equation of state
and plotted against $(s_0/s_c)^{1/3}$, it displays a feature which
turns out to be characteristic for generic $\left<p_T\right>$
distributions: a flattening in the region where the initial entropy
density $s_0$ is chosen to have the critical value $s_c$, while a
linear rise is seen for small ratios $(s_0/s_c)^{1/3}$ (low
temperatures) where $\varepsilon/s$ and $s^{1/3}$ are proportional to
each other.

Unfortunately Eq.\ (\ref{5.41}) is too simple to be true. As the 
QCD plasma fluid undergoes a transverse and a longitudinal 
expansion, the simple relation between $\left\langle 
p_T\right\rangle$ and the initial temperature via $s_0$ 
gets lost. The decoupling temperature $T_d$ at freeze-out 
and the time at which the transverse expansion sets in 
enter as new scales. The essential point is whether 
decoupling and hadronization happen before or after the 
transverse expansion becomes important. The results of 
Blaizot and Ollitrault (1990) are the following. If the 
system decouples before the transverse expansion sets in, 
$\left\langle p_T\right\rangle$ is not sensitive to the 
equation of state, but characterizes a fluid at rest at the 
decoupling temperature. In the opposite case (decoupling 
after transverse expansion), $\left\langle 
p_T\right\rangle$ is roughly that of a thermal distribution 
with a three-dimensional expansion at decoupling. As an 
effect of the longitudinal expansion, $s_0$ has to be 
replaced in the naive expression (\ref{5.41}) in an appropriate 
way. The longitudinal expansion diminishes the available 
energy for transverse expansion.

The value of $\left\langle p_T\right\rangle$ is no longer 
proportional to $\varepsilon/s$. The modified $\left\langle 
p_T\right\rangle$-diagram as function of the initial 
entropy density can be compared with the diagram resulting 
from a bag model equation of state combined with a 
hydrodynamic expansion. The qualitative feature is the 
same: a flattening of $\left\langle p_T\right\rangle/T_c$ 
for initial entropy values lying in the transition region. 
The physical reason is a reduced pressure on individual 
particles, if the entropy contents is distributed among 
more degrees of freedom in a transient plasma. The smaller 
value for the pressure leads to the reduced average 
transverse momentum. Next we come to the relation between 
transverse momentum and pressure.

\paragraph*{Enthalpy and pressure}
The average transverse momentum $\left\langle 
p_T\right\rangle$ can be predicted from solutions of the 
hydrodynamic equations. The transverse velocity component 
is parametrized in terms of rapidity variables. The average 
transverse momentum is written in terms of the average 
transverse rapidity $y_T$ (due to thermal motion) and the 
fluid rapidity $\theta$ according to
\begin{equation}
  \left<p_T\right>=m_\pi\sinh(y_T+\theta)\;.
  \label{5.42}
\end{equation}

{\it Enthalpy measurements} can be performed via 
$\left\langle p_T\right\rangle$, if the fluid rapidity 
$\theta$ is determined as function of the initial enthalpy 
according to (Cleymans et al. 1986)
\begin{equation}
  \theta_R={{c_s}\over{1+c^2_s}}
  \ln{{\varepsilon_0+p_0}\over{\varepsilon+p}}\;,
  \label{5.43}
\end{equation}
where $\theta_R$ is a solution of the set of longitudinal 
hydrodynamic equations, specialized to rarefaction waves 
$\theta_R$. Solving Eq.\ (\ref{5.42}) for $\theta$ with $\theta$ 
given by Eq.\ (\ref{5.43}), we find for large $p_T/m_\pi$ that 
$\left<p_T\right>$ is proportional to $e^{\theta_R}$, thus
\begin{equation}
  \left<p_T\right>\sim{{m_\pi}\over2}\left({{\varepsilon_0+p_0}\over 
      {\varepsilon+p}}\right)^{c_s/(1+c^2_s)}\;.
  \label{5.44}
\end{equation}
It remains the express the r. h. s. of Eq.\ (\ref{5.44}) in terms 
of measurable quantities. Write the initial enthalpy as 
$T \cdot s_0$, eliminate $T_0$ in favor of $\epsilon_0$ 
with an equation of state, replace $s_0$ and 
$\varepsilon_0$ via Eqs.\ (\ref{5.35}) and (\ref{5.39}).

{\it Barometers} are provided by $\left\langle 
p_T\right\rangle$-distributions as function of the 
pressure. The fluid rapidity is now considered as function 
of the final energy density $\epsilon_f$. The analogous 
relation to Eq.\ (\ref{5.43}) reads (Gyulassy, 1984)
\begin{equation}
  \theta(\varepsilon_f)={{c_s}\over{1+c^2_s}}\ln\left({{p_0}\over{p_f}}
  \right)\;,
  \label{5.45}
\end{equation}
leading to
\begin{equation}
  \left<p_T\right>=
  {{m_\pi}\over2}e^{y_T}\left({{p_0}\over{p_f}}\right)^{c_s/(1+c^2_s)}\;.
  \label{5.46}
\end{equation}
Eq.\ (\ref{5.45}) is derived from a set of hydrodynamic equations, 
where the longitudinal expansion has been neglected for 
simplicity. Eq.\ (\ref{5.46}) explains why $\left\langle 
p_T\right\rangle$ distributions are called barometers. If 
the speed of sound and the pressure $p_f$ in the final 
state are known, a measurement of $\left\langle 
p_T\right\rangle$ allows an estimate of the initial pressure 
$p_0$. Eq.\ (\ref{5.46}) gives $\left\langle p_T\right\rangle$ for 
a hadron gas, which is described as a massless ideal pion 
gas with rarefaction waves in transverse direction, but 
without longitudinal expansion.

The transverse flow is overestimated as long as the longitudinal
expansion is neglected. If $p_0/p_f$ or $\epsilon_0/\epsilon_f$ in
Eq.\ (\ref{5.46}) are replaced by the pion rapidity density $dN/dy$,
the resulting $\left\langle p_T\right\rangle$ values are actually too
large as compared to experimentally measured values. The fluid feels
transverse pressure gradients when the rarefaction wave arrives at a
fluid element. The arrival is somewhat delayed, since the wave moves
inwards with the speed of sound. Thus some energy is already gone to
the longitudinal expansion, less energy is available for the
transverse motion. Formally this effect can be included by an
appropriate form factor, which should be multiplied on the r. h. s.'s
of Eqs.\ (\ref{5.46}) and (\ref{5.41}), cf. Gyulassy (1984), Baym
(1984a and b), Blaizot and Ollitrault (1990) for further details. The
terminology to call $\left<p_T\right>$ a ``barometer'' for heavy-ion
collisions goes back to Gyulassy (1984).  Gyulassy further identifies
thermometers and seismometers in the connection of dilepton production
and large multiplicity fluctuations, see sections \ref{sec:V.C.2} and
\ref{sec:V.C.5} below.

\vskip12pt
We summarize. Bulk quantities like $\epsilon(T)$, $p(T)$, 
$s(T)$, derived quantities like the velocity of sound, the 
enthalpy, the energy per degree of freedom $\epsilon/s$ 
(Redlich and Satz, 1986) and QCD's equation of state are 
calculable on the lattice. The equation of state under 
inclusion of dynamical fermions should finally replace the 
bag model equation of state which entered the derivation of 
various relations in the preceding section. Eqs.\ (\ref{5.35}) and 
(\ref{5.44}) are a remarkable example of relations between 
observables which are directly accessible in experiments 
and on the lattice. Initial energy and entropy are related 
to the pion multiplicity per unit rapidity. A variation of 
$dN/dy$ amounts to a variation of $s_0$ or $\epsilon_0$. On 
the other hand, $\left\langle p_T\right\rangle$can be a 
measure for the freeze-out temperature and is sensitive to 
the pressure. Thus a flattening of $\left\langle 
p_T\right\rangle$ in a $\left\langle p_T\right\rangle$ vs 
$dN/dy$ diagram could be nothing else but a reflection of a 
slow change of temperature and pressure during a rapid rise 
in energy and entropy densities, i.e. a rapid crossover or 
a first order transition in a finite volume. The typical 
shape of a $\left\langle p_T\right\rangle$ vs $dN/dy$ 
diagram which is schematically drawn in Fig. \ref{fig:27} has been 
actually seen at Tevatron (Alexopoulos et al., 1990), cf. 
Fig. \ref{fig:28}, which gives $\left\langle p_T\right\rangle$ of 
$K^\pm$ and $\pi^\pm$ in $p-\bar{p}$ collisions as function 
of the charged particle multiplicity. Unfortunately a sharp 
crossover or a phase transition are not the only 
explanations for the shape. The flattening could be just a 
kinematic effect (lack of available energy). Kinematic 
constraints like energy conservation require that 
$\left\langle p_T\right\rangle$ goes to zero at the 
boundaries of the allowed rapidity interval (Schmidt and 
Schukraft, 1993). Rescattering effects or minijet 
production are further alternatives to explain a first 
increase and a subsequent flattening effect (Satz, 1990; 
Eskola and Lindfors, 1990).
The multiplicity increases only slowly with the incident 
kinetic energy. The relevant variable has been identified 
to be $s_0t_0/s_cR_0$, where $t_0$ is the proper time, at 
which the hydrodynamical evolution starts, $s_0$ is the 
entropy density at that time, $R_0$ is the initial radius 
and $s_c$ the entropy density at the critical temperature 
(Blaizot and Ollitrault, 1990). This variable should be 
larger than 1 to see an effect of the transverse expansion. 
Thus it is not sufficient that the initial entropy density 
$s_0$ is larger than the critical value $s_c$, which one 
might naively expect. The inverse initial radius $1/R_0$ is 
proportional to $1/B$, where $B$ is the mass number of the 
smaller of the colliding nuclei. Therefore the use of 
heavier nuclei in the collision may not help, but make 
things worse. For a review about the available data of 
$\left<p_T\right>$-distributions we refer to Schmidt and Schukraft 
(1993).

\subsubsection{Dileptons and real photons}
\label{sec:V.C.2}
\paragraph*{Dileptons}
We assume that the initial energy densities of a collision between
nuclei $A$ of the projectile and nuclei $B$ of the target are high
enough so that a quark gluon plasma is initially generated. Since the
size of the system produced in such a collision is not too large and
the electromagnetic cross section is small, all dileptons can escape,
especially those from the hottest dense state of the plasma. Dileptons
will be emitted from all stages of the evolution, from the initial
plasma phase, an intermediate possible mixture of plasma and hadron
phase till the hadron phase at freeze-out. Thus, differently from
hadronic yields, dileptons are sensitive to the whole spacetime
history of the evolution. The suitable kinematic variables for
dileptonic cross sections are the invariant mass $M$ of the dilepton
pair and its transverse momentum $p_T$. A typical observable is the
multiplicity of dileptons per invariant mass square $M^2$, transverse
momentum $p_T$ and unit rapidity interval $\Delta Y$,
\begin{equation}
  R={{dN}\over{dM^2d^2p_TdY}}=\int_Vd^4x{{dN}\over{d^4xd^4p}}.
  \label{5.47}
\end{equation}
Here $R$ is a spacetime integral over the rate $dN/d^4xd^4p$, which is
the rate at a given four-momentum $\underline p$ of the dilepton pair
at a spacetime point $(\vec x,t)$. The quantity $dN/d^4xd^4p$ depends
on the temperature $T$ via $(\vec x,t)$, and therefore on the phase of
the system. The phase (plasma or hadron) determines the dominant
production mechanisms entering the elementary cross sections and the
structure functions in $dN/d^4xd^4p$. The structure functions reflect
the medium where the elementary processes take place (quarks inside a
pion or in a heat bath of other partons). Collective effects of the
plasma may also influence the rate for a given value of $(\vec x,t)$.

In a given phase, usually several sources are responsible for dilepton
production. We are mostly interested in thermal collisions reflecting
thermodynamic properties of the supposed heat bath environment.
Thermal collisions are by far not the only way to produce dileptons.
Other mechanisms are Drell-Yan production, preequilibrium production
in the plasma phase, and dileptons from pions via $\rho$-resonance
decays in the hadronic phase.  Thermal dileptons result from parton
collisions in a medium which can be characterized by a local
temperature. They are specific for a plasma or a hadron gas which can
be described with thermodynamic concepts; they are absent in
hadron-hadron collisions.

Finally, we want to find out the mass range $\Delta M$, where $R$ is
dominated by thermal dileptons. In this range $R$ should be sensitive
to the kinetics of the phase transition and in particular to its
order. Let us assume $R(\vec x,t)$ has been calculated as a function
of $T(\vec x,t)$.  The spacetime integral in Eq.\ (\ref{5.47}) can be
done if the missing relation $T=T(\vec x,t)$ is known. Such a relation
is provided by the hydrodynamic approach. So far it is the only
framework which is detailed enough to predict particle spectra
according to Eq.\ (\ref{5.47}). Further simplifying assumptions are
made. To these belong Bjorken's scaling ansatz and the neglection of
transverse flow, which lead to entropy conservation as function of
time. As we have seen above, the hadronic multiplicity of the final
state can be related to the initial conditions under the assumption of
entropy conservation.  Once the entropy is known as function of time,
the time dependence of $T$ follows immediately if $s(T)$ is known.
This is the place where the equation of state enters. Further
assumptions about the kinetics of the phase transition are necessary
to justify the application of hydrodynamic concepts throughout all
stages of the evolution.

In the following we will outline the predictions of (Cleymans et
al.,1987) for thermal dilepton spectra which are derived for different
scenarios of the phase transition. The reason why we have selected
this reference although it is not the most recent one in this field is
that it explicitly addresses the influence of the order of the phase
transition on the dilepton production rate. Although the ingredients
do not represent the newest state of the art, the representation is
suited for illustrating which features of dileptonic yields are
sensitive at all to the dynamics of the phase transition. Later we
comment on places of possible improvements and extensions.

Consider the particle rate which is differential in $x$ and $p$. The
general expression for the thermal rate of lepton pairs (here written
for $\mu^+\mu^-$ rather than $e^+e^-$) in the independent particle
approximation of kinetic theory (Kapusta and Gale, 1987) is given by
\begin{eqnarray}
  {{dN}\over{d^4xd^4p}}=&&
  \int{{d^3q_1}\over{2E_1(2\pi)^3}}f(q_1)
  \int{{d^3q_2}\over{2E_2(2\pi)^3}}f(q_2) \nonumber\\
  &&\int{{d^3q_+}\over{2E_+(2\pi)^3}}
  \int{{d^3q_-}\over{2E_-(2\pi)^3}} \nonumber\\
  &&\quad\cdot\vert M(p_1\bar{p_2}\to\mu^
  +\mu^-\vert^2\cdot\delta(p-q_1-q_2)~.
  \label{5.48}
\end{eqnarray}
Here $M$ is the matrix element for the process of leptoproduction from
particles $p_1,\bar{p_2}$ with momenta $q_1,q_2$, $p$ is the momentum
of the lepton pair, $q_+$ and $q_-$ are the momenta of $\mu^+$ and
$\mu^-$.  The statistical distributions $f(q_1)$ and $f(q_2)$ measure
the probability of finding particles $p_1,\bar{p_2}$ with momenta
$q_1$ and $q_2$ in the given medium. The dileptons do not receive such
factors, since we assume that rescattering of the electromagnetically
interacting particles can be neglected. The cross section is small,
and the plasma volume is assumed not to be too large.  In terms of
cross sections $\sigma$, Eq.\ (\ref{5.48}) reads
\begin{eqnarray}
  {{dN}\over{d^4xd^4p}}=\int{{d^3q_1}\over{(2\pi)^3}}{{d^3q_2}\over
    {(2\pi)^3}}v_{q_1q_2}\sigma(p_1\bar{p_2}\to\mu^+\mu^-)
  \nonumber\\
  \cdot f_1(q_1)f_2(q_2)\cdot\delta(p-q_1-q_2)
  \label{5.49}
\end{eqnarray}
upon integration over the lepton momenta, and $v_{q_1,q_2}$ denotes the
relative velocity of particles $p_1$and $\bar{p_2}$
\begin{equation}
  v_{q_1q_2}={{\sqrt{(q_1q_2)^2-m^4_{p_{1,2}}}}\over{E_1\cdot E_2}}.
  \label{5.50}
\end{equation}
In the plasma phase the lowest order process for dilepton production
is the same as in the Drell-Yan production: two quarks annihilate via
a virtual photon to yield a lepton pair
\begin{equation}
  q\bar q\to \gamma*\to\mu^+\mu^-.
  \label{5.51}
\end{equation}
The momentum distribution functions  $f_1$ and $f_2$ of the quarks $q$
and antiquarks $\bar q$ are given by the Fermi-Dirac distributions
\begin{equation}
  f_q={6\over{e^{(\underline u \underline q-\mu)/T}+1}}~, \qquad
  f_{\bar q}={6\over{e^{(\underline u\bar{\underline q}+\mu)/T}+1}} .
  \label{5.52}
\end{equation}
The chemical potential $\mu$ is set to zero in the end, $\underline u$
is the local four-velocity of the plasma fluid element in the fixed
laboratory frame, $\underline q(\bar{\underline q})$ is the
four-momentum of the quark (antiquark). As hydrodynamic scenario a
longitudinal expansion will be used with vanishing transverse velocity
of the plasma.  In view of that, it is convenient to express the
product $\underline u\cdot \underline q$ in terms of the transverse
invariant mass $M$, the rapidity $Y$ of the lepton pair, and the
plasma spacetime rapidity $\theta$ according to
\begin{mathletters}
  \label{5.53}
  \begin{equation}
    \underline u\cdot \underline q=M_T\cosh(\theta-Y)
    \label{5.53a}
  \end{equation}
  with the definitions of the transverse mass $M_T$
  \begin{eqnarray}
    M_T&&=\sqrt{M^2+p^2_T} \\
    Y&&={1\over2}\ln{{E+p_z}\over{E-p_z}} \\
    \theta&&=\hbox{arctanh} v
  \end{eqnarray}
\end{mathletters}
with $v=z/t$ denoting the collective fluid velocity.  If the
transverse mass is large compared to the temperature, the exponentials
are sharply peaked around $\theta=Y$. In this case no final
$\theta$-dependence is left. Upon integration over the dilepton
momentum, the result for the {\it dilepton rate in the plasma phase}
at a spacetime point $\underline x= (\tau,x_T,y)$ ($y$ being the local
spacetime rapidity of a fluid element) and four-momentum $\underline
p=(M_T\cosh Y,p_T,M_T\sinh Y)$ is given by
\begin{eqnarray}
  {{dN_P}\over{d^4xd^4p}}=&&{{\alpha^2}\over{4\pi^4}}
  \left\lbrack 1+{{2m^2}\over{M^2}}\right\rbrack\left\lbrack
    1-{{4m^2}\over {M^2}}\right\rbrack^{1/2} \nonumber\\
  &&e^{-E/T}K_P (\underline p,T,\mu) \cdot\sum_i e^2_i .
  \label{5.54}
\end{eqnarray}
The index $P$ stands for plasma phase, $e_i$ are the charges of the
quarks, $\alpha$ is the electromagnetic coupling, $m$ stands for the
lepton mass and $K_P$ is a function, which depends on $\underline x$
via $T$, which we will not specify further now. It is characteristic
for the plasma phase.

In the {\it hadronic phase} it depends on the invariant mass $M$ of
the dilepton pair which process makes the leading contribution to the
elementary cross section $\sigma$ in Eq.\ (\ref{5.49}).  For small
mass pairs (small compared to the $\rho$-peak), bremsstrahlung-type
emission of soft virtual photons is important. For masses $M$ well
above the order of $T_c$, processes $h\bar h\to \gamma* \to\mu^+\mu^-$
play a role, where $h\bar h$ are hadrons others than pions.  Their
thermal production should be suppressed due to the relatively low
temperatures, unless the in-medium masses are changed due to effects
of chiral symmetry restoration, which may be dramatic! The only
process that has been considered so far in (Cleymans et al., 1987) is
$\pi^+\pi^-\to\rho\to\mu^+\mu^-$.  The electromagnetic cross section
$\sigma(M)$ is modified by the strong interactions of the pions,
leading to
\begin{mathletters}
  \label{5.55}
  \begin{equation}
    \sigma_\pi(M)=F^2_\pi(M)\left(1-{{4m^2_\pi}\over{M^2}}\right)
    ^{1/2}\cdot\sigma(M) ,
    \label{5.55a}
  \end{equation}
  where
  \begin{equation}
    \sigma(M)={{4\pi}\over 3}{{\alpha^2}\over{M^2}}\left\lbrack 1+
      {{2m^2_\mu}\over{M^2}}\right\rbrack\left\lbrack 1-{{4m^2_\mu}\over
        {M^2}}\right\rbrack^{1/2} .
    \label{5.55b}
\end{equation}
\end{mathletters}
The pion form factor $F_\pi$ is treated in the vector-meson-dominance
approximation,  only the $\rho$-pole  is kept
in the sum over all $\rho$-like resonances.

Furthermore the functions $f_1$ and $f_2$ in Eq.\ (\ref{5.49}) have to
be replaced by Bose-Einstein distributions. The result for the {\it
  dilepton rate in the hadron phase} is given by
\begin{eqnarray}
  {{dN_H}\over{d^4xd^4p}}=&&{{\alpha^2}\over{48\pi^4}}
  \left\lbrack 1+2{{m^2_\pi}\over{M^2}}\right\rbrack\left\lbrack
    1-4{{m^2_\pi}\over{M^2}}\right\rbrack^{3/2}\cdot\vert F_\pi(M^2)
  \vert^2 \nonumber\\
  &&\cdot\exp\lbrace-E/T\rbrace\cdot K_H(p,T)~,
  \label{5.56}
\end{eqnarray}
where $K_H$ is a slightly different function from $K_P$ in Eq.\ 
(\ref{5.54}).

\vskip12pt\noindent
{\it Integration over the spacetime history}
\vskip12pt\noindent
The assumption of Bjorken's scaling solution for the hydrodynamical
expansion simplifies the evaluation of the spacetime integration.
The appropriate representation of the volume element is $d^4x=\tau
d\tau dy d^2 x_T$, where $\tau$ is the eigentime in a comoving reference
frame with the fluid, $y$ is the spacetime rapidity (denoted as
$\eta$ above), $x_T$  are the transverse coordinates. The transverse
velocity is neglected, the transverse distributions of thermodynamic
quantities are taken as step functions, $\varepsilon,
p,s$ and $T$  depend then only on the eigentime $\tau$. The dileptonic
yields per  invariant mass, transverse momentum and unit rapidity
interval $dY$ are given by
\begin{equation}
  {{dN_{Ph}(hydro)}\over{dM^2 d^2p_TdY}}={1\over2}\int^{\tau_f}
  _{\tau_i}\tau d\tau\int^{y_{max}}_{y_{min}}dy{{dN_{Ph}}\over{d^4xd^4p}}
  (T(\tau)) .
  \label{5.57}
\end{equation}
The index $Ph$ indicates the dependence on the phase, $\tau_f-\tau_i$
is the eigentime duration of a certain phase or a mixture of phases,
$y_{max}$ is taken to be the rapidity of the incoming beam. The {\it
  equation of state} enters the cooling law $T(\tau)$, the phase
duration is sensitive to the dynamics of the transition.

Cleymans et al. (1987) have discussed the following scenarios
\begin{itemize}
\item a first order transition described by a bag model equation of
  state proceeding in equilibrium or with supercooling and subsequent
  superheating
\item a second order transition with an equation of state taken from
  lattice Monte Carlo results (Redlich and Satz, 1986).
\end{itemize}
As we have seen in section \ref{sec:V.B.2}, the bag model equation of
state leads to the $T^3$ dependence of the entropy density and is
typical for an ideal gas. In the baryonless plasma phase for two
massless flavors, the relation is given by $s_P=4\cdot 37\pi^2
T^3/90$. In the hadron phase described as an ideal gas of massless
pions we have $s_H=2\pi^2 T^3/15$. In the case of the second order
transition $s(T)$ has been read off from numerical data of Redlich and
Satz (1986), which were obtained in lattice QCD. Using the scaling
solution for isentropic flow $(s(\tau)\cdot\tau=$ const., cf. section
\ref{sec:V.B.1}), we find the relations $T(\tau)$. When the scaled
temperature $T/T_0$ is plotted as a function of the scaled eigentime
$\tau/\tau_0$ (where $T_0$ and $\tau_0$ refer to the initial values
chosen as $T_0=284$~MeV and $\tau_0=1$~fm/c), the characteristic
difference between a first order transition proceeding in equilibrium
and a second order transition is that the average temperature is
higher in the quark gluon plasma phase and lower in the hadronic phase
for a second order transition.  Having $T(\tau)$ at hand, the
integration in Eq.\ (\ref{5.57}) can be done.  Depending on the
scenario, four rates are distinguished.
\begin{itemize}
\item The dilepton rate in the pure plasma phase produced in the time
  interval between $\tau_0$ and $\tau_P$ where $T(\tau)$ is given as
  $T_0(\tau_0/\tau)^{1/3}$ for the bag model equation of state.
\item The dilepton rate in the pure hadronic phase produced during the
  interval between $\tau_H$ and freeze-out time (where the application
  of hydrodynamic concepts is already questionable).
\item If there is a coexistence of phases as it is expected in the
  first order case, there is a mixed phase at temperature $T_c$, where
  the plasma (hadron) phase contributes a fraction
  ${{dN_{P/H}\hbox{(mixed)}} \over{dM^2d^2p_T dY}}$, respectively.
  Both rates are calculated separately. Their relative weight is
  described by a factor $f$, which is the fraction of the entropy in
  the plasma phase. Its value follows from $s(\tau)\cdot \tau=$ const
  and Eq.\ (\ref{5.58})
  \begin{equation}
    s(\tau)=f(\tau)s_P+\lbrack 1-f(\tau)\rbrack\cdot s_H.
    \label{5.58}
  \end{equation}
  The entropy densities $s_P$ and $s_H$ remain constant
  for constant $T=T_c$.
\end{itemize}

The ratio of entropy densities $s_P/s_H$ gives also an estimate for
the time scales $\tau_H$ and $\tau_P$, as $\tau_H/\tau_P=s_P/s_H=37/3$
if we assume for a moment that the naive counting of degrees of
freedom is appropriate close to $T_c$. This ratio is frequently quoted
in the literature, although the number 3 in the hadronic phase
(resulting from 3 pions) is most likely incorrect. The large
difference in entropy densities between both phases explains the long
duration time of the mixed phase. It takes time to rearrange the
effective degrees of freedom since they have to be reduced by an order
of magnitude for the above counting when the conversion to the
hadronic phase sets in.

We display the results of Cleymans et al. (1987) only for the cases of
a first-order transition with Maxwell construction and a second-order
transition. Upon integration over the transverse momenta, the
resulting dileptonic rates $dN/dM^2dY$ at $Y=0$~[GeV$^{-2}$] are shown
in Figs.~\ref{fig:29}(a) and (b). The difference is the extension of the
interference region where both phases contribute with dilepton
production. It is considerably smaller for a second order transition.
In this case, the production mechanisms change instantaneously at
$T_c$, while those of both phases would be at work in the coexistence
phase of a first order transition.

The width of the interference region does not yet provide a signal
which is suitable for experiments to infer the phase transition
dynamics. The sudden change in the production mechanism is reflected
in the average transverse momentum $\left<p_T\right>$ of the dilepton
pair or even more clearly in its derivative with respect to $M$. The
average $\left<p_T\right>$ can be derived from the rate $R$ according
to
\begin{equation}
  \left<p_T\right>={{\int{{dN}\over{dM^2dydp_T}}p_Tdp_T}~/
    {}~{\int{{dN}\over{dM^2dydp_T}}dp_T}} .
  \label{5.59}
\end{equation}
A plot of $d\left<p_T\right>/dM$ as a function of $M$ for both
transition scenarios shows a peak in the $d\left<p_T\right>/dM$
distribution for a second-order transition. Thus a peak in the
$d\left<p_T\right>/dM$ distribution is a possible candidate for a
signal of a second-order phase transition (keeping in mind all
approximations that have been applied so far). As outlined by Cleymans
et al. (1987), such a sudden change in the slope of $\left<p_T\right>$
as a function of $M$ can be also produced by a higher mass resonance
like the $\rho'$ (1600). To disentangle the origins of bumps in the
$\left<p_T\right>$ spectra, one should vary the initial temperature.
At higher initial temperature the contribution from the plasma phase
to the dilepton rate increases. A peak structure in the
$\left<p_T\right>$ spectrum due to a 2nd order transition then shows
up already for low initial temperatures, while higher initial
temperatures are necessary to see a similar structure in a first order
scenario.

Yet another possibility may lead to a sudden change in
$\left<p_T\right>(M)$.  Dilepton rates are calculated in a way that
they do not automatically interpolate smoothly at $T_c$, although they
should do so in the case of a second-order transition where no
discontinuities should be seen in thermodynamic observables. Cleymans
et al. (1987) have investigated the influence of smoothly
interpolating the rates between both phases. The bump structure,
however, survived the interpolation between $dN_p/d^4xd^4p$ and
$dN_H/d^4xd^4p$.

In general, it is outstanding to calculate the electromagnetic
current-current correlation $W(x,p)$ (whose spacetime integral is
proportional to $dN/dM^2dp_TdY$) in terms of one effective model which
replaces the bag model in the case of a second order transition. We
have mentioned this point already in section \ref{sec:V.B.2}.

So far we have been concerned with thermal rates and specific features
reflecting the transition dynamics in the rate dependence of the
invariant mass $M$. In heavy-ion experiments it is obviously an
objective of higher priority to guarantee that thermal dileptons can
be identified from variant background sources. In the order of
increasing masses, we have Dalitz pairs, soft gluon bremsstrahlung,
hadronic resonance decays, preequilibrium production, and Drell-Yan
production in the very early stage of the collision.

Ruuskanen (1990) has derived an analytic expression for the integrated
thermal rate $R$. The characteristic feature of this expression
concerns the $T$ and $M_T$-dependence as compared to other production
mechanisms ($M_T=(M^2+p_T^2)^{1/2}$ denotes the transverse mass which
is used instead of the transverse momentum $p_T$). The dependence of
the thermal rates on the initial temperature $T_i$ is strong, i.e.
proportional to $T_i^6$ as compared to the $T_i^3$-dependence for a
Drell-Yan rate. Recall that $T_i$ at equilibration time $\tau_i$ can
be related to the hadronic multiplicity in the final state according
to
\begin{equation}
  (\tau_iT_i^3)^2\propto(dN\pi/dy)^2~,
  \label{5.60}
\end{equation}
if an isentropic expansion scheme is assumed. This allows an
identification of thermal dileptons via their strong (quadratic)
dependence on pion multiplicity fluctuations, Drell-Yan rates show
only a linear dependence. At fixed multiplicity (i.e., fixed
$\tau_iT_i^3$), the thermal rates are expected to show an approximate
power law behavior proportional to $M^{-6}_T$ in a range of
$M_T/T_i<5.5<M_T/T_c$ (Ruuskanen, 1990) with a $\left<p^2_T\right>$
dependence proportional to $M^2$, and an exponential decrease of the
dilepton rate for large $M_T/T_i$, with $\left<p^2_T\right>$ showing a
linear increase in $M$. For an initial temperature of 600~MeV, the
window of the $M^{-6}_T$ dependence of the transverse mass spectrum
lies between 1-3~GeV.  Its lower end has to face a background from the
hadronic phase, its upper end the Drell-Yan background.

Thus the first goal of a measurement of dilepton rates in a collision
of heavy-ion nuclei is to identify a window in $M_T$ of the spectrum
which shows the behavior of a thermal rate produced in a plasma phase.
It is easier to filter out the $M^{-6}_T$ depending part if high
initial temperatures are reached in the experiment. For detailed
calculations of the different backgrounds and their competition to
thermal rates we refer to e.g.  Ruuskanen (1992), Weldon (1991),
Kajantie and Miettinen (1981), McLerran and Toimela (1985). For more
recent predictions of the dilepton spectrum at LHC energies see
Redlich et al. (1992).  The effect of various equations of state on
the transverse expansion and the dilepton production rate is discussed
in (Kataja et al., 1992).  The background of dileptons from
$\rho$-decays is usually considered as undesirable in view of thermal
dilepton identification. Recently it has been shown by Heinz and Lee
(1992) that such dileptons may contain valuable information on the
lifetime of the hadronic phase; that is why they are called ``fast
clocks'' for heavy-ion collisions. The overall judgement is that the
identification of thermal rates in the different phases is difficult
but not completely unrealistic. Yet it seems to be an order of
magnitude more difficult to infer specific features of the transition
itself from structures of $R$ or $\left<p_T\right>$ distributions as
functions of $M$.

\bigskip
We summarize the approximations which enter predictions of dilepton rates
along the lines we have indicated above.
\begin{itemize}
\item Uncertainties in the background to thermal dileptons.  The
  Drell-Yan background is the one which is understood best.
  Corrections due to higher orders in QCD have been assumed to be 
  small, actually they are large in the usual deep-inelastic structure
  function approach. (For recent results on Drell-Yan production at
  collider energies including $O(\alpha_{s}^{2})$ contributions we refer
  to van Neerven (1995).)  
  Deviations from the linear $A$-dependence of the structure
  functions on the mass number $A$ (entering the Drell-Yan cross
  section for colliding nuclei) are assumed to be negligible in the 
  relevant
  mass range ($M=2$-6~GeV). Probably the least understood rates are
  those which are produced shortly after the collision but before the
  onset of thermal equilibrium. Effective models are needed which
  describe nonequilibrium properties of the evolution of the initial
  state. In the fluxtube model of Kajantie and Matsui (1985) and
  Bialas and Czyz (1984, 1985), dilepton rates depend on the strength
  of the color field which is formed after the color exchange between
  the colliding nuclei. The new production mechanism relies on plasma
  oscillations of quarks and antiquarks. In the mass range of a few
  GeV which is of special interest for thermal production, the worst
  perspective is that both the thermal and the Drell-Yan productions
  -- including all their specific signatures -- are buried under the
  yields of plasma oscillations.  For further discussions, see also
  Bialas et al. (1988) and Bialas and Blaizot (1985).
\item The applicability of hydrodynamic concepts made it possible to
  evaluate the integral over the spacetime history of rates at a given
  point in phase space. It is marginal in several aspects. The volumes
  are not so large compared to the mean free paths, at least not close
  to freeze-out, where by definition the system is so diluted that
  interactions can be neglected. The scale of equilibration time is
  estimated in a purely phenomenological way. Thus it cannot really be
  argued, which time $\tau_0$ for the onset of hydrodynamic expansion
  is to be used. The adiabatic expansion guarantees the conservation
  of entropy and leads to a simple relation between entropy and time.
  During the phase conversion this condition may be violated.
\item The bag model equation of state has been used close to $T_c$,
  although strong deviations from the ideal gas behavior are expected
  there. Lattice Monte Carlo calculations should do their best to
  include the effect of dynamical fermions as realistic as possible
  and to derive an equation of state close to $T_c$. In deriving the
  equation of state which was used by Cleymans et al. (1987) and led
  to a second-order transition, the fermions are treated in the
  hopping parameter expansion; the involved parameters are known to be
  not close enough to the continuum limit. A more realistic equation
  of state obtained from lattice calculations with dynamical fermions
  can be found e.g. in (Christ, 1992), although it should not be
  considered as {\it the} realistic equation of state for QCD, see
  section \ref{sec:III.B.2}.
\end{itemize}
Given the hydrodynamic approach and the equation of state, further
simplifications were made in deriving the results we have shown. We
have neglected effects due to
\begin{itemize}
\item a realistic nuclear geometry and
\item the impact parameter dependence. In particular, the initial
  temperature profile depends on the impact parameter $b\not=0$.
\item the transverse coordinate dependence of temperature during all
  stages of the evolution.
\item the transverse flow of the hadron/plasma fluid. When transverse
  flow is included, hadronic rates can be reduced by an order of
  magnitude due to the reduced lifetime of mixed and hadronic phases,
  cf.  Fig.~\ref{fig:10} of Ruuskanen (1990).
\item last but not least the possible temperature and density
  dependence of masses which are involved in the dilepton production.
  Meson masses can be expressed in terms of the chiral condensate
  $\left<\bar\psi\psi\right>$, they should be sensitive to the order
  of the transition. Either $\left<\bar\psi\psi\right>$ vanishes
  smoothly when approaching $T_c$ from below, or two values of
  $\left<\bar\psi\psi\right>$ coexist at $T_c$, a large value and a
  small one.
\end{itemize}
Finally we refer to the work by Geiger and Kapusta (1993), Shuryak and
Xiong (1993), and Ruuskanen (1992) as more recent references in this
field.

\vskip12pt When dileptons are produced from vector meson decays, the
dilepton rates are sensitive to the phase transition dynamics also via
vector meson mass changes in medium both as a function of $T$ and
$\mu$. We have skipped this possible manifestation of the transition
dynamics, as we are not aware of any specific prediction yet.

\paragraph*{Real photons}
Real photons should be understood in contrast to virtual photons in
dilepton production. Like dileptons they give the electromagnetic
response to the plasma evolution.  They could provide clean signals in
the sense that they escape from different stages of the system with
rather small interaction cross sections.  According to their
production conditions, the three major contributions to real photons
are {\it direct photons} from partonic processes in the initial state,
{\it thermal photons} from both phases and {\it decay photons}
produced at a late stage of the evolution. At large transverse momenta
$K_T\geq$ 3~GeV, the lowest order ($\alpha_s$) contributions to direct
photons are the gluon-to-photon compton scattering off (anti)quarks
$q(\bar q)g\to q(\bar q) \gamma$, quark-antiquark annihilation to
photon and gluon processes $q\bar q\to\gamma g$ and bremsstrahlung
(with quark fragmentation into the photon). At very high collision
energies, the main contribution to direct photons comes from the
fragmentation of jets (Gupta, 1990). At low transverse momenta
$K_T\leq$ 3~GeV, decays of neutral mesons $\pi^0,\eta, \eta'$ and
$\omega$ give a large contribution. This contribution is particularly
difficult to estimate, but will be measured (Stachel, 1993).  The
uncertainty in the estimate arises from the lack of knowledge of the
$p_T$-distributions of these mesons in nuclear collisions. For very
small transverse momenta ($K_T\leq10-20$~MeV), hadronic bremsstrahlung
is important (Ruuskanen, 1992).

A possible window for observing thermal photons could be in the
2-3~GeV region (Ruuskanen, 1992). Thermal photons can be produced in
both phases. In earlier calculations, only the plasma contribution has
been considered (see e.g. Hwa and Kajantie, 1985). Thermal photons in
the plasma phase are produced via the same elementary processes as
direct photons, the difference in the final thermal rate as compared
to the direct production rate is a replacement of the product of
structure functions by the entropy squared. The contribution from the
hadronic phase has been calculated by Kapusta, Lichard and Seibert
(1991, 1992) with the result that ``a hadron gas shines as brightly as
a quark gluon plasma'' at temperatures between 150-200~MeV.

In the 1~GeV region, thermal rates are clearly below decay photons.
This does not necessarily imply that thermal rates cannot be measured.
An electromagnetic calorimeter with high energy resolution and fine
granularity has been proposed for RHIC, which aims for measuring
ratios of $(\#\hbox{thermal photons})/(\#\pi_0)$ as function of
$p_T\sim 3$-5\% (Stachel, 1993). In view of section \ref{sec:V.D.3}
we anticipate that an accurate measurement of this ratio is rather
appreciable for verifying certain remnants of a hypothetical second
order chiral transition. At large transverse momenta ($\geq3$~GeV),
direct photons dominate. The chances for thermal photons improve for
very large multiplicities $dN_\pi/dy$ of pions, because thermal
photons are proportional to $(dN_\pi/dy)^2$, while photons from pion
decay vary with $(dN_\pi/dy)$. In the intermediate region, it
sensitively depends on the shape of the hadronic $p_T$-spectra whether
thermal photons have a chance of being identified or not. These
$p_T$-spectra are not well enough understood.  Real photons have
certain advantages over dileptons. Prompt real photons originate from
the Compton scattering of gluons and quarks. Gluons equilibrate very
rapidly compared to quarks. Therefore the prompt photon measurement
provides a direct (and the only direct) access to the very early
initial phase of the collision which is not accessible with dilepton
rates. The signal to noise ratio is expected to be larger than for the
dilepton continuum.

Thus, it seems to us promising in future experiments with large
$dN_\pi/dy$ and high resolution detectors to look for signatures of a
transient plasma and even for certain substructures in thermal photon
distributions which could be conclusive in view of the phase
transition dynamics. The procedure to calculate the thermal rate of
real photons is the same as for dileptons. Once the elementary
processes are identified and emission rates are calculated in the
stationary plasma or hadronic phase, the result has to be folded with
the spacetime expansion.  In a first approximation one can use
Bjorken's similarity flow and correct for transverse flow, nuclear
geometry in further steps. For such calculations we refer to Ruuskanen
(1992) and references therein.  High multiplicities ($\geq4000$) and
short equilibration times favor thermal photon production. We have to
wait and see whether RHIC or LHC energies do reach such
multiplicities.

\subsubsection{Strangeness production}
\label{sec:V.C.3}
In heavy-ion collisions an enhancement of certain particle ratios
involving strange quarks is predicted compared to ratios in hadron
hadron or hadron nucleus collisions. Ratios enhanced by a factor of 2
for $K^+/\pi^+$ have in fact been observed in experiments at the AGS
at Brookhaven and SPS at CERN. A theoretical explanation in terms of a
transient quark gluon plasma is not compulsory. Rescattering processes
in a hot hadronic gas can also change the ratios in the same
direction.

The idea to propose strangeness as a possible signature for a QGP
relies on the following arguments. Strangeness production in the
plasma should be facilitated by two reasons.
\begin{itemize}
\item Independently of the assumed baryon density, the threshold
  energies for strangeness production in the plasma phase are much
  lower than the production of strange baryons or mesons in the hadron
  phase.  In the hadron phase, strange mesons or baryons are typically
  made in a collision of two non-strange hadrons. The reaction with
  the lowest threshold energy requires already 671~MeV
  $(p+n\to\Lambda^0+K^++n)$.\hfill\break In the plasma the threshold
  for $s\bar s$ production is equal to the rest mass of $s\bar s\sim$
  300~MeV.  For a given temperature the density of non-interacting
  strange quarks is higher than the density of non-interacting kaons
  which will be later considered as an example. In a plasma an
  $s$-quark has 2 spin and 3 color degrees of freedom. If it is bound
  in a $K^-$ or a $\bar K^0$ in the hadron phase it has only 2. Also
  the kaon mass is $\sim$ 494~MeV compared to the current $s$-quark
  mass of 180 $\pm$ 30~MeV.
\item The second reason applies to a baryon rich environment, which
  can be found in collisions with high baryon stopping or in the
  rapidity range of fragmentation regions.\hfill\break The Pauli
  principle will prohibit the creation of $u\bar u$ and $d\bar d$
  pairs compared to strange pairs. The light quarks have to supply the
  large Fermi energy represented by the chemical potential $\mu_B$,
  while the $s$-quarks are only suppressed by their finite mass.
  Moreover $\bar u$ and $\bar d$ quarks have a high probability to
  recombine with $u$ and $d$ quarks to form gluons. For strange quarks
  the recombination to gluons is less likely. Due to the volume
  expansion and the decreasing temperature of the fireball, the
  process $g g{\rightarrow\atop\leftarrow} s\bar s$ is soon out of
  thermal equilibrium, that means, it is too slow to proceed in the
  inverse direction. In a background which is rich in $u$ and $d$
  quarks (large $\mu$), the number of (massless) $\bar u$ (or $\bar
  d$) quarks in a free gas is given by
  \begin{equation}
    n(\bar u)=6\int{{d^3 p}\over{(2\pi)^3}}{1\over{\exp\lbrace(p+\mu
        )/T\rbrace+1}}.
    \label{5.61}
  \end{equation}
  This number should be compared to the number of strange quarks
  \begin{equation}
    n(s)=6\int{{d^3 p}\over{(2\pi)^3}}{1\over{\exp\lbrace(p^2
        +m^2_s)^{1/2}/T\rbrace+1}}.
    \label{5.62}
  \end{equation}
\end{itemize}

To lowest order in perturbative QCD, $s\bar s$-pairs are created
in collisions of two gluons and by annihilation of light quark and
antiquark pairs (Rafelski  and  M\"uller, 1982). Rafelski
and M\"uller  have shown that the gluonic production is the dominating
mechanism of strangeness production. For a small window of the quark
mass equal to  (0.5-1.) times the temperature the equilibration time is
similar to the expected lifetime of the plasma phase.

For dileptons it has been sufficient (cf. the previous section
\ref{sec:V.C.2}) to calculate the differential rate per given
spacetime volume. Due to the very small electromagnetic cross section
the dileptons of high invariant mass escape from the plasma without
further interaction.

Strange quarks are kept in the subsequent evolution. That is the
essential difference to dilepton production. In various ways strange
quarks are incorporated as hadronic constituents during the conversion
of the plasma into the hadron phase. After completion of the
conversion strangeness is confined to hadrons.  The strange hadrons
continue to react via strangeness creation, annihilation or exchange
reactions until freeze-out. Thus it is not sufficient to calculate the
formation rate of strangeness per unit time and volume as if the
predicted rate would directly correspond to observed multiplicities
(as for dilepton production).

The change $dn/dt$ in the density $n$ of strange particles as a
function of time is of primary interest now. Later we shall see that
in the case of a static plasma $dn/dt$ is equal to the rate $R$ if $n$
vanishes, that is at the start of the production. The change goes to
zero when the equilibrium value of $n(n_{eq})$ has been reached. It is
a question of time whether $n_{eq}$ can be reached at all. The two
competing time scales are the equilibration time $\tau_{eq}$ defined
as $n(\tau_{eq})=n_{eq}$ and the lifespan of the phase where the
strangeness production is considered.

Already at this point one can suspect that the appropriate ``choice''
of the dynamical scenario for the phase transition is even more
essential here than it was in the case of dilepton production. Time
scale estimates have farther reaching consequences in strangeness
production.  The contribution of a thermal rate is not only a question
of size in the sense that it could be certainly identified if it is
only large enough compared to other backgrounds. Certain rare
multi-strange hadron species may completely fail as candidates for QGP
signatures, if their equilibration time is just too long compared to
the lifetime of the phase where they could have been produced
otherwise.

Below we will describe  the differential equation which determines
the equilibration time for a given species in more detail.

We will not judge the chances for using strangeness enhancement as a
signature for a transient plasma phase. Also we do not see a way of
specifying features which could be conclusive about the underlying
phase transition dynamics. We merely want to give a flavor of the
complexity in calculating equilibration times. The ratio of
$K^-/\pi^-$ serves only as an example. We describe the work of Kapusta
and Mekjan (1986) and indicate places for further extensions or
alternative treatments.

The preconditions which are taken for granted are the following:
The rate constant $R$ for producing strange quarks in the plasma phase
is so large that the density quickly attains its chemical equilibrium
value $n_s^{eq}$. The system expands and cools so rapidly that the
strangeness abundance - once it has been produced - is not lost to a
significant extent due to annihilation into gluons. Both conditions
ensure that the facilitated strangeness production in a plasma
environment has a chance to become manifest in possible plasma
signals.

To get numbers for particle ratios in the very end, we have to specify
\begin{itemize}
\item the kinematical approach and the geometry. Bjorken's ansatz of a
  longitudinal hydrodynamic scaling expansion is used.  Possible
  effects due to viscosity, heat conduction and transverse expansion
  are neglected. Viscosity and heat conduction would prolong the
  decrease of temperature as a function of time, transverse expansion
  will accelerate it. Both effects may or may not approximately
  cancel.
\item the rapidity range. The central rapidity range has been chosen,
  where the baryon number approximately vanishes. Hence the
  calculations are performed at $\mu_B=0$. This may not be a favorable
  choice in view of the different sensitivities of $\bar u,\bar d$ and
  $s,\bar s$ quarks with respect to $\mu_B$, but it simplifies the
  discussion a lot.
\item the equation of state. Once again the bag model equation of
  state is taken.
\item the gross features of the phase transition. Corresponding to the
  bag model the phase transition is assumed to be of first order.  Two
  realizations are considered: a smooth transition via the Maxwell
  construction and a rapid transition with supercooling and subsequent
  reheating.
\item the measure for the strangeness contents of each phase.  For the
  hadron phase $K^-$-mesons will be considered.
\item the elementary reactions entering the cross sections in the rate
  formulas.
\item an ansatz for the differential rate equation that determines the
  equilibration time.
\end{itemize}

Let us start from the end and describe the rate equation as it was
used in (Kapusta and Mekjian, 1986). For an expanding plasma the
change in the density of strange quarks $n_s$ as a function of time
$t$ is given by
\begin{equation}
  {{dn_s(t)}\over{dt}}=R_p(T(t))\left\lbrace 1-\left\lbrack{{n_s(t)}\over
        {n_s^{eq}(T(t))}}\right\rbrack^2\right\rbrace-{{n_s(t)}\over t} .
  \label{5.63}
\end{equation}
Here $n_s^{eq}$ is the equilibrium density of strange quarks which
depends on time via the temperature $T$. The rate $R_p$ is defined as
the number of specific reactions per unit time and volume producing
$s\bar s$-pairs in the plasma phase. It corresponds to $\int d^4p \ 
dN/(d^4xd^4p)$ which entered also the dilepton rate in section
\ref{sec:V.C.2}.  Similarly to Eq.\ (\ref{5.48}) the rate is
calculated in the independent particle limit, that is according to the
ansatz
\begin{eqnarray}
  R(a_1+&&a_2\to X)={1\over{1+\delta_{a_1,a_2}}}\int{{d^3 k_1}\over
    {(2\pi)^3}}f(\vec k_1) \nonumber\\
  &&\int{{d^3k_2}\over{(2\pi)^3}}f(\vec k_2)
  \cdot\sigma(a_1+a_2\to X;\vec k_1,\vec k_2)\cdot v_{rel}~.
  \label{5.64}
\end{eqnarray}
The occupation probabilities $f(\vec k)$ are chosen as Boltzmann weights
\begin{equation}
  f(\vec k)=\exp\lbrace-(\sqrt{k^2+m^2_a})/T\rbrace~,
  \label{5.65}
\end{equation}
$m_a$ is the degenerate mass of particle $a_1$ and $a_2$. This ansatz
neglects possible final state interactions due to the influence of
Pauli blocking. Here it is consistent with considering a baryon free
rapidity region. The prefactor $(1/1+\delta)$ prevents a double
counting when gluon pairs annihilate to $s\bar s$-pairs, $v_{rel}$
denotes the relative velocity between the particles $a_1$ and $a_2$
and is given by the analogous formula to Eq.\ (\ref{5.50}). Finally
$\sigma$ denotes the elementary cross sections which contribute to $R$
in a given phase.  Note also that the spacetime dependence enters $f$
via the temperature $T$.

Some further comments on the rate equation (\ref{5.63}) are in order.
The last term in Eq.\ (\ref{5.63}) is a dilution term due to the
volume expansion. The particle density will roughly decrease like the
entropy density, that is as $1/t$ when the volume increases
proportional to $t$. The first two terms of Eq.\ (\ref{5.63})
characterize the rate equation for a non-expanding source.
Restricting the r.h.s. of Eq.\ (\ref{5.63}) to the static part, the
linearized rate equation implies for small deviations from equilibrium
\begin{equation}
  \tau_{eq}=n_{eq}/(2R)~,
  \label{5.66}
\end{equation}
i.e., a high rate $R$ implies a short equilibration time.

The cooling law as a function of time depends on the chosen scenario. In
the plasma phase the  bag model equation of state implies a decrease
according to
\begin{equation}
  T(t)=T_0(t_0/t)^{1/3}~.
  \label{5.67}
\end{equation}
Further quantities which enter the rate equation (\ref{5.63}) are the
equilibrium density $n_s^{eq}$ of strange particles and the rate
$R_p(gg\ \hbox{or} q\bar q\to s\bar s)$ for producing strange quarks
in the plasma phase. The density $n_s^{eq}$ is approximated by its
Maxwell-Boltzmann limit. The rate $R_p$ is taken from (M\"uller and
Rafelski, 1982), it is based on the lowest order QCD cross section,
where $s\bar s$-pairs are created via $gg\to s\bar s$, $u\bar u\to
s\bar s$ and $d\bar d\to s\bar s$. (It remains to be checked that
suppressions due to higher orders in $\alpha_s$ are not compensated by
large combinatorial factors in the graphical expansion.) The rate
$R_p$ depends on $\alpha_s$ and $m_s$ as parameters, which both vary
with the renormalization scheme, the energy scale, and the
temperature.

Finally for solving the rate equation for $n_s(t)$ in the plasma
phase, one has to specify the initial condition. Since the
net-strangeness of two cold nuclei is zero, it is chosen as
$n_s(t_0)=0$, where $t_0$ denotes the formation time.

The dependence on this initial condition should be weak, if the final
rate is close to its equilibrium value, otherwise it has to be checked
carefully. The conserved charge is the difference of strange and
antistrange particles. Only as long as strange and antistrange
particles do not interact, their charges are conserved separately.
Thus, if the $s$-quark density represented by the $K^-$-mesons is
considered separately in (Kapusta and Mekjian, 1986), it relies on the
treatment of quarks and mesons as effectively ideal gases in both
phases. It is implemented by the choice for the equation of state.  As
lattice results have shown, the ideal gas picture is certainly not
correct in the vicinity of $T_c$, in particular it fails if long range
correlations are enforced by a second order transition. We stress this
point, because in the presence of interactions between $s$ and $\bar
s$- quarks it is not clear what a sensible measure for the strangeness
contents of each phase could look like.

Next we turn to the rate equations in the intermediate phase between
the onset of hadronization and its completion. The terms in the rate
equation depend on the scenario. Kapusta and Mekjian (1986) have
considered two alternatives, the Maxwell scenario and a phase
conversion with superheating and supercooling effects.

In the first case, the transition proceeds adiabatically. The system
stays in local equilibrium, Maxwell's condition for coexisting phases
is assumed to be fulfilled. The nucleation rate of hadronic bubbles
is large compared to the expansion such that a fraction $(1-f(t))$
of the hadron phase is created immediately after  the temperature
has reached its critical value $T_c$. This happens at time $t_1$
\begin{equation}
  t_1=(T_0/T_c)^3 t_0 .
  \label{5.68}
\end{equation}
Eq.\ (\ref{5.68}) follows from entropy conservation and the equation
of state (cf. section \ref{sec:V.C.2} where it has been derived).
Thereafter the temperature stays constant until the conversion to the
hadronic phase is completed. Both phases contribute to the entropy
density according to
\begin{equation}
  s(t)=f(t)s_p(T_c)+(1-f(t))s_h(T_c) ,
  \label{5.69}
\end{equation}
where $s_p$ and $s_h$ are the entropy densities in the plasma and the
hadron phase, respectively.  Combined with the time dependence of
$s(t)$ it implies the time dependence of the fraction $f(t)$
\begin{equation}
  f(t)={{37}\over{34}}(T_0/T_c)^3{{t_0}\over t}-{3\over{34}} .
  \label{5.70}
\end{equation}
All plasma  is converted to the hadron phase at time $t_2$, where
$t_2$ follows from $f(t_2)=0$ as
\begin{equation}
  t_2={{37}\over 3} t_1 .
  \label{5.71}
\end{equation}

The second scenario would be realized, if an interface between the
hadron and the plasma phase costs much excessive energy, i.e. if the
surface tension between the plasma and hadronic bubbles is large. (As
we have shown in section \ref{sec:III}, this is not likely the case.)
The onset of the phase transition is delayed until the temperature has
dropped to a value $T_p<T_c$. At $T_p$ the system is assumed to
transform instantaneously to the hadron phase. The hadron phase is
superheated to a value $T_h>T_c$, while the plasma was supercooled to
$T_p<T_c$.  The entropy should not decrease (during the transition it
may increase), the energy is conserved. These assumptions together
with the bag model equation of state lead to a prediction of the time
dependence of $T$ in the various time intervals.

In view of the final strangeness contents the interesting feature
about the time evolution of temperature is that the system spends
quite a long time in the mixed phase or in the overheated hadron
phase, depending on the scenario and the initial temperature $T_0$.

Even the very ansatz for the rate equations depends on the scenario.
The common part is the one which describes the approach to the
equilibrium density of strange particles in a static situation. When
the matter is expanding, loss terms in the density have to be
included, which depend on the presence and absence of a phase mixture.
In particular, a coexistence of phases as it occurs in Maxwell's
construction leads to an additional complication. For the lifespan of
the mixed phase the volume occupied by the hadron phase expands while
the fraction of plasma phase shrinks. This structure has to be
superimposed on the common volume expansion.

In both scenarios the rate changes $dn/dt$ in $K^-$ mesons are
followed over the period between the onset of the hadron phase until
all plasma has been converted. In Maxwell's scenario the temperature
stays constant at $T_c$, in the supercooling/reheating scenario it
drops from $T_h>T_c$ to $T_c$.

Finally, to solve the rate equations for the density of $K^-$-mesons,
we have to determine the rate constants $R_h(T(t))$ and $R_h(T_c)$ in
the {\it hadronic} phase. Here the possible contributions simplify due
to the restriction of a baryon free environment. The relevant
processes are $\pi^++\pi^-\to K^++K^-,\ \pi^-\pi^0\to K^- K^0$ and
$\pi^0\pi^0\to K^+K^-$. The corresponding cross sections must be
inserted which are inferred from experiments.

After all these specifications we are now ready to solve the rate
equations for $n_s(t)$. In heavy-ion collisions densities are not
directly observable as a function of time.  Instead of densities
particle ratios, in our case $K^-/\pi^-$ ratios are suitable
observables. The volume drops out. Thus we have
\begin{equation}
  K^-/\pi^-\propto n_s/n_{\pi^-}.
  \label{5.72}
\end{equation}
Here $n_s$ is the density of strange particles at freeze-out, hidden
to equal fractions in $K^-$ and $\bar K^0$-mesons and $n_{\pi^-}$ is
the density of $\pi^-$-mesons. For massless pions and under the
assumption of an ideal gas, the pion number density is proportional to
the pion entropy density $s\propto T^3$ which is almost equal to the
total entropy density of the hadronic gas. This leads to
\begin{equation}
  K^-/\pi^-\propto n_s/s_h.
  \label{5.73}
\end{equation}
This ratio should be compared to $n_s\hbox{(plasma)}/s_p$, the number
of strange quarks in the plasma phase and the plasma entropy density.
It has been questioned for some time what a good measure for the
strangeness contents of the plasma is. Rather than $n_s/s$ the ratios
$n_s/n_d$ ($n_d$ being the number of down quarks) have been earlier
considered in (Rafelski, 1982). However, the quantity which should
replace $n_{\pi^-}$ in the plasma phase is $s$ and not $n_d$, since
all $u,d$ and gluonic degrees of freedom (``summed up'' in the
entropy) must hadronize into pions.

Plots of $n_s/s$ and $n_s/n_{eq}$ as a function of time can be found
in (Kapusta and Mekjian, 1986) for three scenarios: the Maxwell
construction, the supercooling/reheating scenario without an extra
entropy production $(\Delta S=0)$ and with doubling of the entropy at
$T_c(\Delta S=S)$. The curves display some expected features: the
higher the initial temperature the faster $n_s$ reaches its
equilibrium value. For the Maxwell scenario $n_s/s$ and $n_s/n_{eq}$
are continuous functions of time. After completion of the phase
transition the final ratio of $K^-/\pi^-$ is shown in
Fig.~\ref{fig:30} as function of the initial temperature $T_0$ for
three scenarios
and three sets of input parameters corresponding to the
dashed, solid and dotted curves. Input parameters are the strong
coupling $\alpha_s$, the strange quark mass $m_s$, two cross sections
entering the rates of $K\bar K$-annihilation and the proper formation
time $t_0$ of the plasma. Obviously in the supercooling/reheating
scenario with $\Delta S=0$ or $\Delta S=S$, the $K^-/\pi^-$ ratios are
little sensitive to the initial temperature $T_0$, and the input
parameters, even to the degree of supercooling and the related entropy
production. The value falls in the range of 0.21-0.25. The value of
chemical equilibrium is in this range, it is given by 0.217.  The
$K^-/\pi^-$ values, predicted here for heavy-ion collisions should be
compared to the same ratios in $\bar p p$-collisions where $K^-/\pi^-$
is 0.07 $\pm$ 0.02.

In the Maxwell scenario such an enhancement by a factor about 3 is
only seen for an ``optimistic'' choice of parameters (dashed curves in
Fig.~\ref{fig:30}) for all values of $T_0\geq $200~MeV. For a
``realistic'' choice (solid lines in Fig.~\ref{fig:30}) it is only
seen for $T_0>250$~MeV, and for a ``pessimistic'' set (dotted lines in
Fig.~\ref{fig:30}) only for large values of $T_0$. The physical reason
for the higher sensitivity of the Maxwell scenario to the input
parameters is likely due to the continuous time evolution throughout
all stages of the evolution. No jumps in thermodynamic quantities at
$T_c$ cut off the history of the plasma evolution.

The enhancement of $K^-/\pi^-$ by a factor of 3 compared to $\bar p
p$-collisions at the same beam energy is to be interpreted as an
indirect sign for a transient plasma phase. Without the first order
phase transition the lifetime of the hadron phase would be shorter, so
short that the time would not suffice to saturate the $K^-/\pi^-$
ratio nearly to its equilibrium value.  Thus, in contrast to
dileptons, $K^-$-mesons are not a direct probe of the high temperature
plasma phase, their abundance is influenced by the phase transition
and the subsequent evolution in the hadron phase.

This example for an estimate of the strangeness enhancement may have
illustrated the urgency of calculating the equation of state, the
latent heat, the surface tension, the temperature dependence of
couplings and masses and the order of the phase transition from first
principles. Theoretical calculations may help to find the most
realistic scenario. In particular we have seen the simplifying power
of the bag model equation of state which does not justify its use. For
strong correlations between particles close to $T_c$, the very
indicator itself may fail as a measure for the strangeness contents of
both phases. Also the gain and loss terms in the rate equations which
take care on the dynamical expansion, are sensitive to {\it details}
of the phase transition scenario.

\bigskip The approach of Mekjian and Kapusta (1986) can be generalized
in various aspects, which we mention rather briefly below. For our
limited aspect of specific signatures for the phase transition a
detailed discussion of a more general framework would only serve to
further support the overall impression: it is extremely difficult to
unfold the underlying phase transition dynamics from strangeness
production.

Koch et al. (1986b) have considered the strangeness production in the
{\it plasma} phase within a framework, where the baryochemical
potential is different from zero. The rate equation is considered in
an alternative form including a linear loss term in $n_s/n_s^{eq}$.
The volume expansion is investigated for different time dependences of
$V$ and $T$, one following from the bag model equation of state and
one chosen by hand to simulate a more rapid expansion. In addition,
Koch et al. have checked the influence of Pauli blocking on final
states by replacing the momentum distributions of strange quarks in a
suitable way.

A more general set of rate equations in the {\it hadronic} phase has
been considered in (Koch et al., 1986a). The underlying physical
picture is similar to the Maxwell scenario of Kapusta and Mekjian
(1986). It is more general in including baryons and antibaryons in
addition to pions and kaons. A wider spectrum of possible reactions in
the hadron phase has to be taken into account: strangeness production
and exchange reactions, annihilation processes. The abundances not
only of $\pi$ and $K$-mesons, but also of $N,Y,\Xi,\Omega$ and their
antiparticles have to be followed in their evolution. An additional
term appears in the rate equation for the mixed phase which is
specific for the chosen hadronization process (here the so called
fragmentation-recombination scenario). Needless to say that the way in
which the strangeness abundance of the plasma is modified during the
phase conversion depends on the choice of the hadronization process.
Unfortunately there is yet a whole variety of fragmentation models at
one's disposal.

We conclude with a short and incomplete confrontation to the
experimental status. The ratios of $K^+/\pi^+$ and $K^-/\pi^-$ have
been measured in the E~802 experiment (Abbott et al., 1990; Abbott et
al., 1991) at the AGS (Brookhaven) and the NA34 experiment at SPS
(CERN) (Van Hecke, 1991). Both experiments are measuring in a baryon
rich region of phase space, thus we cannot compare the results with
the theoretical predictions of (Kapusta and Mekjian, 1986) for a
baryon free environment. In both experiments the same trends are seen:
an enhancement of $K^+/K^-$ ratios by roughly a factor of 2 when going
from hadron hadron to hadron nucleus and nucleus nucleus collisions,
whereas the $K^-/\pi^-$ ratio increases less strongly or stays
approximately constant. It should be mentioned that the ratios
$K^+/\pi^+$ and $K^-/\pi^-$ are not the only measure for strangeness
enhancement. Another quantification is the strangeness suppression
factor $\lambda=2 s\bar s/(u\bar u+d \bar d)$ as evaluated by
Wroblewski (1985) for pp-collisions and applied to S+S-collisions at
200~GeV/n-n (Seyboth, 1992; Foley, 1992). With this criterion one
finds a factor 2 of strangeness enhancement as compared to pp- and
pA-collisions.

This does not mean that a plasma has been observed. For many strange
particles the equilibrium values in a hot extended hadron phase are
similar to those in a quark gluon plasma. Therefore the enhanced
particle ratios do not uniquely signal a transient plasma. If there is
another way to get strangeness into equilibrium by adjusting
parameters in a hadronic cascade model, one will never be able to
distinguish this explanation from a transient plasma. A system in
equilibrium has lost the memory how it got there. There may be
exceptions for finding a unique signal of a plasma (Rafelski, 1991).
Enhanced multistrange (anti)baryon production (antihyperons
$\bar\Lambda,\bar\Xi^-$) have been proposed as special tests for a
plasma environment (Eggers and Rafelski, 1991). A hot hadron gas
cannot overcome their large mass barriers.

\bigskip In Table~\ref{tab:8} we recapitulate the ingredients, which
have led to the prediction of $K^-/\pi^-$ enhancement in the baryon
free case. Some freedom of choice is given for the geometry, the
rapidity range, and the initial conditions.  This freedom corresponds
to the possibility of experimental tuning of certain parameters like
the initial energy density. In contrast to that one is truly not
allowed to choose between various versions for the kinematical
approach, the equation of state, the gross features of the transition,
the rate formula, the contributing elementary processes and the ansatz
for the rate formula or the differential rate equation.  These
ambiguities are due to uncertainties in the theoretical description
and shed some light on the reliability of predicted numbers in the
very end.  For example the choice of a strong supercooling scenario
must be regarded as ad hoc.

\subsubsection{Pion interferometry}
\label{sec:V.C.4}
Interferometry is nowadays a well-known technique. It has been
developed by Han\-bu\-ry-Brown and Twiss (1954) almost 40 years ago in
astrophysics, where it is a tool to measure the size of various
stellar objects in the visible and radio frequency ranges.
Fig.~\ref{fig:31} shows a source which is assumed to emit identical
particles from positions $P_1$ and $P_2$. The particles are later
observed at positions $P_3$ and $P_4$. Both emission points may
contribute to both observation points, even if the particles are
non-interacting, but have small relative momenta. The reason is the
symmetrization (antisymmetrization) of the quantum-mechanical wave
function in the case of bosons (fermions). A correlation function is
constructed from the number of counts at $P_3$ and $P_4$. The
particles which are detected in astrophysical interferometry
experiments are photons.

The overwhelming part of particles which are radiated in heavy-ion 
collisions are pions. Thus we describe pion interferometry 
as a particularly useful tool to estimate the lifetime and 
the final-state size of the source which radiates the 
pions. From the discussion in the previous sections it is 
evident that we need some measurements delivering 
information about the equation of state. Production rates 
of dileptons and strange particles strongly depend on the 
duration of a certain phase. Three extensive quantities 
must be measured to obtain an equation of state for a 
static gas with no conserved charge. In heavy-ion 
collisions the total energy and entropy can always be 
estimated from experiment (cf. section \ref{sec:V.C.1}). If we know 
the volume and the fraction of collective energy at a given 
time of the collision, we have a single value in an 
equation of state corresponding to the conditions at that 
time. The correlation function measured in pion 
interferometry could give an estimate for the size at 
freeze-out. The main reason why we are reporting on this 
experimental tool is the possibility to measure the 
prolongation of lifetime due to a phase transition.

Let us see how we can infer the spacetime structure of the 
source from correlations in momentum space. The 
correlations refer to two-particle correlations in the 
pion's momenta distributions. They are defined as
\begin{equation}
  C(\vec p,\vec q)=P(\vec p,\vec q)/(P(\vec p)P(\vec q))\;,
  \label{5.74}
\end{equation}
where $P(\vec p,\vec q)$ denotes the probability to find two pions with 
three-momenta $\vec q$ and $\vec p$ in the same event. In terms of 
rates, $C(\vec p,\vec q)$ is given as
\begin{equation}
  C(\vec p,\vec 
  q)={{d^6N}\over{d^3pd^3q}}/\left({{d^3N}\over{d^3 p}}
    {{d^3N}\over{d^3 q}}\right)\;.
  \label{5.75}
\end{equation}
Thus $C$ is measurable as a ratio of two-pion to one-pion 
inclusive yields. Its width is a measure for the inverse 
source size. The weaker the correlation the smaller its 
width, the larger is the source which emits the pions.

Two-pion correlations can arise from different origins. The 
only one we will discuss in some more detail are 
Bose-Einstein correlations due to the quantum mechanical 
symmetrization of the outgoing wave functions of identical 
bosons. In the same way electrons in a metal are 
anticorrelated in their spatial distribution due to the 
fact that they occupy only a finite volume in momentum 
space.

Other causes of correlations due to final-state 
interactions are hadronic and Coulomb interactions. Coulomb 
interactions lead to positive correlations between 
particles of opposite charge if the relative momentum is 
small. It has to be checked how far their contribution is 
averaged out due to a comparable number of pairs with equal 
and opposite charges.

A third source for correlations comes from resonance 
decays. Pions are detected that are not directly emitted 
from the hadronic gas at freeze-out,  but result from 
decays of heavier mesons. Such pions may contaminate the
information about the source size.

Bose-Einstein correlations are the only origin for 
correlated emission from the radiating source if the 
source is completely chaotic. If we follow the evolution of 
the plasma through the phase transition to the final state 
of a hadron gas, it consists to $\sim$90\% of pions in 
the baryon free region at freeze-out, where the pion gas is 
already rather dilute. (Present experiments deal with equal 
nucleon and pion numbers (AGS) and a ratio of 1:7 (SPS) for 
S+S, 1:5 (SPS) for S+Ag, the baryon free region in an 
idealization up to now). The assumption of an incoherent 
source for a dilute pion gas seems to be a good 
approximation to start with.

Therefore we focus on the case of a chaotic pion source. 
The probability $P(\vec p,\vec q)$ of measuring two pions 
with momenta $\vec p$ and $\vec q$ in the same event is 
given as (Pratt, 1984)
\begin{mathletters}
  \label{5.76}
  \begin{eqnarray}
    P(\vec p,\vec q)=P(\vec p)P(\vec q)+
    \int && d^4xd^4y\ g(\vec K,x)g(\vec K,y) \nonumber\\
    &&\cdot \exp\{ik(x-y)\}\;,
    \label{5.76a}
  \end{eqnarray}
  where
  \begin{equation}
    P(\vec p)=\int d^4x\ g(\vec p,x)\;,
    \label{5.76b}
  \end{equation}
  furthermore
  \begin{equation}
    \vec K=(\vec p+\vec q)/2
    \label{5.76c}
  \end{equation}
  is the average momentum of the pion pair and
  \begin{equation}
    k=p-q
    \label{5.76d}
  \end{equation}
\end{mathletters}
denotes the relative four-momentum. The function $g(\vec K,x)$ gives
the probability of emitting a pion of momentum $\vec K$ from a
spacetime point $x,\ g(\vec K,x)$ is called the emission function.
Thus, in the simplest case, where the source is incoherent and only
Bose-Einstein correlations relate the wave functions of emitted
particles, the correlation function essentially measures the
expectation value of $\cos(k r)$ weighted by the product of emission
functions, $r=x-y$ is the relative distance of the pions at their
source). Stated differently, the measurable correlation in momentum
space is the Fourier transform of the emission functions. (A general
ansatz for the two-particle correlation function due to final state
interactions in rapidity space can be written in a similar form.) The
probability $P(\vec p,\vec q)$ depends on six degrees of freedom.
Under a further constraint on the spacetime evolution of the source or
on its symmetry properties, it may be possible to determine the
emission function $g$ from a measurement of $P$, $g$ depends
originally on seven degrees of freedom. The final spacetime
distribution described by the product of $g()$'s is a result of the
spacetime evolution of the plasma passing a possible phase transition
to the hadronic phase.  The spatial distributions of the pions at the
moment of freeze-out constitute the radiating pion source we are
interested in. Hence the ansatz for $g(\vec K,x)$ depends on two
ingredients: the {\it equation of state} and the {\it dynamical
  concept} relating the spacetime dependence to thermodynamic
quantities.

Let us recall why a phase transition is expected to delay the
break-up. The moment of freeze-out or break-up $\tau_B$ is determined
by the degree of dilution.  Equivalently, it is given by the value of
its entropy density $s(\tau_B)$. For fixed initial energy density and
equilibration time $\tau_0$, the initial entropy density for a plasma
is considerably higher than for a hot pion gas to start with. The
total entropy is assumed to be approximately conserved during the
expansion. Thus it takes longer to reach the same low entropy density
at freeze-out if one starts in the plasma phase. A phase transition
must have occurred at an intermediate stage.

In Fig.~\ref{fig:32} (from Bertsch, 1989) we visualize how the
prolongated lifetime is expected to influence the effective spatial
dimension of the source. The beam axis is chosen in $z$-direction and
labelled $x_L$. The two transverse directions are $x_{sidewards}(x_s)$
and $x_{outwards}(x_o)$, $\vec x_o$ points in the direction where the
detector is placed. This coincides only with the direction of the main
pion stream if the pions are viewed in a frame in which they are
emitted perpendicular to the cylinder. Fig.~\ref{fig:32} shows a
snapshot of the pion sources at freeze-out for both scenarios: a
plasma at the beginning or a hot and dense pion gas. In the latter
case, the high density of pions in the formation zone leads to a
rescattering. Thus the transverse dimensions at freeze-out would
clearly exceed the size of the formation zone if the formation zone
were at rest. The longitudinal expansion of the source modifies the
picture in the following way. The collective flow of the pions singles
out a preferred direction. Recall that in interferometry measurements
pions are only detected which have almost the same momentum. For a
moving source the probability to detect such pairs in a certain
direction is reduced. Only pions which come from the same side of the
cylinder as the detector have a chance to arrive in the detector with
nearly the same momenta.  This reduces the transverse dimensions of
the pion cloud considerably in spite of the rescattering effects
(Pratt, 1984; Bertsch, 1989).

A {\it first-order phase transition} which proceeds via nucleation of
the plasma to the hadronic phase will alter the picture. The plasma
lives only a very short time, thus its transverse expansion may be
neglected. In the intermediate phase, a possible scenario is that the
plasma breaks up into droplets of hadrons. The conversion to the
hadronic phase is slow, because the rearrangement of degrees of
freedom takes time. The pions are emitted over a longer period at a
smaller rate, hence they have less chance to rescatter. The result is
a {\it small sideward} and a {\it large outward-dimension} of the
cloud at the moment of last interactions. This is illustrated in
Fig.~\ref{fig:32}. The hope to identify a (strong) first order phase
transition via pion interferometry is essentially based on this
picture: to measure a large source size in the outward direction at
freeze-out.

Pratt (1986) has derived a functional dependence of 
the emission function $g()$ for two equations of state (the 
bag model equation of state to simulate a phase transition 
and an ultrarelativistic pion gas without a phase 
transition). As dynamical concept he has used the 
hydrodynamical equations for two geometries, a spherically 
symmetric exploding plasma and Bjorken's scaling solution. 
For Bjorken's scenario, an analytic expression can be 
obtained, for the spherically exploding plasma the 
hydrodynamic equations have been solved for different 
initial conditions. The change in the dynamics which is 
caused by a phase transition is more pronounced the longer 
the mixed phase lasts where the pressure stays constant 
while the energy is absorbed in latent heat. Qualitatively 
the following features can be seen in a plot of $C(K,k)$ as 
a function of $k$, the relative momentum. Higher average 
momentum $K$ leads to a broader correlation corresponding 
to a smaller effective source size. This reflects the 
collective expansion of the system. One effect of a phase 
transition is a reduction of the explosive velocity. The 
other effect is only apparent in $C$ when the relative 
momentum is chosen parallel to the average momentum. For 
larger values of $K$, the correlations drop to smaller 
values indicating a larger size or a longer lifetime of the 
source. The strength of the signals depend, however, on the 
initial conditions which have been chosen for the 
spherically exploding fireball. They may fade away under 
realistic conditions. In order to get a hint on a transient 
phase transition in a cylinder geometry, the orientation of 
the relative momentum $k$ with regard to the average 
momentum $K$ has to be chosen in an appropriate way. The 
large outward dimension shows up only in correlations with 
$\vec k$ parallel to $\vec K_{per}$ ($\vec K_{per}$ being 
the transverse component of the average momentum).

\vskip12pt
The difference between sideward and outward dimensions in a 
cylinder geometry has been numerically investigated by 
Bertsch et al. (1988). This calculation is based on a 
cascade description of the hadronic phase starting at a 
time where the plasma begins to convert to hadronic matter. 
At a first view such a treatment seems to be more 
appropriate for the hadronic phase, especially in the 
period close to freeze-out, where the system is too diluted 
to justify a hydrodynamic approach. On the other hand it 
turns out that a hydrodynamic expansion in the sense of a 
collective transverse and longitudinal expansion velocity 
may be reobtained from a more microscopic starting point, 
if realistic cascade descriptions are used (Stachel, 
1993). In the cascade approach of Bertsch et al.  
individual pions are followed along their classical 
trajectories. The plasma is assumed to break up into 
``blobs'' which emit and reabsorb pions at a certain rate. 
Entropy conservation is no longer an ingredient, its 
production during the phase conversion comes out as a 
result. Also the freeze-out time is predicted in this 
approach. The break-up is described to occur 
instantaneously, i.e. without
supercooling delay. 

Again one has to specify an ansatz for the emission 
function. The classical treatment of pion motions leads to 
an emission function which is too singular to display 
interference effects. To implement interference effects, 
the momentum dependence of the source function is smoothed. 
Bertsch et al. (1988) have derived the following 
parametrization
\begin{equation}
  g(x,\vec K)=f(\vec K_{per})\tau\ 
  e^{-(\tau/\tau_0)^2} 
  e^{-(r_{per}/R)^2}e^{-(y_r 
    -y_{\vec K})^2/y^2_0}\;.
  \label{5.77}
\end{equation}
The spacetime coordinates are $x$, the proper time
$\tau=\sqrt{t^2-z^2},r_{per}$ denotes the transverse position,
$y_r\equiv\tanh^{-1}(z/t)$ is the spacetime rapidity, the average
three-momentum $\vec K$ is written in terms of the transverse
components $\vec K_{per}$ and the momentum rapidity $y_{\vec
  K}\equiv\tanh^{-1} (\vec K_z/E)$. The values for $\tau_0, R$ and
$y_0$ are chosen as $\tau_0=9$~fm/c, $R=3.3$~fm and $y_0=0.76$.
Arguments for this choice can be found in the original reference. The
result for the rate $C(\vec p,\vec q)$ to find a pion pair with
momenta $\vec p$ and $\vec q$ is obtained from Eq.\ (\ref{5.77}) as an
integral over the transverse positions. The integral which is left is
given as
\begin{eqnarray}
  C(\vec p,&&\vec q) = 1 + e^{ - (k_{per} R )^2 / 2} 
  \int d\tau d\tau' dy\ dy' {{4 \tau \tau' } \over {\tau^4_0 \pi 
      y^2_0 }} \nonumber\\
  && \quad\cdot \exp \left\{ - {{ \tau^2 + \tau^{'2} } \over 
      {\tau^2_0} } \right\} \exp \{ - [ ( y_{\vec K} - y )^2
  \nonumber\\
  && + ( y_{\vec K} - y' )^2 ] / y^2_0 \} \nonumber\\
  && \quad\cdot \cos [ ( \vec 
  p_z - \vec q_z ) ( z - z' ) - ( E_{\vec p} - E_{\vec q} ) 
  (t - t' ) ] \;.
  \label{5.78}
\end{eqnarray}
\vskip12pt
Here $k_{per}\equiv|(\vec p-\vec q)_{per}|$ has a sideward and an
outward component. The result is a Gaussian in sideward direction, $C$
measures the transverse size of the source which is the same in
sideward and outward direction as long as the time dependence of the
distribution is neglected. If the transverse relative momentum is
chosen in the outward direction (i.e. $\vec k_{perp\ s}=0,\vec
k_{\perp\ 0}\not=0$), the Gaussian in $\vec k_{perp\ 0}$ is modified
by the $\vec k_{perp\ 0}$-dependence of $E_{\vec p}$ and $E_{\vec q}$,
an effective size in outward direction is sensitive to the time
correlation as it is seen in Eq.\ (\ref{5.78}). The numerical results
for sideward and outward correlations are displayed in
Figs.~\ref{fig:33}(a) and (b).

The reduced correlation for outward (compared to sideward) 
relative momenta reflects the effectively larger  size of 
the source in this direction, as has been anticipated above 
(cf. Fig.~\ref{fig:32}).

Expression (\ref{5.78}) for the pion correlation partly motivates an
ansatz which is sometimes used in real interferometry experiments (see
e.g. Schmidt and Schukraft, 1993). That is a threefold Gaussian in
$\vec k_{perp\;0}$, $\vec k_{perp\;s}$ and $\vec k_L$ (the
longitudinal component of $\vec k$ instead of the rapidity $y_{\vec
  K}$). The possibility is implemented to have different length scales
$R_{T_{out}},R_{T_{side}}$ and $R_{T_L}$ in all three directions, cf.
Eq.\ (\ref{5.79}) below.

\vskip12pt
So far we have reported on pion interferometry as a tool to 
estimate the {\it lifetime} of a pion radiating source. 
Naturally the question arises as to whether it is also 
possible to resolve spatial inhomogeneities of bubble 
structures in the mixed phase. This question has been 
recently addressed by Wieand et al. (1992). The answer is 
in principle positive. Pion interferometry may be a tool to 
resolve the grain structure of the hadronic phase. Clearly 
the bubble structure should not be too fine-grained, 
otherwise it is not resolvable for the detectors. In 
practice it does not seem to be a realistic possibility. 
The theoretical predictions have not yet reached a stage 
where it is clear what should be verified experimentally.

\vskip12pt\noindent {\it The status of experiments.}
\vskip12pt\noindent A few aspects of the experimental status of
interferometry measurements are described which are restricted to
applications in heavy-ion experiments. The primary observables are the
double-pion and single-pion inclusive distribution functions
$N(p_1,p_2)$ and $N(p)$ which determine the correlation $C$ according
to Eq.\ (\ref{5.74}). In Bjorken's geometry, $C$ depends on five
independent variables, the transverse and longitudinal components of
the average momenta $K_{transv}$, $K_L$ or the rapidity $Y$, and the
sideward, outward and longitudinal components of the relative momenta
($k_{per\;s}$,$k_{per\;0}$,$k_L$, respectively). An ansatz which is
frequently used by experimentalists is a simple Gaussian in the
relative momentum of the pion pair. A comparison to the analytic
expression Eq.\ (\ref{5.78}) may indicate the gap between theory and
experiment. The parametrization which has been used in the NA35
experiment at CERN-SPS (Baechler et al., 1992) is a Gaussian in all
three components of the relative momentum
\begin{eqnarray}
  C(\vec k)=&&1+\lambda\cdot\exp\{-k^2_{per\ s}R^2_{per\ s}/2\}
  \nonumber\\
  &&\cdot\exp\{-k^2_{per\ 0}R^2_{per\ 0}/2\}
  \cdot\exp\{-k^2_{L}R^2_{L}/2\}\;,
  \label{5.79}
\end{eqnarray}
where $k_{per\;s}$ is orthogonal to $K_{transv}$ and $K_L$,
$k_{per\;0}$ is parallel to $K_{transv}$ and orthogonal to $K_L$, and
$k_L$ is orthogonal to $K_{transv}$ and parallel to $K_L$.
Furthermore, $\lambda$ is the coherence or {\it chaoticity parameter}
which is assumed to absorb the Coulomb corrections of final state
interactions. It is 1 for an incoherent boson source. The parameters
$R_{per\ s},\ R_{per\ 0}$ and $R_L$ are the observables which are
determined from a fit. Note that the simple Gaussian ansatz of
Eq.\ (\ref{5.79}) describes only {\it spatial} correlations.  The time
correlation is more complicated if a phase transition occurs. The
parameters $R$ include some information on the spacetime evolution of
the system (Ferenc et al., 1992). The data for $R_L$ [fm] as a
function of rapidity $Y$ as seen by an observer at fixed $Y_{obs}=2.5$
are consistent with a longitudinally expanding source. This result is
not trivial. Predictions for a static source are quite different
(Makhlin and Sinyukov, 1988). The longitudinal radius $R_L$ can be
related to the freeze-out or decoupling time $\tau_f$. The relation
between $R_L$ and $\tau_f$ for a scaling expansion is given as
(Makhlin and Sinyukov, 1988)
\begin{equation}
  R_L=(2T_f/<m_T>)^{1/2}\tau_f
  \label{5.80}
\end{equation}
in a comoving frame, $T_f$ is the freeze-out temperature,
$\left\langle m_T\right\rangle$ the average transverse pion mass. From
the measured values of $R_L,\ \tau_f$ is estimated to be
$4.5<\tau_f<6.3$~fm/c for $100<T_f<200$~MeV (Ferenc et al., 1992). The
difference between the outward and sideward extensions $R_{per\ 
  0}-R_{per\ s}$ should be a measure for the duration of pion emission
as was argued above. The source dimension in the outward direction has
two contributions (cf. Fig.~\ref{fig:32}): the effective geometrical
depth (which is comparable to the sideward extension and can be larger
than the geometrical size of the formation zone due to rescattering
effects) and the effective prolongation due to a reduced particle
emission rate in case of a phase transition. In Fig.~\ref{fig:34} from
Ferenc (1992) it is seen that the difference in the transverse
dimensions is negligible.

For a specification of the rapidity intervals, see Ferenc 
(1992). A more careful estimate for the duration of 
particle emission is $\Delta\tau_f\leq2$~fm/c. This rules 
out a long-lived source ($\geq10$~fm/c) as a result of a 
first-order phase transition with supercooling delay. 
Finally, the transverse size at decoupling time 
$R_{\tau_f}$ turns out to be 1.5-2 times larger than the 
geometrical size if the geometrical size is estimated as a 
transverse size of the projected transverse density 
distribution. This is compatible with a high density 
formation zone, from which the pions first rescatter before 
they freeze out. The freeze-out volume becomes independent 
of the formation volume in this case, it merely depends on 
the particle density.

The statistics of the data has still to be improved to check the
reliability of the projections of $R_{per}$ on $R_{per\ s}$ and
$R_{per\ 0}$. One should also keep in mind that the results for the
different length scales could be an artifact of the simplified ansatz
(\ref{5.79}) for $C$. It is unlikely that just one parameter $\lambda$
can absorb the full complexity of different origins of pion
correlations.  Another complication may arise from rescattering with
the surrounding target spectator matter (Schmidt and Schukraft, 1993),
a ``dirt effect'' which we have not mentioned so far. The relative
weights of the contributions to pion correlations of different origins
depend on the impact parameter. One would like to ignore these
effects, unfortunately one is not allowed to do so.

For comparison we also mention results of pion-interferometry
measurements with the AGS-machine at BNL (Stachel, E814 collaboration,
1994; Xu, E814 collaboration, 1994). The data have been measured for
central 14.6~GeV/$n-n$ Si+Al- and Si+Pb-collisions. At freeze-out the
result is the following. In the center-of-mass frame the pions are
emitted from a source with transverse radius $R_{per}=6.7$ fm and
longitudinal radius $R_L=5.0$ fm (on average). The source is nearly
spherically symmetric. If here the transverse radius is compared to
the initial transverse size of the system $R_{per}(Si)=2.9$ fm, one
finds a transverse expansion of the system by a factor of 2.3.
Assuming an expansion velocity of $v/c=0.3$ $(0.2)$, the timescale of
the expansion follows as 10 (15) fm/$c$. The uncertainty in the
expansion velocity is due to resonance decays. The pion-interferometry
results lead to a freeze-out volume of $V=2400$fm$^3$. A kind of
consistency check is provided by the estimate for the volume, if
freeze-out is defined to occur when the average distance from a given
nucleon to the nearest pion is $d=\sqrt{\sigma_{\pi n}/\pi}$, where
$\sigma_{\pi n}$ is the pion-nucleon cross section. For a freeze-out
temperature of 140~MeV and $\sigma_{\pi n}=62$~mb one finds
$V=2750$fm$^3$ at freeze-out in reasonable agreement with the
interferometry estimate.

In these measurements the estimates for the radii are not obtained
from a direct fit of a certain functional form to the data, but by use
of a dynamical model (realized in the event generator RQMD (Sorge et
al., 1989)) with known spacetime characteristics of the source.
Two-particle correlations are constructed from the RQMD-generated
single-particle distributions in a way that the Bose-Einstein effect
is imposed (Xu, E814 collaboration, 1994). The realistic experimental
parameters are imposed as input data in the RQMD-code.

Although there is no indication for a first order transition in these
experiments, the important conclusion is that the hadron gas does have
time enough to be in thermal (and chemical) equilibrium at freeze-out.
This supports one of the basic underlying assumptions for applying
thermodynamic concepts.

\vskip12pt
Signals of a strong first order transition should not be 
missed by pion-interferometry. A prolongation of lifetime 
by an order of magnitude should be visible for a certain 
projection of the relative pair's momentum. Non-appearance  
of such signals is less conclusive. Specific signals for a 
second order transition with a large correlation length 
have not been invented so far.
Experimental groups which are at present dealing with pion, 
kaon and rho-in\-ter\-fe\-ro\-met\-ry are the E802 
collaboration (Abbot et al., 1992), (Akiba et al., 1993), 
the E814 collaboration (see Xu, 1994, and Stachel, 1994) 
at Brookhaven, and the NA35 collaboration (Seyboth, 1992; 
Ferenc, 1992), the NA44 collaboration (Sarabura et al., 
1992), and the WA80 collaboration (Peitzmann, 1993; 
Peitzmann et al., 1992 and 1993a and b) at CERN.
A chance for future high statistics experiments may lie in 
kaon interferometry. The yield of kaons is considerably 
lower than that of pions, but kaons are less plagued by 
contamination due to resonance decays (Ferenc, 1992b).

\subsubsection{Multiplicity fluctuations}
\label{sec:V.C.5}
The possible signatures for a first-order transition 
which we have discussed in the preceding sections are based 
on a prolongated duration of the matter evolution. 
Multiplicity fluctuations rely on additional entropy 
production which might be generated in explosive processes 
during the plasma conversion to hadronic matter. Several 
candidates for hadronization processes have been under 
discussion: nucleation in the bulk, ``boiling by cooling'' 
(van Hove, 1983), or an adiabatic procedure, where the 
system stays in local thermal equilibrium, pressure and 
temperature remain constant during the conversion. These 
scenarios do not lead to strong multiplicity fluctuations.

In this section, we describe alternatives to the adiabatic scenarios.
These are ha\-dro\-ni\-za\-tion processes via {\it deflagrations} or
{\it detonations}. They are compatible with a strong first order phase
transition and a liberation of a large amount of latent heat,
supercooling and/or superheating effects during the transition. One of
the proposed physical pictures is the following (van Hove, 1985).
Energy and entropy densities change drastically over a small
temperature and pressure interval in the vicinity of $T_c$. Therefore
the proposed hadronization mechanism has to provide a fast liberation
of the large entropy and energy amount of the plasma, a large energy
has to flow into the hadron phase in an efficient way. The main part
of hadronization occurs at the phase transition. Hadronization via
evaporation of hadronic bubbles before the transition is negligible in
this picture. Close to the transition temperature, the color fields
(induced by the color charges of the plasma) collapse to flux tubes
creating something like a network of strings (cf. Patel, 1984 a and b,
and section \ref{sec:IV.B.1}). Due to string formation, stretching and
breaking, plasma droplets will be stopped in their longitudinal
expansion or break up in a few more droplets. In the next stage, the
main part of the hadronization is assumed to proceed via ejection of
hadrons through the surface of the plasma blobs with the kinematics
and thermodynamics of a deflagration. Deflagration is said to occur
only at the outer surface, but not in the interior (in the ``bulk'')
of the plasma droplets. The plasma blobs themselves are small in size
(a few fermi at most).

Let us see which constraints from energy momentum and 
positive entropy production are obtained for deflagration 
or detonation scenarios. Again we use a hydrodynamic 
description and follow Cleymans et al. (1986) to sketch the 
derivation of kinematic constraints. Deflagrations start at 
the surface of the plasma, detonations in its interior and 
form hadronic bubbles. In the case of deflagrations the 
front separating both phases moves more slowly inwards than 
the hadrons escape outwards. In the rest frame of quark 
matter the hadrons and the front move in opposite 
directions. In detonations the velocities of the front and 
the hadrons are both directed outwards, but at a different 
speed. Usually the speed of the hadrons is assumed to be 
lower than the speed of the front. In the case that the 
speed of the hadrons is higher, the relative velocity of 
the hadrons with respect to the front is negative. This 
explosion is called an {\it eruption} and is very unlikely 
as we will see.

Let us go to the rest frame of the front. Pictorially, in 
this frame the front ``eats up'' the plasma which wants to 
pass the front. In the case of deflagrations the hadron 
velocity exceeds the velocity of quark matter
leading to
\begin{equation}
  \tanh\theta_h>\tanh\theta_q\;.
  \label{5.81}
\end{equation}
Here $\theta_{h,q}$ are the fluid rapidities in the hadron 
and quark phases, respectively. In usual detonations the 
quark matter is ``eaten up'' faster than the hadronic 
bubbles are formed ($\theta_h<\theta_q$), in eruptions 
quark and hadronic matter velocities have opposite sign in 
the rest frame of the front, cf. Fig. \ref{fig:35}(a)-(c).

Thus the difference between deflagrations and (usual) 
detonations lies in the relative magnitude of their 
velocities in the rest frame of the separating front. To 
derive constraints for pressure, energy and entropy 
densities, we go to the rest frame of the front, where the 
flow of quark and hadronic matter is steady. Derivations 
$\partial/\partial\tau$ vanish. Further, we specialize 
the interface separating both phases to a (1+1)-dimensional 
planar interface. The interface is chosen to be 
infinitesimally thin. Energy momentum conservation 
$\partial_\mu T^{\mu\nu}=0$ across the front leads to
\begin{eqnarray}
  T^{zz}&=const \nonumber\\
  T^{z0}&=const
  \label{5.82}
\end{eqnarray}
in $(t,z)$ coordinates. The energy-momentum tensor is taken 
as that of a perfect fluid (apart from the place of 
discontinuity). The velocity is parametrized in terms of 
the fluid rapidity $\theta$ as usual according to 
$u_0=\cosh\theta$ and $u_z=\sinh\theta$. Continuity of 
$T^{zz}$ and $T^{z0}$ across the front leads to
\begin{eqnarray}
  (\varepsilon_p+p_h)\sinh^2\theta_h+p&=(\varepsilon_q+p_q) 
  \sinh^2\theta_q+p_q \nonumber\\
  (\varepsilon_h+p_h)\sin\theta_h\cos\theta_h&
  =(\varepsilon_q+p_q)\sin\theta_q 
  \cos\theta_q
  \label{5.83}
\end{eqnarray}
for a perfect fluid. Solving for $\tanh^2\theta_{h,q}$ and 
implementing the unequalities for $\theta_h$ and $\theta_q$ 
Eq.~(\ref{5.83}) gives
\begin{equation}
  \varepsilon_q-p_q>\varepsilon_h-p_h\ \qquad \hbox{for deflagrations}
  \label{5.84}
\end{equation}
and
\[
  \varepsilon_q-p_q<\varepsilon_h-p_h\ \qquad \hbox{for 
    detonations}\;.
\]
Imposing the bag model equation of state (section \ref{sec:V.B.2}) 
with $c_s=1/\sqrt3$, this means
\begin{eqnarray}
  \varepsilon_q+2B>\varepsilon_h\ &\qquad\hbox{for 
    deflagrations} \nonumber \\
  \varepsilon_q+2B<\varepsilon_h\ &\qquad\hbox{for 
    detonations}\;.
  \label{5.85}
\end{eqnarray}
The unequality in Eq.\ (\ref{5.85}) for detonations shows that a
strong supercooling is necessary for $\varepsilon_q$ being small
enough to satisfy Eq.\ (\ref{5.85}). (The gap between the energy
densities of both phases may be unrealistically large in the bag
model, but it is obvious that some supercooling is necessary.) The
next constraint on the allowed values of $\varepsilon$ and $p$ comes
from the second law of thermodynamics. The entropy flux into the front
may not be larger than out of the front. The condition of entropy
increase across the discontinuity leads to ${{s^2_h} / {s^2_q}} \geq
{{(\varepsilon_h + p_h ) ( \varepsilon_h + p_q)} / { ( ( \varepsilon_q
    + p_q ) ( \varepsilon_q + p_h ) ) } }$. For a perfect fluid, where
dissipation effects are absent, entropy is strictly conserved inside
each phase $(\partial_\mu(su^\mu))=0)$. It is also approximately
conserved across the front (equality sign above), if the discontinuity
in $\varepsilon$ is infinitesimally small (van Hove, 1983).

In general, however, deflagrations and detonations are 
irreversible processes and do produce entropy. The equation 
for $s_h^2 / s_q^2$ can be exclusively expressed in terms 
of $\varepsilon_h$ and $\varepsilon_p$ by using the 
thermodynamic relation $T^.s=\varepsilon+p$ and the 
equation of state. Finally the velocities must be smaller 
than the velocity of light. This means
\begin{equation}
  0\leq\tanh^2\theta_{h,q}\leq1\;.
  \label{5.86}
\end{equation}
The combined constraints for $s_h^2 / s_q^2$ and (\ref{5.86}) limitate
the values for $\varepsilon_q$ and $\varepsilon_h$ which are
compatible with detonations or deflagrations to a small window. Thus
reliable results for $\varepsilon_q$ and $\varepsilon_h$ and the
equation of state could lead to an exclusion of these scenarios.

Deflagration und detonation solutions of the hydrodynamic 
equations have been constructed in a numerical analysis by 
Gyulassy et al. (1984). The plasma expansion is neglected 
in the initial state, the baryon number is set to zero. The 
constants from energy conservation and entropy increase are 
evaluated as indicated above, the equation of state is 
taken from the bag model. The results are the following.

Detonations and deflagrations use the latent heat in a 
different way. Detonations need strong supercooling and 
subsequent strong superheating of the hadron phase. 
Deflagrations are possible under less extreme conditions. 
They are compatible with small supercooling. The 
calculations have been extended from discontinuities along 
a single surface to several ones, that means bubble 
deflagration rather than surface deflagration.
A surface deflagration shock had been considered by van 
Hove (1985). It seems to move too slowly into the plasma 
for providing a realistic possibility of plasma conversion 
to the hadron phase.

Entropy producing processes can lead to enhanced rapidity 
density fluctuations, peaks in $dN/dy$ of the final state 
hadrons. These peaks may be isolated or overlapping, 
depending on the resolution $\delta y$ and the size of a 
typical scale $\Delta y$ over which $N$ fluctuates. Large 
$\left\langle p_T\right\rangle$-momenta are another 
possible effect. Multiplicity fluctuations are sometimes 
called seismometers (Gyulassy, 1984). Experimental hints 
for ``crests'' of plasma bubbles are taken from cosmic ray 
events (Iwai et al., 1976 and 1982). The observations are 
at least compatible with an explanation of deflagration 
processes in QCD-plasma conversions (Gyulassy et al, 
1984; van Hove, 1985).

Let us assume that the QCD transitions are truly sharp 
crossover phenomena. The question arises as to whether 
there is still a chance for observing deflagrations. The 
answer seems to be positive, if we look at the derivation 
of deflagrations e.g. in van Hove (1985). The essential 
ingredients are a large variation of $\varepsilon$ and $s$ 
over a small interval of $T$ and $p$. An ingredient is the 
small value of $p$ as compared to $T^.s=\varepsilon+p$. In 
case of a crossover phenomenon the transition interval is 
defined as a region of rapid variation of $dp/dT$. Van Hove 
(1985) has studied the implications for deflagrations in 
both cases, a first and a second order transition. The 
results are quite similar in both cases. Deflagrations may 
be realized in heavy-ion collisions in spite of a smooth 
crossover if it is only sharp enough. The result is not 
surprising. Nucleus-nucleus collisions are performed in a 
finite space-time volume. From numerical simulations we 
know that a strictly first order transition and a sharp 
crossover phenomenon are practically indistinguishable in a 
finite volume, if the measurement is performed for a single 
size. Similarly deflagrations may be practically the same 
for first order transitions and a crossover phenomenon in a 
finite volume. The finite volume constitutes part of the 
physical boundary conditions in heavy-ion collisions. 
Experimental results will not expose finite volume 
roundings as artifacts of approximations to a hypothetical 
infinite volume world.

\subsubsection{Intermittency analysis}
\label{sec:V.C.6}
Experimental manifestations of the QCD transitions are frequently
discussed under the assumption that they are of first order. As we
have seen in sections \ref{sec:III} and \ref{sec:IV}, there are
indications that the finite temperature transition resembles more a
crossover phenomenon. The realistic set of quark masses may be close
to a ``critical'' set in mass parameter space, which could lead to a
second order transition. A standard physical picture of critical
phenomena in spin systems is a diverging correlation between spins
leading to clusters of aligned spins of arbitrary length scales. The
$3-d-Z(2)$-spin system is supposed to share the universality class of
an $SU(2)$ pure gauge theory, an $O(4)$-ferromagnet is assumed to
share the class of 2-flavor QCD.

Thus the question arises how a picture of clusters of aligned spins in
color or isospin space can be transferred to typical observables in
heavy-ion collisions, e.g. to particle multiplicities. One of the
possible answers is {\it intermittent behavior} in rapidity
distributions. (For a particular realization of a second order
transition in case of a hot pion gas see also section \ref{sec:V.D.3}
below).

The concept of intermittency was originally introduced in 
studies of turbulent behavior of fluids (Zel'dovich et al., 
1987). In general, intermittency can be defined as the 
appearance of structure in random media. In the context of 
particle physics, it has to deal with large fluctuations of 
charged particle density in small regions of phase space. 
In heavy-ion collisions intermittency refers to certain 
moments of rapidity distributions. Other distributions like 
energy, pseudo-rapidity and azimuthal angles can be 
considered as well (Ochs and Wosiek, 1988).

Let us first consider a toy model that is universal enough 
for applications in different areas. Consider a set of $N$ 
balls distributed in a box of total size $R$ into $M$ cells 
of size $L(M=R/L)$. In the case of rapidity distributions, 
they correspond to $N$ particles per given unit of some 
kinematic range. A cell in a box is a rapidity interval in 
a certain available rapidity range. For an Ising model, the 
balls are the spins on a spacetime lattice $R$ subdivided 
into cells of size $L$. We are interested in the ball 
distributions if the resolution of the lattice is made 
finer and finer $(M\to\infty)$. There are many ways to 
realize this limit. One possibility is to keep $R$ constant 
and let $L$ going to zero. This limit is of interest for 
rapidity distributions, where $L$ stands for the bin size 
$\delta y$ of the rapidity interval. Another realization is 
the thermodynamic limit with $R\to\infty$ for fixed $L$. 
This limit is usually taken when critical phenomena are 
discussed in the infinite volume limit.

The same type of limit is of interest in a second order 
finite temperature transition in spacetime continuum, which 
is described by a model of lattice gauge theory. One may 
expect that intermittency in the limit of ($R\to\infty$, 
$L$ fixed) has a correspondence in the limit of ($R$ fixed, 
$L\to0$).

We define the $\ell$th normalized moment $f_\ell$ for a 
given distribution of $N$ balls as
\begin{equation}
  f_\ell(M)=\left[{1\over M}\sum^M_{m=1}K^\ell_m\right]/
  \left[\left({1\over M}\right)\sum^M_{m=1}K_m\right]^\ell\;.
  \label{5.87}
\end{equation}
Here $K_m$ denotes the number of balls in the $m$th cell. 
First, we keep $N$ fixed and vary $M$. In the extreme case 
of an equidistribution of $N/M$ balls in each box, it is 
easily seen that $f_\ell(M)$ is independent of the grain 
size, i.e.
\begin{equation}
  f_\ell(M)=1\ \forall\ell\;.
  \label{5.88}
\end{equation}
In the other extreme, a strong fluctuation (where all balls 
are concentrated in one box), we find a logarithmic 
dependence of $\ln f_\ell$ on the ``resolution size'' $L$
\begin{equation}
  \ln f_\ell(M)=-(\ell-1)\ln L+(\ell-1)\ln R\;.
  \label{5.89}
\end{equation}
Such a behavior of a given distribution with $N$ balls is 
called {\it intermittent}.

More generally, intermittent behavior is attributed to 
average values of $f_\ell$. The weighted average 
$\left\langle \dots\right\rangle$ of the $\ell$th moments 
runs over an ensemble of configurations in a 
$d$-dimensional volume $R^d$ divided into $M=(R/L)^d$ cells 
of equal size $L^d$. In a multiplicity measurement it could 
be an average over all events, in the Ising model it is the 
thermodynamic average with Boltzmann weights.

Intermittency is called the property of fluctuations around 
some average distribution which lead to a power law 
behavior of $\left<f_\ell\right>$ in the number of cells $M$, or 
equivalently if
\begin{equation}
  \ln <f_\ell(L)>=-\lambda_\ell\ln L+g_\ell(R)\;.
  \label{5.90}
\end{equation}
Here $\lambda_\ell>0$ are constant and $g_\ell(R)$ is 
independent of $L$, $L$ being the linear cell size. The 
constants $\lambda_\ell$ are called intermittency indices, 
they are a measure of the strength of intermittent behavior.

Results for $\lambda_\ell$ and $g_\ell$ in case of a 
d-dimensional Ising model can be found in (Satz, 1989). The 
essential point where criticality enters the derivation of 
intermittency in an Ising model is scale invariance at 
$T_c$. While the balls of the toy model are the same 
independently of the partition of the lattice, one has 
first to argue, why one may choose the same type of 
variables on all length scales in an Ising model. This is 
justified by selfsimilarity as $T\to T_c$, if $1\ll L \ll 
\xi$ is satisfied. For a given $\ell$ the intermittency 
indices can be expressed in terms of the more familiar 
critical exponents. Like critical exponents they are 
universal for all models belonging to the same universality 
class.

The mere divergence of a correlation length $\xi$ is not 
sufficient for introducing intermittent behavior. In a 
1-$d$-Ising model, where $\xi$ has an {\it essential} 
singularity rather than a power-law singularity as in the 
2- and 3-dimensional cases, the normalized momenta $f_\ell$ 
are bounded from above for all $\ell$ (Hajdukovi\'c and 
Satz, 1992).

We turn now to intermittent behavior in heavy-ion 
experiments. To identify this behavior in rapidity 
distributions, we consider moments $\cal C_\ell$ which have 
been introduced by Bialas and Peschanski (1986) according to
\begin{equation}
  {\cal C}_\ell={1\over M}\sum^M_{m=1}(Mp_m)^\ell\;.
  \label{5.91}
\end{equation}
Here $M$ is the number of intervals of size $\delta y$ in a 
given rapidity interval $\Delta y=M\cdot\delta y$, 
$p_m(m=1,\dots,M$) denotes the probability for finding 
particles in any of these rapidity intervals $dp_M$. The 
total distribution $P(p_1,\dots,p_M)dp_1 \dots dp_M$ of 
probabilities for finding particles in the intervals 
$dp_1 \dots dp_M$ is normalized such that
\begin{eqnarray}
  p_1+...+p_M&&=1 \nonumber\\
  \int dt\int dp_1...dp_M\ P(p_1...p_M)&&=1\;.
  \label{5.92}
\end{eqnarray}
The variable $t$ stands for a collection of kinematic cuts 
like the energy of the collision. The average moment 
$\left\langle \cal C_\ell\right\rangle$ is obtained as the 
sum over all configurations in rapidity space weighted by 
the probability distribution $P(\;)dp$
\begin{equation}
  \left\langle {\cal C}_\ell\right\rangle = \int dt \int 
  dp_1 \dots dp_M P(p_1...p_M;t) {1 \over M} \sum^M_{m=1} 
  (Mp_m)^\ell\;.
  \label{5.93}
\end{equation}
In terms of rapidity variables genuine intermittent 
behavior is signalled, if
\begin{equation}
  \ln \left\langle {\cal C}_\ell\right\rangle = \phi_\ell 
  \ln (\Delta y/\delta y) = \phi_\ell \ln M\
  \label{5.94}
\end{equation}
in the limit of $\delta y\to0$, i.e. a logarithmic 
dependence on the resolution size in the limit of 
increasing resolution, which is equivalent to the 
definition in Eq.\ (\ref{5.90}). Small values for $\delta y$ 
correspond to a fine resolution in rapidity space, $\Delta 
y$ is the full considered rapidity interval. The lower 
bound on $\delta y$ is given by the experimental 
resolution. The intermittency indices $\phi_\ell$ vary 
between $O<\phi_\ell\leq\ell-1$. They are a measure for the 
strength of the intermittent behavior. Intermittency occurs 
if selfsimilar fluctuations exist on all scales $\delta y$. 
In the case  of rapidity distributions intermittency should 
be seen in contrast to dynamical fluctuations.

In {\it real experiments} we have to deal with finite size 
systems. Statistical fluctuations around the probability 
distribution $dp_1 \dots dp_MP(p_1,\dots,p_M;t)$ will 
always provide a noisy background for fluctuations of 
dynamical origin that we are trying to identify. Unless the 
multiplicity in the events is very high, the probability 
$p_m$ for finding particles in the rapidity interval $m$ is 
different from the {\it measured fraction} $K_m/N$ of the 
total multiplicity $N$. As filter of dynamical fluctuations 
Bialas and Peschanski (1986) have proposed to consider 
scaled factorial moments $\left\langle F_\ell\right\rangle$ 
of the distribution $Q$ which is actually measured in the 
experiment
\begin{mathletters}
  \label{5.95}
  \begin{eqnarray}
    \left\langle F_\ell\right\rangle = M^\ell 
    &&\sum_{K_1\dots K_M} Q (K_1...K_M) {1 \over M} \nonumber\\
    &&\sum^M_{m=1}
    {{K_m(K_m-1)\dots(K_m-i+1)}\over{N(N-1)\dots(N-i+1)}}\;,
    \label{5.95a}
  \end{eqnarray}
  where
  \begin{eqnarray}
    Q(K_1...K_M)=\int dt
    \int &&dp_1...dp_MP(p_1...p_M;t) \nonumber\\
    &&B(p_1...p_M;K_1...K_M)\;,
    \label{5.95b}
  \end{eqnarray}
\end{mathletters}
and $B$ is the Bernoulli distribution.The statement is that 
the scaled moments $\left\langle {\cal C}_\ell 
\right\rangle$ of a probability distribution $P$ are equal 
to the scaled factorial moments $\left\langle 
F_\ell\right\rangle$ of the experimental distribution $Q$. 
It can be shown that $\left\langle 
F_\ell\right\rangle=\left\langle {\cal 
C}_\ell\right\rangle$. Notice the advantage, if one uses 
$\left\langle F_\ell\right\rangle$ instead of $\left\langle 
{\cal C}_\ell\right\rangle$. In Eq.\ (\ref{5.95a}) $\left\langle 
F_\ell\right\rangle$ is exclusively expressed in quantities 
which are directly measured. The average in Eq.\ (\ref{5.95a}) can 
be evaluated by absorbing the weights $Q(...)$ in the 
selection of events.

Further information about the origin of intermittency is 
contained in the $\ell$-dependence of $\phi_\ell$. 
Equivalently, one can use the $\ell$-dependence of 
anomalous fractal dimensions $d_\ell$ which are related to 
$\phi_\ell$ according to
\begin{equation}
  d_\ell=\phi_\ell/(\ell-1)\;.
  \label{5.96}
\end{equation}
At least two physical mechanisms are known which may lead 
to real intermittent behavior in heavy-ion collisions (real 
in contrast to fluctuations which are induced by 
two-particle correlations). One of these mechanisms is 
{\it the QCD transition} if its correlation length diverges 
at $T_c$, or, adapted to the finite volume, if its 
correlation length extends the typical volume in the 
collision. In this case, $d_\ell$ should be approximately 
independent of $\ell$. Other origins are {\it selfsimilar 
cascades} which are attempts to model the evolution of the 
plasma to the hadronic phase. Here $d_\ell$ will depend on 
$\ell$ in general. Unfortunately, the $\ell$-dependence of 
$d_\ell$ does not allow a unique identification of the 
origin of intermittency. Neither does the independence on 
$\ell$ guarantee a phase transition nor does its dependence 
exclude a transition. According to the work by Bialas and 
Peschanski (1986), it may be possible to tune the 
parameters of a selfsimilar cascade in such a way that 
$d_\ell$ does not depend on $\ell$. On the other hand, 
multifractal phase transitions are known which lead to 
$\ell$-dependent $d_\ell$ (although the latter possibility 
is not predicted by QCD).

We add a remark on dynamical fluctuations which are due to 
two-particle correlations. In this case the scaled moments 
first rise with decreasing bin size, then they saturate to 
some constant value below a certain resolution size $\delta 
y_0,\, \delta y_0$ gives also the typical size of the 
dynamical fluctuations. Two-particle correlations can 
result from resonance decays or Bose-Einstein correlations. 
For resolutions $\delta y>\delta y_0$ two-particle 
correlations are mainly responsible for apparent 
intermittency effects. At least in $e^+e^-$-reactions the 
observed intermittent behavior can be fully explained by 
two-particle correlations within a certain particle 
production scheme. For heavy-ion collisions the particle 
production processes in the matter evolution are less well 
understood, at least the contribution from Bose-Einstein 
correlations can be tested experimentally, see e.g. Schmidt 
and Schukraft (1993).

We summarize what a measurement of the anomalous fractal 
dimension $d_\ell$ can tell us. Assuming that the 
contribution from two-particle correlations can be 
subtracted, the following alternatives can be 
distinguished.
\begin{itemize}
\item If $d_\ell\sim0$, the fluctuations are of purely statistical
  origin. A first order transition with a correlation length smaller
  than the size of the system would be compatible with such a
  scenario.
\item If $d_\ell$ strongly depends on $\ell$, it seems rather unlikely
  that the QCD transition itself is responsible for inducing such a
  multifractal behavior. More likely is a cascade mechanism at work
  whose vertices should be chosen to reproduce the observed
  $\ell$-dependence of $d_\ell$. For a cascading process one expects
  $d_\ell$ to depend linearly on $\ell$ (Bialas and Peschanski, 1988).
\item If $d_\ell$ depends weakly on $\ell$, but is clearly different
  from zero, a second order transition provides an
  explanation for intermittency, but not the only one.
\end{itemize}

\vskip12pt
We conclude with a short glance at the experimental status. 
Intermittency has been found in $e^+e^-$-reactions, hadron 
hadron- and heavy-ion collisions, see e.g. (Abreu et al., 
1989; Buschbeck, 1989; {\AA}kesson et al., 1990, 
respectively). Qualitatively, different experiments agree 
about the result that intermittency has been observed and 
different mechanisms are at work, as the observed behavior 
is not uniform. Quantitatively, they disagree. In 
Fig.~\ref{fig:36} we display the dependence of the anomalous fractal 
dimension on $\ell\equiv p$ at $\sqrt s=20$~GeV taken from 
Bialas and Hwa (1991). Obviously, the $p$-dependence of 
$d_p$ weakens when passing from lepton-hadron to 
hadron-hadron and nucleus-nucleus collisions. In (Bialas 
and Hwa, 1991) this is interpreted as a result of 
selfsimilar cascade mechanisms in the first two cases, but 
as a phase transition in the $S-Ag,Br$ central collisions. 
The data for $Ag,Br$ collisions which are consistent with 
an almost constant nonzero value of $d_\ell$ suffer from 
low statistics (Schmidt and Schukraft, 1993). Another 
tendency which is seen in Fig.~\ref{fig:36} is a decrease of $d_p$ 
when the process becomes more complex, but this decrease is 
less strong for small values of $p$.
Thus it may be more than an artifact of generally small 
values of $d_p$ that $d_p$ is almost independent of $p$ for 
heavy-ion collisions.

Less encouraging are the high statistics data taken at CERN 
SPS (Bloomer et al., 1992) for $^{32}S+S$ and $^{32}S+Au$ 
collisions. The slope of $\ln\left\langle F_2\right\rangle$ 
vs $(-\ln\delta\eta)$ ($\eta$ being the pseudorapidity) is 
consistent with zero or less than zero. Thus one of the 
conclusions of the authors is that there is ``no need for 
new physics'' in the sense of fluctuations of unknown 
dynamical origin.

Another systematic trend seems to be well established by 
the data. As we have mentioned above, two-particle 
correlations may look like intermittency signals as long as 
the resolution is not high enough. In fact, for lighter 
projectiles like oxygen or silicon  the data for 
intermittency  indices $\phi_2$ can be well fitted as a 
function of the average particle density $\left\langle 
\rho\right\rangle$. Such a scaling law has been proposed by 
 Seibert (1990) and is based on the assumption that only 
two-particle correlations are responsible for the observed 
intermittent effects. The optimistic news are that for 
heavier projectiles like sulfur-induced reactions such an 
explanation is not sufficient. For heavier projectiles, 
there seems to be a chance for seeing collective 
effects. These could be due to a nearly second-order 
transition or a cascade mechanism during the matter 
evolution. Last but not least detector effects should be 
kept in mind as possible source for contaminating the data. 
For further details on the experimental status we refer to 
the review of Schmidt and Schukraft (1993) and references 
therein.

Thus the main task in an intermittency analysis is to disentangle
fluctuations of different origins: geometry (impact parameter
dependence), statistics (finite number of particles), resonance decays
or Bose-Einstein correlations, hadronization along cascade mechanisms,
and a phase transition with a large correlation length.  Recently work
by Campi and Krivine (1995) questions any connection between the
observation of intermittent behavior in mass distributions of finite
nuclei and a second order phase transition.  In any case, one has to
face a superposition of several backgrounds on the type of
multiplicity fluctuations one is primarily interested in. Once an
interesting structure is filtered from the backgrounds, the
$\ell$-dependence of intermittency indices may indicate the specific
dynamical origin of this structure. The situation resembles that for
dilepton rates, where the main practical problem was an identification
of thermal rates. Once the thermal rates are identified, the remaining
task is to look for conclusive structures of the transition scenario
(cf. section \ref{sec:V.C.2}).

\subsection{Theoretical concepts for off-equilibrium situations}
\label{sec:V.D}
\subsubsection{Nucleation rate of hadronic bubbles}
\label{sec:V.D.1}
Many phenomenological implications assume a scenario of a first-order
transition which is often described by a bag model equation of state.
The scenarios differ in assumptions about the entropy production
during the transition, the amount of supercooling, the production of
latent heat.

In this section, we describe an approach of Csernai and Kapusta (1992a
and b) for calculating the nucleation rate of hadronic bubbles in the
plasma phase.  In principle, such a calculation can select a realistic
scenario. For example, it should tell us whether a
Maxwell-construction of two-phase equilibrium is realistic in view of
the total duration of the plasma evolution to the final hadronized
state, which is of the order of $10-20$~fm/c.  It requires a
nucleation rate large enough that the latent heat is conducted fast
enough from the bubble surfaces for the bubbles to grow.  Naturally
transport properties of the latent heat are determined by properties
of the ``conducting'' medium surrounding the bubble. Such properties
are the thermal conductivity, the shear and bulk viscosity, if a
hydrodynamical description is used. In our case of an expanding plasma
on its way to hadronization, the plasma viscosity will turn out to
determine the nucleation rate of the hadronic bubbles.

The prediction of a definite value for the nucleation rate and the
time delay due to a first order transition is rather ambitious and
involves unavoidably a number of assumptions and approximations.
Deviations from equilibrium have to be admitted during the phase
transition, which need not be perturbatively small. One has to deal
with an expanding system of many particles interacting according to
the excitation spectrum of QCD.

Quite generally, the nucleation rate of one phase out of another (here
the hadronic phase out of the plasma phase) has the form
\begin{equation}
  I=I_0\ e^{-\Delta F/k_BT}.
  \label{5.97}
\end{equation}
This is the usual expression for the rate of a thermally activated
process.  More precisely, it is the probability to form a (hadronic)
bubble of critical size per unit time and unit volume. The difference
$\Delta F$ is the activation energy to form a bubble of critical size.
More generally, it is the difference in the free energies between two
configurations with and without a condensed part of the new phase (it
need not have the form of a droplet). The more delicate part is the
prefactor, which describes some sort of quantum and/or thermal
fluctuations. It sensitively depends on the context and the
approximation scheme.

A specific realization of Eq.\ (\ref{5.97}) is the famous formula of
Callan and Coleman (1977) for the decay rate of the `false' vacuum.
The $\Delta F$ in the exponent is given by the Euclidean action $S_4$
in four dimensions of the considered field theory, evaluated at a
solution $\phi$ of the Euclidean equation of motion satisfying
$\phi\to 0$ for $x^2+t^2\to\infty$ as boundary condition. The
prefactor depends on $S_4$ and a ratio of determinants depending on
some zero-temperature (effective) potential $V$. The occurrence of
determinants in the prefactor is a common feature of formulas for
decay rates when fluctuations are taken into account in the Gaussian
approximation.  Although Callan and Coleman's formula is sometimes
mentioned in the context of phase transitions in the early universe,
it is not appropriate for calculating a decay rate at finite
temperature in a situation out of equilibrium, which is encountered in
heavy-ion collisions.

A suitable framework is provided by Langer's approach (Langer, 1969),
which has been generalized to relativistic field theory and applied to
QCD by Csernai and Kapusta (1992a). We briefly summarize the main
ingredients in Langer's formalism. The zero-temperature infinite
volume potential in Callan and Coleman's formula is replaced by a so
called coarse-grained free energy, which is the appropriate quantity
for describing a phase coexistence in a finite volume. Langer
considers an effectively classical system of $N$ degrees of freedom,
represented by a set of $N$ coordinates and their conjugate momenta
$\eta_i$, $i=1,\ldots,2N$. These variables need not describe the
system on a microscopic level, but enter the coarse-grained free
energy as some kind of collective coordinates. Whereas the Euclidean
action $S_4$ (mentioned above) depends on microscopic variables, the
coarse-grained free energy is formulated in terms of `mesoscopic'
variables, which are collective coordinates.

The $\Delta F$-term in Eq.\ (\ref{5.97}) keeps its meaning as the
change in (coarse-grained) free energy due to a formation of one
droplet of critical radius $R_*$.

The prefactor $I_0$ of the general formula accounts for
(thermal) fluctuations in a specific way. It is written as a product
of two terms
\begin{equation}
  I_0={\kappa\over{2\pi}}\Omega_0,
  \label{5.98}
\end{equation}
a statistic factor $\Omega_0$, which is a measure of phase space
volume around the saddle point, and a dynamical factor $\kappa$. While
$\Omega_0$ will be seen to depend on equilibrium quantities, $\kappa$
knows about off-equilibrium features of the dynamical expansion. The
dynamics may be described by hydrodynamical equations of motion. Thus
it is no longer restricted to the regime of $O(4)$-invariant Euclidean
equations of motion in equilibrium situations. A direct manifestation
of off-equilibrium features is $\kappa$'s dependence on the thermal
conductivity in a liquid-gas system or on the shear and bulk
viscosities ($\rho$ and $\eta$ respectively) in the case of QCD, cf.
Eq.\ (\ref{5.117}) below. This is clearly a generalization as compared
to the approach of Callan and Coleman, but also Langer's formulation
does not apply to generic off-equilibrium systems.  The {\it basic
assumption} is that the conversion proceeds via nucleation.  The
nucleation rate comes out as time-independent by construction. The
phase transition is described as motion of a probability current
$\rho$ in phase space. It starts at $\{\eta_0\}$, the metastable
phase, passes most likely a saddle point $\{\bar\eta\}$, where the
current is assumed to be stationary
\begin{equation}
  \partial_t\rho = 0 ,
  \label{5.99}
\end{equation}
and ends in the stable phase configuration $\{\eta_f\}$. The
configuration $\{\bar\eta\}$ differs from $\{\eta_0\}$ just by one
critical droplet of new phase. To allow a stationary current across
the saddle in phase space, there has to be a heat bath in the
background, which replenishes the metastable phase at the same rate as
it is lost in the phase conversion process. The assumption in Langer's
derivation is that such a heat bath exists, it is in equilibrium prior
to each interaction with the converting phase. The probability for a
transition between two configurations $\{\eta\}$ and $\{\eta'\}$,
induced by thermal fluctuations of the heat bath, is determined by the
same Hamiltonian $F$ as it enters the dynamical part of the system
which transforms into the new phase.

The dynamical part is assumed to be governed by equations of motion of
the form
\begin{equation}
  {{\partial\eta_i}\over{\partial t}}=-\sum^{2N}_{j=1} M_{ij}
  {{\partial F}\over{\partial \eta_j}}.
  \label{5.100}
\end{equation}
In general, $F$ plays the role of a Hamiltonian, in an application to
a hydrodynamical system $F$ has the specific form of a coarse grained
free energy. Later $F$ will contain the information which is specific
for the underlying QCD. The matrix $M$ is the mobility matrix, in our
application the coefficients $M_{ij}$ can be read off from a
comparison with the equations of relativistic fluid dynamics. Eq.\ 
(\ref{5.100}) enters the time development of the distribution function
$\rho(\{\eta\},t)$ which is associated with the variables $\eta_i$ and
gives the probability density over configurations $\{\eta\}$. For
equilibrium configurations $\rho$ is proportional to the Boltzmann
weight $\exp[-F\{\eta\}/T]$, where $F$ are the costs in (coarse
grained) free energy to generate such a configuration.
Skipping some steps in the derivation, the time evolution of $\rho$
can be finally written in the form of a continuity equation
\[
  {{\partial\rho\lbrace\eta\rbrace}\over{\partial t}}=
  -\sum_i{{\partial J_i}\over{\partial\eta_i}}
\]
with
\begin{equation}
  J_i=-\sum_j M_{ij}\left({{\partial F}\over{\partial\eta_j}}
    \rho+T{{\partial\rho}\over{\partial\eta_j}}\right)~,
  \label{1.101}
\end{equation}
where $F$ is the Hamiltonian and $T$ is the temperature. The
probability current across the saddle is identified with the decay
rate. The prefactor $\kappa$ gives the initial exponential growth rate
of a bubble which has just exceeded its critical size $R_*$. It is
defined as
\begin{equation}
  \kappa = {d \over dt}{\ln[R(t)-R_*]} .
  \label{5.102}
\end{equation}
To obtain $\kappa$ from the time evolution of $\rho$, Eq.\ 
(\ref{5.102}) is solved for {\it small} deviations
$\nu_i=\eta_i-\bar\eta_i$ from the saddle $\{\bar\eta\}$. Note that
this is a further assumption entering the derivation of $\kappa$. The
deviations from the saddle must be small for the saddle point
approximation to make sense. For details about the statistical
prefactor $\Omega_{0}$ we refer to the original literature (Langer,
1969; Langer and Turski, 1973 and 1980).

Let us turn next to the application in QCD (Csernai and Kapusta, 1992a
and b).  We have to specify the collective variables $\eta_i$, the
coarse grained free energy $F$, the hydrodynamic equations of motion
and their energy momentum tensor $T_{\mu\nu}$. The mobility matrix $M$
and the stationary configuration $\{\bar\eta\}$ must be identified.

{\it The collective variables} $\eta_i$ are chosen as energy density
$e(\vec r)$ and flow momentum $\vec P(\vec r)$ at positions $\vec r$
in the system.

The choice of a {\it coarse grained free energy} $F$ for QCD is more
subtle.  The hydrodynamic approach is suitable for describing
nucleation rates in a fluid, since droplet formation occurs on a
semi-macroscopic level where a large number of particles contributes
to a single droplet. The concept of the coarse grained free energy is
introduced when hydrodynamics is derived from microscopic kinetic
theory (Van Kampen, 1964; Jackson 1960). The macroscopic system is
divided into cells of a given volume. Specific densities ($e$ and
$\vec{\cal P}$ in our case) are assigned to each cell. If the
partition function of the microscopic system is evaluated under these
cellular constraints on the microscopic variables, one obtains an
effective action (in a more field theoretical language), which is
called the coarse grained free energy (over $k_B T$). (Note that this
concept is very similar to the {\it constrained effective potential}
used in section \ref{sec:IV.A.4}.  In the constrained effective
potential the entire system is considered as one cell subjected to a
constraint on the vacuum expectation value of the order parameter
field.)

The underlying QCD dynamics enters the interaction part $f(e)$ of the
coarse grained free energy $F$ for a small volume of nonuniform energy
density $e$.  The grain size of $F$ is obviously essential for the
approximation to make sense. If the grain size is too small, the use
of hydrodynamic variables is not meaningful, if it is too large, the
coarse grained free energy cannot resolve the substructure of the
fluid. This happens if phase separation occurs within single cells,
$F$ would be a convex function of the local average energy density $e$
in this case. Similarly the constrained effective potential becomes
convex in the large volume limit. Thus an ansatz for $f(e)$ which
leads to a non-convex shape of $F$ as function of $e$, will indirectly
implement the right grain size which is sensitive to nonequilibrium
properties like droplet growth and phase separation. Note that $F$ is
non-convex, but real, the non-convex shape is not an artifact of a
perturbative approximation scheme, in which a thermodynamic potential
is evaluated in the infinite volume limit. The very grain structure is
implemented in the choice of hydrodynamic variables, implicitly
involving averages over cells.

As ansatz for $f(e)$ Csernai and Kapusta have chosen a fourth order
polynomial in $(e-e_{0})$, where $e_{0}$ is the location of the top of
the potential barrier. The coefficients depend on the energy densities
in the hadron and plasma phase, and on the curvature $f''_{0}$ of
$f(e)$ at $e_{0}$. In the full ansatz for $F$ one is left with two
free parameters, $f''_{0}$, and the coefficient $K$ of the derivative
term $(\nabla e)^{2}$. Both parameters can be expressed in terms of
the interface tension $\sigma$ and the interface thickness
$2\xi_{0}$. Here $\xi_{0}$ denotes the correlation length defined at
the top of the barrier by $\xi_{0}^{2}\equiv -K/f''_{0}$. Note that
both quantities ($\sigma$ and $\xi_{0}$) can be calculated from first
principles, i.e. from the underlying QCD Lagrangian.

The {\it equations of motion} are the equations of relativistic fluid
dynamics, $\partial_\mu T^{\mu\nu}=0$.  In the baryon-free region of
the quark gluon plasma we have to deal with a relativistic system.
Without a net baryon number there is no distinguished reference frame
to define the flow four velocity $u_\mu$ of the matter. Also the
pressure is not small compared to the energy.  Therefore $u_\mu$ is
defined as velocity of the collective energy flow. The {\it
  stress-energy-momentum tensor} for a relativistic dissipative fluid
depends on the shear and bulk viscosities $\eta$ and $\zeta$,
respectively. It should be noticed that for a vanishing dissipation
part the growth rate of bubbles would vanish. Similarly it has been
noticed by Langer and Turski (1973) that the growth rate in a
liquid-gas transition vanishes near the critical point, if the thermal
conductivity is set to zero. The bubbles can only grow, if the latent
heat is conducted from the bubble surface.

We skip further details of the derivation and come to the results.
The exponential suppression factor in Eq.\ (\ref{5.97}) depends on the
change in free energy $\Delta F$, if a hadronic bubble has formed in
the QCD plasma at $T<T_c$ due to a thermal fluctuation. It is given as
a sum of a volume and a surface term according to
\begin{equation}
  \Delta F = {4\pi \over 3} (f_h-f_p) R^3+4\pi R^2\sigma~,
  \label{5.103}
\end{equation}
where the indices $h$ and $p$ indicate the corresponding phases. The
critical radius $R_*$ is given by
\begin{equation}
  R_*(T) = {2\sigma \over p_h(T)-p_p(T)}~,
  \label{5.104}
\end{equation}
where $p_h$ and $p_p$ denote the pressure in the hadron and the plasma
phase, respectively. (Bubbles with $R<R_*$ collapse, those with
$R>R_*$ expand.)

{\it The statistical prefactor} $\Omega_0$ can be expressed in terms
of the surface tension $\sigma$, the bubble radius $R$, the
correlation length $\xi_p$ in the plasma phase and the volume $V$
according to
\begin{equation}
  \Omega_0={2\over{3\sqrt3}}\left({\sigma\over T}\right)^{3/2}
  \left({R\over{\xi_P}}\right)^4 V~.
  \label{5.105}
\end{equation}
{\it The dynamical prefactor} $\kappa$ which determines the
exponential growth of a fluctuation $\nu$ of a droplet of critical
size is derived as
\begin{equation}
  \kappa={{4\sigma(\zeta+4\eta/3)}\over{(\Delta w)^2 R^3_*}}~,
  \label{5.106}
\end{equation}
where $\Delta w$ is the enthalpy difference between both phases. Note
that $\kappa$ is determined in terms of equilibrium quantities
$\sigma$, $R_*$ and $w$, which are in principle accessible in lattice
calculations, and nonequilibrium quantities, which are the transport
coefficients $\eta$ and $\zeta$. For an ultrarelativistic gas the bulk
viscosity $\zeta$ is much smaller than the shear viscosity $\eta$,
which has been estimated to leading order in perturbative QCD by Baym
et al. (1990). For two flavors the result is
\begin{equation}
  \eta={{2.11 T^3}\over{\alpha^2_s\ln(1/\alpha_s)}} ,
  \label{5.107}
\end{equation}
where $\alpha_s$ is the strong coupling constant at finite temperature.

Eqs.\ (\ref{5.98}), (\ref{5.105}) and (\ref{5.106}) with $\Delta F$
from Eq.\ (\ref{5.103}) imply for the nucleation rate
\begin{equation}
  I={4\over{\pi}}\left({\sigma\over{3T}}\right)^{3/2}
  {{\sigma(\zeta_P+4\eta_P/3)\cdot R_*}\over{\xi^4_P(\Delta w)^2}}e^{-
    \Delta F/T} .
  \label{5.108}
\end{equation}
The input parameters are chosen as follows:
$\sigma=50~\hbox{MeV/fm}^2$, $\alpha_s=0.23$, $\xi_P=0.7$~fm,
$\eta_P=14.4 T^3$ and $B^{1/4}=235$~MeV for the bag constant. The
pressure and the enthalpy difference $\Delta w$ between the hadronic
and the plasma phase are calculated from the bag model equation of
state. We would like to add a comment on the input. More recent
results on the surface tension in pure $SU(3)$ gauge theory show that
the estimate of $50~\hbox{MeV/fm}^2$ which has been used by Csernai
and Kapusta is too large by an order of magnitude, see e.g. section
\ref{sec:III.B.3} or (Kajantie, 1992). It would be also quite
interesting to update the equation of state, replacing the bag model
values by more recent lattice results, since the bag model equation of
state fails in particular in the transition region.

The nucleation rate $I$ naturally defines a {\it nucleation time} if
$I$ is multiplied by the volume of a bubble of critical size
\begin{equation}
  \tau^{-1}_{nucl}(T)={{4\pi}\over3}R^3_*(T)I(T)~.
  \label{5.109}
\end{equation}
This time scale is a measure for the delay due to supercooling. It
should be distinguished from the total time duration of the phase
transition. It just refers to bubbles of critical size, but neglects
bubble growth. When $\tau$ is plotted as a function of ($T/T_c$), one
sees that 5\% of supercooling are necessary for the nucleation to set
in, when the above input parameters are used. A use of updated input
parameters may considerably reduce the degree of supercooling.

A hint on the self-consistency of the approximation scheme is provided
by the temperature dependence of the critical bubble radius $R_*$,
which is determined by Eq.\ (\ref{5.104}). The radius diverges at
$T_c$, but rapidly reduces to typical hadronic scales and even below.
At $T/T_c=90\%$, $R_*\sim 0.73$~fm. This is comparable to the
correlation length in the plasma phase $\xi_P$ and of the order of the
surface thickness. For such values of $R_*$ the approximation scheme
breaks down. Therefore the question arises as to whether the phase
transition has been completed before or after these low temperatures
are reached. The quantity of interest is the {\it time dependence of
  temperature}.  In the early universe $T$ drops so slowly with time
that the phase transition is most likely over before the
approximations loose their validity.

For heavy-ion collisions the time dependence of $T$ has been estimated
by Csernai and Kapusta (1992b). The $T(t)$-dependence in the bulk
phase is known, once an equation of state is combined with an ansatz
for the expansion scenario, e.g. Bjorken's scaling ansatz. The time
dependence of $T$ {\it during} the phase conversions is a more subtle
problem. In the Maxwell construction $T$ is independent of $t$ right
from the beginning. To find $T(t)$ for a given nucleation rate $I$,
one starts from a dynamical equation which couples $T(t)$ to the
fraction of space which has been converted to the hadronic phase.

The result for $T(t)$ clearly shows supercooling, superheating effects
and a time delay in the temperature decrease as compared to a Maxwell
scenario.  When the temperature has decreased to $\sim 0.8 T_c$, the
nucleation is sufficient for reheating the system up to $\sim 0.95
T_c$, where further nucleation is stopped. From that temperature on
the previously created bubbles grow further until the conversion to
the hadronic phase is complete. Compared to the Maxwell scenario the
time of completion is delayed by 11~fm/c leading to an increase in
volume at that completion time. The increase is linear in $t$ in
Bjorken's ansatz. If the entropy density at completion is only
determined by the equation of state in the pure hadronic phase, but
independent of the conversion procedure, a delay of 11~fm/c implies
30\% of extra entropy generation to keep the entropy density fixed.
This is non-negligible in an extrapolation from the final pion
multiplicity to the initial temperature. It has to be checked, whether
the amount of extra entropy generation is an artifact of the bag model
equation of state.

Finally we note that the critical bubble radius in the late period of
the phase conversion is about 1~fm (not too large in view of the
finite size of the system and not too small for the approximation to
hold).

Coming back to the question posed in the beginning of this section: Is
a Maxwell construction a realistic scenario ? The time dependence of
temperature during the transition based on the {\it derived}
nucleation rate $I$ (Eq.\ (\ref{5.108})) comes out not so different
from a Maxwell construction, which is usually taken as an ad hoc
assumption. It remains to be checked whether the similarity can be
maintained, if one improves on various approximations, and most recent
input data are used for the surface tension, the surface thickness,
and the enthalpy difference.

\paragraph*{Time dependent nucleation rates}
In Langer's formalism for deriving the nucleation rate an
embedding heat bath has been assumed to guarantee a stationary flow in
probability space or a time independent nucleation rate. Langer's
description does not apply to a situation far out of equilibrium. The
remnants of off-equilibrium are contained in the viscosity
coefficients of the nucleation rate. It may well be that such a heat
bath is not realized during all stages of phase conversion in
heavy-ion collisions. Some time dependence of the nucleation rate is
natural, at least at the onset and the completion of nucleation. A
calculation of time dependent rates requires a different formalism.

Boyanovsky and Aragao de Carvalho (1993 and 1994) have analyzed
the thermal activation via sphaleron transitions in a
(1+1)-dimensional field theory, where they use a real time formalism.
They obtain a time-dependent nucleation rate from the real time
evolution of the initial density matrix along the unstable direction
in configuration space. The rate is rather sensitive to the initial
state, which is typical for an off-equilibrium process. The initial
state is a metastable state that has been formed after a period of
{\it rapid supercooling}.

One may think of the phase transition in heavy-ion collisions in a
similar way.  The initial state for the conversion process is a
metastable state arising from a rapid cooling rather than being in
local equilibrium with a heat reservoir. The time evolution of the
conversion may come out quite differently, if the heat bath is not
replenished during the conversion process. A comparison of nucleation
rates calculated in one or the other way would be quite useful for
assessing the estimates of time delay and extra entropy production,
which should be visible in heavy-ion collisions.

Recent results from the lattice and studies in effective models
suggest that the transition is weakly first order -- if it is of first
order at all. In this case nucleation may completely fail as
conversion mechanism from the plasma to the hadron phase. Large domain
coarsening might replace nucleations as $T$ approaches $T_{c}$ from
above. Nonequilibrium aspects in weak first order transitions are
discussed in the next section.

\subsubsection{Large domain coarsening}
\label{sec:V.D.2}
Different nonequilibrium aspects may be relevant for the transition
dynamics. Here we distinguish only between {\it nucleation} and {\it
  large domain coarsening} or {\it spinodal decomposition}. Nucleation
of hadronic bubbles in a first order QCD transition has been the topic
of the preceding section. The physical assumption there was that the
system is still in a homogeneous plasma phase, as $T_{c}$ is reached
from above. Some supercooling below $T_{c}$ is necessary, until
bubbles of critical size can be formed and the phase conversion
proceeds via nucleation. This scenario is likely a realistic
description for a {\it strong first order} transition. The formal
manifestation for the underlying physical assumption is the saddle
point evaluation of the path integral in Langer's formulation, as we
have seen above. The saddle point approximation is justified as long
as the fluctuations about the homogeneous metastable ground state are
small. The system should be well localized in the metastable minimum
as the temperature drops below $T_{c}$.

The nonequilibrium dynamics of the phase conversion may drastically
change if the transition is {\it weakly first order} (Borrill and
Gleiser, 1995). In spite of a potential barrier the system may no
longer be well localized in the metastable minimum, before the
temperature drops below $T_{c}$. Large amplitude fluctuations are
allowed, subcritical bubbles can form {\it above} $T_{c}$. {\it At}
$T_{c}$ the system is found in a completely mixed phase of high and
low temperature ``components''. The phase mixing is assumed {\it not}
to be an artifact of the finite volume. (Recall that spontaneous
symmetry breaking strictly occurs only in the infinite volume limit.
Here the volume is chosen large enough that the tunneling rate
between the different phases is negligible.) If the system starts at
$T_{c}$ from a mixed phase and then cools below $T_{c}$, the phase
separation may evolve by domain coarsening in a way that resembles
spinodal decomposition rather than nucleation.

Borrill and Gleiser (1995) have studied the phase mixing {\it at}
$T_{c}$ as a function of the strength of the first order transition.
The framework are discretized scalar field theories in $3+1$
dimensions. The homogeneous part of the free energy density contains a
quadratic, cubic, and quartic term in a real scalar field. The
couplings in the quadratic and cubic part are assumed as temperature
dependent. The precise form is motivated by an effective potential for
the electroweak phase transition, but the results are applicable to
QCD as well. The strength of the first order transition has been tuned
by the quartic coupling $\lambda$, keeping the cubic coupling fixed.
Small values of $\lambda$ correspond to a high barrier in the
effective potential and a strong first order transition, large values
of $\lambda$ to a weak first order transition. The system is always
considered at fixed temperature $T_{c}$. It is prepared to be
initially well localized in the high temperature phase. It is then
evolved by a discretized Markovian Langevin equation with white
stochastic noise to mimic the coupling with a thermal bath.

Let us denote the fractions of the total volume in the high (low)
temperature phase divided by the total volume $f_{0}$ ($f_{+}$),
respectively. The fraction of the total volume in each phase,
$f_{0,+}(t)$, is then measured as a function of time, until the final
ensemble-averaged equilibrium values $f_{0,+}^{\rm{eq}}$ are
reached. The qualitative results of Borrill and Gleiser are the
following. For small values of $\lambda$ (strong first order
transitions) the system remains localized in the initial metastable
minimum, $f_{0}^{\rm{eq}}\sim 1$. For large enough $\lambda$
(sufficiently weak first order transitions), $f_{0}^{\rm{eq}}$
approaches 0.5, corresponding to a complete phase mixing at $T_{c}$.

A further interesting observation is made, if the equilibrium
fractional population difference at $T_{c}$
\begin{equation}
  \label{5.110}
  \Delta F_{\rm{eq}}=f_{0}^{\rm{eq}}-f_{+}^{\rm{eq}}
\end{equation}
is plotted as a function of $\lambda$. The curve suggests a second
order phase transition as function of $\lambda$ (the control parameter
of the strength of the first order finite-$T$ transition). There seems
to exist a critical strength of the first order transition (obtained
for $\lambda=\lambda_{c}$), so that for larger values of $\lambda$
(weaker transitions) $\Delta F$ soon approaches zero, corresponding to
a perfect thermal mixing $f_{0}^{\rm{eq}}=0.5=f_{+}^{\rm{eq}}$. For
smaller values of $\lambda$ (stronger transitions), $\Delta
F_{\rm{eq}}$ remains close to 1, the system is trapped in its
homogeneous initial state. Therefore Borrill and Gleiser proposed,
$\Delta F_{\rm{eq}}$ to be used as order parameter for the
`transition' from strong to weak first order transitions, $\Delta
F_{\rm{eq}}$ is a quantitative measure for how strong the finite-$T$ 
transition is.

The prediction of complete thermal mixing at $T_{c}$ relies on a
cooling process, which is slow compared to other equilibration scales
in the system. Keeping $T$ fixed at $T_{c}$, this assumption was
implicitly built in the derivation. The narrow transition region
between the two areas of strong and weak first order transitions may
be related to the fact that the order parameter for the finite $T$
transition $\left<\phi(t)\right>$ is not conserved. For conserved
order parameters in binary mixtures and long internal equilibration
times the transition between the two regimes is smooth (Borrill and
Gleiser, 1995, and references therein).

\vskip12pt
To conclude we summarize some possible implications for QCD. The
results of Borrill and Gleiser have shown that the nonequilibrium
transition dynamics may change from nucleation to large domain
coarsening in passing from strong to weak first order transitions. The
interpolation between both schemes need not be smooth, but may be a
phase transition itself. In QCD it is likely that the finite
temperature transitions are rather weakly first order (if they are of
first order at all), or rapid crossover phenomena, which may be
indistinguishable from a practical point of view. The role of the
control parameter of the strength of the transition will be played by
the current quark masses. If the cooling of the plasma is slow
compared to equilibration time scales of amplitude fluctuations about
the metastable ground state, the QCD transition may proceed via large
domain coarsening rather than supercooling with subsequent nucleation.

Observable signatures in heavy-ion collisions will in general depend
on the dominant process during the phase conversion. Large domain
coarsening may result in similar effects as they are expected for
second order transitions, while strong supercooling would lead to
explosive effects in the other extreme case.

In view of the results of Borrill and Gleiser, a good quantitative
estimate of the strength of the first order/crossover phenomenon does
no longer appear as a minor detail, once the order of the transition is
determined. It is particularly important, if the regimes of nucleation
and spinodal decomposition are separated by a narrow range in the
control parameter of the potential barrier. Thorough estimates of time
scales for the cooling process and relaxation phenomena during the
phase conversion are needed to turn the speculations of today into
predictions for observable signatures.

\subsubsection{Transport coefficients}
\label{sec:V.D.3}
As we have seen in section \ref{sec:V.D.1}, dissipation effects like
shear and bulk viscosities of the plasma will influence the nucleation
rate of hadronic bubbles, if the transition proceeds via nucleation.
Viscosity and thermal conductivity are further sources of entropy
production during the evolution from the initial to the final state.
Although it is sometimes claimed that the amount of extra entropy is
negligible compared to the entropy generated during a hypothetical
supercooling scenario, it is worthwhile being calculated; it may well
be that there is no supercooling at all.

Corresponding to this extra entropy, dissipation leads to an increase
in the final rapidity density of observed particles. The entropy as a
function of time enters various production rates, thus an extra
entropy will influence predictions of final rates as well.

Transport processes result from gradients in thermodynamical
parameters.  Gradients in the temperature lead to a thermal current,
in the velocity field to a friction force, in color distributions to a
color current. They are irreversible and typical nonequilibrium
phenomena which have to be treated with nonequilibrium methods. At
length we have discussed results from equilibrium thermodynamics.
Typically they refer to the existence of phase transitions, critical
parameters and ingredients for the equation of state. In order to
incorporate all this knowledge in nonequilibrium effects, the
equilibrium has to play the role of an expansion point, at which small
perturbations are allowed. Thus it is not surprising that transport
coefficients are calculated within some linear response approach. The
linear response refers to the response of the energy momentum tensor
to the thermodynamical forces (friction forces, temperature gradients,
etc.).

Two approaches are known which we call the {\it kinetic} and the {\it
  phase space approach}. The kinetic approach emphasizes particle
aspects. Transport processes arise as the result of collisions between
particles (partons or hadrons, depending on the phase). In the
vicinity of $T_c$, kinetic theory is expected to fail. A hydrodynamic
treatment seems to be more adequate, which is incorporated in the
phase-space approach. In principle, the phase-space approach applies
to all temperatures. The transport coefficients are expressed in terms
of expectation values of retarded Green's functions. Following Zubarev
(1974), Hosoya et al. (1984) elaborated the phase-space approach for
the hydrodynamic regime (although it is not necessarily restricted to
that). The hydrodynamic description is apparent in the choice of
variables, locally defined temperature and velocity fields. As usual
this assumes that the mean free time of constituent particles is much
shorter than the typical relaxation time of the system.

Only linear terms in gradients of inhomogeneous distributions are
kept. The system has to be in a late stage of preequilibrium expansion
rather close to equilibrium to satisfy this condition. Under these
restrictions one obtains as an expression for the shear viscosity
\begin{eqnarray}
  \eta=-&&\int d^3x'\int^t_{-\infty}dt_1\exp\{\varepsilon(t_1-t)\}
  \nonumber\\
  &&\int^{t_1}_{-\infty}dt'
  \left<T_{12}(\vec x,t)T_{12}(x',t')\right>_{ret}
  \label{5.111}
\end{eqnarray}
in the limit of $\varepsilon\to0$, $T_{\mu\nu}$ denote the components
of the energy momentum tensor. The formulas for the heat conductivity
$\chi$ and the bulk viscosity $\zeta$ are very similar, they mainly
differ in the components of $T_{\mu\nu}$. The expectation value
$\left<\ldots\right>_{ret}$ stands for a retarded Green's function in
{\it Minkowski} time. It is related to the thermal expectation value
$\left<\ldots\right>_0$ with respect to the thermal equilibrium
distribution according to
\begin{eqnarray}
  &&\left<T_{\mu\nu}(\vec x,t)T_{\mu\nu}(\vec x',t')\right>_{ret}~
  \nonumber\\
  &&\qquad
  =-i\theta(t'-t)
  \left<[T_{\mu\nu}(\vec x,t),T_{\mu\nu}(\vec x',t')]\right>_0.
  \label{5.112}
\end{eqnarray}
Eq.\ (\ref{5.112}) is the expression we are finally interested in. On
the other hand, expectation values which are easily accessible on the
lattice are thermal Green's functions in {\it Euclidean} time $\tau$.
They are expressed via thermal expectation values
\begin{equation}
  \left<T_{\mu\nu}(\vec x,\tau)T_{\mu\nu}(0,0)\right>_0,\quad
  0<\tau<\beta={1\over T}~.
  \label{5.113}
\end{equation}
The procedure of analytic continuation from Euclidean to Minkowski time is
standard and can be found in textbooks, see e.g. (Kadanoff and Baym, 1962).
Here we only sketch the logics.

In order to relate the retarded Green's function $G_R(\vec x,t)$ to the
thermal function $G_\beta(\vec x,\tau)$, one proceeds via their Fourier
transforms $\tilde G_R(\vec p,p_0)$ and $\tilde G_\beta(\vec p,\omega_n)$,
respectively. Both have almost identical representations in terms of the
spectral functions $\rho(\vec p,\omega)$. We have
\begin{equation}
  \tilde G_\beta(\vec p,\omega_n)=\int d\omega{{\rho(\vec p,\omega)}\over
    {i\omega_n-\omega}}~.
  \label{5.114}
\end{equation}
Here $\tilde G_\beta(\vec p,\omega_n)$ is only defined on a discrete
(but infinite) set of Matsubara frequencies $\omega'=\omega_n$ (due to
the periodicity in Euclidean time). Next one performs an analytic
continuation of the Fourier coefficients $\tilde G_\beta(\vec
p,\omega_n)$ for all (non real) $\omega'$. If there is no essential
singularity at $\omega'=\infty$, the unique continuation is
\begin{equation}
  \tilde G_\beta(\vec p,\omega')=\int d\omega{{\rho(\vec p,\omega)}\over
    {\omega'-\omega}}~.
  \label{5.115}
\end{equation}
The spectral function is then given by the discontinuity of $\tilde
G_\beta (\vec p,\omega')$ across the real axis. Once we know $\rho$,
the Fourier transform of the retarded Green's function can be
calculated according to
\begin{equation}
  \tilde G_R(\vec p,p_0)=\int d\omega{{\rho(\vec p,\omega)}\over{p_0+i
      \varepsilon-\omega}}
  \label{5.116}
\end{equation}
with the same $\rho$ as in Eq.\ (\ref{5.114}). The retarded Green's
function $G_R(\vec x,t)$ then follows as a Fourier transform of Eq.\ 
(\ref{5.116}). Thus in the ideal case we are done if we are able to
extract the spectral function from certain correlators in Euclidean
space. The spectral function is not only essential for analytic
continuation. It also allows an identification of the physical degrees
of freedom. In a given channel $C$, $\rho_c$ has a pole for each
physical degree of freedom with quantum numbers $C$, and a brunch cut
for every allowed multiparticle state with the same $C$.

Practically, it is very difficult to determine $\rho$ in lattice
calculations. $G_\beta(\vec x,\tau)$ is only known for a finite (and
even small) number of points $\tau_i$. Therefore $\tilde G_\beta(\vec
p,\omega_n)$ cannot be calculated as integral $\int^\beta d\tau
e^{i\omega_n\tau}$ (...), but the spectral representation of $\tilde
G_\beta(\vec p,\omega_n)$ is just the quantity of interest.

What can be done in spite of these difficulties has been shown by
Karsch and Wyld (1987). Karsch and Wyld have pointed out the
possibilities and limitations of calculating transport coefficients in
lattice gauge theory. The approximation for extracting the spectral
function is to make an ansatz for $\rho$ involving certain unknown
parameters, to calculate $\tilde G_\beta(\vec p,\omega_n)$ and
$G_\beta(\vec p,\tau)$, and to fix the unknown parameters by fitting
$G_\beta(\vec p,\tau)$ with Monte Carlo data.

Kubo-type formulas (cf.~Eq.\ (\ref{5.111})) have been used with an
energy momentum tensor of pure $SU(3)$ lattice gauge theory. The
ansatz for the zero-momentum spectral function is given as
\begin{equation}
  \rho(\omega)={{A(1-e^{-\beta m})}\over\pi}
  \left({\gamma\over{(m-\omega)^2
        +\gamma^2}}-{\gamma\over{(m+\omega)^2+\gamma^2}}\right).
  \label{5.117}
\end{equation}
It involves three parameters $A$, $\gamma$ and $m$. The spectral
function of a free field theory contains just two, $A$ and $m$, thus
$\gamma$ represents the interactions. Let us label the different
transport coefficients $\alpha_i$ with index $i$.  Correspondingly the
spectral functions $\rho_i(\omega)$ needed for the retarded Green's
functions $G_{R_i}(t)$ in the associated Kubo formulas depend on
parameters $A_i$, $\gamma_i$ and $m_i$. The transport coefficients
read in terms of $(A_i,\gamma_i,m_i)$ (Karsch and Wyld (1987))
\begin{equation}
  \alpha_i=
  2A_i(1-e^{-\beta m_i}){{2\gamma_im_i}\over{(\gamma_i^2+m_i^2)^2}}.
  \label{5.118}
\end{equation}
This expression can be easily reformulated in dimensionless quantities
which are directly measurable on the lattice: $m_\ell= m\cdot
a,\gamma/m,1/T=N_\tau a$ and $A_\ell=Aa^5$, $a$ being the lattice
constant. The goal is to determine $A_\ell, m_\ell$ and $\gamma/m$
from fits of $G_\beta^{(i)}(\tau)$. For example $G_\beta^{(i)}(\tau)$
for the shear viscosity is given in terms of space-space-off-diagonal
correlations according to
\begin{equation}
  G_\beta^{(\eta)}(\tau)=\sum_x 
  \left<T_{12}(0,0)T_{12}(\vec x,\tau)\right>~.
  \label{5.119}
\end{equation}
This quantity has been evaluated on an $8^3\times4$ lattice. Thus the
time distance $\tau$ in lattice units is restricted to three values 0,
1, 2, it cannot be larger than 2. In fact, only $G^{(\eta)} (\tau)$
could be measured up to distance 2. For the other correlators
$G^\zeta$ and $G^\chi$, the signal is already lost in the noise at
$\tau=2$.

The measurements have been performed above and below the critical
coupling of the deconfinement transition, which is at $6/g^2=5.68$ for
the given lattice size. The mass parameters $m_i$ are easier
accessible than $A_i$ and $\gamma_i$. They follow from the exponential
decay of the correlations. The interesting result is that the
correlations drop rapidly above and below $T_c$.  Especially above
$T_c$, there still seem to be massive modes in the system with a large
effective mass. Thus there is no indication for an abrupt change from
a free glueball gas below $T_c$ to a free gluon gas above $T_c$. Only
at low temperatures, the effective masses approach the lowest glueball
masses obtained from independent calculations (Berg and Billoire,
1983), in the vicinity of $T_c$ they differ. Yet one should keep in
mind the typical caveats for any lattice calculation. In particular,
the extraction of masses from short distance properties could lead to
artificially large values. For the interaction parameter $\gamma$,
Karsch and Wyld obtain an upper bound, resulting in an upper bound for
the shear viscosity $\eta$, $\eta/T^3<9.5$. The bound is in agreement
with analytic estimates in the collision time approximation of the
kinetic approach, cf. Hosoya and Kajantie (1985).

\medskip\noindent We summarize. The original objection to derive
nonequilibrium properties in Minkowski space from Euclidean field
theory in thermal equilibrium is divided into two steps. Based on a
linear response approach, first the nonequilibrium properties are
expressed in terms of equilibrium quantities in Minkowski space. The
second step is the analytic continuation between Minkowski and
Euclidean spacetime. The essential ingredient is the spectral function
$\rho$. A basic obstacle to derive it from Euclidean time correlators
measured on the lattice is the rather limited set of time points.  At
best, one can only guess an ansatz for the spectral function that is
compatible with the time correlator. In addition, the correlators
$G_\beta$ of the matrix elements of $T_{\mu\nu}$ turned out to drop so
rapidly that the small values for $G_\beta$ are lost in the noise for
large distances. Large distances are needed to project on lowest
excitation states, to justify the very ansatz for $G_\beta$. Thus the
values for $\tau$, the time distance in lattice units, are too small
both in number and size. From a realistic point of view, it seems to
be impossible to unfold the rich analytic structure of the spectral
function from a lattice approach. The main part of it will be hidden
in the Monte Carlo noise. Further discussions of transport
coefficients in the lattice approximation and comparisons to the
results of kinetic theory can be found in (Horsley and Schoenmaker,
1986) and (Schoenmaker and Horsley, 1988).

\bigskip Heat conductivity or viscosities are {\it relaxation
  phenomena} of the {\it energy-momentum} with respect to temperature
gradients, friction forces etc.  Their impact on the phase transition
concerns the phase conversion (cf. the preceding section). A direct
manifestation of the phase transition should be visible in a different
relaxation phenomenon associated with color degrees of freedom. It is
noteworthy that one of the most challenging properties of a quark
gluon plasma - its {\it color conductivity} - poses itself as a
transport phenomenon. In analogy to the Mott transition from an
electric insulator to a conductor, the transition from the hadronic
color insulator to the plasma color conductor should be manifest in
the transport coefficient of color conductivity, which drops to zero
either smoothly or suddenly, when $T_c$ is approached from above.

Selikhov and Gyulassy (1993) have found a surprisingly small value for
the color conductivity, which is determined by the ratio of the static
color electric and magnetic screening masses, $m_{e}$ and $m_{m}$,
respectively. (Their assumption is $m_{m}\ll m_{e}$. This assumption
is not borne out by lattice simulations in an $SU(2)$ and $SU(3)$
gauge theory, where $m_{m}\sim m_{e}$ is found from $T_{c}$ up to 2-3
$T_{c}$ (Boyd, 1995).)  The color conductivity coefficient is
calculated from the linear response of the system to a weak external
field. The reason for the rapid damping of collective color modes is
rapid color diffusion. Color diffusion is characterized by a specific
{\it relaxation time scale} $t_c$, which measures the precession
frequency of the color of a parton in a fluctuating background field.
This time scale turns out to be much smaller than a typical {\it
  momentum relaxation scale} $t_p$, which has been used in (classical)
non-abelian transport theory (see e.g. Heinz, 1986; Baym et al., 1990;
Eskola and Gyulassy, 1993). Its small value explains the poor
conductivity.

Color transport phenomena are still in their infancy and deserve
further attention in future work. One may think about constructing a
device for a `color amp\`eremeter'.

\bigskip We conclude with some remarks on a promising new direction in
transport theory, which is suited for applications in heavy-ion
collisions. It has been studied by Zhang and Wilets (1992) and is
based on a closed time-path Green's function technique (Schwinger,
1961; Keldysh, 1964 and 1965) combined with the loop expansion scheme
of Cornwall, Jackiw and Tomboulis (1974). The framework leads to
transport equations which include the generalized Boltzmann equations
in the special case of the quasiclassical limit. Its virtue is a
unified description of equilibrium and off-equilibrium systems. The
deviations from equilibrium in the considered off-equilibrium
situations are not restricted to a small size, whereas small
deviations are necessary for justifying a linear response approach.

In particular the framework is well suited for incorporating the phase
transition. The order parameter (here chosen as the chiral condensate
of the Nambu-Jona-Lasinio model) is a {\it dynamical variable} that
enters the generalized Boltzmann equations. The Boltzmann equations
reduce to the Vlasov equation in the Hartree approximation. The Vlasov
dynamics controls the collision dynamics, in particular medium effects
of chiral symmetry breaking in transverse flow. The transverse flow is
directly sensitive to the chiral condensate.

A further numerical elaboration of these microscopic transport
equations could lead to predictions of observable signatures for
chiral symmetry restoration.  Such predictions should be more reliable
(in the sense of more realistic) than any predictions based on
equilibrium thermodynamics. Ultimately full QCD should be substituted
for the NJL model used in the analysis of Zhang and Wilets. A less
ambitious extension would incorporate three rather than two flavors,
or vary the effective microscopic model for QCD to include gluonic
degrees of freedom as well.

\subsubsection{Dynamical universality and disoriented chiral condensates}
\label{sec:V.D.4}
Most of the proposed experimental consequences of a QCD transition at
finite temperature are based on the assumption of a first order
transition scenario.  As manifestation of a second order transition we
have discussed intermittent behavior in section \ref{sec:V.C.6}. In
this section we describe further consequences of a second order
transition under an additional assumption of rapid cooling from the
high to the low temperature phase. The underlying theory is the
3-dimensional $O(4)$-model of section \ref{sec:V.A},
which is supposed to
describe 2-flavor QCD in the vicinity of $T_c$ and predicts a second
order transition in the chiral limit. At criticality long range order
in an $O(4)$-model leads to large clusters of aligned spins, or
translated to the particle contents of the sigma model, large clusters
in three-dimensional space, where the ratio of charged to neutral
pions is fixed. The question arises as to whether the cluster size is
sufficiently large to have observable effects. Due to the finite quark
masses the second order of the transition is an idealization anyway,
(unless the masses take tricritical values). Although the up- and down
masses are only a few MeV, the induced zero temperature pion mass is
$\sim 135$~MeV and further increases with temperature (Rajagopal and
Wilczek, 1993).  As lightest particles in the spectrum the pions lead
to the largest correlation length. Thus the largest correlation length
is smaller than $(135~\hbox{MeV})^{-1}$ which is not large as compared
to $T_c^{-1}$ (if we use for $T_c \sim 140$~MeV for two flavors
(Bernard et al., 1992)). The number of pions arising from a typical
cluster of ``aligned $O(4)$-spins'' in isospin space can be estimated
from the energy which is stored in a typical correlation volume in the
vicinity of $T_c$ (Rajagopal and Wilczek, 1992). The estimate is based
on lattice results and says that not more than one or two pions belong
to the same correlation volume. Thus the hope of seeing clusters of
hundreds of pions with a fixed ratio of neutral to charged pions seems
to be gone. The upper bound on the maximal size of a correlation
volume and the energy stored in this volume are based on equilibrium
thermodynamics. They apply to a phase transition in heavy-ion
collisions, if the system remains always close to local equilibrium,
in particular during the phase transition.

The situation may drastically change, if the transition proceeds far
out of equilibrium via a {\it quench}. Let us recall what a quench is
in spin systems.  The system starts at high temperature in a
completely disordered state. Now the temperature is instantaneously
turned to zero. An example is a bar of iron which is plunged into ice
water rather than slowly cooled through the transition point. The high
temperature spin configuration is far out of equilibrium, when the
thermal fluctuations are suddenly switched off. The equilibrium
configuration at zero temperature would be a well ordered ground
state. In condensed matter physics it is well known what happens to
high temperature spin configuration after a quench (Hohenberg and
Halperin, 1977; Halperin et al., 1974). The system evolves according
to the zero temperature equations of motion, which determine the
universal dynamics of the long wavelength modes near $T_c$. Theories
which are described by the same equations of motion for the long
wavelength modes define a {\it dynamical universality class}. In
contrast to {\it static} universality classes dynamical universality
classes are not only specified by the space dimension and the number
of order parameter components, but also by those quantities which are
conserved. The important result of condensed matter physics (Bray,
1990; Newman et al., 1990) is that the linear size of a correlation
volume $L(t)$ grows with time according to $L(t)\sim t^p$, where $p$
depends on the characteristic parameters of the dynamical universality
class. It is crucial in view of the application to QCD that $L(t)$
does not depend on equilibrium correlation lengths. From the
digression to condensed matter physics let us return to QCD.

The interesting suggestion of Rajagopal and Wilczek (1993) is that QCD
belongs to the dynamical universality class of an
$O(4)$-anti-ferromagnet in the vicinity of $T_c$. The implications of
the hypothesis are discussed in the following.

The question of interest is the specific way, how different regions of
space relax to the ground state at low temperature. In case of an
$O(4)$-model the ground state at low $T$ is characterized by small
oscillations around the $\sigma$-direction in isospin space.  The
vacuum expectation value of the $O(4)$-field is
$\left<\phi\right>=\left<\sigma,\vec\pi\right>=(f_{\pi},\vec 0)$.  If
the cooling is adiabatically slow, it is likely that different
clusters of aligned spins relax independently to the ground state
configuration. In contrast, if the cooling proceeds via a quench, the
long-wavelength modes relax much more slowly to the equilibrium
configuration. This allows the formation of large clusters of
misaligned spins characterized by a `misaligned' condensate value
$\left<\vec\pi\right> \neq \vec 0$. Such regions of misaligned
condensates have been called {\it Disoriented Chiral Condensates}
(DCC) by Bjorken et al. (1993). Bjorken et al. had speculated about
events, in which the pion yield comes in clusters of pions aligned in
a single direction in isospin space over a large fraction of the
collision volume.  When these regions relax {\it coherently} to the
true ground state with $\left<\sigma\right>\neq 0$, a specific
radiation of pions is emitted. This radiation is proposed as
observable signature in heavy-ion collisions. The `misaligned' high
temperature vacuum is determined by oscillations around some fixed
direction $(\sigma,\vec{\pi})$ in classical field space. The radiated
pions correspond to a coherent configuration in that direction. Pions
originating from one such cluster of misaligned condensate have a
fixed ratio of charged to neutral pions. The probability distribution
for a given charge ratio $R$ is predicted as (Rajagopal and Wilczek,
1993)
\begin{equation}
  \rm{Prob}(R)={1 \over 2} R^{-1/2}
  \label{5.120}
\end{equation}
with $R=\sharp\pi^0/(\sharp(\pi^0+\pi^++\pi^-))$.

Correlations in space are not directly accessible in heavy-ion
collisions. Thus a further relation is needed to translate
correlations in space to measurable correlations in rapidity space.
Such a relation is provided by Bjorken's scaling ansatz. In Bjorken's
description different positions in space become different positions in
rapidity space. Given $z$ and $t$ we know the rapidity
\hbox{$y={1\over 2}\ln((1+v)/(1-v))$}, cf.~section \ref{sec:V.B.1}).

Thus an observable effect of a quenched QCD transition could be
constant ratios of neutral to total number of pions over large volumes
in rapidity space.  Fluctuations in pion multiplicities of this type
should be distinguished from fluctuations originating from exploding
plasma blobs, where the criterion to form a blob is a genuine
nucleation process rather than a fixed direction in isospin-space,
cf.~section~V.C.5.

Generic spin systems and an $O(4)$-model as effective description for
QCD share similarities in view of dynamical universality, but also
differ in several aspects, which we list in the following.

\begin{itemize}
\item At high temperature the $O(4)$-vector will not really fluctuate
  in arbitrary directions from cluster to cluster due to the `external
  field'. The non-vanishing pion mass singles out a preferred
  direction even at high temperatures assuming that the model has not
  lost its validity at high temperatures. The analysis of Bray(1990)
  and Newman et al. (1990) has to be extended to account for the
  effect of finite quark masses (see Rajagopal and Wilczek, 1993).
\item The succeeding quench as instantaneous cooling is certainly an
  idealization of the realistic cooling process. In Bjorken's scaling
  ansatz, $T$ drops according to $T=T_0(\tau_0/\tau)^{c_s^2}$, more
  slowly than the corresponding entropy density if transverse
  expansion is neglected.
\item Strictly speaking it is not possible to talk about a
  well-ordered ground state at $T=0$. The plasma is expanding until
  freeze-out, where the temperature is larger than zero. After
  freeze-out it does no longer make sense to call some parameter
  `temperature'.
\item Furthermore a description in terms of field configurations is
  not convenient when individual particle aspects become more
  relevant. In the very end detectors register single particles. Even
  before freeze-out the language of kinetic theory is more adequate.
\item The equations of motion at zero temperature have to be Lorentz
  invariant in the QCD case. The proposal is to use the $T=0$-$O(4)$
  linear sigma model to guarantee Lorentz invariant equations of
  motion.
\item The plasma is expanding, whereas the ferromagnet is static.
\end{itemize}

\bigskip In spite of the complications in heavy-ion collisions there
is at least a chance that a fast cooling amplifies the role of
long-wavelength pionic modes during the phase conversion of the quark
gluon plasma. However, for realistic coupling strengths in the sigma
model, their relaxation to the low $T$-equilibrium configuration goes
too fast to allow the formation of large regions of `misaligned'
vacuum (Gavin et al., 1994b). Gavin et al. performed a numerical
simulation of the zero-temperature $O(4)$-linear sigma model in the
presence of an external field $H$ with couplings in the Lagrangian
that induce physical values for $f_{\pi}=92.5$~[MeV],
$m_{\pi}=135$~[MeV] and $m_{\sigma}=600$~[MeV]. In particular the
quartic coupling is strong ($\lambda=20$). In contrast, simulations at
weak couplings corresponding to unrealistic light masses
($m_{\pi}=0.3$~[MeV] and $m_{\sigma}=1.8$~[MeV]) reproduce the
formation of domains with a field slowly varying about some nonzero
value $\left<\vec\pi\right>$. The `collapse' of these domains to the
true ground state goes along with a characteristic pion radiation,
where the number of pions in a typical clump could be large enough for
detection.

Under realistic conditions the cooling in heavy-ion collisions may not
be fast enough to sufficiently amplify the long-wavelength modes, and
the wavelengths of the lightest modes may not be long enough for the
formation of disoriented chiral condensates.
For further references to
disoriented chiral condensates see (Krzywicki, 1994; Huang, 1995),
(Rajagopal, 1995, and references therein).

A useful tool which helps in assessing the chance for observing the
predicted pion radiation, is an energy budget for the chiral and the
deconfinement transition (Wilczek, 1994). The energy density which is
stored in the chiral vacuum (associated with chiral condensation)
should be compared with the energy density of the quark gluon plasma
for temperatures $T\geq T_c$. The decimation of QCD's degrees of
freedom to pions and the description of the cooling process as a
quench are certainly idealizations. Within the idealized limits we
have seen the predictive power of the renormalization group approach.
As we have argued in sections \ref{sec:IV.A}.1 and 2, the basic open
question is, how important the heavier modes are in the relative to
the light modes associated with spontaneous chiral symmetry breaking.
One measure for their importance is their contribution to the total
energy density.  The accumulating effects of heavier modes can bury
the interesting (predictable) structures arising from the
singularities in the chiral 2-flavor transition. A quench does not
guarantee, but improves the chance for seeing relics of the chiral
transition.

Support from `experiment' is taken from a particular class of cosmic
ray events called Centauros (Rajagopal and Wilczek, 1993). 
In Centauros isospin invariance seems indeed to be violated. The 
number of charged pions strongly exceeds the number of neutral ones. 
For Centauros see e.g. Lattes et al., 1980.

%
%
\section{SUMMARY AND CONCLUSIONS}
\label{sec:VI}
The ultimate question of the order of the finite temperature QCD
transition for three colors, two light and one heavier flavor with
physical values for the quark masses is open. Closest to this physical
case come lattice simulations with two light and one heavier flavor in
the fermion schemes with staggered and Wilson fermions. At present
their results are in disagreement, predicting a crossover phenomenon
in the staggered formulation (section \ref{sec:III.C.5}) and a first
order transition in the Wilson formulation (section
\ref{sec:III.C.7}). Both results may be still artifacts of the strong
couplings inherent in the simulations.

Partial answers about the order of the QCD transitions are known in
limiting cases. The first (second) order of the deconfinement
transition in the $SU(3)$ ($SU(2)$) Yang Mills theory seems to be well
established (section \ref{sec:III.B.1} and \ref{sec:II.B.2}). General
agreement is also found about the second order of the chiral
transition in the limit of two massless flavors when dynamical
fermions are included. An open question is the universality class of
two-flavor QCD (whether it is $O(2)$, $O(4)$ or mean-field).
Consistent results in different fermion formulations are found for the
first order of the chiral transition in case of $N_{f}\geq 3$ flavors
with sufficiently light masses (section \ref{sec:III.C.5} and
\ref{sec:III.C.7}).

An interesting topic for future studies is a determination of
the critical strange quark mass $m_{s}^{*}$ at which the first order
of the chiral transition changes into second order. For larger values
of $m_{s}$ it stays of second order for $m_{u,d}=0$ and disappears
otherwise. The location of the `critical' phase boundary between mass
regions of first order transitions and crossover phenomena could tell
us, how far the physical quark masses are from `critical' quark
masses. It would be appealing both from a theoretical and an
experimental point of view, if they are close by. The powerful
renormalization group approach becomes then applicable for finite mass
scaling predictions. Remnants of a nearby second order transition
should lead to large correlation lengths and pronounced effects in
RHIC-experiments (section \ref{sec:III.C.5}, \ref{sec:III.C.7} and
\ref{sec:IV.A.4}).

Lattice calculations start from first principles as they
simulate QCD in its discretized version. They are {\it the}
nonperturbative tool for studying the phase transition region, where
many expansion schemes break down. But they are plagued with artifacts
from the finite volume, the finite lattice constant and unphysical
mass parameters. IR-artifacts may amount to a harmless rounding of an
infinite volume singularity or a misleading double-peak structure,
suggesting a first order transition though it is truly of second order
(section \ref{sec:II.A.2}). UV-artifacts may result in a `harmless'
change of critical exponents ($O(2)$ rather than $O(4)$) (section
\ref{sec:III.C.3}), but also in bulk transitions being a {\it pure}
lattice artifact without any impact on continuum physics (section
\ref{sec:III.C.4}).

Progress on the lattice is not only a question of powerful machines
(like special purpose computers). Progress in the past came also from
a better control of UV- and IR-artifacts with finite size and finite
mass scaling analysis, using refined criteria which have been
developed in statistical physics. Extrapolations to the infinite
volume and zero mass limits are unavoidable in numerical simulations.
Therefore a precise measurement of critical exponents is worth the
effort assuring that the right extrapolation formulas are used
(section \ref{sec:III.C.5}).

The earlier discrepancies between predictions from staggered and
Wilson fermions shrink. Wilson fermions seem to have a
chiral limit in the confinement phase with broken chiral symmetry at
low temperatures, if the lattice is large enough. The lattice
size must be even larger than for staggered fermions to reach the
continuum limit, too large for practicable simulations, unless
improved actions accelerate the approach to the continuum limit. Work
in this direction is in progress (section \ref{sec:III.C.7}).

\bigskip The main contributions from the Euclidean lattice formulation
of QCD to physical applications in RHIC-experiments should be reliable
predictions of {\it in-equilibrium} properties. To these belong the
equation of state, the tension of interfaces between coexisting
phases, and the velocity of sound. A derivation of the equation of
state under the inclusion of dynamical fermions is still in an
exploratory stage. The dependence on the lattice size has not yet been
analyzed, and the simulations are performed for 2 rather the 2+1
flavors, but the ingredients are fully nonperturbative (section
\ref{sec:III.C.6}). This is an important progress compared to earlier
calculations using perturbative and nonperturbative input data in an
inconsistent way (cf. also section \ref{sec:III.B.2}).

From results in the pure gauge sector it has become clear that strong
deviations from the phenomenological bag model equation of state must
be expected in the transition region. Deviations from the ideal gas
behavior are pronounced up to temperatures of the order of 2-3 $T_{c}$
(section \ref{sec:III.B.2}). So far many phenomenological applications
are based on the bag model equation of state. An update is called for,
once the lattice results are well established.

With a reliable calculation of interface tensions one could exclude
certain scenarios, which are under discussion for RHIC-experiments,
right from the beginning. Thus a small value for the interface tension
is incompatible with a strong supercooling scenario (section
\ref{sec:III.B.3}).

Studies of the phase structure of QCD in effective models in the
spacetime continuum may give useful hints about the underlying
physical mechanisms. Results in dual Ginzburg-Landau models suggest
that abelian $U(1)\times U(1)$ monopoles are the most important field
configurations in the confinement mechanism (section
\ref{sec:IV.B.2}) in the maximal abelian gauge. Confinement would then 
have an explanation as a
dual Meissner effect. The type of most relevant topological field 
configurations does depend on the gauge. 

{\it Nonequilibrium} aspects enter the evolution of the initially
created hot and dense `fireballs' in RHIC-experiments and the
transition dynamics during the phase conversion. Most of the
nonequilibrium approaches in the past have studied {\it small}
deviations from in-equilibrium properties. Field theoretical
descriptions of nonequilibrium phenomena are still in its infancy.
Recently they became increasingly important for the electroweak
transition in the early universe. These tools may have fruitful
applications to QCD as well. For example it is yet unclear, whether
the conversion from the plasma to the hadron phase proceeds via
nucleation or large domain coarsening (section \ref{sec:V.D.1} and
\ref{sec:V.D.2}).

For phenomenological discussions of QCD transitions in lab-experiments
theoretical concepts are sometimes missing. Thus it seems to be rather
difficult to derive the equilibration time of the hot initial state
from first principles. Formulas are missing for particle production
rates in media with {\it long-range} correlations. Such correlations
exist at $T_{c}$, if the transition is almost of second order. An
interpolation between nuclear structure functions and thermal
Bose-Einstein or Fermi-Dirac distributions would be desirable to close
the gap between the initial off-equilibrium and the final
in-equilibrium medium.

Altogether the judgement of the experimental situation is the
following. Even from a very conservative point of view, extended and
dense systems have been observed which display some collective
features. Some of them can be explained with a transient plasma, but
alternative mechanisms explain the observed features as well. To date
the question of the order of the transition has been of secondary
importance in heavy-ion experiments. Once the creation of a plasma
phase may be taken for granted, one can focus on substructures
exhibiting details of the transition dynamics (section \ref{sec:V}).

Our aim was to list the most important signatures which are at least
in principle sensitive to the transition, and in case it is hopeless
to see an effect to argue why it is hopeless (see e.g. section
\ref{sec:V.C.3}).

In Table~\ref{tab:9} we summarize signatures in various experimental
tools which may be indicative for a first or second order transition
and display alternative explanations in a separated column. Crosses
(x) stand for an affirmative answer which generally should be
taken with care, ``/'' stands for ``not known''. Explanations of
further notations can be found in the corresponding sections.

We conclude with the following {\it conjecture}. For physical quark
masses the deconfinement and the chiral transitions are replaced by
rapid crossover phenomena. The rapid change is seen in the energy and
entropy density over a temperature interval of the order of 10~MeV,
and in variations of the order parameter expectations values like the
Polyakov loop and the light quark condensate. It would be interesting
to investigate on the phenomenological side when a change in one of
these quantities is considered as rapid enough for producing
multiplicity fluctuations in rapidity space beyond the statistical
noise or clear deviations in dilepton production rates from the
standard rates without a crossover. Is a temperature interval of 10
MeV resolvable from an experimental point of view~? Finding a
quantitative answer to these questions is a challenge for further
studies.

%
%
\acknowledgments
The author wishes to thank J. Kapusta and E. Shuryak for an invitation
to the workshop on Strong Interactions at Finite Temperature at the
ITP in Santa Barbara in August 1993 and for discussions with the
participants of this workshop. I would also like to thank J. Cleymans,
A. Patk\'os, H.-J. Pirner, K. Redlich and H. Satz for discussions. In
particular I am indebted to G. Boyd, F. Karsch, and J. Stachel for
critical comments on parts of the manuscript.

%
%
\section*{REFERENCES}
\begin{description}
\item[] Abbott, T., {\it et al.}, E802 Collaboration, 1990, Phys.
  Rev.  Lett. {\bf 64}, 847.
  
\item[] Abbott, T., {\it et al.}, E802 Collaboration, 1991, Phys.
  Rev. Lett. {\bf 66}, 1567.
  
\item[] Abbott, T., {\it et al.}, E802 Collaboration, 1992, Phys.
  Rev. Lett.  {\bf 69}, 1020.
  
\item[] Abreu, P., {\it et al.}, DELPHI Collaboration, 1990, Phys.
  Lett.  B {\bf 247}, 137.
  
\item[] Aizu, K., 1970, Phys. Rev. B {\bf 2}, 754.
  
\item[] \AA kesson, T., {\it et al.}, NA34 Collaboration, 1990, Phys.
  Lett.  B {\bf 252}, 303.
  
\item[] Akemi, K., M. Fujisaki, M. Okuda, Y. Tago, Ph. de Forcrand,
  T. Hashimoto, S. Hioki, O. Miyamura, T. Takaishi, A. Nakamura, and
  I. O. Stamatescu, 1993, Phys. Rev. Lett., 3063.

\item[] Akiba, Y., {\it et al.}, E802 Collaboration, 1993, Phys. Rev.
  Lett.  {\bf 70}, 1059.
  
\item[] Albanese, M., {\it et al.}, APE Collaboration, 1987, Phys.
  Lett. B {\bf 192}, 163.
  
\item[] Alexopoulos, T., {\it et al.}, E735, 1990, Phys. Rev. Lett.
  {\bf64}, 991.
  
\item[] Altmeyer, R., K. D. Born, M. G\"ockeler, R. Horsley,
  E. Laermann, and G. Schierholz, 1993, Nucl. Phys. B {\bf 389}, 445.

\item[] Appelquist, T., and J. Carazzone, 1975, Phys. Rev. D {\bf
    11}, 2856.
  
\item[] Applegate, J.H., C.J. Hogan, and R.J. Sherrer, 1987, Phys.
  Rev. D {\bf 35}, 1151.
  
\item[] Applegate, J.H., 1991, Nucl. Phys. A {\bf 527}, 195c.
  
\item[] Asakawa, M., and K. Yazaki, 1989, Nucl. Phys. A {\bf 504},
  668.
  
\item[] Bacilieri, P., {\it et al.}, 1988, Phys. Rev. Lett. {\bf 61},
  1545.
  
\item[] Baechler, J., {\it et al.}, NA35 Collaboration, 1992, Nucl.
  Phys.  A {\bf 544}, 531c.
  
\item[] Bak, P., S. Krinsky, and D. Mukamel, 1976, Phys. Rev. Lett.
  {\bf 36}, 52.
  
\item[] Baker, G., B. Nickel, D. Meiron, 1978, Phys. Rev. D {\bf 17},
  1365.
  
  
\item[] Banerjee, B., and R. V. Gavai, 1992, Phys. Lett. B {\bf 293},
  157.

\item[] Banks, T., R. Myerson, and J. B. Kogut, 1977, Nucl. Phys., B {\bf
    129}, 493.

\item[] Banks, T., and A. Casher, 1980, Nucl. Phys. B {\bf 169}, 103.

\item[] Banks, T., and A. Ukawa, 1983, Nucl. Phys. B {\bf 225} [FS9],
  145.
  
\item[] Barber, M.N., 1983, {\it Phase Transitions and Critical
    Phenomena}, vol. 8, edited by C. Domb and J.L. Lebowitz (Academic,
  New York), p. 145.
  
\item[] Barbour, I., C. Davies, and Z. Sabeur, 1988, Nucl. Phys. B
  {\bf 215}, 567.
  
\item[] Barbour, I., and Z. Sabeur, 1990, Nucl. Phys. B {\bf 342},
  269.
  
\item[] Barbour, I., and A. Bell, 1992, Nucl. Phys. B {\bf 372}, 385.
  
\item[] Barbour, I., 1992, Nucl. Phys. B (Proc. Suppl.) {\bf 26}, 22.
  
\item[] Barrette, J. et al., E814 Collaboration, 1992, Phys. Rev. C
  {\bf 45}, 819; 1992, Phys. Rev. C {\bf 46}, 312.
  
\item[] Barrette, J., 1994, Nucl. Phys. A {\bf 566}, 411c.

\item[] Barrette, J. et al., E877 Collaboration, 1995, Phys. Rev. C
     {\bf 51}, 3309.

\item[] Batrouni, G.G., G.R. Katz, A.S. Kronfeld, G.P. Lepage, B.
  Svetitsky, and K.G. Wilson, 1985, Phys. Rev. D {\bf 32}, 2736.
  
\item[] Baumann, B., and B. Berg, 1985, Phys. Lett. B {\bf 164}, 131.
  
\item[] Baym, G., J.P. Blaizot, W. Czyz, B.L. Friman, and M. Soyeur,
  1983, Nucl. Phys. A {\bf 407}, 541.

\item[] Baym, G., 1984a, Nucl. Phys. A {\bf 418}, .

\item[] Baym, G., 1984b, Phys. Lett. {\bf 138}B, 18.
  
\item[] Baym, G., H. Monien, C.J. Pethick, and D.G. Ravenhall, 1990,
  Phys. Rev. Lett. {\bf 64}, 1867.
  
\item[] Berezin, F.A., 1966, {\it The Method of Second Quantization,
    Pure and Applied Physics}, Vol. 24 (Academic, New York).
  
\item[] Berg, B.A., and A. Billoire, 1983, Nucl. Phys. B {\bf 221},
  109.
  
\item[] Berg, B.A., and T. Neuhaus, 1991, Phys. Lett. B {\bf 267},
  249.
  
\item[] Berg, B.A., and T. Neuhaus, 1992, Phys. Rev. Lett. {\bf 68},
  9.
  
\item[] Bernard, C., 1974, Phys. Rev. D {\bf 9}, 3312.
  
\item[] Bernard, C., M.C. Ogilvie, M.A. DeGrand, C. DeTar, S.
  Gottlieb, A. Krasnitz, R.L. Sugar, and D. Touissant, 1992, Phys.
  Rev. D {\bf 45}, 3854.
    
\item[] Bernard, C., et al., 1995a, Nucl. Phys. B (Proc. Suppl.) {\bf
    42}, 448.

\item[] Bernard, C., T. Blum, T.A. DeGrand, C. DeTar, S. Gottlieb, U.Heller,
      L. K\"arkk\"ainen, K. Rummukainen, B. Sugar, and D. Toussaint,
      1995b, Nucl. Phys. B (Proc. Suppl.) {\bf 42}, 451.

\item[] Bernard, V., and U.G. Meissner, 1988, Phys. Rev. D {\bf 38},
  1551.
  
\item[] Bertsch, G., M. Gong, and M. Tohyama, 1988a, Phys. Rev. C
  {\bf 37}, 1896.
  
\item[] Bertsch, G., M. Gong, L. McLerran, V. Ruuskanen, and E.
  Sarkkinen, 1988b, Phys. Rev. D {\bf 37}, 1202.
  
\item[] Bertsch, G., 1989, Nucl. Phys. A {\bf 498}, 173c.

\item[] Bervillier, C., 1976, Phys. Rev. B {\bf 14}, 4964.
  
\item[] Bhanot, G., and M. Creutz, 1981, Phys. Rev. D {\bf 24}, 3212.
  
\item[] Bialas, A., and W. Czyz, 1984, Phys. Rev. D {\bf 30}, 2371;
  1985, Z. Phys.  C {\bf28}, 255.
  
\item[] Bialas, A., and J.P. Blaizot, 1985, Phys. Rev. D {\bf 32},
  2954.
  
\item[] Bialas, A., and R. Peschanski, 1986, Nucl. Phys. B {\bf 273},
  703.
  
\item[] Bialas, A., W. Czyz, A. Dyrek, and W. Florkowski, 1988, Nucl.
  Phys. B {\bf 296}, 611.
  
\item[] Bialas, A., and R. Peschanski, 1988, Nucl. Phys. B {\bf 308},
  857.
  
\item[] Bialas, A., and R.C. Hwa, 1991, Phys. Lett. B {\bf 253}, 436.
  
\item[] Bijnens, J., C. Bruno, and E. de Rafael, 1993, Nucl. Phys. B
  {\bf 390}, 501.
  
\item[] Bili\'c, N., K. Demeterfi, and B. Petersson, 1992a, Nucl.
  Phys. B {\bf 377}, 651.
  
\item[] Bili\'c, N., F. Karsch, and K. Redlich, 1992b, Phys. Rev.  D
  {\bf 45}, 3228.
  
\item[] Billoire, A., R. Lacaze, and A. Morel, 1990a, Nucl. Phys. B
  {\bf 340}, 542.
  
\item[] Billoire, A., S. Gupta, A. Irb\"ack, R. Lacaze, A. Morel, and
  B. Petersson, 1990b, Phys. Rev. B {\bf 42}, 6743.
  
\item[] Billoire, A., 1991, ``A Monte Carlo study of the Potts model
  in two and three dimensions,'' preprint SPHT/91/014.
  
\item[] Billoire, A., 1992, Int. J. Mod. Phys. C {\bf 3}, 913.
  
\item[] Billoire, A., R. Lacaze, and A. Morel, 1992, Nucl. Phys. B
  {\bf 370}, 773.
  
\item[] Binder, K., 1981, Z. Phys. B {\bf 43}, 119.
  
\item[] Binder, K., 1982, Phys. Rev. A {\bf 25}, 1699.
  
\item[] Binder, K., and D.P. Landau,1984, Phys. Rev. B {\bf 30},
  1477.
  
\item[] Binder, K., 1987, Rep. Prog. Phys. {\bf 50}, 783.
  
\item[] Bitar, K.M., {\it et al.}, 1991, Phys. Rev. D {\bf 43}, 2396.
  
\item[] Bitar, K.M., A.D. Kennedy, and P. Rossi, 1990, Phys. Lett. B
  {\bf 234}, 333.
  
\item[] Bitar, K.M., et al., 1993, Nucl. Phys. B (Proc. Suppl.)
  {bf 30}, 315.
  
\item[] Bjorken, J.D., 1983, Phys. Rev. D {\bf27}, 140.
  
\item[] Bjorken, J. D., K. L. Kowalski, and C. C. Taylor, 1993,
  Observing Disoriented Chiral Condensates, Proceedings of the
  workshop on Physics at Current Accelerators and the Supercollider,
  Argonne (1993), hep-ph/9309235, and SLAC preprint SLAC-PUB-6109
  (April 1993).

\item[] Blaizot, J.-P., and J.-Y. Ollitrault, 1989, Phys. Lett. B
  {\bf217}, 386.

\item[] Blaizot, J.-P., and J.-Y. Ollitrault, 1990, in {\it
    Quark-Gluon Plasma}, edited by R.C. Hwa (World Scientific,
  Singapore), p. 393.

\item[] Bl\"ote, H.W.J., and M.P. Nightingale, 1982, Physica {\bf
    112}A, 405.
  
\item[] Bloomer, M.A., {\it et al.} WA80 Collaboration, 1992, Nucl.
  Phys. A {\bf544}, 543c.

\item[] Blum, T., T. A. DeGrand, C. DeTar, S. Gottlieb, A. Hasenfratz,
  L. K\"arkk\"ainen, D. Touissant, R. L. Sugar, 1994, Phys. Rev. D
  {\bf 50}, 3377.

\item[] Blum, T., S. Gottlieb, L. K\"arkk\"ainen, and D. Toussaint,
  1995a, Phys. Rev. D {\bf 51}, 5153.

\item[] Blum, T., S. Gottlieb, L. K\"arkk\"ainen, and D. Toussaint,
  1995b, Nucl. Phys. B (Proc. Suppl.) {\bf 42}, 460.

\item[] Bochicchio, M., L. Maiani, G. Martinelli, G. Rossi, and M.
  Testa, 1985, Nucl. Phys. B {\bf 262}, 331.
  
\item[] Borgs, C., and R. Koteck\'y, 1990, J. Stat. Phys. {\bf 61},
  79.
  
\item[] Borgs, C., R. Koteck\'y, and S. Miracle-Sole, 1991, J. Stat.
  Phys.  {\bf 62}, 529.
  
\item[] Born, K. D., E. Laermann, N. Pirch, T.F. Walsh, and
   P.M. Zerwas, 1989, Phys. Rev. D {\bf 40}, 1653.

\item[] Borrill, J., and M. Gleiser, 1995, Phys. Rev. D {\bf 51},
  4111.
  
\item[] Boyanovsky, D., and C. Aragao de Carvalho, 1993, Phys. Rev. D
  {\bf 48}, 5850.

\item[] Boyanovsky, D., and C. Aragao de Carvalho, 1994, Phys. Rev. D
  {\bf 49}, 2769.

%
  
\item[] Boyd, G., J. Fingberg, F. Karsch, L. K\"arkk\"ainen, and B.
  Petersson, 1992, Nucl. Phys. B {\bf 376}, 199.

\item[] Boyd, G., 1995, private communication.
  
\item[] Boyd, G., J. Engels, F. Karsch, E. Laermann, C. Legeland,
  M. Lutgemeier, and B. Petersson, 1995, preprint BI-TP-95-23, and
  hep-lat-9506025.

\item[] Braaten, E., and R. D. Pisarski, 1990, Phys. Rev. Lett. {\bf
    64}, 1338; Phys. Rev. D {\bf 42}, 2156; Nucl. Phys. B {\bf 337},
  569; ibid. B {\bf 339}, 310.

\item[] Braaten, E., and R. D. Pisarski, 1992, Phys. Rev. D {\bf 45},
  1827; ibid. D {\bf 46}, 1829.
  
\item[] Brahm, D.E., and S.D.H. Hsu, 1991a and b, Caltech preprints
  CALT-68-1762, HUTP-91-A064, Dec. 1991, and CALT-68-1705,
  HUTP-91-A063, Dec. 1991.
  
\item[] Bravina, L., L.P. Csernai, P.  Levai, and D. Strottman, {\it
    Proc. 2nd Int. Workshop Heavy Ion Physics at the} AGS, MIT Jan.
  93, MITLNS-2158, p. 329.

\item[] Bray, A.J., 1990, Phys. Rev. B {\bf41}, 6724.
  
\item[] Br\'ezin, E., D.J. Wallace, and K.G. Wilson, 1973, Phys. Rev.
  B {\bf 7}, 232.

\item[] Br\'ezin, E., J.C. Le Guillou, and J. Zinn-Justin, 1976, in
  {\it Phase Transitions and Critical Phenomena}, Vol. 6, edited by C.
  Domb and M.S. Green (Academic Press, New York).

\item[] Brower, R., S. Huang, J. Potvin, and C. Rebbi, 1992, Phys.
  Rev. D {\bf 46}, 2703.

\item[] Brown, F.R., N.H. Christ, Y. Deng, M. Gao, and T.J. Woch,
  1988, Phys. Rev. Lett. {\bf 61}, 2058.

\item[] Brown, F.R., F.P. Butler, H. Chen, N.H. Christ, Z. Dong, W.
  Schaffer, L. Unger, and A. Vaccarino, 1990a, Phys. Rev. Lett. {\bf
    65}, 2491.

\item[] Brown, F.R., F.P. Butler, H. Chen, N.H. Christ, Z. Dong, W.
  Schaffer, L. Unger, and A. Vaccarino, 1990b, Phys. Lett. B {\bf251},
  181.
  
\item[] Brown, F.R., H. Chen, N.H. Christ, Z. Dong, R.D. Mawhinney, W.
  Schaffer, and A. Vaccarino, 1992, Phys. Rev. D {\bf 46}, 5655.

\item[] Brown, G.E., J. Stachel, and G.M. Welke, 1991, Phys. Lett. B
  {\bf 253}, 19.
  
\item[] Bunatian, G., and J. Wambach, 1994, Phys. Lett. B {\bf 336},
  290.
  
\item[] Buschbeck, B., 1989, in {\it Festschrift L\'eon Van Hove},
  edited by A. Giovanni and W. Kittel (World Scientific, Singapore).
  
\item[] Cahn, J.W., and J.E. Hilliard, 1958, J. Chem. Phys. {\bf28},
  258.
  
\item[] Callan, C., R. Dashen, and D.J. Gross, 1976, Phys. Lett. B
  {\bf 63}, 334.
  
\item[] Callan, C.G., Jr., and S. Coleman, 1977, Phys. Rev. D
  {\bf16}, 1762.
  
\item[] Callaway, D.J.E., and A. Rahman, 1982, Phys. Rev. Lett. {\bf
    49}, 613.
  
\item[] Callaway, D.J.E., and A. Rahman, 1983, Phys. Rev. D {\bf 28},
  1506.
  
\item[] Campbell, B.A., J. Ellis, and K.A. Olive, 1990, Nucl.  Phys.
  B {\bf 345}, 57.
  
\item[] Campi, X. and H. Krivine, 1995, preprint IPNO/TH 95-13, IPN
  Orsay, Paris.
  
\item[] Celik, T., J. Engels, and H. Satz, 1983a, Phys. Lett. B {\bf
    125}, 411.
  
\item[] Celik, T., J. Engels, and H. Satz, 1983b, Phys. Lett. B {\bf
    129}, 323.
  
\item[] Challa, M.S., D.P. Landau, and K. Binder, 1986, Phys. Rev. B
  {\bf 34}, 1841.
  
\item[] Chan, L.-H., and R. W. Haymaker, 1973, Phys. Rev. D {\bf 7},
  415.

\item[] Chernodub, M. N., M. I. Polikarpov, and A. I. Veselov, 1995,
  Phys. Lett. B {\bf 342}, 303.

\item[] Chodos, A., R.L. Jaffe, K. Johnson, C.B. Thorn, and V.F.
  Weisskopf, 1974, Phys. Rev. D {\bf9}, 3471.
  
\item[] Christ, N., 1992a, Nucl. Phys. A {\bf 544}, 81c.
  
\item[] Christ, N., 1992b, Nucl. Phys. B (Proc. Suppl.) {\bf26}, 217.
  
\item[] Clapeyron, B.P.E., 1834, J. l'Ecole Polytechnique (Paris)
  {\bf 14}, 153.
  
\item[] Clausius, R., 1850, Ann. Phys. (Leipzig) {\bf 79}, 368, 500.
  
\item[] Cleymans, J., R.V. Gavai, and E. Suhonen, 1986, Phys. Rep.
  {\bf 130}, 217.
  
\item[] Cleymans, J., J. Fingberg, and K. Redlich, 1987, Phys. Rev. D
  {\bf 35}, 2153.
  
\item[] Coleman, S., R. Jackiw, and H.D. Politzer, 1974, Phys. Rev. D
  {\bf 10}, 2491.

\item[] Cooper, F., and G. Frye, 1974, Phys. Rev.  D {\bf 10}, 186.
  
\item[] Cornwall, J., R. Jackiw, and E. Tomboulis, 1974, Phys. Rev. D
  {\bf 10}, 2428.

\item[] Creutz, M., 1981, Phys. Rev. Lett. {\bf 46}, 1441.
  
\item[] Creutz, M., L. Jacobs, and C. Rebbi, 1983, Phys. Reports {\bf
    93}, 201.
  
\item[] Csernai, L.P., and J. Kapusta, 1992a, Phys. Rev. Lett.
  {\bf 69}, 737.
  
\item[] Csernai, L.P., and J. Kapusta, 1992b, Phys. Rev. D {\bf 46},
  1379.
  
\item[] Damgaard, P.H., N. Kawamoto, and K. Shigemoto, 1986, Nucl.
  Phys. B {\bf 264}, 1.
  
\item[] Davies, C.T.H., and E.G. Klepfish, 1991, Phys. Lett. B {\bf
    256}, 68.
  
\item[] DeGrand, T.A., R.L. Jaffe, K. Johnson, and J. Kiskis, 1975,
  Phys. Rev.  D {\bf 12}, 2060.
  
\item[] DeGrand, T. A., and D. Touissant, 1980, Phys. Rev. D {\bf 22},
  2478.

\item[] DeGrand, T.A., and C.E. DeTar, 1983, Nucl. Phys. B {\bf 225},
  590.

\item[] DeGrand, T. A., and C. E. DeTar, 1986, Phys. Rev. D {\bf 34},
  2469.

\item[] Deng, Y., 1989, Nucl. Phys. B (Proc. Suppl.){\bf 9}, 334.

\item[] DeTar, C.E., 1985, Phys. Rev. D {\bf 32}, 276.
  
\item[] DeTar, C.E., and T. Kunihiro, 1989, Phys. Rev. D {\bf 39},
  2805.
  
\item[] Dong, Z., and N.H. Christ, 1992, Nucl. Phys. B (Proc. Suppl.)
  {\bf26}, 314.
  
\item[] Duane, S., 1985, Nucl. Phys. B {\bf 257} [FS14], 652.
  
\item[] Duane, S., A.D. Kennedy, B.J. Pendleton, and D. Roweth, 1987,
  Phys. Lett. B {\bf 195}, 216.
  
\item[] Dyakonov, D.I., and A.D. Mirlin, 1988, Phys. Lett B {\bf
    203}, 299.
  
\item[] Eggers, H.C., and J. Rafelski, 1991, Int. J. Mod. Phys. {\bf
    6}, 1067.

\item[] Ejiri, S., S. Kitahara, Y. Matsubara, and T. Suzuki,
    1995, Nucl. Phys. B (Proc. Suppl.) {\bf 42}, 481.

\item[] Engels, J., J. Fingberg, and M. Weber, 1990a, Nucl. Phys. B
  {\bf 332}, 737.
  
\item[] Engels, J., J. Fingberg, F. Karsch, D. Miller, and M. Weber,
  1990b, Phys. Lett. B {\bf 252}, 625.
  
\item[] Engels, J., F. Karsch, and K. Redlich, 1995, Nucl. Phys. B
  {\bf 435}, 295.

\item[] Eskola, K.J., and J. Lindfors, 1990, Z. Phys. C {\bf 46},
  141.
  
\item[] Eskola, K.J., and M. Gyulassy, 1993, Phys. Rev. C {\bf 47},
  2329.
  
\item[] Espinosa, J.R., M. Quiros and F. Zwirner, 1992, Phys. Lett. B
  {\bf 291}, 115.
  
\item[] F\"aldt, G. and B. Petersson, 1986, Nucl. Phys. B {\bf 264},
  197.
  
\item[] Ferenc, D., et al., NA35 collaboration, 1992, in {\it Proc.
    Quark Matter '91}, Nucl. Phys. A {\bf 544}, 531c.
  
\item[] Ferenc, D., 1992, private communication.
  
\item[] Ferrenberg, A.M., and R.H. Swendsen, 1988, Phys. Rev. Lett.
  {\bf 61}, 2635.
  
\item[] Ferrenberg, A.M., and R.H. Swendsen, 1989, Phys. Rev. Lett.
  {\bf 63}, 1658.
  
\item[] Feynman, R.P., 1981, Nucl. Phys. B {\bf 188}, 479.

\item[] Fisher, M.E., 1974, Rev. Mod. Phys. {\bf 46}, 597.
  
\item[] Fisher, M.E., and A.N. Berker, 1982, Phys. Rev. B {\bf 26},
  2507.

\item[] Flory, P.J., 1941a, J. Amer. Chem. Soc. {\bf 63}, 3083.

\item[] Flory, P.J., 1941b, J. Amer. Chem. Soc. {\bf 63}, 3091.
  
\item[] Foley, K.Y., E810 Collaboration, 1992, in {\it Proc. Quark
    Matter '91}, Nucl. Phys. A {\bf 544}, 335c.
  
\item[] Frei, Z., and A. Patk\'os, 1989, Phys. Lett. B {\bf 229},
  102.
  
\item[] Frei, Z., and A. Patk\'os, 1990, Phys. Lett. {\bf 247}B, 381.
  
\item[] Fucito, F., and A. Vulpiani, 1982, Phys. Lett. {\bf 89}A, 33.
  
\item[] Fukugita, M., and A. Ukawa, 1985, Phys. Rev. Lett. {\bf 55},
  1854.
  
\item[] Fukugita, M., S. Ohta, and A. Ukawa, 1986, Phys. Rev. Lett.
  {\bf 57}, 1974.
  
\item[] Fukugita, M., S. Ohta, and A. Ukawa, 1988, Phys. Rev. Lett.
  {\bf60}, 178.
  
\item[] Fukugita, M., M. Okawa, and A. Ukawa, 1989, Phys. Rev. Lett.
  {\bf 63}, 1768.
  
\item[] Fukugita, M., M. Okawa, and A. Ukawa, 1990a, Nucl. Phys. B
  {\bf337}, 181.
  
\item[] Fukugita, M., H. Mino, M. Okawa, and A. Ukawa, 1990b, Phys.
  Rev. Lett.  {\bf 65}, 816.
  
\item[] Fukugita, M., H. Mino, M. Okawa, and A. Ukawa, 1991, Nucl.
  Phys. B (Proc. Suppl.) {\bf 20}, 258.
  
\item[] Fulde, P., and H. Wagner, 1971, Phys. Rev. Lett. {\bf 27},
  1280.
  
\item[] Fuller, G.M., G.J. Mathews, and C.R. Alcock, 1988, Phys. Rev.
  D {\bf 37}, 1380.
  
\item[] Gasser, J., and H. Leutwyler, 1984, Ann. Phys. {\bf 158},
  142.
  
\item[] Gavai, R. V., and F. Karsch, 1985, Nucl. Phys. B {\bf 261},
  273.

\item[] Gavai, R.V., J. Potvin, and S. Sanielevici, 1988, Phys. Rev.
  D {\bf 37}, 1343.
  
\item[] Gavai, R.V., F. Karsch, and B. Petersson, 1989, Nucl. Phys. B
  {\bf 322}, 738.
  
\item[] Gavai, R.V., and F. Karsch, 1989, Phys. Lett. B {\bf 233}, 417.
  
\item[] Gavai, R.V., S. Gupta, A. Irb\"ack, F. Karsch, S. Meyer, B.
  Petersson, H. Satz, and H.W. Wyld, 1990, Phys. Lett. B {\bf 241},
  567.
  
\item[] Gavai, R. V., M. Grady, and M. Mathur, 1994, Nucl. Phys. B
  {\bf 423}, 123.

\item[] Gavai, R., and M. Mathur, 1995, Nucl. Phys. B (Proc. Suppl.)
  {\bf 42}, 490.

\item[] Gavai, R. V., J. Potvin, and S. Sanielevici, 1987,
  Phys. Rev. Lett. {\bf 58}, 2519,

\item[] Gavin, S., and M. Gyulassy, 1988, Phys. Lett. B {\bf 214},
  241.
  
\item[] Gavin, S., A. Gocksch, and R. D. Pisarski, 1994a, Phys. Rev. D
  {\bf 49}, R3079.
  
\item[] Gavin, S., A. Gocksch, and R. D. Pisarski, 1994b, Phys. Rev.
  Lett. {\bf 72}, 2143.

\item[] Geiger, K., 1992a, Phys. Rev. D {\bf 46}, 4965.

\item[] Geiger, K., 1992b, Phys. Rev. D {\bf 46}, 4986.
  
\item[] Geiger, K., and J. Kapusta, 1993, Phys. Rev. Lett {\bf 70},
  1920.

\item[] Gell-Mann, M., and M. Levy, 1960, Nuovo Cimento {\bf 16}, 705.
  
\item[] Gell-Mann, M., R.J. Oakes, and B. Renner, 1968, Phys. Rev.
  {\bf 175}, 2195.
  
\item[] Gerber, P., and H. Leutwyler, 1989, Nucl. Phys. B {\bf 321},
  387.

\item[] Gibbs, P., 1986, Phys. Lett. B {\bf 182}, 369.

\item[] Ginsparg, P., 1980, Nucl. Phys. B {\bf 170} [FS1], 388.
  
\item[] Gocksch, A., U. Heller, and P. Rossi, 1988, Phys. Lett. B {\bf
    205}, 334.

\item[] Gocksch, A., 1991, Phys. Rev. Lett. {\bf 67}, 1701.

\item[] Goldberg, H., 1983, Phys. Lett. B {\bf 131}, 133.
  
\item[] Gonzalez-Arroyo, A., and M. Okawa, 1987, Phys. Rev. Lett.
  {\bf 58}, 2165.
  
\item[] Gottlieb, S., W. Liu, R.L. Renken, R.L. Sugar, and D.
  Touissant, 1987a, Phys. Rev. D {\bf 35}, 2531; 1987b, ibid., 3972.
  
\item[] Gottlieb, S., W. Liu, D. Toussaint, R.L. Renken, and
   R.L. Sugar, 1987c, Phys. Rev. Lett. {\bf 59},
  1881.
  
\item[] Gottlieb, S., W. Liu, R. Renken, R. Sugar, and D. Touissant,
  1989, Phys. Rev.  Lett. {\bf 40}, 2389.
  
\item[] Gottlieb, S., W. Liu, R.L. Renken, R.L. Sugar, and D.
  Touissant, 1991, Phys. Rev. D {\bf 41}, 622.
  
\item[] Gottlieb, S., 1991, Nucl. Phys. B (Proc. Suppl.) {\bf 20},
  247.
  
\item[] Gottlieb, S., A. Krasnitz, U.M. Heller, A.D. Kennedy, 
   J.B. Kogut, R.L. Renken, D.K. Sinclair, R.L. Sugar, D. Toussaint, 
   and K.C. Wang, 1993a, Phys. Rev. D {\bf 47}, 3619.

\item[] Gottlieb, S., et al., 1993b,
  Nucl. Phys. B (Proc. Suppl.) {\bf 30}, 315.

\item[] Griffiths, R.B., 1965, J. Math. Phys. {\bf 8}, 478.
  
\item[] Gross, D.J., R.D. Pisarski, and L.G. Yoffe, 1981, Rev. Mod.
  Phys.  {\bf 53}, 43.
  
\item[] Grossmann, B., M.L. Laursen, T. Trappenberg, and U.J. Wiese,
  1992, Phys. Lett. B {\bf 293}, 175.
  
\item[] Grossmann, B., and M.L. Laursen, 1993, Nucl. Phys. B {\bf
    408}, 637.

\item[] Gupta, S., 1990, Phys. Lett. B {\bf248}, 453.

\item[] Gupta, S., and H. Satz, 1992, Phys. Lett. B {\bf283}, 439.

\item[] Gyulassy, M., 1984, Nucl. Phys. A {\bf418}, 59c.
  
\item[] Gyulassy, M., K.  Kajantie, H. Kurki-Suonio, and L. McLerran,
  1984, Nucl. Phys. B {\bf 237}, 477.

\item[] Hagedorn, R., 1983, Riv. Nuovo Cim. {\bf 6}, 1.
  
\item[] Hagedorn, R., 1985, in {\it Quark Matter'84}, Lecture Notes
  in Physics {\bf 211} (Springer, Berlin).
  
\item[] Hajdukovi\'c, D., and H. Satz, 1992, ``Does the
  one-dimensional Ising model show intermittency?,'' preprint
  CERN-TH-.6674/92 and BI-TP 92/43.
  
\item[] Halperin, B.I., P.C. Hohenberg, and S.-K. Ma, 1974, Phys.
  Rev. B {\bf10}, 139.
  
\item[] Hanbury-Brown, R., and R.Q. Twiss, 1954, Philos. Mag. {\bf
    45}, 633.
  
\item[] Hasenfratz, A., and D. Toussaint, 1992, Nucl. Phys. B {\bf
    371}, 539.
  
\item[] Hasenfratz, P., and F. Karsch, 1983, Phys. Lett. {\bf 125}B,
  308.
  
\item[] Hasenfratz, P., F. Karsch, and I.O. Stamatescu, 1983, Phys.
  Lett.  B {\bf 133}, 221.
  
\item[] Hatsuda, T., and T. Kunihiro, 1985, Progr. Theor. Phys. {\bf
    74}, 765.

\item[] Hatsuda, T., and T. Kunihiro, 1987, Phys. Lett. B {\bf 198}, 126.
  
\item[] Hatsuda, T., 1992, in {\it Proc. Quark Matter '91}, Nucl.
  Phys. A {\bf544}, 27c.
  
\item[] van Hecke, H., NA34 Collaboration, 1991, Nucl. Phys. A {\bf
    525}, 227c.
  
\item[] Heinz, U. 1986, Ann. Phys. (N.Y.) {\bf 168}, 148.
  
\item[] Heinz, U., and K.S. Lee, 1992, Nucl. Phys. A {\bf544}, 503c.
  
\item[] Herrmann, H.J., and F. Karsch, 1991, Eds., {\it Fermion
    Algorithms, Workshop on Fermion Algorithms} (World Scientific,
  Singapore).

\item[] Hioki, S., S. Kitahara, S. Kiura, Y. Matsubara, O. Miyamura,
    S. Ohno, and T. Suzuki, 1991, Phys. Lett. B {\bf 272}, 326.

\item[] Hoek, J., 1990, Nucl. Phys. B {\bf 339}, 732.
  
\item[] Hohenberg, P.C., A. Aharony, B.I. Halperin, and E. Siggia,
  1976, Phys. Rev. B {\bf 13}, 2986.
  
\item[] Hohenberg, P.C., and B.I. Halperin, 1977, Rev. Mod. Phys.
  {\bf49}, 435.
  
\item[] 't Hooft, G., 1981, Nucl. Phys. B {\bf 190}, 455.

\item[] Horsley, R., and W. Schoenmaker, 1986, Phys. Rev. Lett.
  {\bf57}, 2894.
  
\item[] Hosoya, A., M.-A. Sakagami, and M. Takao, 1984, Ann. Phys.
  {\bf154}, 229.
  
\item[] Hosoya, A., and K. Kajantie, 1985, Nucl. Phys. B {\bf250},
  666.

\item[] van Hove, L., 1982, Phys. Lett. {\bf118} B, 138.

\item[] van Hove, L., 1983, Z. Phys. C {\bf 21}, 93.

\item[] van Hove, L., 1985, Z. Phys. C {\bf 27}, 135.
  
\item[] Huang, S., J. Potvin, C. Rebbi, and S. Sanielevici, 1990,
  Phys. Rev.  D {\bf42}, 2864; 1991, errata D {\bf 43}, 2056.

\item[] Huang, Z., 1995, hep-ph/9501366.

\item[] Hubbard, J., 1959, Phys. Rev. Lett. {\bf 3}, 77.
  
\item[] H\"ufner, J., Y. Kurihara, and H.J. Pirner, 1988, Phys. Lett.
  B {\bf215}, 218.
  
\item[] Hwa, R.C., and K. Kajantie, 1985, Phys. Rev. D {\bf 32},
  1109.
  
\item[] Ilgenfritz, E.M., and E.V. Shuryak, 1989, Nucl. Phys. B {\bf
    319}, 511.

\item[] Imry, Y., 1980, Phys. Rev. B {\bf 21}, 2042.

\item[] Iwai, J., N. Suzuki, and Y. Takahashi, 1976, Prog. Theor. 
    Phys. {\bf 55}, 1537.

\item[] Iwai, J., I. Ohta, Y. Muraki, T. Ogata, S. Toyoda, R. Ihara,
    F. Fumuro, T. Yanagita, Y. Takahashi, and J.N. Capdevielle,
     1982, Nuovo Cimento {\bf 69}A, 295.
  
\item[] Iwasaki, Y., 1983, preprint UTHEP-118 (December 1983),
  unpublished.

\item[] Iwasaki, Y., Y. Tsuboi, and T. Yoshi\'e, 1989, Phys. Lett. B
  {\bf 220}, 602.
  
\item[] Iwasaki, Y., K. Kanaya, T. Yoshi\'e, T. Hoshino, T. Shirakawa,
  Y. Oyanagi, S. Ichii, and T. Kawai, 1991, Phys. Rev. Lett. {\bf 67},
  3343.

\item[] Iwasaki, Y., K. Kanaya, T. Yoshie, T. Hoshino, T. Shirakawa,
   Y. Oyanagi, S. Ichii, and T. Kawai, QCDPAX collaboration, 1991, Phys.
  Rev. Lett.  {\bf 67}, 3343.

\item[] Iwasaki, Y., K. Kanaya, T. Yoshi\'e, T. Hoshino, T. Shirakawa,
  Y. Oyanagi, S. Ichii, and T. Kawai, 1992, Phys. Rev. D {\bf 46},
  4657.

\item[] Iwasaki, Y., K. Kanaya, T. Yoshie, T. Hoshino, T. Shirakawa,
   Y. Oyanagi, S. Ichii, and T. Kawai, QCDPAX collaboration, 1992a, Phys.
  Rev. D {\bf 46}, 4657.

\item[] Iwasaki, Y., K. Kanaya, S. Sakai, and T. Yoshi\'e, 1992b,
  Phys. Rev.  Lett.  {\bf 69}, 21.

\item[] Iwasaki, Y., K. Kanaya, S. Sakai, and T. Yoshi\'e, 1993,
  Nucl. Phys. B (Proc. Suppl.) {\bf 30}, 327.

\item[] Iwasaki, Y., K. Kanaya, L. K\"arkk\"ainen, K. Rummukainen,
  and T. Yoshi\'e, 1994, Phys. Rev. D {\bf 49}, 3540.

\item[] Iwasaki, Y., K. Kanaya, S. Kaya, S. Sakai, and T. Yoshie,
  1995a, Nucl. Phys. B (Proc. Suppl.) {\bf 42}, 499.

\item[] Iwasaki, Y., K. Kanaya, S. Sakai, and T. Yoshie, 1995b,
  Nucl. Phys. B (Proc. Suppl.) {\bf 42}, 502.

\item[] Iwasaki, Y., 1995, Nucl. Phys. B (Proc. Suppl.) {\bf 42}, 96.

\item[] Jackson, J.L., 1960, Phys. Fluids {\bf 3}, 928.
  
\item[] Jacob, M., and H. Satz, 1982, Eds., {\it Quark Matter
    Formation and Heavy Ion Collisions} (World Scientific, Singapore).

\item[] Jacobsen, H.H., and D.J. Amit, 1981, Ann. Phys. (N.Y.) {\bf
    133}, 57.

\item[] Jain, V., 1993, Nucl. Phys. B {\bf 394}, 707.

\item[] Janke, W., 1992, Int. J. Mod. Phys. C {\bf 3}, 1137.

\item[] Johnson, K., 1975, Acta Phys. Polonica B {\bf 6}, 865.
  
\item[] Kadanoff, L.P., and G. Baym, 1962, {\it Quantum Statistical
    Mechanics} (Benjamin/Cummings Publ. Comp.), London).
  
\item[] K\"arkk\"ainen, L., P. Lacock, D.E. Miller, B. Petersson, and
  T. Reisz, 1992, Phys. Lett. B. {\bf 282}, 121.

\item[] K\"arkk\"ainen, L., P. Lacock, B. Petersson, and T. Reisz,
  1993, Nucl. Phys. B {\bf 395}, 733.

\item[] K\"arkk\"ainen, L., P. Lacock, D. E. Miller, B. Petersson, and
  T. Reisz, 1994, Nucl. Phys. B {\bf 418}, 3.

\item[] Kajantie, K., and H.I. Miettinen, 1981, Z. Phys. C {\bf 9}, 341.
  
\item[] Kajantie, K., R. Raitio, and P.V. Ruuskanen, 1983, Nucl.
  Phys. B {\bf 222}, 152.
  
\item[] Kajantie, K., and T. Matsui, 1985, Phys. Lett. B {\bf 164},
  373.
  
\item[] Kajantie, K., and H. Kurki-Suonio, 1986, Phys. Rev. D {\bf
    34}, 1719.
  
\item[] Kajantie, K., L. K\"arkk\"ainen, and K. Rummukainen, 1990,
  Nucl. Phys. B {\bf 333}, 100.
  
\item[] Kajantie, K., L. K\"arkk\"ainen, and K. Rummukainen, 1991,
  Nucl. Phys. B {\bf 357}, 693.

\item[] Kajantie, K., 1992, Int. J. Mod. Phys. C {\bf 3}, 1137.

\item[] van Kampen, N.G., 1964, Phys. Rev. A {\bf 135}, 362.

\item[] Kanki, T., 1988, Osaka University preprint  OSGE 86-24.

\item[] Kapusta, J., 1979, Nucl. Phys. {\bf148}B, 461.

\item[] Kapusta, J., and A. Mekjian, 1986, Phys. Rev. D {\bf 33}, 1304.

\item[] Kapusta, J., and C. Gale, 1987, Phys. Rev. C {\bf35}, 2107.
  
\item[] Kapusta, J., 1989, in {\it Finite Temperature Field Theory}
  (Cambridge University Press, Cambridge).
  
\item[] Kapusta, J., P. Lichard, and D. Seibert, 1991, Phys. Rev. D
  {\bf44}, 2774; 1992, Nucl. Phys. A {\bf544}, 485c.
  
\item[] Kapusta, J., L. McLerran, and D.K. Srivastawa, 1992, Phys.
  Lett. B {\bf 283}, 145.

\item[] Karsch, F., 1982, Nucl. Phys. B {\bf 205}[FS5], 285.

\item[] Karsch, F., and H.W. Wyld, 1987, Phys. Rev. D {\bf35}, 2518.
  
\item[] Karsch, F., and K.-H. M\"utter, 1989, Nucl. Phys. B {\bf313},
  541.

\item[] Karsch, F., 1990, Nucl. Phys. B (Proc. Suppl.) {\bf 15}, 157.
  
\item[] Karsch, F., 1992, private communication.
  
\item[] Karsch, F., 1993, in Proc. QCD 20 Years Later, ed. by P. M.
  Zerwas and H. A. Kastrup, Vol. 2 (World Scientific, Singapore, 1993)
  p. 717 and references therein.
  
\item[] Karsch, F., 1994, Phys. Rev. D {\bf 49}, 3791.

\item[] Karsch, F., 1995, Nucl. Phys. A {\bf 590}, 367C.

\item[] Karsch, F., and E. Laermann, 1994, Phys. Rev. D {\bf 50},
  6954.

\item[] Kataja, M., J. Letessier, P.V. Ruuskanen, and A. Tounsi,
  1992, Z.  Phys. C {\bf 55}, 153.

\item[] Keldysk, L.V., 1964, Zh. Eksp. Teor. Fiz. {\bf 47}, 1515; 
1965, Sov. Phys. JETP {\bf 20}, 1018.

\item[] Kennedy, A.D., J. Kuti, S. Meyer, and B.J. Pendleton,
     1985, Phys. Rev. Lett. {\bf 54}, 87.
  
\item[] Kitahara, S., Y. Matsubara, and T. Suzuki, 1995,
  Progr. Theor. Phys. {\bf 93}, 1; Nucl. Phys. B (Proc. Suppl.) {\bf
    42}, 511.

\item[] Klaetke, J.-U. and K.-H. M\"utter, 1990, Nucl. Phys. B {\bf
    342}, 764.
  
\item[] Klimt, S., M. Lutz, and W. Weise, 1990, Phys. Lett. B. {\bf
    249}, 386.

\item[] Ko, C.M., and M. Asakawa, 1994, Nucl. Phys. A {\bf 566}, 447c.

\item[] Ko, C.M., 1995, private communication.
  
\item[] Koch, P., B. M\"uller, and J. Rafelski, 1986a, Z. Phys. A
  {\bf 324}, 453.
  
\item[] Koch, P., B. M\"uller, and J. Rafelski, 1986b, Phys.Rep. {\bf
    142}, 167.
  
\item[] Koci\'c, A., J. B. Kogut, and M.-P. Lombardo, 1993, Nucl.
  Phys. B {\bf 398}, 376.
  
\item[] Koci\'c, A., and J. B. Kogut, 1995,
     Phys. Rev. Lett. {\bf 74}, 3109.
  
\item[] Kogut, J., and L. Susskind, 1975, Phys. Rev. D {\bf 11}, 395.

\item[] Kogut, J.B., 1983, Rev. Mod. Phys. {\bf 55}, 775.
  
\item[] Kogut, J.B., H. Matsuoka, M. Stone, H.W. Wyld, S. Shenker, J.
  Shigemitsu, and D.K. Sinclair, 1983, Phys. Rev. Lett. {\bf 51}, 869.
  
\item[] Kogut, J.B., J. Polonyi, H.W. Wyld, and D.K. Sinclair, 1985,
  Phys.  Rev. Lett. {\bf 54}, 1475.
  
\item[] Kogut, J.B., and D.K. Sinclair, 1988a, Nucl. Phys. B {\bf295}
  [FS21], 465.
  
\item[] Kogut, J.B., and D.K. Sinclair, 1988b, Phys. Rev. Lett. {\bf
    60}, 1250.
  
\item[] Kogut, J.B., and D.K. Sinclair, 1989, Phys. Lett. B {\bf
    229}, 107.

\item[] Kronfeld, A. S., M.L. Laursen, G. Schierholz, and U.J. Wiese,
   1987a, Phys. Lett.{\bf 198} B, 516;
 
\item[] Kronfeld, A.S., G. Schierholz, and U.J. Wiese, 1987b, 
  Nucl. Phys. B {\bf 293}, 461.

\item[] Krzywicki, A., 1994, Disoriented Chiral Condensates, talk at
  29th Rencontres de Moriond, LPTHE-ORSAY-94-43, and hep-ph/9405244.

\item[] Kunihiro, T., 1989, Phys. Lett. B {\bf 219}, 363.

\item[] Kunihiro, T., 1991, Nucl. Phys. B {\bf 351}, 593.
  
\item[] Kusaka, K., and W. Weise, 1992, Z. Phys. A {\bf 343}, 229.
  
\item[] Laermann, E., G. Boyd, J. Engels, F. Karsch, C. Legeland, 
  M. L\"utgemeier, and B. Petersson, 1995, Nucl. Phys. B 
  (Proc. Suppl.) {\bf 42}, 120.

\item[] Landau, L.D., and E.M. Lifshitz, 1958, {\it Statistical
    Physics} (Pergamon, Oxford).
  
\item[] Landau, L.D., and E.M. Lifshitz, 1959, {\it Course of
    Theoretical Physics Vol. 6, Fluid Mechanics} (Pergamon, London).

\item[] Landsman, N.P., 1989, Nucl. Phys. B {\bf 322}, 498.

\item[] Langer, J.S., 1969, Ann. Phys. {\bf54}, 258.

\item[] Langer, J.S., and L.A. Turski, 1973, Phys. Rev. A {\bf8}, 3230.
  
\item[] Lattes, C.M.G., Y. Fujimoto, and S. Hasegawa, 1980, Phys.
  Rep.  {\bf65}, 151.

\item[] Lee, J., and J.M. Kosterlitz, 1991, Phys. Rev. B {\bf 43}, 3265.
  
\item[] van Leeuwen, J.M.J., 1975, in {\it Fundamental Problems in
    Statistical Mechanics}, Vol. 3, E.D.G. Cohen (ed.), North Holland,
  Amsterdam, p.81.
  
\item[] Leutwyler, H., 1988, Nucl. Phys. B (Proc. Suppl.) {\bf 4},
  248.
  
\item[] Leutwyler, H., 1992, Phys. Lett. B {\bf 284}, 106.

\item[] Lipatov, A., 1977, JETP72, 411.

\item[] Ludlam, T., and N.P. Samios, 1988, Z. Phys. C {\bf38}, 353.
  
\item[] Luttinger, J. M., and J. C. Ward, 1960, Phys. Rev. {\bf 118},
  1417.
  
\item[] Lutz, M., S. Klimt, and W. Weise, 1992, Nucl. Phys. A {\bf
    542}, 521.
  
\item[] Ma, S.-K., 1976, {\it Modern Theory of Phase Transitions}
  (Reading, MA: Benjamin).
  
\item[] Makhlin, A.N., and Yu.M. Sinyukov, 1988, Z. Phys. C {\bf 39},
  69.
  
\item[] Malmstr\"om, G., and D. Geldart, 1980, Phys. Rev. B {\bf 21},
  1133.
  
\item[] Manohar, A., and H. Georgi, 1984, Nucl. Phys. B {\bf 234},
  189.

\item[] Margetis, S. et al., NA49 Collaboration, 1995, Nucl. Phys. A
      {\bf 590}, 335c.
   
\item[] Matsubara, Y., S. Ilyar, T. Okude, K. Yotsuji, and T. Suzuki,
    1995, Nucl. Phys. B (Proc. Suppl.) {\bf 42}, 529.

\item[] Matsubara, Y., S. Ejiri, and T. Suzuki, 1994, Nucl. Phys. B
  (Proc. Suppl.) {\bf 34}, 176.

\item[] Matsui, T., and H. Satz, 1986, Phys. Lett. {\bf 178}B, 416.

\item[] Mawhinney, R.D., 1993, Nucl. Phys. B (Proc. Suppl.) {\bf 30},
  331.
  
\item[] McLerran, L.D., and T. Toimela, 1985, Phys. Rev. D {\bf 31},
  545.

\item[] McLerran, L.D., 1986, Rev. Mod. Phys. {\bf 58}, 1021.
  
\item[] Metropolis, N.A., M.N. Rosenbluth, A.H. Rosenbluth, E.
  Teller, and J. Teller, 1953, J. Chem. Phys. {\bf 21}, 1087.
  
\item[] Meyer, B.S., C.R. Alcock, G.J. Mathews, and G.M. Fuller,
  1991, Phys.  Rev. D {\bf 43}, 1079.

\item[] Meyer-Ortmanns, H., 1989, in {\it Proceedings of the XXIII.
    International Symposium on the Theory of Elementary Particles},
  edited by E. Wieczorek (Ahrenshoop, GDR).

\item[] Meyer-Ortmanns, H., H.-J. Pirner, and A. Patk\'os, 1992,
  Phys. Lett.  B {\bf 295}, 255.

\item[] Meyer-Ortmanns, H., and B.-J. Schaefer, 1995, submitted to
  Phys. Rev. D.
  
\item[] Miyamura, O., 1995, Nucl. Phys. B (Proc. Suppl.) {\bf 42},
  538.
  
\item[] Michel, L., 1980, Rev. Mod. Phys. {\bf 52}, 617.
  
\item[] Miller, J.C., and O. Pantano, 1989, Phys. Rev. D {\bf40},
  1789; Phys. Rev. D {\bf42}, 3334.

\item[] Miyamura, O., 1995, Nucl. Phys. B (Proc. Suppl.) {\bf 42},
  538.
  
\item[] Montvay, I., and G. M\"unster, 1994, {\it Quantum Fields on
    the Lattice}, Cambridge Monographs on Mathematical Physics
  (Cambridge University Press, Cambridge).

\item[] Moriarty, K.J.M., 1981, Phys. Lett. {\bf106B}, 130.

\item[] Mott, N.F., 1968, Rev. Mod. Phys. {\bf40}, 677.
  
\item[] M\"uller, B., and J. Rafelski, 1982, Phys. Rev. Lett. {\bf
    48}, 1066.
  
\item[] Narison, S., 1989, {\it QCD Spectral Sum Rules}, Lecture
  Notes on Physics, Vol. 26 (World Scientific, Singapore).

\item[] van Neerven, W.L., 1995, Int. J. Mod. Phys. A {\bf 10},2921.
  
\item[] Newman, T.J., A.J. Bray, and M.A. Moore, 1990, Phys. Rev. B
  {\bf42}, 4514.
  
\item[] Nielsen, H.B., and M. Ninomiya, 1981a, Nucl. Phys. B {\bf
    185}, 20.
  
\item[] Nielsen, H.B., and M. Ninomiya, 1981b, Phys. Lett. {\bf
    105}B, 219.
  
\item[] Nienhuis, B., and M. Nauenberg, 1975, Phys. Rev. Lett. {\bf
    35}, 477.
  
\item[] Nowak, M.A., J.J.M. Verbaarschot, and I. Zahed, 1989, Nucl.
  Phys. B {\bf 325}, 581.

\item[] Ochs, W., and J. Wosiek, 1988, Phys. Lett. B {\bf 214}, 617.

\item[] Ohta, S., and S. Kim, 1991, Phys. Rev. D {\bf 44}, 504.

\item[] Parisi, G., and Y.-S. Wu, 1981, Sci. Sin. {\bf 24}, 483.

\item[] Patel, A., 1984a, Nucl. Phys. B {\bf 243}, 411.

\item[] Patel, A., 1984b, Phys. Lett. B {\bf 139}, 394.

\item[] Paterson, A.J., 1981, Nucl. Phys. B {\bf 190} [FS3], 188.

\item[] Patk\'os, A., 1991, Nucl. Phys. B {\bf 365}, 243.
  
\item[] Peitzmann, T., {\it et al.}, WA80 collaboration, 1992, Z.
  Phys. C {\bf 53}, 225.
  
\item[] Peitzmann, T., {\it et al.}, WA80 collaboration, 1993a, {\it
    Proceedings on Physics and Astrophysics of the Quark-Gluon
    Plasma}, Calcutta, India, Jan. 1993.
  
\item[] Peitzmann, T., {\it et al.}, WA80 collaboration, 1993b, Z.
  Phys.  C {\bf 59}, 127.
  
\item[] Petcher, D.N., and D.H. Weingarten, 1981, Phys. Lett. {\bf
    99}B, 333.
  
\item[] Pfeuty, P., and G. Toulouse, 1977, {\it Introduction to the
    Renormalization Group and Critical Phenomena} (Wiley, New York).
  
\item[] Pisarski, R.D., and F. Wilczek, 1984, Phys. Rev. D {\bf 29},
  338.
  
\item[] Polonyi, J., and K. Szlachanyi, 1982, Phys. Lett. B {\bf
    110}, 395.
  
\item[] Polyakov, A. M., 1975, Phys. Lett. B {\bf 59}, 82.

\item[] Potvin, J., and C. Rebbi, 1989, Phys. Rev. Lett. {\bf 62},
  3062.
  
\item[] Potvin, J., and C. Rebbi, 1991, Nucl. Phys. B (Proc. Suppl.)
  {\bf 20}, 317.

\item[] Pratt, S., 1984, Phys. Rev. Lett. {\bf 53}, 1219.

\item[] Pratt, S., 1986, Phys. Rev. D {\bf 33}, 1314.
  
\item[] Privman, V., and M.E. Fisher, 1983, J. Stat. Phys. {\bf 33},
  385.

\item[] Rafelski, J., 1982, Phys. Rep. {\bf 88}, 331.
  
\item[] Rafelski, J., and B. M\"uller, 1982, Phys. Rev. Lett. {\bf
    48}, 1066.

\item[] Rafelski, J., 1991, Phys. Lett. {\bf262}B, 333.


\item[] Rajagopal, K., and F. Wilczek, 1993, Nucl. Phys. B {\bf 399},
  395; ibid. B {\bf 404}, 577.

\item Rajagopal, K., 1995, The chiral phase transition in QCD:
  Critical phenomena and longwavelength pion oscillations, to appear
  in Quark-Gluon Plasma 2, edited by R. Hwa, World Scientific, 1995,
  and hep-ph- 9504310.

\item[] Redlich, K., and H. Satz, 1986, Phys. Rev. D {\bf33}, 3747.
  
\item[] Redlich, K., H. Satz, and J. Cleymans, 1992, Nucl. Phys. A
  {\bf544}, 509c.
  
\item[] Redlich, K., R. Baier, H. Nakkagawa, and A. Ni\'egawa, 1992,
  Nucl. Phys. A {\bf544}, 511c.
  
\item[] Reisz, T., and B. Petersson, 1991, Nucl. Phys. B {\bf 353},
  757.

\item[] Reisz, T., 1992, Z. Phys. C {\bf 53}, 169.

\item[] Rothe, H., 1992, {\it Lattice Gauge Theories} (World
  Scientifique, Singapore).
  
\item[] Rummukainen, K., 1995, Nucl. Phys. B (Proc. Suppl.) {\bf 42},
  505.

\item[] Ruuskanen, P.V., 1990, in {\it Quark-Gluon Plasma}, edited by
  R. Hwa (World Scientific, Singapore), p. 519.
  
\item[] Ruuskanen, P.V., 1992, Nucl. Phys. A {\bf544}, 169c.
  
\item[] Sarabura, M., et al., NA44 Collaboration, 1992, {\it Proc.
    Quark Matter '91}, Nucl. Phys. A {\bf544}, 125c.

\item[] Satz, H., 1989, Nucl. Phys. B {\bf326}, 613.
  
\item[] Satz, H., 1990a, in {\it Proceedings Large Hadron Collider
    Workshop}, edited by G. Jarlskog and D. Rein, Vol. 1, p. 188.

\item[] Satz, H., 1990b, Phys. Lett. B {\bf242}, 107.

\item[] Schechter, J., 1980, Phys. Rev. D {\bf 21}, 3393.
  
\item[] Schmidt, H.R., and J. Schukraft, 1993, J. Phys. G {\bf 19},
  1705.

\item[] Schmidt, M., G. R\"opke, and H. Schulz, 1990, Ann. Phys. {\bf
    202}, 57.
  
\item[] Schoenmaker, W., and R. Horsley, 1988, Nucl. Phys. B (Proc.
  Suppl.) {\bf 4}, 318.
  
\item[] Schramm, D.N., B. Fields, and D. Thomas, 1992, Nucl. Phys. A
  {\bf 544}, 267c.

\item[] Schwinger, J., 1961, J. Math. Phys. {\bf 2}, 407.  

\item[] Seibert, D., 1990, Phys. Rev. D. {\bf 41}, 3381.
  
\item[] Seiler, E., 1982, {\it Gauge Theories as a Problem of
    Constructive Quantum Field Theory and Statistical Mechanics},
  Springer-Verlag, Heidelberg.
  
\item[] Selikhov, A.V., and M. Gyulassy, 1993, Phys. Lett. B {\bf
    316}, 373.
  
\item[] Seyboth, P., {\it et al.}, NA35 Collaboration, 1992, in {\it
    Proc. Quark Matter '91}, Nucl. Phys. A {\bf 544}, 293c.
  
\item[] Shaposhnikov, M.E., 1992, Phys. Lett B {\bf 277}, 324;
  Erratum-ibid. B {\bf282}, 483.

\item[] Shen, Y., 1993, Phys. Lett. B {\bf 315}, 146.

\item[] Shiba, H., and T. Suzuki, 1994, Phys. Lett. B {\bf 333}, 461;
  1994, Nucl. Phys. B (Proc. Suppl.) {\bf 20}, 236.

\item[] Shiba, H., and T. Suzuki, 1995, Phys. Lett. B {\bf 343}, 315.

\item[] Shuryak, E.V., 1978, Phys. Lett. {\bf 78}B,  150.

\item[] Shuryak, E.V., 1979, Phys. Rep. {\bf 61}, 71.

\item[] Shuryak, E.V., 1984, Phys. Rep. {\bf 115}, 151.
  
\item[] Shuryak, E.V., 1988, {\it The QCD Vacuum, Hadrons and the
    Superdense Matter} (World Scientific, Singapore).

\item[] Shuryak, E.V., 1992,  Phys. Rev. Lett. {\bf 68}, 3270.
  
\item[] Shuryak, E.V., and L. Xiong, 1993, Phys. Rev. Lett. {\bf 70},
  2241.
  
\item[] Sorge, H., H. St\"ocker, and W. Greiner, 1989, Nucl. Phys. A
  {\bf 498}, 567c.

\item[] Stachel, J., and G.R. Young, 1992, Annu. Rev. Nucl. Part. Sci.
   {\bf 42}, 537.
  
\item[] Stachel, J., E814 collaboration, 1994, Nucl. Phys. A {\bf
    566}, 183c.

\item[] Stack, J. D., and R. J. Wensley, 1992, Nucl. Phys. B {\bf
    371}, 597.

\item[] Stanley, H.E., 1971, {\it An Introduction to Phase
    Transitions and Critical Phenomena} (Oxford University Press,
  Oxford).

\item[] Stratonovich, R.L., 1958, Sov. Phys.  Dokl. {\bf 2}, 416.

\item[] Susskind, L., 1977, Phys. Rev. D {\bf 16}, 3031.

\item[] Suzuki, T., 1993, Nucl. Phys. B (Proc. Suppl.) {\bf 30}, 176.

\item[] Suzuki, T., and I. Yotsuyanagi, 1990, Phys. Rev. D {\bf 42},
  4257.
  
\item[] Suzuki, T., and I. Yotsuyanagi, 1991, Nucl. Phys. B
  (Proc. Suppl.) {\bf 20}, 236.
  
\item[] Suzuki, T., and H. Shiba, 1995, Nucl. Phys. B (Proc. Suppl.)
  {\bf 42}, 282.

\item[] Svetitsky, B., and L.G. Yaffe, 1982a, Phys. Rev. D {\bf 26},
  963.

\item[] Svetitsky, B., and L.G. Yaffe, 1982b, Nucl. Phys. B {\bf 210}
[FS6], 423.

\item[] Svetitsky, B., 1986, Phys. Rep. {\bf 132}, 1.

\item[] Toledano, J.-C., 1981, Ferroelectrics {\bf 35}, 31.
  
\item[] Toussaint, D., 1988, ``Introduction to algorithms for Monte
  Carlo simulations and their application to QCD''; lectures presented
  at the Symposium on New Developments in Hardware and Software for
  Computational Physics (Buenos Aires).
  
\item[] Toussaint, D., 1990, Nucl. Phys. B (Proc. Suppl.) {\bf 17},
  248.
  
\item[] Trappenberg, T., and U.J. Wiese, 1992, Nucl. Phys. B {\bf
    372}, 703.

\item[] Trinchero, R.C., 1983, Nucl. Phys. B {\bf 227}, 61. 
  
\item[] Turski, L.A., and J.S. Langer, 1980, Phys. Rev. A {\bf22},
  2189.
  
\item[] Venugopalan, R., M. Prakash, M. Kataja, and V. Ruuskanen,
  1994, Nucl. Phys. A {\bf 566}, 473c.

\item[] Vladikas, A., 1988, Nucl. Phys. B (Proc. Suppl.) {\bf 4}, 322.

\item[] Waldhauser,R., D.H. Rischke, U. Katscher, J.A. Maruhn, 
  H. St\"ocker, and W. Greiner, 1992, Z. Phys. C {\bf 54}, 459.
  
\item[] Wegner, F. J., 1972, Phys. Rev. B {\bf 5}, 4529.

\item[] Weinberg, S., 1966, Phys. Rev. Lett. {\bf 17}, 616; 1967,
  Phys. Rev. Lett. {\bf 18}, 188; 1979, Physica A {\bf 96}, 327.

\item[] Weldon, H.A., 1991, Phys. Rev. Lett. {\bf 66}, 293.
  
\item[] Welke, G.M., R. Venugopalan, and M. Prakash, 1990, Phys.
  Lett. B {\bf 245}, 137.

\item[] Werner, K., 1989, Phys. Rev. Lett. {\bf 62}, 2460.
  
\item[] Wieand, K.L., S.E. Pratt, and A.B. Balantekin, 1992, Phys.
  Lett.  B {\bf 274}, 7.

\item[] Wilczek, F., 1992, Int. J. Mod. Phys. A {\bf 7}, 3911.
  

\item[] Wilczek, F., 1994, Nucl. Phys. A {\bf 566}, 123C.

\item[] Wilson, K., and J. Kogut, 1974, Phys. Rep. C {\bf 12}, 75.

\item[] Wilson, K., 1974, Phys. Rev. D {\bf 14}, 2455.

\item[] Witten, E., 1979, Nucl. Phys. B {\bf 156}, 269.

\item[] Wroblewski, A., 1985, Acta Phys. Pol. B {\bf 16}, 379.
  
\item[] Xu, N., E814 collaboration, 1994, Nucl. Phys. A {\bf 566},
  585c.

\item[] Zel'dovich, Ya.B., {\it et al.}, 1987, Usp. Fiz. Nauk {\bf
    152}, 3 [Sov. Phys. Usp. {\bf 30}, 353 (1987)].

\item[] Zhang, W.-M., and L. Wilets, 1992, Phys. Rev. C {\bf 45}, 1900.
  
\item[] Zubarev, D.N., 1974, {\it Nonequilibrium Statistical
    Thermodynamics} (Plenum, New York).

\end{description}

%
%
\narrowtext
\begin{table}
  \caption{Criteria to distinguish between 1st
    and 2nd order transitions in a large but finite volume}
  \label{tab:1}
  \begin{tabular}{ccc}
    criterion & 1st order & 2nd order \\
    \tableline
    $P_L(E)$     & double peak     & single peak  \\
    $c_{L max}$  & {\hbox{$\propto L^d$}}  
    & {\hbox{$\propto L^{\alpha/\nu}$}} \\
    $\chi_{max}$ & {\hbox{$\propto L^d$}}    & {\hbox{$\propto
        L^{\gamma/\nu}$}} \\
    $\beta(c_{L max})-\beta_c$ & {\hbox{$\propto L^{-d}$}} &
    {\hbox{$\propto L^{-1/\nu}$}} \\
    $B_{min}$    & {\hbox{$>0$}} & {\hbox{$\rightarrow 0$}} \\
    $U4_{min}$   & {\hbox{$\rightarrow 1$}} & {\hbox{$>1$}}
  \end{tabular}
\end{table}

\begin{table}
  \caption{Dictionary between statistical physics and QCD.}
  \label{tab:2}
\end{table}

\begin{table}
  \caption{Order of the chiral transition
    as function of the  number of flavors ($N_f$) and the anomaly
    strength ($g$), partly conjectural}
  \label{tab:3}
  \begin{tabular}{cccc}
    $N_f$  & g=0          & g=const    & g=g(T) \\
    & \hbox{$N_c=\infty$} & \hbox{of order 1} & \hbox{ g(T)$\sim d_I(T)$} \\
    \noalign{\hrule}      
    1 & \hbox{2nd order} & \hbox{no transition } & \hbox{no transition} \\
    & O(2)-exponents &  &  \\
    \noalign{\hrule}
    2 & \hbox{1st order} & \hbox{2nd order} & \hbox{1st order} \\
    & & O(4)-exponents & \hbox{O(2)$\times$ O(4)-symmetry} \\
    \noalign{\hrule}
    3 & \hbox{1st order} & \hbox{1st order} & \hbox{1st order} \\
    &           &           &
    $U_A(1)\times SU(3)\times SU(3)$ \\
    & & & symmetry \\
    \noalign{\hrule}
    $\ge 4$ & \hbox{1st order} & \hbox{1st order} & \hbox{1st order} \\
    \noalign{\hrule}
  \end{tabular}
\end{table}

\mediumtext
\begin{table}
  \caption{Infinite volume behavior of the mass gap, the 
    tunneling mass $m^{\rm(unp)}$ and their corresponding 
    correlation lengths.}
  \label{tab:4}
  \begin{tabular}{ccc}
    phase & 1st transition & 2nd transition \\
    \tableline
    \\
    symmetric 
    & $m^{(p)}$ stays finite as $\beta\to\beta_c^-$ 
    & $\matrix{m^{(p)}\to(\beta-\beta_c)^\nu\to 0~
      \rm{as}~\beta\to\beta_c^-~, \cr
      \xi^{(p)}\to\infty~ }$
    \medskip\\
    broken 
    & $m^{(p)}$ and $\xi^{(p)}$ are finite
    & $m^{(p)}$ and $\xi^{(p)}$ are finite
    \medskip\\
    symmetric
    & $m^{\rm(unp)}$ stays finite as $\beta\to\beta_c^-$
    & $\matrix{ m^{\rm(unp)}\sim (\beta_c-\beta)^\nu\to 0~
      \rm{as}~\beta\to\beta_c^-~, \cr
      \xi^{(t)}\to\infty}$
    \medskip\\
    broken
    & $m^{\rm(unp)}=0$, $\xi^{(t)}=\infty$
    & $m^{\rm(unp)}=0$, $\xi^{(t)}=\infty$
    \\
  \end{tabular}
\end{table}
\narrowtext

\begin{table}
  \caption{Critical exponents $1/\beta\delta$ and $1/\delta$ for
    2-flavor QCD in comparison with $O(4)$, $O(2)$ and mean-field
    critical exponents.}
  \label{tab:5}
  \begin{tabular}{cllll}
    & 2-flavor QCD & $O(4)$ & $O(2)$ & mean-field \\
    \tableline
    $1/(\beta\delta)$ & $0.77\pm 0.14$ & 0.55(2) & 0.60(1) & 0.67 \\
    $1/\delta$ & $0.21<1/\delta<0.26$ & 0.208(2) & 0.2080(3) & 0.33 \\
  \end{tabular}
\end{table}

\begin{table}
  \caption{Mass parameters and results for 2 and 3 flavors.}
  \label{tab:6}
  \begin{tabular}{cllc}
    $N_f$ & $m_{u,d}\cdot a$ & $m_s\cdot a$ &
    results of Brown et al. (1990) \\
    \tableline
    2 & 0.01 & $\infty$ & no transition \\
    2 & 0.025 & $\infty$ & no transition \\
    3 & 0.025 & 0.025 & 1st order \\
    3 & 0.025 & 0.1 & no transition \\
  \end{tabular}
\end{table}

\mediumtext
\begin{table}
  \caption{Tree level parametrization of the $SU(3)\times 
SU(3)$ linear sigma model (input data taken from experiment)}
  \label{tab:7}
  \smallskip
  Input
  \begin{tabular}{ccccccc}
    $\mu_{0}^{2}~\rm{[GeV]}^{2}$ & $f_{1}$ & $f_{2}$ & $g$~[MeV] &
    $f_{\pi}$~[MeV] & $\epsilon_{0}~\rm{[GeV]}^{3}$ &
    $\epsilon_{8}~\rm{[GeV]}^{3}$ \\
    \tableline
    $5.96\cdot 10^{-2}$ & 4.17 & 4.48 & -1,812.0 & 94 & 0.0265 & -0.0345
    \\
  \end{tabular}
  \medskip
  Output (all masses are understood in units of [MeV])
  \begin{tabular}{ccccccccc}
    & $m_{\pi}$ & $m_{K}$ & $m_{\eta}$ & $m_{\eta'}$ &
    $m_{\sigma_{\pi}}$ & $m_{\sigma_{K}}$ & $m_{\sigma_{\eta}}$ &
    $m_{\sigma_{\eta'}}$ \\
    \tableline
    $\matrix{\rm{realistic}\cr \rm{mass~values}}$
    & 129.3 & 490.7 & 544.7 & 1045.5 & 1011.6 &
    1031.2 & 1198.0 & 749.5 \medskip\\
    $\matrix{\rm{experimental}\cr \rm{mass~values}}$
    & 138.0 & 495.7 & 547.5 & 957.8
    & $\matrix{980~~\rm{if}\cr \sigma_{\pi}\equiv a_{0}}$
    & $\matrix{1322.0~~\rm{if}\cr \sigma_{K}\equiv K_{0}^{*}}$
    & $\matrix{1476.0~~\rm{if}\cr \sigma_{\eta}\equiv f_{0}(1476)}$ 
    & $\matrix{975~~\rm{if}\cr \sigma_{\eta'}\equiv f_{0}(975)}$ \\
  \end{tabular}
\end{table}
\narrowtext

\begin{table}
  \caption{Flow diagram of `ingredients' to predict the $K^{-}/\pi^{-}$
    enhancement in the very end (for special notations see the text).}
  \label{tab:8}
\end{table}

\begin{table}
  \caption{Summary of chapter V.}
  \label{tab:9}
\end{table}


%
%
\section*{FIGURE CAPTIONS}
\begin{figure}
  \caption{Signature for a second order phase transition
    (a) Landau's free energy as function of the scalar order
    parameter field $\phi$ (b) Order parameter $\phi$ and
    associated susceptibility above ($\chi_+$) and below
    ($\chi_-$) $T_c$ as function of the temperature $T$.}
  \label{fig:1}
\end{figure}

\begin{figure}
  \caption{Hypothetical phase
    diagrams of QCD in the $(m,T)$-plane, $T$ being the temperature,
    $m$ standing for generic current quark masses; (a) the transitions
    survive the external fields and coincide, (b) both transitions are
    washed out for intermediate (realistic~?) mass values, (c) a
    $(H,T)$-diagram for a ferromagnet and (d) a $(p,T)$-diagram for a
    liquid/gas system. For further explanations see the text.}
  \label{fig:2}
\end{figure}

\begin{figure}
  \caption{Double peak structure of the probability 
    distribution $P_L(S)$ of the order parameter $S$ for $L=30$ at
    $\beta=0.36675$, $L=36$ at $\beta=0.367$ and $L=48$ at
    $\beta=0.367025$, from Gavai et al. (1989).}
  \label{fig:3}
\end{figure}

\begin{figure}
  \caption{Energy density (a) and pressure (b) in an $SU(3)$ pure gauge
    theory on $16^{3}\times 4$, $32^{3}\times 6$ and $32^{3}\times 8$
    lattices. The dashed horizontal lines indicate the corresponding
    results for an ideal gas on lattices of the same size. The solid
    horizontal line shows the Stephan-Boltzmann result in the
    continuum limit, from Karsch (1995).}
  \label{fig:4}
\end{figure}

\begin{figure}
  \caption{Interaction measure $\epsilon-3p$ normalized to $T^{4}$ 
    for a pure $SU(3)$ gauge theory for different lattice sizes, from
    Karsch (1995).}
  \label{fig:5}
\end{figure}

\begin{figure}
  \caption{Auxiliary multicanonical probability distribution 
    as function of the action density $s$, from Berg and Neuhaus
    (1991).}
  \label{fig:6}
\end{figure}

\begin{figure}
  \caption{Time history of the Wilson line Re~$\Omega$ for $N_f=2$
    and $m_f=0.025$ on an $N_\sigma^3\times4$ lattice with
    $N_\sigma=4,6,8$ and 12, from Fukugita et al., 1990b.}
  \label{fig:7}
\end{figure}

\begin{figure}
  \caption{same as Fig.~\ref{fig:7}, but $N_f=4$ and
    $N_\sigma=4,6,8,10$ and 12, from Fukugita et al., 1990b.}
  \label{fig:8}
\end{figure}

\begin{figure}
  \caption{Generic phase diagram in the $N_\tau-\beta$- plane for 
    8-flavor QCD, $\beta=6/g^2$, $N_\tau$ is the temporal extent of
    the lattice. The full line locates the bulk transition, the dashed
    line suggests a possible finite temperature transition. Solid
    squares label parameter values where simulations have been
    actually performed, open squares locate critical couplings, from
    Brown et al., 1992.}
  \label{fig:9}
\end{figure}

\begin{figure}
  \caption{Hot (upper curve) and cold (lower curve) start
    evolution of $\left<\bar\chi \chi\right>$ on a $16^4$ lattice at
    $\beta=4.65$, from Brown et al. (1992).}
  \label{fig:10}
\end{figure}

\begin{figure}
  \caption{Evolution of $\left<\bar\chi \chi\right>$
    from a cold start on a $16^3\times 32$ lattice with $\beta=4.60$.
    The jump at $\tau\approx 250$ is interpreted as tunneling from the
    metastable, weak coupling phase to the stable, strong coupling
    phase at this $\beta$-value, from Brown et al.  (1992).}
  \label{fig:11}
\end{figure}

\begin{figure}
  \caption{(a) $N_f=2$ with $m a=0.01$ and $\beta=5.265$. 
    Evolutions of $\left<\bar\chi \chi\right>$ without clear tunneling
    events, ordered start (solid), disordered start (dotted) curves,
    (b) $N_f=2$ with $ma=0.01$. Overlapping
    histograms from left to right $\beta=5.275$, 5.265, 5.25. Counts
    are in units of trajectories in phase space,
    (c) $N_f=3$ with $m a=0.025$, $\beta=5.132$.  Evolutions for an
    ordered start (solid) and a disordered start (dotted curve) signal
    a first order transition, (d) $N_f=3$,
    $\beta=5.171$ with nearly physical masses: $m_{u,d} a=0.025$, $m_s
    a=0.1$. The evolution of $\left<\bar\chi \chi\right>$ signals no
    transition, from Brown et al. (1990).}
  \label{fig:12}
\end{figure}

\begin{figure}
  \caption{Generic phase diagram, partly conjectural, for 2 
    and 3 flavors. Solid lines indicate supposed 2nd order
    transitions, shaded areas 1st order transitions, solid circles
    correspond to mass parameters, where the transition is seen, solid
    squares, where it is absent. The dashed circle locates the
    physical mass point, from Brown et al. (1990).}
  \label{fig:13}
\end{figure}

\begin{figure}
  \caption{Energy density (upper two curves) and three times the
    pressure (lower curve) vs temperature in the staggered fermion
    scheme of QCD with two light flavors ($a\cdot m_{q}=0.025$
    (octagons) and $a\cdot m_{q}=0.1$ (squares)), from Blum et al.
    (1995a).}
  \label{fig:14}
\end{figure}

\begin{figure}
  \caption{Line of critical hopping parameters $\kappa_{c}(\beta)$ and
    thermal transition line $\kappa_{T}(\beta)$ vs $\beta=6/g^{2}$. The
    arrows indicate the shift of $\kappa_{T}(\beta)$ towards weaker
    couplings for increasing $N_{\tau}$ and stronger couplings for
    increasing $N_{f}$. The qualitative shape of the curves remains
    unchanged for $N_{f}\leq 6$.}
  \label{fig:15}
\end{figure}

\begin{figure}
  \caption{Qualitative temperature dependence of the critical chemical
    quark potential $\mu_{q}^{\rm{crit}}$~[MeV].}
  \label{fig:16}
\end{figure}

\begin{figure}
  \caption{Quark condensate as a 
    function of temperature in the chiral limit, from Gerber 
    and Leutwyler (1989).}
  \label{fig:17}
\end{figure}

\begin{figure}
  \caption{Energy density of the pion gas in the
    chiral limit, from Gerber and Leutwyler (1989).}
  \label{fig:18}
\end{figure}

\begin{figure}
  \caption{Quark condensate for 
    nonzero quark mass ($m_{\pi}=140$~MeV), from Gerber 
    Leutwyler (1989).}
  \label{fig:19}
\end{figure}

\begin{figure}
  \caption{Temperature 
    dependence of the quark condensate for nonzero quark 
    masses. The shaded area represents the superposition of the 
    contributions generated by the pions and by the massive
    states, from Gerber and Leutwyler (1989).}
  \label{fig:20}
\end{figure}

\begin{figure}
  \caption{The light quark condensate normalized to its value at zero
    temperature $\left<\bar q q\right>_{T}/\left<\bar q
      q\right>_{0}$ as a function of $T$ in the $SU(3)$-symmetric
    case. The weakening of the first order transition is obvious, when
    $\epsilon_{0} [\rm{GeV}^{3}]$ is varied between
    $\epsilon_{0}=0(\times), 2\cdot 10^{-4}~(\diamond), 2.5\cdot
    10^{-4}~(+)$ and $6.6\cdot 10^{-4}~(\Box)$.}
  \label{fig:19X}
\end{figure}

\begin{figure}
  \caption{Light ($\left<\bar q q\right>$) and strange 
    ($\left<\bar s s\right>$) quark condensates normalized to their
    corresponding values at zero temperature as a function of
    temperature. The crossover behavior is most rapid between
    $181.5\leq T\leq 192.6$~[MeV].}
  \label{fig:20X}
\end{figure}

\begin{figure}
  \caption{Entropy density $s$ over $T^{3}$, energy density $\epsilon$
    over $T^{4}$, and pressure $p$ over $T^{4}$ in the
    large-$N_{f}$-approximation for physical meson masses.}
  \label{fig:21}
\end{figure}

\begin{figure}
  \caption{Flux tubes between static
    quarks in pure $SU(2)$ gauge theory (a) at low temperatures, (b)
    at high temperatures.}
  \label{fig:22}
\end{figure}

\begin{figure}
  \caption{String bifurcation for a baryon (a) and an antibaryon (b).}
  \label{fig:23}
\end{figure}

\begin{figure}
  \caption{Allowed vertices for the $SU(3)$ gauge theory.}
  \label{fig:24}
\end{figure}

\begin{figure}
  \caption{$SU(3)$-vacuum at low temperatures filled with glueballs.}
  \label{fig:25}
\end{figure}

\begin{figure}
  \caption{(a) disallowed structure
    below $T_c$, (b) allowed structure above $T_c$.}
  \label{fig:26}
\end{figure}

\begin{figure}
  \caption{Hypothetical correlation between the average transverse 
    momentum and the multiplicity of final state particles in a given
    rapidity interval. The vertical line separates areas of the
    hadronic phase (left) and the transition region (right).}
  \label{fig:27}
\end{figure}

\begin{figure}
  \caption{The average $p_T$ of $K^{\pm}$ and $\pi^{\pm}$ in $p-\bar p$
    collisions as function of the charged particle multiplicity, from
    Satz (1990a).}
  \label{fig:28}
\end{figure}

\begin{figure}
  \caption{Cross section for the
    production of muon pairs vs the invariant mass $M$ of the dilepton
    pair calculated (a) for a first order transition and (b) for a
    second order transition. Dashed-dotted line -- hadron
    contribution; dashed line -- quark contribution; solid line -- sum
    of quark and hadron contributions.  The two vertical lines
    separate regions where different phases give the dominant
    contribution, from left to right: the hadronic phase, the
    interference region, the quark gluon plasma phase, from Cleymans
    et al.  (1987).}
  \label{fig:29}
\end{figure}

\begin{figure}
  \caption{Sensitivity of the ratios $K^-/\pi^-$ at $T_c$ (when the
    phase conversion is completed) to the chosen scenario $(\Delta
    S=S,0,$ or $M$) and the input parameters. For further explanations
    see the text, from Kapusta and Mekjian (1986).}
  \label{fig:30}
\end{figure}

\begin{figure}
  \caption{Source emitting photons at points $P_1$ and $P_2$ which
    are registered at $P_3$ and $P_4$.}
  \label{fig:31}
\end{figure}

\begin{figure}
  \caption{Illustration of two pion formation zones (shaded part) 
    emitting pions in the sideward and outward directions; upper part:
    pion clouds (dotted area) without a phase transition; lower part:
    with a transition.}
  \label{fig:32}
\end{figure}

\begin{figure}
  \caption{(a) Sideward transverse momentum correlation between pions 
    with $k_{perp\ 0}$ $=$ $0$ and $k_{longitudinal}=0$. (b) Outward
    transverse momentum correlation between pions, with $k_{perp\ 
      s}=0$ and $k_{longitudinal}=0$. Solid lines are the cascade
    results, dashed lines correspond to a semianalytic parametrization
    (from Bertsch et al., 1988).} 
  \label{fig:33}
\end{figure}

\begin{figure}
  \caption{$R_{per\ 0}-R_{per\ s}$ [fm] for the indicated collisions,
    from Ferenc et al.  (1992).}
  \label{fig:34}
\end{figure}

\begin{figure}
  \caption{(a)-(c) Velocities of quark and hadronic matter for
    three possible scenarios viewed in the rest frame of the front.}
  \label{fig:35}
\end{figure}

\begin{figure}
  \caption{Anomalous dimensions of pseudorapidity spectra at 
    $s^{1/2}\sim 20$~GeV,
    from Bialas and Hwa (1991).}
  \label{fig:36}
\end{figure}

\end{document}